\documentclass[aps,rmp,reprint,amsmath,amssymb,graphicx,longbibliography,floatfix]{revtex4-1}
\usepackage{float}

\usepackage{graphicx}
\graphicspath{{figs/}}
\usepackage{bm}
\usepackage{color}

\usepackage{xspace}
\def\mcfm{\texttt{MCFM}\xspace}
\def\sherpa{\texttt{Sherpa}\xspace}
\def\powhegbox{\texttt{Powheg Box}\xspace}
\def\alpgen{\texttt{ALPGEN}\xspace}
\def\vbfnlo{\texttt{VBFNLO}\xspace}
\def\mg{\texttt{MadGraph5\_aMC\@NLO}\xspace}
\def\pt{\ensuremath{p_{\mathrm{T}}}\xspace}
\def\et{\ensuremath{E_{\mathrm{T}}}\xspace}
\newcommand*{\ifb}{\mbox{fb$^{-1}$}\xspace}
\begin{document}

\title{Multi-Boson Interactions at the LHC}

\author{D. R. Green}
\affiliation{Particle Physics Division, Fermi National Accelerator Laboratory, 
PO Box 500, Batavia, IL 60510, USA}
\author{P. Meade}
\affiliation{C.N. Yang Institute for Theoretical Physics, Stony Brook University, Stony Brook, NY 11794, USA}
\author{M.-A. Pleier}
\affiliation{Physics Department, Omega Group, Brookhaven National Laboratory,
P.O. Box 5000, Upton, NY 11973, USA}

\date{\today{}}

\begin{abstract}
  This review covers results on the production of all possible
  electroweak boson pairs and 2-to-1 vector boson fusion at the
  CERN Large Hadron Collider (LHC) in proton-proton collisions at a
  center of mass energy of 7 and 8~TeV. The data were taken between
  2010 and 2012. Limits on anomalous triple gauge couplings (aTGCs)
  then follow. In addition, data on electroweak triple gauge boson
  production and 2-to-2 vector boson scattering yield limits on
  anomalous quartic gauge boson couplings (aQGCs). The LHC hosts two
  general purpose experiments, ATLAS and CMS, which have both reported
  limits on aTGCs and aQGCs which are herein summarized.
  The interpretation of these limits in terms of an effective field
  theory is reviewed, and recommendations are made for testing other
  types of new physics using multi-gauge boson production.
\end{abstract}

\maketitle

\tableofcontents{}

\section{Introduction}
\label{intro}
The Standard Model (SM) of particle physics is based on the
$SU(3)_C \otimes SU(2)_L \otimes U(1)_Y$ gauge symmetry group and
describes the interactions among all the elementary particles. With
the discovery of a light Higgs boson, the SM is a complete and
self-consistent theory which can and should be tested as closely as
possible.

Because the electroweak gauge bosons carry weak charge the SM predicts
interaction vertices which contain three bosons (triple gauge
coupling) or four bosons (quartic gauge coupling). These interactions
contribute to the inclusive production of pairs and triplets of gauge
bosons as expected in the SM.

Previous experiments have studied the production of pairs of gauge
bosons. The Large Electron-Positron (LEP) collider experiments studied $WW$ and $WZ$ production as a
function of center of mass energy. Indeed, the triple vertices
were found to be critical in limiting the growth of the cross sections
with energy giving strong confirmation of the correctness of the
SM. Limits were set by the LEP experiments on anomalous triple gauge
couplings (aTGCs) for the first time and these
limits~\cite{Schael:2013ita} have remained the most stringent until
the advent of the Large Hadron Collider (LHC) at CERN.

Experiments at the Tevatron (CDF and D0) also measured exclusive gauge
pair production extending the data on final states to $WW$, $WZ$,
$ZZ$, $W\gamma$, $Z\gamma$ and $\gamma\gamma$. In these final states
the dynamics of the process, especially at large diboson center of mass energy
could be used to further test the predictions of the SM. For a recent
review of the relevant Tevatron results, the reader is referred 
to~\cite{Kotwal:2014mna}.

The LHC experiments ATLAS and CMS have begun to exploit the increased
center of mass energy of the LHC and the associated large increase in cross
section to expand the gauge coupling studies. In particular, the
energy at the triple and quartic vertices has been pushed into the TeV
range. As the energy and luminosity of the LHC continue to increase,
ever more incisive studies will open up.

This review covers the LHC proton-proton data taking up to the end of
2012, which occured at 7 and 8~TeV and is referred to as LHC
Run~I. The diboson states herein covered consist of all gauge boson
pairs, $\gamma\gamma$, $W\gamma$, $Z\gamma$, $WW$, $WZ$ and $ZZ$. In
each case limits on aTGCs could be set and they have now surpassed the
previous LEP and Tevatron limits. A unique feature of the LHC data is
the first exploration of triple gauge boson production with
$W\gamma\gamma$, $Z\gamma\gamma$ and $WWW$ final states, compiled in
this review. The corresponding first limits on anomalous quartic gauge
boson couplings (aQGCs) which have been reported are herein
summarized.

A second set of limits on aQGCs arise from the studies of exclusive
final states in the Vector Boson Scattering (VBS) topology. In that
case the initial proton-proton state, due to the virtual emission of
two gauge bosons, contains two remnant, forward going jets and a more
centrally produced final state with the resulting VBS dibosons. In
this review, aQGC limits are derived for the VBS states, $W\gamma jj$, $WV jj$,
$W^\pm W^\pm jj$, $WZ jj$ and $\gamma\gamma\to WW$, where the
symbol $j$ refers to the remnant jet. In the particular case where the
protons emit soft photons in the VBS initial state
($\gamma\gamma\to VV$), remnant jets are not part of the VBS
signature.

In the presentation of experimental results the distributions of
kinematic quantities which are well measured experimentally and which
also serve as a proxy for the energy at a triple or quartic gauge
boson vertex for a specific final state are shown. Where available,
predicted deviations from the SM due to anomalous couplings are also
shown in order to give an idea of the sensitivity of the measurement
to deviations from the SM.

This article is organized as follows.  In Section~\ref{s.theory} the
theory of multi-gauge boson interactions in the SM and the modern
treatment of deviations being described by an effective field theory (EFT) are reviewed.
Additionally, the impact of multi-gauge boson physics beyond simple
shifts in aTGC and aQGC measurements is emphasized, and a model-independent 
recommendation for experiments is made. 
In Section~\ref{sec:expsetup}
a brief description of the relevant experimental issues is given with
references to the corresponding experimental aspects of the
ATLAS and CMS experiments. In Section~\ref{diboson} the published LHC
diboson studies are presented while in Section~\ref{triboson} the
triboson results are shown. In Section~\ref{VBF} the vector boson
fusion (VBF) data for $W$ and $Z$ bosons are shown as a proof of principle
that this electroweak process can be extracted from the experimental
backgrounds. Armed with those analyses, the data on VBS are presented
in Section~\ref{VBS}. The existing limits on gauge couplings are
collected in Section~\ref{aTGC} for aTGCs and Section~\ref{aQGC} for
aQGCs. Finally, Section~\ref{HL-LHC} explores the prospects for gauge
coupling studies with the increased luminosity planned for the LHC as
set out by CERN.

\section{Theory}
\label{s.theory}

There are a variety of theoretical motivations for testing the structure of multi-boson interactions at the LHC.  Given the non-Abelian nature of the ElectroWeak (EW) sector of the SM, this allows one to test non-Abelian gauge theories directly.  While this of course had already been done in other ways with QCD, the weakly coupled non-confining nature of the EW gauge symmetry allows for its investigation in unprecedented detail, at higher energies, and with larger data sets.   Even more important is the connection between the study of multiple EW gauge bosons and the structure of ElectroWeak Symmetry Breaking (EWSB).  

The $W^\pm,Z$, and $\gamma$ (through mixing) represent the SM particles most strongly coupled to EWSB other than the top quark.  Since the discovery of a Higgs boson by ATLAS~\cite{Aad:2012tfa} and CMS~\cite{Chatrchyan:2012xdj}, we have definitive proof that the ultimate mechanism of EWSB must look very much like the simple ad hoc Higgs mechanism.  However, this results in many more theoretical problems than answers.  In particular, the appearance of spontaneous symmetry breaking without a dynamical origin associated with a scalar field brings the hierarchy problem to the fore.  Since EW gauge bosons can be cleanly identified at the LHC, they provide one of the best ways to seek {\em any} structure to EWSB beyond the Higgs.

Both the non-Abelian nature of EW gauge bosons and their connection to EWSB were used in past phenomenological studies that have spurred decades-long experimental programs at different colliders.  Historically these two threads, EWSB and non-Abelian couplings, were studied independently despite their intertwined nature.  

\begin{figure}[htbp]
  \centering
\includegraphics[width=0.25\textwidth]{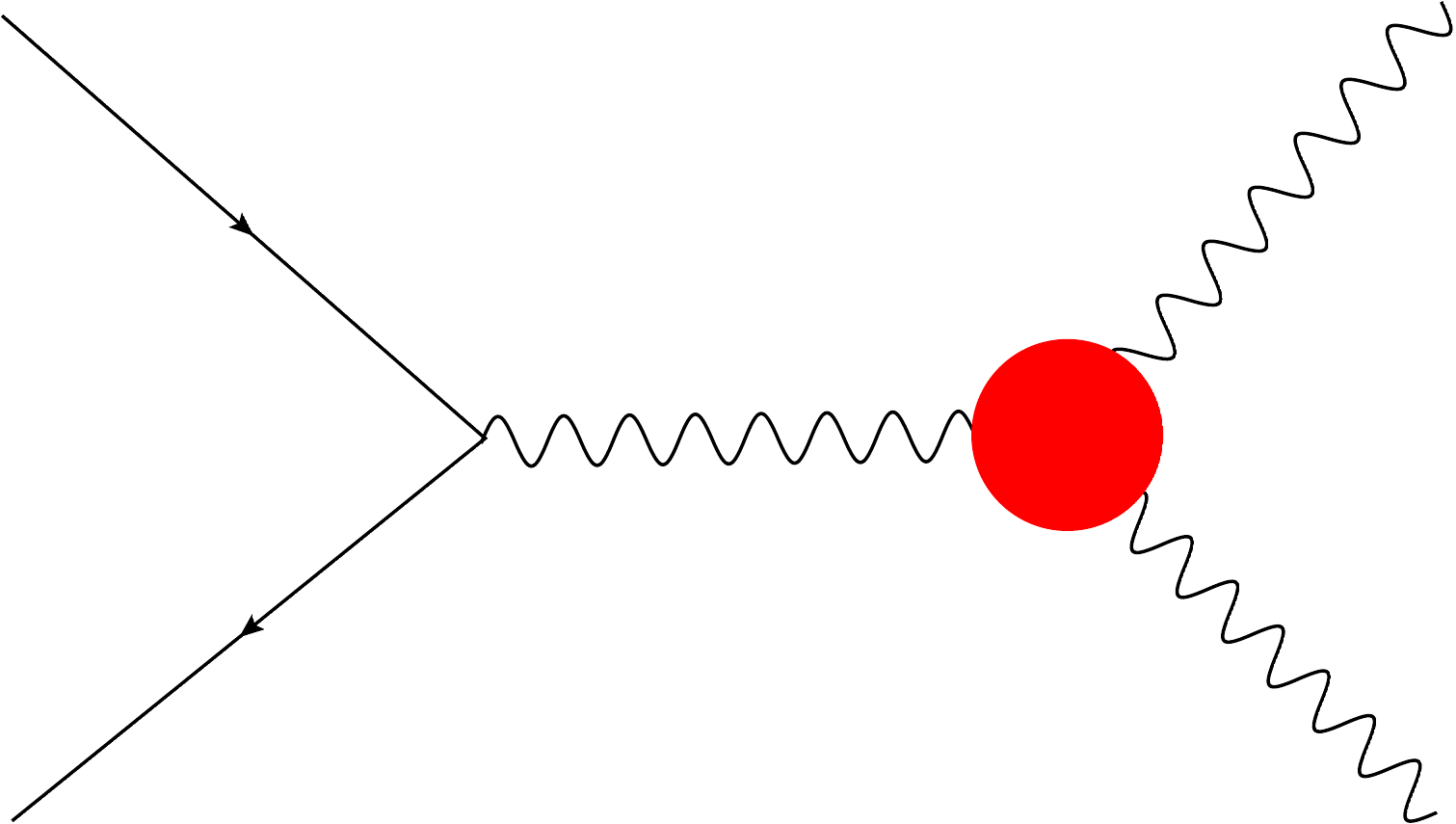}
  \caption{Diboson production via Drell-Yan at a lepton or hadron collider. The red insertion represents using a term from the parametrized Lagrangian in Equation~(\ref{eqn:hagiwara}). }
  \label{fig:aTGC}
\end{figure}

The origin of testing the non-Abelian structure using EW gauge bosons goes back to~\cite{Hagiwara:1986vm}. In that paper a parametrization of possible triple gauge boson couplings consistent with Lorentz invariance and charge conservation was given:
\begin{eqnarray}\label{eqn:hagiwara}
&&\mathcal{L}_{WWV}= ig_1^V \left(W^\dagger_{\mu\nu}W^\mu V^\nu-W_\mu^\dagger V_\nu W^{\mu\nu}\right) \nonumber\\
	&&+\frac{i\lambda_V}{m_W^2} W^\dagger_{\lambda\mu}W^\mu_\nu V^{\nu \lambda}-g_4^V W_\mu^\dagger W_\nu \left( \partial^\mu V^\nu+\partial^\nu V^\mu\right) \nonumber\\
	&&+ g_5^V \epsilon^{\mu\nu\rho\sigma} \left(W_\mu^\dagger \overset{\leftrightarrow}{\partial} W_\nu\right) V_\sigma+i \tilde{\kappa}_V W_\mu^\dagger W_\nu \tilde{V}^{\mu\nu}\nonumber \\
	&&+\frac{i \tilde{\lambda}_V}{m_W^2}W^\dagger_{\lambda\mu}W^\mu_\nu \tilde{V}^{\nu\lambda}+i\kappa_V W_\mu^\dagger W_\nu V^{\mu\nu},
\end{eqnarray}
where $W^\mu$ is the $W^-$, $V$ represents either the $Z$ or $\gamma$, the two-index $V$ or $W$ tensors are Abelian field strengths, and $\tilde{V}$ is the result of contracting two indices with the four-index epsilon tensor.   Historically, this was a very relevant parametrization since it preceded the experimental $WW$ production studies at LEP~II and large deviations from the non-Abelian structure had not yet been ruled out.   Once energies sufficient to produce dibosons were achieved, the effective Lagrangian (\ref{eqn:hagiwara}) could lead to deviations in processes such as those shown in Figure~\ref{fig:aTGC}, or constraints placed on the various couplings.
This parametrization was then carried forward and has been used as the basis for experimental studies of aTGCs for approximately the last three decades.  

\begin{figure}[htbp]
  \centering
\includegraphics[width=0.33\textwidth]{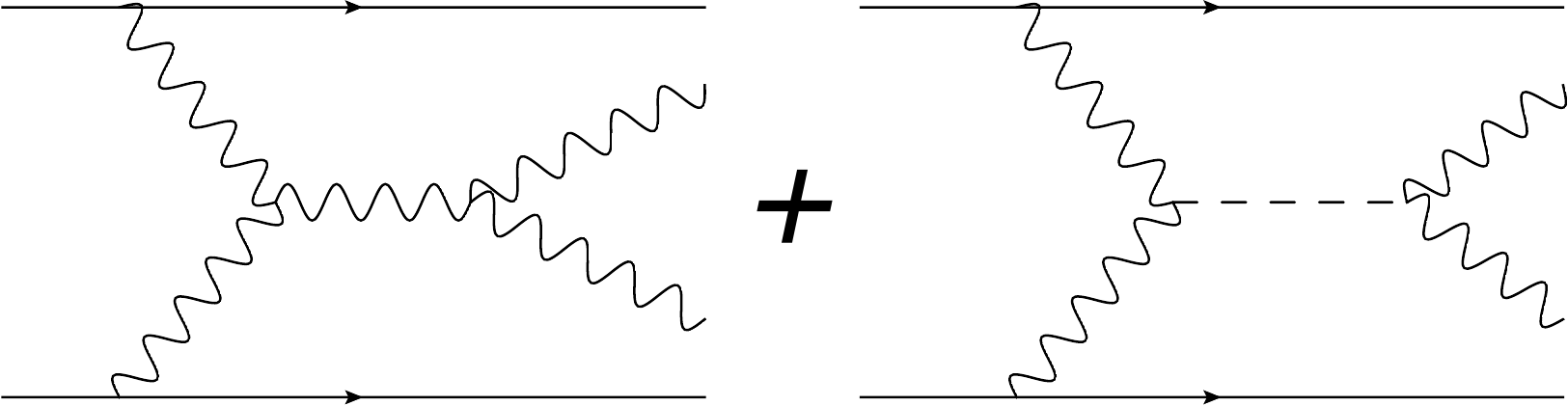}
  \caption{VBS in the SM with the exchange of gauge bosons on the left-hand side and the Higgs on the right-hand side needed to preserve perturbative unitarity in the SM. }
  \label{fig:vbshiggs}
\end{figure}

\begin{figure}[htbp]
  \centering
\includegraphics[width=0.2\textwidth]{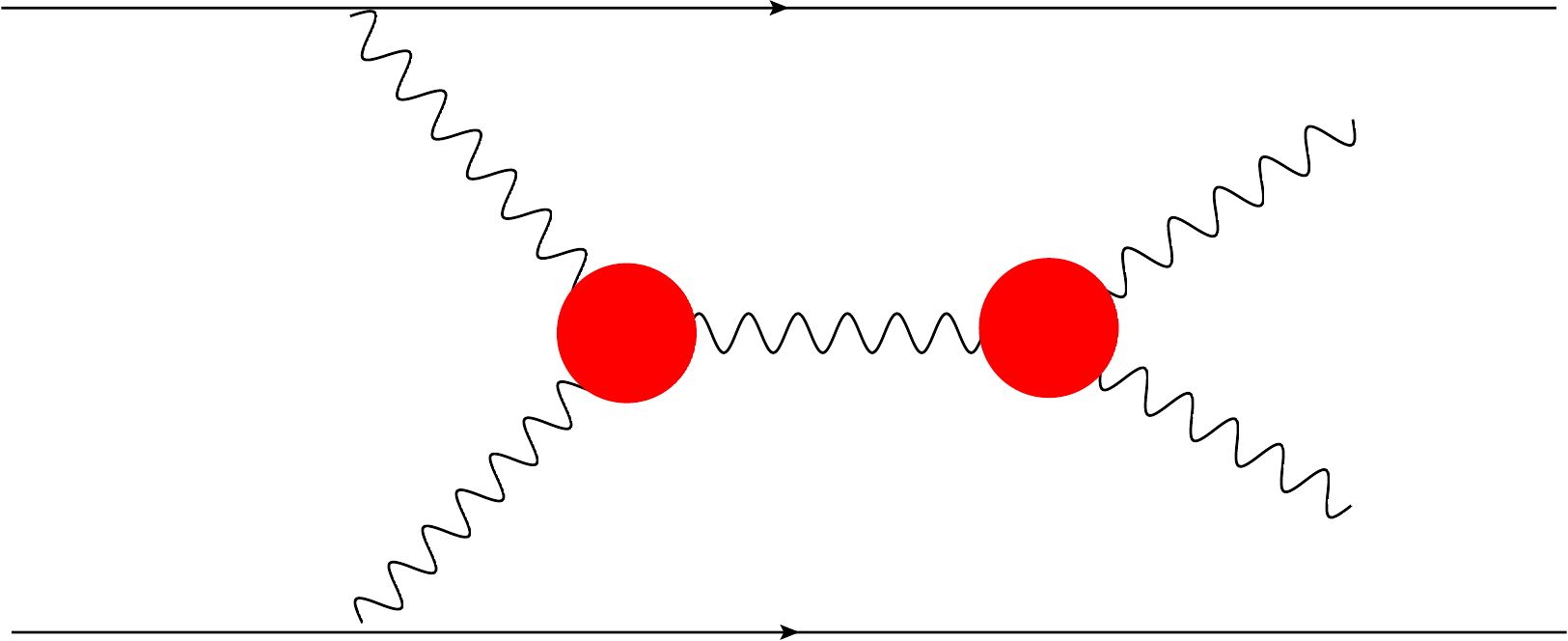}\hspace{0.5cm}
\includegraphics[width=0.2\textwidth]{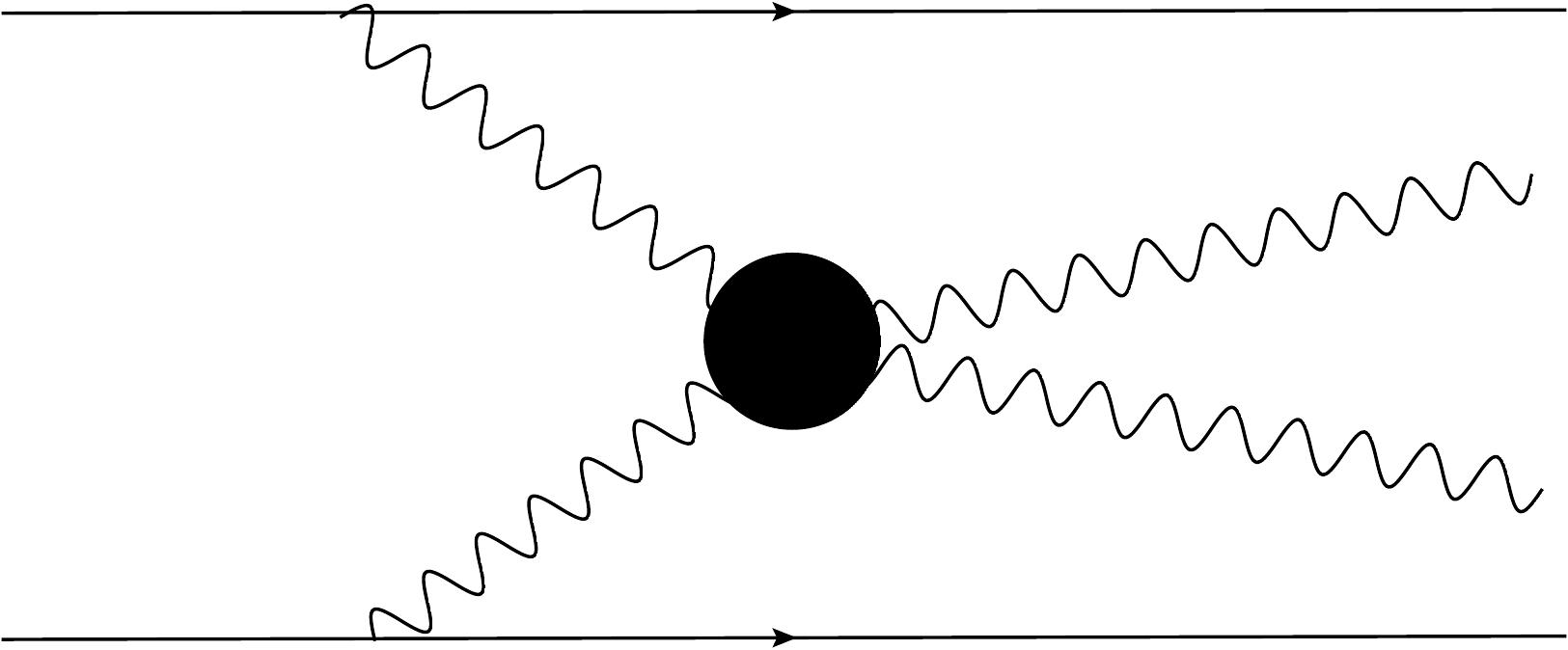}
  \caption{Examples of VBS contributions from aTGCs and aQGCs.}
  \label{fig:vbsanomalous}
\end{figure}

The historical connection between multiple vector boson production and EWSB is the role of the Higgs in unitarizing VBS~\cite{LlewellynSmith:1973yud,Dicus:1992vj,Cornwall:1973tb,Cornwall:1974km,Lee:1977yc,Chanowitz:1985hj}.  Well before the discovery of the Higgs, it was known that the scattering of massive vector bosons without a Higgs-like state has amplitudes that grow as $\sim E^2$.  Naively, if the SM EW gauge bosons were scattered at energies $\sim 4\pi m_W/g$, tree-level unitarity would appear to be violated.  Of course this did not mean that unitarity would actually have been violated, it simply meant that the theory of EWSB and massive gauge bosons would become strongly coupled and unpredictive at these scales.  If the Higgs existed, the growth with energy would be canceled by the Higgs contribution, and perturbative unitarity would have been manifest and calculable within this framework.  To test VBS, the simplest process one can study experimentally is shown in Figure~\ref{fig:vbshiggs}.
As in the case of aTGCs, the proposal to use VBS to test EWSB preceded the experimental observation of a Higgs boson.  At that point there were promising alternatives to the ad hoc Higgs mechanism which could explain EWSB dynamically, such as Technicolor~\cite{Farhi:1980xs} and composite Higgs models~\cite{Kaplan:1983fs}.  In these models unitarity was not violated either: instead of invoking the Higgs, VBS would be unitarized by massive Beyond-the-SM (BSM) states which couple to SM gauge bosons, as shown in the right-hand side diagram of Figure~\ref{fig:vbshiggs}.  If the energy of the collider is too low to directly produce the new states responsible for perturbative unitarity, an indirect way of studying this is again through anomalous couplings.  For instance, if one introduces both aTGCs as in Equation~(\ref{eqn:hagiwara}) and anomalous quartic gauge bosons couplings as shown in Figure~\ref{fig:vbsanomalous}, both will have effects on VBS measurements.
Regardless of how deviations from new sources of EWSB are parametrized, the connection between EWSB and VBS has been viewed as a window into the nature of EWSB since the early days of planning for the Superconducting Super Collider (SSC)~\cite{Chanowitz:1985hj}.   Since the discovery of a Higgs-like state, the direct connection to perturbative unitarity studies has been reduced; nevertheless, it will be an important validation of the SM to show the effects of the Higgs on vector boson scattering at the LHC.  Additionally, there could still be small deviations in the EWSB that would manifest themselves in VBF or VBS either as obvious deviations in the differential cross section or in searches for aTGCs or aQGCs in these channels.

These two independent threads, testing non-Abelian gauge boson couplings and unitarity in massive vector boson scattering were both originally very well motivated to search for large deviations in the EW sector. However, with the advancement of knowledge from LEP, the Tevatron and the LHC it is important to understand their failings in our modern understanding of the SM including the Higgs.  In Section~\ref{ss:bsm} we review the breakdown of historical methods for studying multiple production of EW gauge bosons.  These methods are however still used today, including in the experimental sections of this review. We also discuss how attempts to improve on testing for deviations in coupling constants have been done through EFT methods.  This is a useful tool to understand where to look for deviations in experimental results and how to parametrize them, but only if used correctly. We attempt to delineate these efforts both theoretically and experimentally, as there are failings on both sides with respect to the application of EFTs for multi-gauge boson production. In Section~\ref{sss:bsm} we discuss the important role of multi-gauge boson production in searches for BSM physics which has no connection to EFTs whatsoever. This is an important and often overlooked or factorized result given the structure of the ATLAS and CMS Collaborations, which typically relegate these processes to BSM groups that try to avoid regions of SM-like kinematics, or use them as control regions.  Nevertheless, the relation that massive gauge bosons have to EWSB dictates that searches for BSM in SM EW-like kinematic regions are as important as any other topic in studying multi-gauge boson production in the SM.  Finally, in Section~\ref{sss:conventions} we collect all conventions for anomalous couplings and effective operators used in the experimental results that are covered in this review.

Making any connection to BSM physics using multiple production of EW gauge bosons requires a precise understanding of the theoretical predictions for the SM.  There has been rapid progress in this field over the past few years; notably, several new Next-to-Next-to-Leading-Order (NNLO) QCD calculations have become available.  Simultaneously the LHC has entered into an era where there are sufficient statistics in multiple gauge boson production channels that NNLO and higher-order corrections are required to explain the data well. In addition, to go beyond the first implications of the Higgs and delve into possibilities for EWSB, the Higgs itself has become inextricably intertwined in current and future predictions for the LHC.  Therefore, before turning to BSM possibilities, we briefly review in Section~\ref{ss:sm}  the current theoretical understanding of SM predictions for multiple EW gauge boson production.

\subsection{Current Theoretical Understanding of SM cross sections}
\label{ss:sm}
The precision of theoretical calculations over the past decade has grown by leaps and bounds, particularly, in the last few years.  Prior to the LHC, the state of the art for many calculations was Next to Leading Order (NLO) in $\alpha_s$, and even that was not fully developed.  In particular, in 2005 there was an ``experimentalists NLO wish-list'' developed at Les Houches~\cite{Buttar:2006zd} for many processes relevant for the LHC.  Since then, this wish-list has essentially been completed, and now Monte Carlo (MC) programs are available to calculate at NLO in QCD automatically.  This amazing progress of course has been matched experimentally by the exquisite high-statistics measurements done at the LHC.   This has necessitated at least three important new developments in theory.  

The first is simply improving the theoretical precision of  inclusive cross sections from NLO to NNLO in $\alpha_s$, ultimately reaching this accuracy in fully differential cross sections as well.  There has been much progress on this front that we discuss further in the next section.  Increasing the order of the calculations can also introduce new production channels.  At lowest order all production processes that we discuss in this review are quark initiated, however, for instance at NNLO in $\alpha_s$, $pp\rightarrow VV$ includes both $q\bar{q}\rightarrow VV$ and $gg\rightarrow VV$. This implies that when reaching NNLO accuracy defined for the quark initiated process we have only reached LO in gluon initiated processes.  Therefore it is also important to advance to NLO for gluon initiated processes to learn the size of the first correction.  Here there is some recent progress that we will discuss for the channels where it has been calculated.

The second necessary development is the inclusion of NLO EW corrections.  If we parametrize the cross section as going from LO to higher in powers of $\alpha_W$ and $\alpha_s$ (keeping in mind the caveat of new channels at higher order) as
\begin{eqnarray}
d\sigma\sim d\sigma_{LO} \biggl( 1+ \textstyle \sum\limits_{i} \alpha_s^i &d\sigma_{N^{i}LO}+\sum\limits_{i} \alpha_W^i d\sigma_{N^{i}LO_{EW}} \nonumber\\
&+ \mathrm{mixed\,corrections}\biggr),
\end{eqnarray}
reaching NNLO QCD accuracy implies the need for NLO EW as well, since at the EW scale $\alpha_s^2\sim\alpha_W$. This of course is just a rough estimate, as there are many factors that enter besides the coupling constant.  However, EW corrections typically have the opposite sign as QCD corrections, especially in the high invariant mass and high-\pt regions, and are thus very important in searching for new physics. 

The third new development is due to the nature of the measurements performed at the LHC.  In attempting to isolate multi-boson processes, one has to deal with many QCD background processes.  Reducing the QCD background by exclusively looking in the zero-jet final state of a leptonic diboson decay is experimentally advantageous.  However, this introduces a new scale into the problem which is typically disparate from the hard scale.  The existence of two very different scales requires one to resum the large logarithms which arise to make accurate predictions.

In the following Sections~\ref{ss:sm:vv}--\ref{ss:sm:vvv} we outline the current status of theoretical calculations for three distinct types of processes at the LHC. First, we discuss the inclusive diboson processes, whose large cross sections and potentially clean final states can provide a standard candle for many measurements and searches at the LHC.  We then discuss the exclusive VBF and VBS processes which represent a subset of those for inclusive single or diboson production.  Finally, we briefly discuss the theoretical status of triboson production. These new measurements go beyond those at previous colliders and will become more important with the High Luminosity (HL)-LHC run, both as a signal and as a background to searches.  For a more complete status of SM theoretical calculations beyond those of just multi-boson production we refer the reader to~\cite{Badger:2016bpw, Rauch:2016pai} and~\cite{Campanario:2015vqa}.  We also note that there are a number of multipurpose event generators used for the various multi-boson processes, such as \vbfnlo~\cite{Arnold:2008rz,Arnold:2011wj, Baglio:2014uba}, \mg~\cite{Alwall:2014hca}, \powhegbox~\cite{Nason:2004rx,Frixione:2007vw,Alioli:2010xd, Melia:2011tj, Nason:2013ydw}, \sherpa~\cite{Gleisberg:2008ta, Hoeche:2009rj, Gleisberg:2008fv, Schumann:2007mg}, and \mcfm~\cite{Campbell:1999ah,Campbell:2011bn,Campbell:2015qma, Boughezal:2016wmq}. In Sections~\ref{ss:sm:vv}--\ref{ss:sm:vvv} we concentrate on the current status of theoretical calculations rather than comparing the different MC capabilities.

\subsubsection{Diboson Production}
\label{ss:sm:vv}
For diboson production, the state-of-the-art QCD calculation is NNLO for $W^+W^-$~\cite{Gehrmann:2014fva,Grazzini:2016ctr}, $W^\pm\gamma$~\cite{Grazzini:2015nwa,Denner:2014bna}, $W^\pm Z$~\cite{Grazzini:2016swo}, $ZZ$~\cite{Cascioli:2014yka,Grazzini:2015hta}, $Z\gamma$~\cite{Grazzini:2015nwa,Denner:2015fca} and $\gamma\gamma$~\cite{Campbell:2016yrh}.  There has been recent rapid progress on this front using $q_T$ subtraction techniques and in the near future public codes, such as MATRIX~\cite{Wiesemann:2016eiw}, should be available to automate event generation.  To get to this accuracy, $VV'$ with an additional jet has also been calculated to NLO accuracy in QCD.  It will also be important to understand how to combine NNLO cross sections with parton showers to simulate fully differential events~\cite{Alioli:2013hqa}.  It is important to also push $gg\rightarrow VV$ to NLO because when computing formally at NNLO, this is only the lowest order $gg\rightarrow VV$  process.  Recently there has been progress in this, with $gg\rightarrow W^+W^-$ being calculated at NLO~\cite{Caola:2015rqy} as well as $gg\rightarrow ZZ$~\cite{Caola:2015psa}.  Finally, the matching of NLO gluon initiated processes to a parton shower must also be included and was recently done for $ZZ$~\cite{Alioli:2016xab}.

The NLO EW corrections have also been computed for a subset of the processes for which the NNLO QCD corrections are known.   The NLO EW corrections were calculated in~\cite{Biedermann:2016yvs} for $ZZ$ production including decay. For the case of $W^+W^-$ this was carried out in~\cite{Biedermann:2016guo}. The $Z\gamma$ and $W\gamma$ processes were calculated at the NLO EW order in~\cite{Denner:2015fca} and \cite{Denner:2014bna}, respectively. In the case of $W\gamma$ and $Z\gamma$ this was done in combination with the NLO QCD corrections. First calculations of NLO EW corrections to off-shell vector-boson scattering have also been performed~\cite{Biedermann:2016yds}. The next frontier is the joint calculation to NNLO in QCD and NLO in EW, as well as including the decays in the calculations.

For diboson production, once NNLO in $\alpha_s$ is reached, there is also the possibility to evaluate the interference between $gg\rightarrow VV$ and $gg\rightarrow H\rightarrow VV$. This was pointed out and calculated in~\cite{Campbell:2011cu} where a non-negligible effect was demonstrated.
 
There are also various types of resummation that have been carried out for diboson production such as threshold resummation, \pt resummation, and, in certain cases, jet-veto resummation.  Threshold resummation can give a good approximation for higher-order calculations, for instance the $W^+W^-$ cross section was approximated to NNLO using threshold resummation in~\cite{Dawson:2013lya}.  However, given that all diboson channels are now computed at fixed order to NNLO, these calculations would have to be pushed further to compete.  
 
The resummation of \pt is useful for all diboson channels, given that in these colorless final states it provides a roughly universal prediction.  The prediction for the \pt spectrum of dibosons can now be tested in a new regime, as done previously for single gauge boson production. It is also important to get the correct kinematic distributions since dibosons are important backgrounds for many other processes including Higgs boson production.  The current state of the art is NNLO+Next-to-Next-to-Leading-Log (NNLL) which for $W^+W^-$ and $ZZ$ is computed in~\cite{Grazzini:2015wpa}.  Given that the $W^\pm Z$ final state was only recently computed at NNLO, the current state of the art for this channel is NLO+NNLL as in~\cite{Wang:2013qua}, but this should change in the near future.

For the $W^+W^-$ channel,  a jet-veto is used by the experiments to control the background coming from top quark pair production.  More generally an exclusive measurement is made in different jet-multiplicities.   In this case jet-veto resummation is also needed since there is a large difference of scales between the jet-veto scale and the invariant mass of the diboson system.  In fact, not including this effect led to early measurements of the $W^+W^-$ cross section being significantly overestimated when experiments extrapolated from fiducial to inclusive measurements.  The effect of the jet-veto is also correlated with \pt resummation and its impact on extrapolating to the total cross section was first pointed out for \pt resummation in~\cite{Meade:2014fca} and for jet-veto resummation in~\cite{Jaiswal:2014yba}.  These results naively disagreed, but after taking into account scale choices and adopting a uniform approach, they agreed at NLO+NNLL~\cite{Jaiswal:2015vda}.  Currently the state of the art for jet-veto resummation for this channel is NNLO+NNLL as performed in~\cite{Dawson:2016ysj}.  This slightly reduces the effect of the jet-veto on the total cross section compared to NLO+NNLL.  Additionally one must include the NLO effects of $gg\rightarrow VV$ in the calculation, as done in~\cite{Caola:2015rqy} where it was shown to be large, but this needs to be resummed as well.  Hopefully a more complete theoretical picture for this channel will be developed in the next few years and the same level of scrutiny will be applied to all diboson channels simultaneously. 

\subsubsection{Vector Boson Scattering and Vector Boson Fusion}
\label{ss:sm:vbs}
From the experimental point of view, the separation between VBF and VBS comes down to whether a single gauge boson is produced from two ($VV\rightarrow V$), or whether two gauge bosons come out ($VV\rightarrow VV$).   They are of course related as shown in the representative VBS diagrams shown in Figure~\ref{fig:vbsanomalous}, as the VBF fusion process can also contribute to VBS.  However, experimentally VBF, where only one gauge boson is produced, can be tagged separately from VBS allowing the TGC and QGC vertices to be tested separately in principle.  The current theoretical state of the art is NLO in QCD corrections, and this is implemented in the MC generator \vbfnlo.  Additionally, the NLO EW corrections are also known for these processes~\cite{Badger:2016bpw}. It is important to combine all effects at this order in the future. 

\subsubsection{Triple-boson production and beyond}
\label{ss:sm:vvv}
The process $pp\rightarrow VV'V''$ is interesting for a variety of reasons..  Leptonic $V$ decays represent some of the most relevant multi-lepton backgrounds to new physics.  Additionally, they represent a new and independent avenue for testing TGCs and QGCs beyond those from diboson production, and offer consistency conditions that must be satisfied once these processes are observed with sufficient statistics.  The process $W^\pm\gamma\gamma$ was calculated at leading order in \cite{Baur:1997bn}, and can be used as a test of the QGC.  This process was then calculated at NLO in QCD~\cite{Lazopoulos:2007ix}. By now, general triboson processes are available at NLO in QCD, for example, implemented in the generator \vbfnlo. The effects of EW corrections have also been calculated at NLO accuracy for instance in~\cite{Yong-Bai:2016sal} for $WWW$ and in~\cite{Yong-Bai:2015xna} for $WZZ$.  Higher multiplicity EW gauge boson production will also be observable in the future and can be computed with existing MC generators at NLO in QCD.

\subsection{Beyond the Standard Model Interplay}
\label{ss:bsm}

As discussed earlier, the study of multiple gauge boson production is an important avenue for searching for new physics at the LHC due to its connection to non-Abelian gauge theories and EWSB.  In particular, before the EW sector was tested at high precision by LEP or the evidence of the Higgs mechanism was directly found, large deviations were possible.  However, we are now in an era where the EW structure $SU(2)\times U(1)_Y$ of the SM is established, and the measured Higgs boson mass and couplings closely resemble those of the SM Higgs.  This in turn has led to a modernization of our theoretical and experimental understanding of how to use multi-gauge boson production to probe new physics. 

For instance, the parametrization of aTGCs given in Equation~(\ref{eqn:hagiwara}) manifestly breaks the gauge-invariance that we know to be true and was reformulated in a ``gauge-invariant'' manner to fully incorporate LEP results (see~\cite{Gounaris:1996rz} for a review). This reduced the general parametrization of Equation~(\ref{eqn:hagiwara}) to a subset of related couplings and better formulated the search for aTGCs as deviations from the SM values  $g_1^Z=g_1^\gamma=\kappa_Z=\kappa_\gamma=1$ (appropriately rescaled by the coupling constants $g$ of $SU(2)$ and $g'$ of $U(1)_Y$) while all other terms are non-existent at tree-level.  We discuss this further in Section~\ref{sss:conventions}. However, to truly make Equation~(\ref{eqn:hagiwara}) gauge-invariant requires the introduction of new fields that transform under $SU(2)\times U(1)$ which requires a {\em model-dependent} choice.

Up until the discovery of a Higgs boson, there were many competing models for EWSB.  The reason for this proliferation of models was that the Higgs mechanism in the SM has EWSB put in by hand and cannot explain why the symmetry is broken.   Additionally the Higgs mechanism on its own suffers from extreme fine-tuning unless new physics occurs around the TeV scale.  Models such as Technicolor~\cite{Weinberg:1975gm,Susskind:1978ms} where EWSB occurs dynamically, and similarly to other examples of spontaneous symmetry breaking in nature, offered an attractive alternative.   In the extreme case of strongly coupled EWSB such as Technicolor, or other incarnations of Higgsless models~\cite{Csaki:2003dt,Csaki:2003zu}, there is no Higgs field and the extra-modes required for gauge-invariance come from the ``pions'' of a larger broken symmetry.  There are also models of strongly coupled EWSB which include a mode that resembles the SM Higgs, but the Higgs is also a pseudo Goldstone boson of a larger symmetry, for instance in composite Higgs~\cite{Georgi:1984af} or little Higgs models~\cite{ArkaniHamed:2001nc,ArkaniHamed:2002qy}.  In both of these cases, gauge invariance is parametrized through a non-linear representation of the modes, similar to one used for chiral Lagrangians that describe the breaking of global symmetries in QCD.  In weakly coupled models with fundamental scalar fields, e.g. the Minimal Supersymmetric SM (MSSM), there is a Higgs field that can be used directly to make gauge-invariant contributions to aTGCs and is often described in the literature as a linear representation.   Regardless of the choice of ``new physics'' parametrization (or even simply the Higgs itself) that restores gauge-invariance for aTGCs, accounting for deviations such as those parametrized in Equation~(\ref{eqn:hagiwara}) requires the introduction of new physics beyond the SM.  However, the parametrization does have implications for the size of deviations expected and the interpretation of experimental results.  Once a Higgs boson was discovered (and there were many hints for this from prior EW precision tests that this would be true), a linear representation is highly favored and makes any other starting point almost as contrived as assuming $SU(2)\times U(1)$ is not a good symmetry.  This of course does not preclude the fact that the Higgs could be a composite from dynamical symmetry breaking, but it does restrict the form of corrections, as we will see.
 
 A useful method for looking at the effects of new physics that incorporates all the previous ideas in a ``model-independent'' framework is to use an EFT description of the SM.  This is in fact what {\em all} quantum field theories are in our modern understanding of Wilsonian renormalization.  In practice, this means defining a scale, $\Lambda$, of new physics higher than the energy scale being probed in the experiment and using the fields {\em of the SM} to write higher dimension operators in addition to the dimension $\Delta \leq 4$ operators of the SM
 \begin{equation}
 \mathcal{L}_{EFT}=\mathcal{L}_{SM}+\sum_i \frac{g_i \mathcal{O}_i}{\Lambda^{\Delta_i-4}},
 \end{equation}
 where $g_i$ are called Wilson coefficients.
 Given that $\Lambda$ is much higher than all the scales involved, the contributions to observables are well described by a perturbative series in momenta/energy $(E/\Lambda)^{\Delta_i-4}$ provided that the dimensionless Wilson coefficients are $\mathcal{O}(1)$.  This series then allows experiments to search for the effects of the lowest dimension operators which contribute the most to observables.  At a given dimension $\Delta$ there are always a finite number of operators that can contribute to any observable.  In fact through $\Delta=6$ all operators are known and have been reduced from a general set~\cite{Buchmuller:1985jz} to an irreducible basis~\cite{Grzadkowski:2010es}.  Given that there is only one gauge-invariant operator at dimension 5, the SM neutrino mass operator, the dominant effects of new physics describable by an EFT occur at $\Delta=6$ unless they are forbidden by an additional symmetry assumption.  Given that the EFT includes within it all the symmetries of the SM, this serves as the best starting point for describing small deviations to the SM from physics occurring at higher mass scales.  We describe these EFT methods in more detail in Section~\ref{sss:eft}, their relation to previous aTGC studies, and where they should and should {\em not} be used.  It is important to note though that an EFT manifestly does {\em not} describe physics at a scale $\Lambda$ accessible to the LHC.   Given that one of the most important reasons for studying multiple EW gauge boson production is its strong coupling to the EWSB sector, the possibility that there may be new physics at the EW scale that affects these measurements is a logical possibility.  In fact in almost any model of new physics that explains EWSB naturally, there are new particles near the EW scale with EW quantum numbers that would contaminate the same final states used for the measurements of cross sections.  The EFT formalism cannot be used for this possibility, and currently there is almost no experimental effort in this direction where the kinematics are very SM-like.  In Section~\ref{sss:bsm} we discuss some possible uses of multi-gauge bosons to search for new physics in this region and to test the SM in ways other than what is utilized by EFT, aTGC, aQGC, and vector boson scattering measurements.

\subsubsection{EFT interpretation of SM measurements}
\label{sss:eft}

Given our current experimental and theoretical understanding, treating the SM as an EFT is an incredibly well-motivated starting point.  It incorporates all the symmetries and fields we know and by definition matches the current data in the limit $\Lambda\rightarrow \infty$, since we have no current evidence of BSM physics.  As mentioned earlier, this is not a truly model-independent description of all new physics -- for instance those with a scale directly accessible at an experiment cannot be analyzed effectively in this manner.  However, it is a useful description for those models which are well described through an EFT. Given a model that is well described through an EFT one can then perform a matching calculation of Wilson coefficients in the full model and EFT to set bounds on all such applicable models.  In particular this formalism can describe both ``linear'' representations for BSM and non-linear representations that are still viable when the compositeness scale is large. For example, in the Strongly Interacting Light Higgs (SILH) model~\cite{Giudice:2007fh} if the scale of composite resonances $m_{*}$ is well above the scale we have currently probed, the standard non-linear representation can be expanded and matched onto the EFT description.   In Section~\ref{sss:conventions}  a list of all EFT operators and conventions typically used will be given, as well as their relation to anomalous couplings (a more extensive discussion can be found in ~\cite{Degrande:2013rea}).  Before going into the conventions, it is important to understand that despite EFTs being a well-motivated framework that can apply to many different models, there are also drawbacks depending on how they are used experimentally and theoretically.  The drawbacks arise for two ``different'' reasons, unitarity and model dependence; however, they are both related to the range of validity of the EFT formalism. 

The power of EFTs to describe new physics in a model-independent manner comes explicitly from the expansion $(E/\Lambda)^{\Delta_i-4}\ll 1$.  However, this means that the effects on SM observables are also small.  If one introduces an operator into an effective Lagrangian and naively calculates the experimental limits, the most discriminating power comes from the opposite regime $(E/\Lambda)^{\Delta_i-4}\sim 1$, where the EFT is not valid and an infinite set of operators would be needed to describe the physics.  Beyond invalidating the nature of the EFT expansion, naively calculating with a given operator, with a contribution $(E/\Lambda)^{\Delta_i-4}$ to a matrix element, will also give an apparent unitarity violation at some energy.   This is different from the motivations based on tree-level unitarity violation in vector boson scattering studies back when the nature of EWSB was unknown (although the concept of unitarity violation is just as meaningless there once understood properly as strong coupling). Unitarity violation from the SM EFT is {\em completely unphysical} and simply reflects an incorrect use of an EFT.  Apparent unitarity violation is simply just another guise for the EFT becoming strongly coupled and unable to make predictions.  This point is theoretically well understood, however, experiments still refer to unitarization methods when they use an EFT framework for multi-gauge boson measurements (due to these inconsistent limits $(E/\Lambda)^{\Delta_i-4}\ll 1$ and $(E/\Lambda)^{\Delta_i-4}\sim 1$ for setting the most powerful bounds).  This is understandable given that the implementation of a higher dimension operator at the MC level is always just included as an extra interaction term and thus can be used outside of the physically sensible region if additional constraints are not imposed.  In practice an additional form factor is included to avoid apparent unitary violations in the MC predictions (this is also the case for the use of anomalous gauge couplings as in Equation~(\ref{eqn:hagiwara})).  This typically takes the form
\begin{equation}\label{eqn:ff}
F(\hat{s})\sim\frac{1}{\left(1+\frac{\hat{s}}{\Lambda_{FF}^2}\right)^n},
\end{equation}
where $\hat{s}$ is the invariant mass of the system, $\Lambda_{FF}$ is an arbitrary scale unrelated to $\Lambda$ in practice, and $n$ is some positive power.  The $n$ used depends on the type of EFT operator or anomalous coupling of interest.  This is due to the fact that as the operator dimension $\Delta_i$ grows there is naively a larger growth in energy that would have to be dampened by an insertion of a form factor with a sufficiently large $n$ to make the amplitude convergent in this setup.  There are also other methods used for unitarization such as K-Matrix unitarization (see for instance~\cite{Kilian:2014zja}) which directly deforms the S-matrix of the theory to enforce unitarity, instead of putting a form factor into the action.  

The form factor approach used for unitarization can be related to the physical intuition from matching a UV theory onto an IR EFT.  For instance in the case of Fermi's theory of weak interactions, the dimension-6 charged-current (CC) four-fermion operator arises as an expansion from integrating out the $W$ at tree-level.  This corresponds to an expansion of the $W$ propagator in a geometric series of $p^2/m_W^2$ and keeping the lowest order term.  The expansion is given by
\begin{equation}\label{eqn:fermi}
\frac{g^2}{p^2-m_W^2}=\frac{g^2}{m_W^2}\frac{-1}{1-\frac{p^2}{m_W^2}}=\frac{-g^2}{m_W^2}\sum_{k=0}^{\infty} \left(\frac{p^2}{m_W^2}\right)^k, 
\end{equation}
for $\vert p^2/m_W^2\vert < 1$. If {\em only} the $k=0$ term is kept, this gives the usual relation that the amplitudes for SM CC interactions are well reproduced by a dimension-6 four-fermion operator when $p^2/m_W^2 \ll 1$:
\begin{equation}
\mathcal{A}_{SM}\sim \mathcal{A}_{\bar{\psi}\psi\bar{\psi}\psi}.
\end{equation}
However, $\mathcal{A}_{\bar{\psi}\psi\bar{\psi}\psi}\sim E^2$ which would make it appear that unitarity was violated by CC interactions in Fermi theory, which of course is not the case in the full SM.  There is no {\em actual} violation of unitarity; the Fermi theory with only $\Delta_i=6$ operators is simply incomplete when $E\sim m_W$.  Moreover, to even give an approximately correct answer as $E$ approaches $m_W$ would require keeping more and more terms in the infinite sum, i.e. many more higher dimension operators.    Furthermore, above the mass of the $W$ it is simply {\em impossible} to capture the correct scaling of the amplitude even with an infinite number of terms, since it is outside the domain of convergence of the series.  This leads to the usual overstatement that unitarity is violated in the EFT above a scale that can be predicted.   This is incorrect.  To make a prediction for this scale implies that we can trust perturbation theory at this scale with a finite number of terms, and this is simply not true.  While unitarity is normally treated as a separate problem for EFTs compared to strong coupling, in reality they are one and the same. To go further, a particular UV completion of the EFT is needed and one is then no longer using the EFT formalism as at the start.   In this particular case where the UV completion is the inclusion of  the $W$ gauge boson and its propagator, it can motivate a form for the choice of $\Lambda$ and $n$ in Equation~(\ref{eqn:ff}).  If larger $\Delta_i$ operators are included then $n$ would have to be increased.  In the case of K-matrix unitarization there is not a good physical model since it corresponds to an infinitely heavy resonance of infinite width~\cite{Degrande:2013rea}.  However, it cannot be stressed strongly enough, if one ``unitarizes'' an EFT one defeats the model-independent purpose of using an EFT description.   Once a unitarization method is chosen, there is an explicit UV model dependence introduced, and different UV models make different predictions for the region $(E/\Lambda)^{\Delta_i-4}\sim 1$ or for lower energies as we will see.

The second drawback to using EFTs is again related to their use in an invalid region, and comes from the careful application of matching Wilson coefficients to underlying theories.  Naively LEP, LHC, or other experiments can set bounds on the dimensionful Wilson coefficients $c_i\sim g_i/\Lambda^{\Delta_i-4}$, and these can be compared between experiments.  In fact this is often done to show the increased sensitivity of the LHC relative to previous experiments, including in this review.  However, it is important to keep in mind that the dimensionful Wilson coefficient $c$ always arises from some matching calculation where new physics at a scale $M$ is integrated out.   For example, $G_F$ is the Wilson coefficient of the four-fermion operator that arises from integrating out the massive $W$ and $Z$.  In a general case there can be a new state with coupling $g$ to SM particles and a mass $M$ which, if integrated out at tree-level to form a $\Delta=6$ operator, gives a Wilson coefficient
\begin{equation}\label{eqn:tree}
c\sim\frac{g^2}{M^2}.
\end{equation}
While naively one could use this EFT up to energy scales $c^{-1/2}$, if $g<1$ one would reach the scale of the mass of the new physics $M$ much earlier,  thereby invalidating the EFT description of this model at such an energy scale.    This is the case with our familiar four-fermion operator where $G_F^{-1/2}>m_W$.  If one attempted to use the operator up to the scale $G_F^{-1/2}$, the predictions would be completely wrong.  The resonance behavior would be missed and one would continue to wrongly assume that the operator's importance was still growing with $E$ rather than decreasing after passing through the resonance.  Furthermore, the on-shell production of $W$ bosons in the final state would be unaccounted for if the EFT was still the description being used.  Alternatively though if $g>1$,  this implies the true mass scale $M>c^{-1/2}$.  This illustrates why an underlying understanding of how  power counting the couplings of new physics and matching to Wilson coefficients can vastly affect whether a ``bound'' on an EFT operator has any meaning, or in what class of theories it has relevance.  In particular in weakly coupled theories, the range of validity can be much reduced, and by definition the underlying effects should be small.  Furthermore, it is quite possible that new physics does not generate SM EFT operators at tree-level as in Equation~(\ref{eqn:tree}), and the leading order contributions to Wilson coefficients arise at loop level (this can easily be the case if, for instance, there is a symmetry forbidding interactions between certain SM and BSM states, such as R-parity or T-parity).  In this case 
\begin{equation}\label{eqn:loop}
c\sim\frac{g^2}{16 \pi^2 M^2},
\end{equation}
and even if $g\sim \mathcal{O}(1)$, the scale where the EFT becomes invalid is now order $c^{-1/2}/4\pi$.  In such a theory, the conclusions drawn from using a bottom up EFT description would be even more misleading than the usual tree-level caveats.   In strongly coupled theories, these numerical factors can naively be overcome, but of course at strong coupling there is no theoretical control.  Therefore using experimental bounds on EFT operators to match to these strongly coupled theories and constrain them is an empty step unless augmented by an additional non-generic argument that provides theoretical control.  In addition, now that a Higgs boson has been discovered, we know that there cannot be a parametrically large shift in the physics of EWSB implying that new physics must appear weakly coupled at the scales we are probing at the LHC.  Therefore we {\em must} be careful about the power counting of Wilson coefficients when comparing experimental results; otherwise, we are led to possibly misleading conclusions as we now illustrate. 

\begin{figure}[htbp]
  \centering
\includegraphics[width=0.33\textwidth]{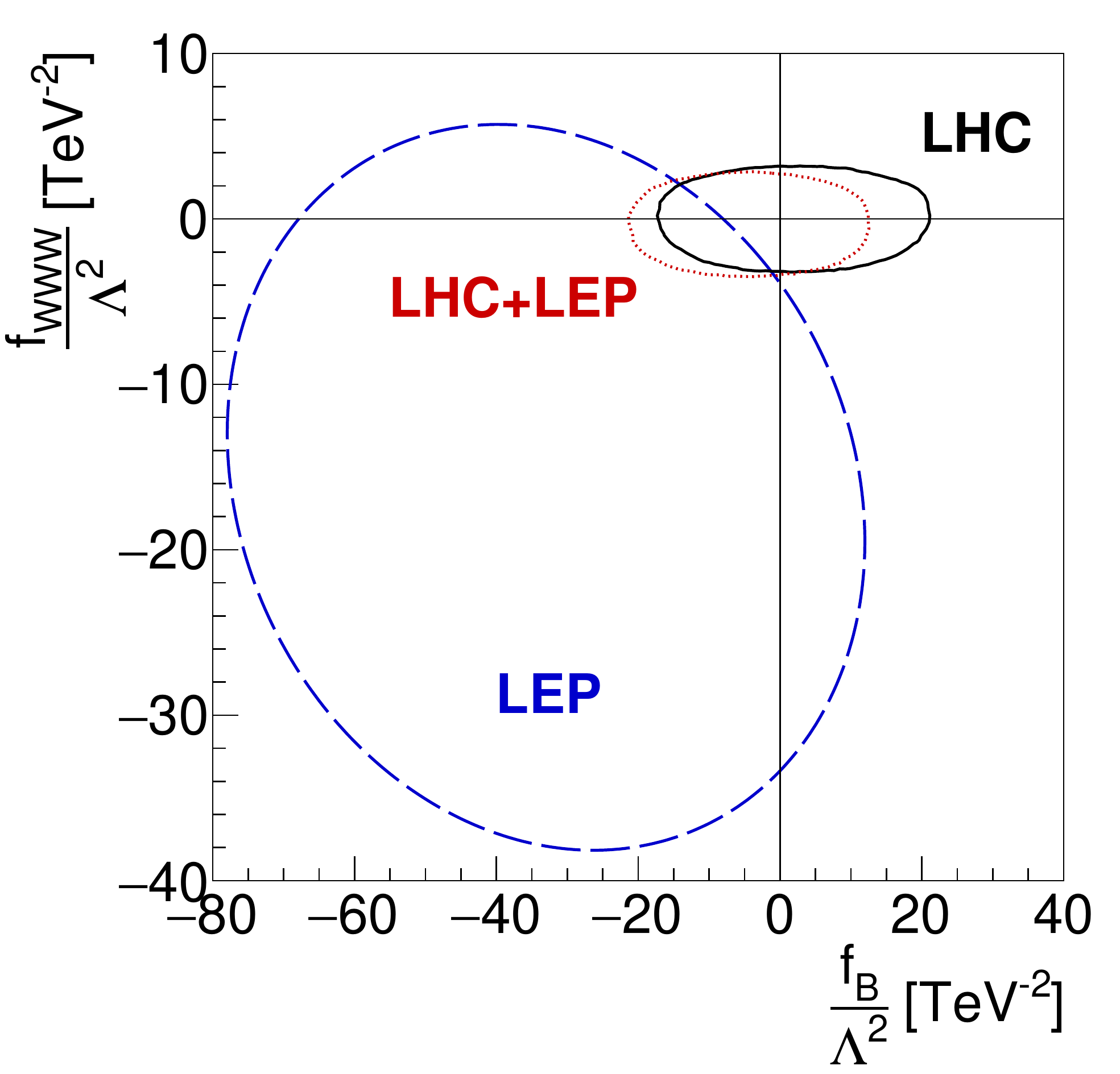}
  \caption{Figure from~\cite{Butter:2016cvz} demonstrating increased sensitivity of LHC over LEP.  The naming conventions are such that our $c_{WWW}$ is their f$_{WWW}/\Lambda^2$ and our $c_B$ is their f$_B/\Lambda^2$, the Wilson coefficients of the operators given in Equation~(\ref{eqn:atgc}).}
  \label{fig:lepLHCeft}
\end{figure}

If one takes the bounds set by different experiments on the same SM EFT operators, naively one could conclude that one experiment has increased sensitivity over another.  For instance in the recent theoretical analysis of ~\cite{Butter:2016cvz} it was concluded that diboson measurements at the LHC set better bounds on operators that contribute to aTGCs than LEP.  The analysis of~\cite{Butter:2016cvz} is not incorrect.  The LHC can indeed measure VBF and diboson production at high \pt enormously better than LEP.  Additionally, in the aforementioned analysis they also check the first caveat discussed in this section about unitarity.  However, a question still remains when using the EFT framework to set bounds:  based on the scales involved and operators analyzed, are there generic statements that can come from the EFT description?  Or are the results useful only to a small subset of strongly coupled models which lack predictive power?  Typically these questions are not investigated in as much detail as the unitarity questions, but as we will show they can be just as important.  We use the TGC as an example of how one can be misled~\cite{continotalk}.   In Section~\ref{sss:conventions} we go into more detail about our full set of EFT operators, but for TGCs the comparison between LEP and the LHC is straightforward because there are only three operators at $\Delta=6$ that contribute to aTGC measurements:
\begin{eqnarray}\label{eqn:atgc}
\mathcal{O}_{W}&=&D_\mu h^\dagger W^{\mu\nu} D_\nu h \nonumber\\
\mathcal{O}_{B}&=&D_\mu h^\dagger B^{\mu\nu} D_\nu h \nonumber\\
\mathcal{O}_{WWW}&=& \mathrm{Tr} \left( W_{\mu\nu} W^{\nu\rho} W^\mu_\rho\right).
\end{eqnarray}
In~\cite{Butter:2016cvz} a fit was performed demonstrating the increased sensitivity that the LHC had from Run~I compared to LEP, with an example shown in Figure~\ref{fig:lepLHCeft}.
However, this increased sensitivity as described in~\cite{Butter:2016cvz} comes from the high-\pt regions available at the LHC.  Therefore, one must ask whether the operators used in Equation~(\ref{eqn:atgc}) correspond to theories in which the EFT description is valid, or whether the increased sensitivity is an artifact of the high-\pt tail of the EFT.  For example, one potentially viable model with alternative EWSB is the SILH model.  In this model there are two  parameters which describe the new physics, a coupling $g_*$ and a mass scale $m_*$.  As with any model, a matching calculation to a SM EFT can be performed, leading to specific predictions for the Wilson coefficients.  In this case there are different power countings of couplings and masses for the different operators of Equation~\ref{eqn:atgc}, and when the one-dimensional bounds on the operators in~\cite{Butter:2016cvz} are recast in terms of the $m_*$ and $g_*$ one reaches a contradiction.  For the case of $c_{HW,HB}$  this leads to
\begin{equation}
c_{W,B}\sim \frac{g}{m_*^2}\left(\frac{g_*^2}{16\pi^2}\right) \rightarrow m_*\gtrsim 300\,\mathrm{GeV}\left(\frac{g_*}{4\pi}\right),
\end{equation}
where at strong coupling the mass scale needs to be $m_*\gtrsim 300\,\mathrm{GeV}$, but the LHC has already probed this territory.  In the case of $c_{3W}$,  in Figure~\ref{fig:lepLHCeft}, we naively see large gains compared to LEP, while with the SILH power counting we have
\begin{equation}
c_{WWW}\sim \frac{g}{m_*^2}\left(\frac{g^2}{16\pi^2}\right) \rightarrow m_*\gtrsim 20\,\mathrm{GeV},
\end{equation}
which shows that it is invalid to bound this type of new physics through EFTs with current data.  While this is only for the SILH power counting, it is part of a more generic set of consequences for aTGCs noted in~\cite{Arzt:1994gp}. In~\cite{Arzt:1994gp} it was shown that the operators which lead to aTGCs must be generated at loop level, and therefore one will always be fighting the loop-factor just as in the SILH power counting.    Now this example of course does not invalidate the use of EFTs at the LHC. However, it illustrates the limitations in an EFT operator analysis, i.e., there could be large swaths of motivated models that cannot always be described or tested consistently in an EFT framework at the LHC.  Does this mean that all channels and interpretations suffer this difficulty when using EFTs to parametrize new physics at the LHC? No, it simply reflects that for aTGCs, given that the Wilson coefficients of operators are typically suppressed, until a higher precision is reached by the LHC the EFT analysis may not be self-consistent.  Once a sufficiently high precision has been reached, these bounds will be generic and useful.  

EFTs have been pursued by experimentalists because of their generic character, but using them to compare to different experimental data has to be done with caution and theory prejudices in mind.  For instance, if one takes the correct LEP bounds on dimension-6 operators, they are quite constrained, and there may not be increased sensitivity at the LHC as of yet unless the high-\pt behavior is exploited.  As a way around this, ATLAS and CMS moved forward with a program that looked at the effects on aQGCs by ignoring all dimension-6 operators and including only dimension-8 operators in their analysis (the operators in question are listed in Section~\ref{sss:conventions}).   This defeats the original motivation for using EFTs, as it is focusing solely on extremely non-generic models where dimension-6 Wilson coefficients vanish or are highly suppressed and the new physics generates leading dimension-8 operators.  While not impossible \cite{Arzt:1994gp,Liu:2016idz}, it is not model-independent at all and requires specific mechanisms to override the standard power counting.  Using the dimension-6 operators may not show improvement compared to LEP for aTGCs for instance {\em yet}, but nevertheless the bounds will apply to a much larger set of models.  

EFTs are a robust theoretical tool and a welcome addition to the experimentalists arsenal.  When used with the SM, they account for the Higgs and known symmetries which helps greatly when organizing search strategies for multi-boson physics.  However, as discussed there are many potential drawbacks as well, and they are not a panacea for model-dependent statements in experimental measurements.  It simply is a fact that at this point, for many channels, the LHC is not better suited to bounding models where an EFT description is applicable.  To realize this, it is not as simple as using a MC and setting a bound on the dimensionful Wilson coefficient and then comparing different colliders.  One must also check whether it is consistent with unitarity/strong coupling and whether there is a self-consistent description of the coefficients of the operators and the scales being probed.  While this is taught in graduate lectures (e.g., TASI~\cite{Skiba:2010xn}: ``If one cannot reliably estimate coefficients of operators then the effective
theory is useless as it cannot be made systematic.''), this point has not been sufficiently stressed in the recent years where EFTs have become more and more used in the experimental communities.  This does not mean that the LHC does not have enormous capabilities for searching for new physics and constraining a wide variety of models that LEP could never dream of constraining.  It is simply a question of how the experimentalists choose to parametrize the constraints.   In the next section we make a recommendation of a generic procedure that applies to situations where EFTs are both applicable or not applicable.

\subsubsection{Fiducial Cross Sections and BSM recommendations}
\label{sss:bsm}

As discussed in Section~\ref{sss:eft}, EFTs provide useful ways to search for new physics, but they also have inherent disadvantages at hadron colliders.  On top of the drawbacks associated with EFTs, they are by definition useless for describing physics at scales directly accessible to the LHC.  However, multi-boson processes still are one of, if not, the most important channels to search for new physics due to their connection to EWSB.  In principle, new physics accessible at LHC energies could be discovered or constrained by direct searches in groups other than the SM groups.  However, in many scenarios of BSM physics there are difficult kinematic regions which direct searches in other groups have trouble accessing.  In this section we demonstrate examples where SM measurements can provide powerful discriminating power for BSM physics even when the EFT description is invalid.  Most importantly, the measurements we propose are equally powerful in searching for BSM physics as EFTs, but avoid all the issues of EFT searches associated with unitarization, strong coupling, power counting, and spurious symmetry arguments. 

Before discussing generalities it is useful to look at an interesting example from Run~I, that came about, not originally from a theoretical effort, but from a series of measurements by ATLAS and CMS.  The $W^+W^-$ cross section as measured by ATLAS and CMS was systematically higher than the predicted NLO cross section at both 7 and 8~TeV.  This eventually led to the theoretical developments involving higher fixed order calculations as well as higher-order resummed calculations that brought theory into good agreement with the measurements  (see \cite{Dawson:2016ysj} for the state of the art which still is slightly low compared to the measured value when jet-veto resummation effects are theoretically included).  However, an intriguing possibility before the higher-order SM calculations were available was that this could have also been caused by a new BSM contribution to the  $W^+W^-$ cross-section measurement.   An example of this was provided in~\cite{Curtin:2012nn} where the  supersymmetric (SUSY) pair-production of Charginos would lead to a final state $pp\rightarrow \chi^+\chi^-\rightarrow W^+W^-\chi^0\chi^0$, with the same $l^+l^- +$ missing transverse energy (MET) final state.  Typically such a process is sought in direct SUSY searches, but if the spectrum is such that the kinematics is similar to that of the SM background it is very difficult to disentangle and could be missed.  Kinematics in a SUSY process similar to multi-boson final states naturally arises if EW BSM states are similar to the EW scale.  However, this also holds true if the mass splittings between the initially produced states and their decay products are similar to the EW scale.  In~\cite{Curtin:2013gta} it was realized that the $W^+W^-$ cross-section measurement itself could be used to bound a number of these scenarios.  In particular, by using this measurement the first bounds on right-handed sleptons that exceeded LEP limits were found.  This was applied to other SM channels as well, for instance the $t\bar{t}$ final state in~\cite{Czakon:2014fka}.  

Having BSM physics which mimics SM final states is a very generic phenomenon.  For example, many different types of models were written to attempt to explain the $W^+W^-$ cross-section excess~\cite{Curtin:2012nn,Curtin:2013gta,Rolbiecki:2013fia,Jaiswal:2013xra,Curtin:2014zua}.  Some of these did not even directly rely on partners of EW gauge bosons for production, but nevertheless led to final states that potentially contaminated the SM measurement.  Almost all Exotic/SUSY searches have gaps when a SM background and BSM signals become kinematically similar.  Dedicated search strategies can be set up to try to close these gaps, but it is very model dependent and takes much effort to understand the SM background.  Naturally, as demonstrated in~\cite{Curtin:2013gta}, a SM measurement is already an incredibly powerful place to search for this type of generic BSM physics.   However, this has been carried out only by theorists and the methods could be pushed further by those making the measurements.  Unfortunately, as discussed so far, multi-gauge boson cross-section measurements are only used by the experiments to search for EFTs, aTGCs and aQGCs, none of which are relevant for the processes described here.  Fortunately, there is a way out already adopted by BSM groups within ATLAS and CMS, which recently has also been adopted by the SM groups and should be extended to all channels.

We recommend that for all multi-gauge boson measurements, the experiments place bounds via upper limits on fiducial cross sections as an alternative to EFT  and anomalous coupling interpretations.  The  ATLAS SUSY group began giving limits like this.  In addition to interpreting their signal regions through models, they included $95\%$ Confidence Level (C.L.) upper cross-section limits on signal regions~\cite{ATLAS:2011ad} independent of interpretation.  ATLAS and CMS have given fiducial cross sections in multi-gauge boson production measurements, and in a few cases $95\%$ C.L. upper limits on signal regions as well, which we strongly endorse. By giving upper limits on cross sections in different fiducial regions, any model can be interpreted whether or not an EFT approach is valid or a model must be used.  There is no loss in discriminating power compared to previous studies of SM cross sections.  For instance signal regions used for aTGCs or aQGCs	 based on high \pt or invariant mass can be kept, and theorists can easily recast the bounds.  However, it avoids the interpretation issues for the experiments on the validity of EFTs, aTGCs, or aQGCs.  In particular, the theoretical statement of when a certain model or approach is theoretically valid resides with the theorists.  Additionally,  it allows for the direct comparison with models that are not describable in the theoretical approaches implemented by the experimental groups, for instance the $W^+W^-$ example given earlier.  Furthermore, by reducing the time spent on theoretical interpretation, it allows for more ``signal'' regions to be investigated.  We emphasize that this is not what has been done at the LHC when moving from EFTs of Dark Matter (DM)~\cite{Fox:2011pm} to Simplified Models~\cite{Abdallah:2015ter} because of concerns with the EFT approach.  In the case of DM at the LHC, it was realized that having an EFT description of DM was often not valid due to the unitary/strong coupling or Wilson coefficient and power counting arguments and another approach was needed.  To couple DM to SM charged particles generically requires new physics that is charged under the SM gauge symmetry which we call messenger particles (there are notable exceptions but this is quite common).   Therefore it is typically more straightforward and theoretically consistent to search for these messenger particles directly, rather than searching for EFT operators via radiative processes such as mono-jets that may not be self-consistent.   For example, this is why SUSY bounds on neutralinos were never set via {\em direct} production of neutralinos tagged from an initial-state radiation jet.  In principle one could attempt to identify simplified models for EW processes relevant for multi-boson physics as an alternative to EFTs.  However, there are always drawbacks to simplified models as well, and searches in BSM experimental groups typically are not nearly as sophisticated in the SM theory prediction as for a SM measurement.  Rather than duplicate effort that may exist elsewhere and run into issues of theoretical interpretations such as whether or not simplified models provide sufficient coverage, it is much more useful and direct to have the SM groups of ATLAS and CMS provide upper limits on fiducial cross sections. This does not have to be motivated solely from the BSM perspective.  Having more differential distributions in fiducial regions that are well understood by the experiments can point to where more SM theoretical effort is needed, e.g., NNLO QCD, NLO EW, or various resummations.  

\subsubsection{Theoretical Conventions used in Experimental Results}
\label{sss:conventions}
Despite the caveats presented in the previous sections, it is useful to understand what the current measurements are based on and therefore we review the common conventions used for EFT operators that are pertinent for multi-boson processes as well as the anomalous coupling parametrizations.  In addition we give the dictionary that translates between these approaches, although this does {\em not} mean they are equivalent.  The EFT parametrization is theoretically sound when used correctly, while anomalous couplings as in Equation~(\ref{eqn:hagiwara}) are not relevant nor sensible post-Higgs.  Most of the conventions used here are explicitly given in the excellent Snowmass white papers~\cite{Degrande:2013rea} and \cite{Degrande:2012wf}, but we give a succinct version here for completeness.

We begin with our description of the EFT operators that will be used in the experimental sections.   As discussed, the operators of interest are those that include gauge fields and are of dimension $\Delta_i=6$ or possibly $\Delta_i=8$.  The $\Delta_i=6$ are the most important operators when the EFT is valid {\em unless} there is a systematic power counting due to a particular UV interpretation that would suppress the dimensionless Wilson coefficients~\cite{Arzt:1994gp,Liu:2016idz}.  At $\Delta_i=6$ there are already 59 operators in the SM~\cite{Buchmuller:1985jz,Grzadkowski:2010es}, while for $\Delta_i=8$ an exhaustive list of 535 operators was finally classified in~\cite{Lehman:2015coa}.  While there are slight differences in the number of operators at a fixed dimension in the literature depending on what assumptions are chosen, the operator basis has now been extended through $\Delta_i=12$ in the SM using more sophisticated mathematical techniques~\cite{Henning:2015alf}.   However, the important and simple to understand point is that as $\Delta_i$ increases the number of operators greatly proliferates.  Therefore even though in this review we are only interested in operators which can modify multiple vector boson production, there will be a much larger number of operators than can contribute at larger $\Delta_i$.  One final point to keep in mind, when using an EFT of a particular set of fields (in this case the SM fields): there is inherently a basis choice that one must make, as operators can be related to one another through various identities, integration by parts, or equations of motion.   In this review we focus on operators that affect multi-gauge boson production, but one must keep the basis choice in mind when comparing to bounds on other operators involving  the gauge boson and Higgs fields not surveyed here.  

At $\Delta_i=6$ there are three independent operators, given in Equation~(\ref{eqn:atgc}) and reproduced below, which affect diboson production by giving new contributions to triple gauge boson and quartic gauge boson couplings,
\begin{eqnarray}\label{eqn:atgc2}
\mathcal{O}_{W}&=&D_\mu h^\dagger W^{\mu\nu} D_\nu h \nonumber\\
\mathcal{O}_{B}&=&D_\mu h^\dagger B^{\mu\nu} D_\nu h \nonumber\\
\mathcal{O}_{WWW}&=& \mathrm{Tr} \left( W_{\mu\nu} W^{\nu\rho} W^\mu_\rho\right).
\end{eqnarray}
The Wilson coefficients for the operators in Equation~(\ref{eqn:atgc2}) are given by $c_{W}/\Lambda^2$, $c_{B}/\Lambda^2$, and $c_{WWW}/\Lambda^2$.  While there are only three operators that contribute at this dimension to diboson production, there are many other operators at $\Delta_i=6$ that involve the Higgs and gauge fields.  These can be shown to affect the propagators, as for instance in the case of the Peskin-Takeuchi S, T, U parameters~\cite{Peskin:1991sw} which all have $\Delta_i=6$ operator definitions.  While these operators do not contribute to diboson production, their Wilson coefficients are already highly constrained.  Therefore it is important to keep in mind that when studying the operators in Equation~(\ref{eqn:atgc2}), a generic UV completion may already be strongly constrained leading to suppressed Wilson coefficients for these operators as well.  

At $\Delta_i=8$ there are 18 operators divided into three classes that can modify multiple vector boson production by generating additional contributions to quartic gauge boson couplings.   Gauge fields, in a gauge covariant setup, can appear in the operators either in covariant derivatives or field strengths and therefore the operators are classified by their contributions from these basic building blocks.  We use the naming conventions found in~\cite{Eboli:2006wa} that have become standard in this community~\cite{Degrande:2013rea}: $S$-type operators involve only covariant derivatives of the Higgs (listed in Table~\ref{tab:stype}), $M$-type operators include a mix of field strengths and covariant derivatives of the Higgs (listed in Table~\ref{tab:mtype}), and $T$-type operators include only field strengths (listed in Table~\ref{tab:ttype}).  Note that not all operators in~\cite{Eboli:2006wa} are listed here.  Some of the original operators in this notation vanish identically or can be related to others. For a more detailed discussion see~\cite{Rauch:2016pai}.

\begin{table}[htbp]
   \centering
   \begin{tabular}{|c|c|} 
   \hline
   \multicolumn{2}{|c|}{$S$-type operators} \\
   \hline 
   Operator name & Operator \\ \hline
  $ {\cal O}_{S,0}$ & $\left [ \left ( D_\mu \Phi \right)^\dagger D_\nu \Phi \right ] \times\left [ \left ( D^\mu \Phi \right)^\dagger D^\nu \Phi \right ] $ \\ \hline
$ {\cal O}_{S,1}$ &  $\left [ \left ( D_\mu \Phi \right)^\dagger D^\mu \Phi  \right ] \times \left [ \left ( D_\nu \Phi \right)^\dagger D^\nu \Phi \right ] $\\ \hline
$ {\cal O}_{S,2}$ &  $\left [ \left ( D_\mu \Phi \right)^\dagger D_\nu \Phi  \right ] \times \left [ \left ( D^\nu \Phi \right)^\dagger D^\mu \Phi \right ] $\\ \hline
   \end{tabular}
   \caption{Each operator $\mathcal{O}_i$ is parametrized by a Wilson coefficient $f_i/\Lambda^4$. $ {\cal O}_{S,2}$ was introduced in~\cite{Eboli:2016kko}.}
   \label{tab:stype}
\end{table}

\begin{table}[htbp]
   \centering
   \begin{tabular}{|c|c|} 
      \hline
   \multicolumn{2}{|c|}{$M$-type operators} \\
   \hline 
   Operator name & Operator \\ \hline
  $  {\cal O}_{M,0} $ & $\hbox{Tr}\left [ {W}_{\mu\nu} {W}^{\mu\nu} \right]\times  \left [ \left ( D_\beta \Phi \right)^\dagger D^\beta \Phi \right ] $ \\ \hline
$ {\cal O}_{M,1} $ & $\hbox{Tr}\left [ {W}_{\mu\nu} {W}^{\nu\beta} \right ]\times  \left [ \left ( D_\beta \Phi \right)^\dagger D^\mu \Phi \right ]$ \\ \hline
$ {\cal O}_{M,2} $ & $ \left [ B_{\mu\nu} B^{\mu\nu} \right ]\times  \left [ \left ( D_\beta \Phi \right)^\dagger D^\beta \Phi \right ]$ \\ \hline
$ {\cal O}_{M,3} $ &  $\left [ B_{\mu\nu} B^{\nu\beta} \right ]\times  \left [ \left ( D_\beta \Phi \right)^\dagger D^\mu \Phi \right ]$ \\ \hline
$ {\cal O}_{M,4} $ & $\left [ \left ( D_\mu \Phi \right)^\dagger {W}_{\beta\nu} D^\mu \Phi  \right ] \times B^{\beta\nu}$ \\ \hline
$ {\cal O}_{M,5} $ & $ \left [ \left ( D_\mu \Phi \right)^\dagger {W}_{\beta\nu} D^\nu \Phi  \right ] \times B^{\beta\mu} $ \\ \hline
$ {\cal O}_{M,7} $ & $\left [ \left ( D_\mu \Phi \right)^\dagger {W}_{\beta\nu} {W}^{\beta\nu} D^\mu \Phi  \right ] $ \\ \hline
   \end{tabular}
   \caption{Each operator $\mathcal{O}_i$ is parametrized by a Wilson coefficient $f_i/\Lambda^4$.}
   \label{tab:mtype}
\end{table}

\begin{table}[htbp]
   \centering
   \begin{tabular}{|c|c|} 
      \hline
   \multicolumn{2}{|c|}{$T$-type operators} \\
   \hline 
   Operator name & Operator \\ \hline
  $  {\cal O}_{T,0} $ & $ \hbox{Tr}\left [ {W}_{\mu\nu} {W}^{\mu\nu} \right ] \times   \hbox{Tr}\left [ {W}_{\alpha\beta} {W}^{\alpha\beta} \right ] $ \\ \hline
$ {\cal O}_{T,1} $ & $\hbox{Tr}\left [ {W}_{\alpha\nu} {W}^{\mu\beta} \right ] \times   \hbox{Tr}\left [ {W}_{\mu\beta} {W}^{\alpha\nu} \right ] $ \\ \hline
$ {\cal O}_{T,2} $ & $\hbox{Tr}\left [ {W}_{\alpha\mu} {W}^{\mu\beta} \right ]\times   \hbox{Tr}\left [ {W}_{\beta\nu} {W}^{\nu\alpha} \right ] $ \\ \hline
$ {\cal O}_{T,5} $ & $  \hbox{Tr}\left [ {W}_{\mu\nu} {W}^{\mu\nu} \right ]\times   B_{\alpha\beta} B^{\alpha\beta} $ \\ \hline
$ {\cal O}_{T,6} $ & $ \hbox{Tr}\left [ {W}_{\alpha\nu} {W}^{\mu\beta} \right ]\times   B_{\mu\beta} B^{\alpha\nu}  $ \\ \hline
$ {\cal O}_{T,7} $ & $ \hbox{Tr}\left [ {W}_{\alpha\mu} {W}^{\mu\beta} \right ]\times   B_{\beta\nu} B^{\nu\alpha} $ \\ \hline
$ {\cal O}_{T,8} $ & $ B_{\mu\nu} B^{\mu\nu}  B_{\alpha\beta} B^{\alpha\beta} $ \\ \hline
$ {\cal O}_{T,9} $ & $ B_{\alpha\mu} B^{\mu\beta}   B_{\beta\nu} B^{\nu\alpha} $ \\ \hline
   \end{tabular}
   \caption{Each operator $\mathcal{O}_i$ is parametrized by a Wilson coefficient $f_i/\Lambda^4$.}
   \label{tab:ttype}
\end{table}

The parametrization of the higher dimension operators in Equation~(\ref{eqn:atgc2}) and Tables~\ref{tab:stype}-\ref{tab:ttype} are the most relevant and sensible for the LHC and for searching for physics beyond the SM because they are inherently gauge-invariant under $SU(2)\times U(1)$ and incorporate EWSB by a SM-like Higgs.   There are also the analogs that are CP-violating operators at $\Delta_i=6$ obtained by inserting a dual field strength in the place of one of the field strengths listed.  In this review experimental limits on $S$-, $T$-, and $M$-type operators are presented, although limits using older conventions are also given.

While the operators listed are the recommended best choice for future studies, anomalous coupling measurements existed long before this modern EFT point of view and therefore there are many legacy parametrizations still used by experiments.  For instance, before the Higgs was confirmed experimentally there were many other possibilities for EWSB as reviewed in earlier sections.  Because of this there were other parametrizations of ``higher-dimensional operators''~\cite{Alboteanu:2008my,Reuter:2013gla}, where an effective chiral Lagrangian was used to describe EWSB and the interactions of the longitudinal modes of gauge bosons.  While this parametrization is not as good a starting point post-Higgs there are still some experimental results that use it. In particular, the $\alpha_4$ and $\alpha_5$ parameters are used, that provide new contributions to quartic gauge boson couplings and can be mapped to a Higgs-like theory straightforwardly.  Assuming a $\Sigma$ field describing the longitudinal degrees of freedom, one can define the longitudinal vector field as ${\bf V}=\Sigma (D\Sigma)^\dagger$.  The $\alpha_4$ and $\alpha_5$ parameters are given as the coefficients of the operators
\begin{eqnarray}
\mathcal{O}_4=\hbox{Tr}\left[ \bf V_\mu V_\nu \right] \hbox{Tr} \left[ \bf V^\mu V^\nu \right],\\
\mathcal{O}_5=\hbox{Tr}\left[ \bf V_\mu V^\mu \right] \hbox{Tr} \left[ \bf V_\nu V^\nu \right].
\end{eqnarray}
We strongly recommend using the parametrizations of the $\Delta_i=6$ and $8$ operators previously given instead of $\alpha_4$ and $\alpha_5$.  If necessary one could translate results in a model of weakly coupled EWSB, i.e., a Higgs-like theory, to this parametrization and the $\alpha$'s would be of order $v^2/\Lambda^2$ up to a dimensionless coefficient.

Another example of pre-Higgs higher dimension operators are the quartic gauge boson coupling operators in the Lagrangian given in~\cite{Stirling:1999ek}:
\begin{equation}
\mathcal{L}=-\frac{e^2 a_0^W}{16\pi\Lambda^2}  F_{\mu\nu}F^{\mu\nu} \vec{W}^\alpha \vec{W}_\alpha-\frac{e^2 a_c^W}{16\pi\Lambda^2}  F_{\mu\alpha}F^{\mu\beta} \vec{W}^\alpha \vec{W}_\beta,
\end{equation}
where $\vec{W}_\beta$ is a three-dimensional vector of the $W$ and $Z$ gauge bosons.   Again the gauge symmetry of the SM is not manifest, but such an operator could be generated at $\Delta_i=8$ in a gauge-invariant way and then mapped to this operator when the Higgs acquires a VEV.  In particular, one can map from all the $M$-type operators in Table~\ref{tab:mtype} to these $a$'s as
\begin{equation}
\frac{f_{M,j} v^2}{\Lambda^4}\sim \frac{a^W_{0,c}}{\Lambda^2}.
\end{equation}
The exact numerical mapping depends on the normalizations and can be found in~\cite{Degrande:2013rea}.

There are also higher dimension operators in the outdated anomalous gauge boson coupling Lagrangian as in Equation~(\ref{eqn:hagiwara}).  For example, the $\lambda_V$ and $\tilde{\lambda}_V$ terms are dimension-6 operators.  However, this is not a consistent EFT expansion given the symmetries we know, but they are gauge invariant and can be mapped directly as 
\begin{equation}
\frac{c_{WWW}}{\Lambda^2} \sim \frac{\lambda_V}{m_W^2},
\end{equation}
or its CP-violating analog, which then allows for a consistent power counting in the EFT.

Finally we must review the ``modern'' anomalous coupling parametrizations as given for instance in Equation~(\ref{eqn:hagiwara}) reduced to the LEP scenario~\cite{Gounaris:1996rz} discussed earlier.  As stressed many times, this parametrization should not be used and we recommend that the EFT basis from  Equation~(\ref{eqn:atgc2}) and Tables~\ref{tab:stype}-\ref{tab:ttype} be used {\em if} one insists on a theory interpretation rather than fiducial cross sections.   Nevertheless, anomalous coupling searches existed long before the modern EFT point of view and therefore they have remained as a legacy that still remains in the experimental community.  The original parametrization of anomalous triple gauge boson couplings and quartic gauge boson couplings, is given in Equation~(\ref{eqn:hagiwara}).  As mentioned earlier, in an attempt to make Equation~(\ref{eqn:hagiwara}) more relevant in the LEP era the general parametrization was reformulated in a gauge-invariant manner where $g_1^Z=g_1^\gamma=\kappa_Z=\kappa_\gamma=1$ (appropriately rescaled by the coupling constants $g$ of $SU(2)$ and $g'$ of $U(1)_Y$) while all other terms do not exist at tree-level in the SM.   Deviations from this limit are parametrized as $\Delta g_1^V \equiv (g_1^V-1)$, $\Delta \kappa_V\equiv (\kappa_V-1)$, and $\lambda_V$, which are the experimentally bounded quantities in a search for BSM contributions to anomalous couplings~\cite{Gounaris:1996rz}. This standard aTGC parametrization has long been used; however it manifestly violates unitarity and lacks a systematic program for renormalization unlike an EFT~\cite{Degrande:2012wf}.   As a kludge, form factors were introduced to parametrize vertex functions for triple gauge boson couplings in momentum space.   This is not sensible nor gauge invariant, but has nevertheless propagated into modern measurements.   The choice of parametrization~\cite{Gaemers:1978hg,Hagiwara:1986vm} used is
\begin{eqnarray}
\Gamma_V^{\alpha\beta\mu}&=&f_1^V(q-\bar q)^\mu g^{\alpha\beta}-\frac{f_2^V}{M_W^2}(q-\bar q)^\mu P^\alpha P^\beta\nonumber\\&&
+f_3^V(P^\alpha g^{\mu\beta}-P^\beta g^{\mu\alpha})+if_4^V(P^\alpha g^{\mu\beta}+P^\beta g^{\mu\alpha})\nonumber\\&&
+if_5^V\epsilon^{\mu\alpha\beta\rho}(q-\bar{q})_\rho-f_6^V\epsilon^{\mu\alpha\beta\rho}P_\rho
\nonumber\\&&
-\frac{f_7^V}{m_W^2}(q-\bar{q})^\mu
\epsilon^{\alpha\beta\rho\sigma}P_\rho (q-\bar{q})_\sigma,
\end{eqnarray}
where two of the gauge bosons are $W$'s and $V$ is a $Z$ or $\gamma$, while $q,\bar{q}$, and $P$ are the respective four-momenta.  A similar approach was undertaken in~\cite{Baur:1992cd} for a triple neutral vertex
\begin{eqnarray}
\Gamma^{\alpha\beta\mu}_{Z\gamma V} (q_1,q_2,P) &=& \frac{P^2-q_1^2}{m_z^2}[ h_1^V(q_2^\mu g^{\alpha\beta}-q_2^\alpha g^{\mu\beta} \nonumber\\
&+&\frac{h_2^V}{m_z^2}P^\alpha ((P\cdot q_2)g^{\mu\beta}-q_2^\mu P^\beta) \nonumber \\
+h_3^V \epsilon^{\mu\alpha\beta\rho} q_{2\rho}&+&\frac{h_4^V}{m_z^2}P^\alpha\epsilon^{\mu\beta\rho\sigma}P_\rho q_{2\sigma}].
\end{eqnarray}
The vertex function approach is particularly opaque compared to the EFT operator treatment and because there is not a straightforward mapping, given that the form factors are undetermined functions (although they could be taken to have a fixed value if one wanted to treat this as a Fourier transform of some position space operators). Again, this manifestly does not include gauge-invariance and does not deal with strong coupling/unitarity in a systematic way.  This parametrization should not be used in the future.  Given the systematic gauge-invariant parametrization of the EFT, once the Higgs acquires a VEV, the Wilson coefficients can be mapped to the anomalous couplings approach.  For example,
\begin{equation}
\Delta g_1^Z=c_W \frac{m_z^2}{2\Lambda^2},
\end{equation}
but this is only a one-way mapping and does not mean these two approaches are equivalent.  The EFT can be extended systematically, and with a full mapping of Wilson coefficients to anomalous couplings, it enforces certain correlations that would otherwise not exist in an anomalous couplings approach.  While we maintain our recommendation to simply measure fiducial cross sections, if a theory interpretation must be made, use the EFT approach.  However, the self-consistency of the EFT approach must also be verified as explained in previous sections or the interpretations can be misleading or wrong.  For further relations between parameters or connections to MC generator parameters see~\cite{Degrande:2013rea}.

\section{Experimental Setup}
\label{sec:expsetup}

Detailed descriptions of the Large Hadron Collider, the ATLAS, and the
CMS detectors are available elsewhere~\cite{Evans:2008zzb,
  Aad:2008zzm, Chatrchyan:2008aa}. The definitions of the physics
objects used in the described analyses vary in both efficiency and
purity and are selected based on the needs of the specific physics
process under study. CMS makes extensive use of particle flow
algorithms which use all the CMS subsystems~\cite{Chatrchyan:2011ds, Beaudette:2014cea}.

The triggers selecting the final states of interest to be recorded for
offline analysis are generally based on the selection of energetic
electrons or muons if present in the final state, with thresholds
depending on the data taking period under study and its instantaneous
luminosity. The trigger thresholds for electrons and muons are
efficient for $W$ and $Z$ boson leptonic decays, and reconstruction
thresholds also maintain high efficiency. In the absence of charged
leptons in the signature, other characteristics such as the presence
of energetic photons or large MET are utilized. The hadronic decay
products of $W$ or $Z$ bosons are not required to satisfy a trigger.
The performance of the ATLAS trigger system is described in more
detail elsewhere~\cite{Aad:2012xs, ATLAS-CONF-2012-048, Aad:2014sca},
and a detailed description of the CMS system is given 
in~\cite{Adam:2005zf, Chatrchyan:2009ab, Khachatryan:2016bia}.

The performance of ATLAS and CMS for
photons~\cite{Aaboud:2016yuq, ATL-PHYS-PUB-2011-007, ATLAS-CONF-2012-123, Aad:2014nim,
  Khachatryan:2015iwa}, electrons~\cite{Aaboud:2016vfy,
  Aad:2014fxa, Khachatryan:2015hwa}, muons~\cite{Aad:2014rra,
  Aad:2014zya, Chatrchyan:2013sba}, 
MET~\cite{Aad:2012re, ATLAS-CONF-2013-082, ATLAS-CONF-2014-019,
  Khachatryan:2014gga, Aad:2016nrq}, and jets~\cite{ATL-LARG-PUB-2008-002,
  Aad:2011he, Aad:2014bia, ATLAS-CONF-2015-037, Aad:2012ag,
  Chatrchyan:2011ds, Khachatryan:2016kdb} is well-documented.

\begin{figure}[htbp]
  \begin{tabular}{c}
\includegraphics[width=0.4\textwidth]{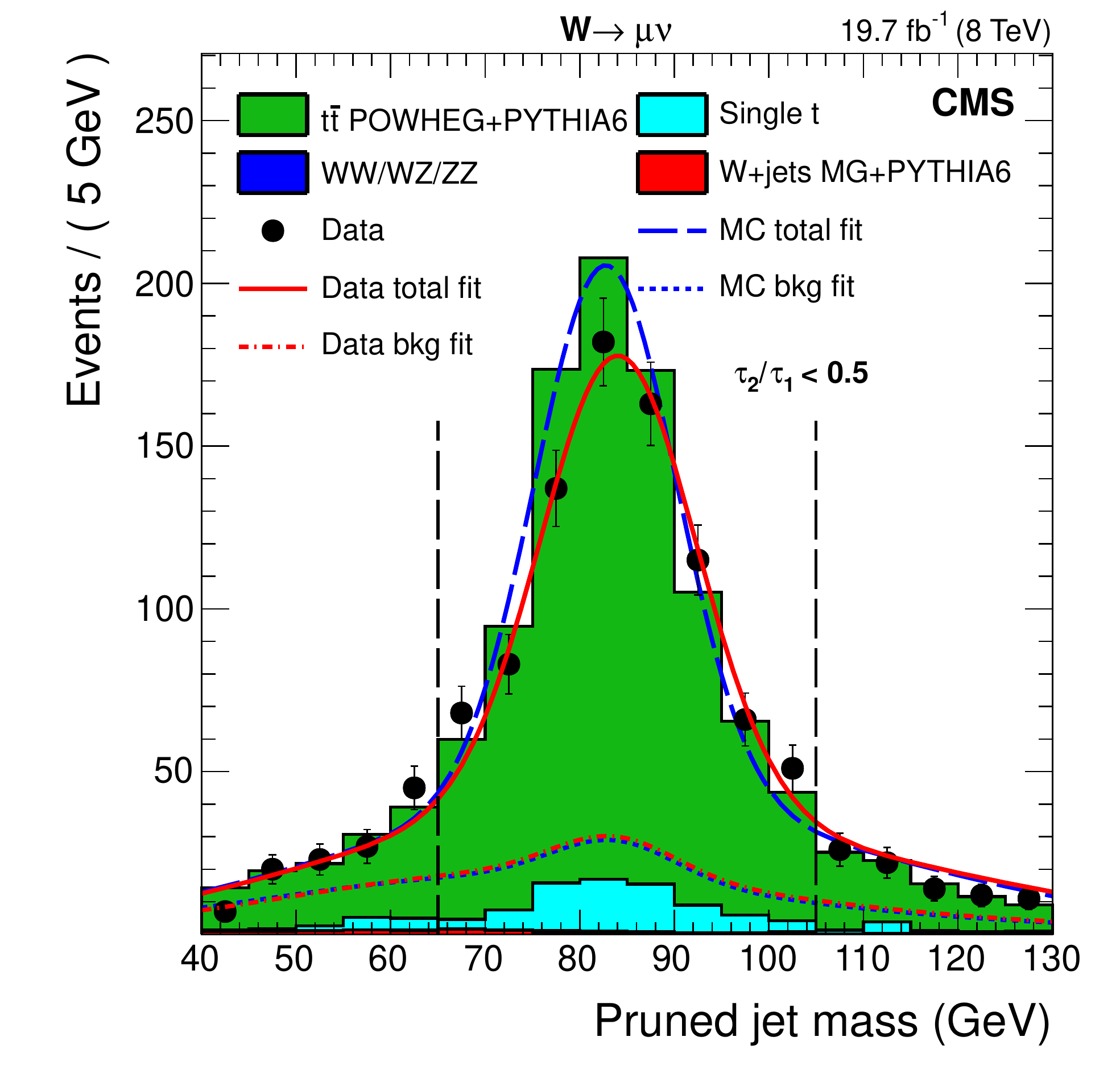}\\
    (a)\\
\includegraphics[width=0.4\textwidth]{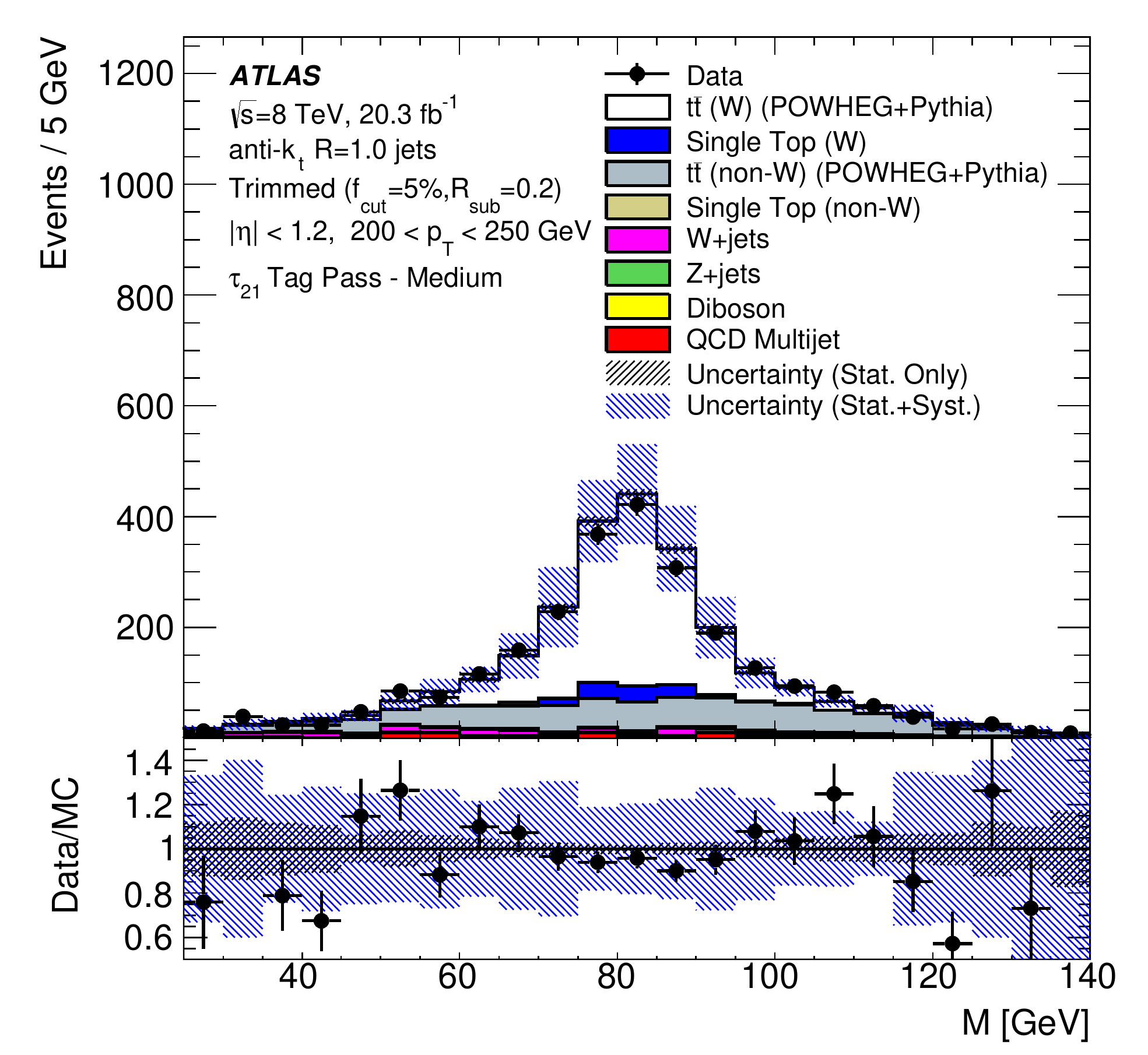}\\
    (b)\\
\end{tabular}
\caption{Jet mass distributions in events with a leptonic $W$ boson
  decay and jets with cuts appropriate for boosted top jets: (a)
  CMS~\cite{Khachatryan:2014vla} for the muon channel and (b)
  ATLAS~\cite{Aad:2015rpa} for the combined electron and muon
  channels.\\
  A wide variety of MC generators is used to model SM signal and
  background processes in the figures of this review. We give an
  overview of the commonly used generators in Section~\ref{ss:sm}, but
  for the specific details of each analysis see the
  provided analysis references.}
\label{fig:jetmW}
\end{figure}
The sensitivity to anomalous gauge couplings is greatest at high mass,
when the hadronic decay products of the gauge bosons are merged into a
single unresolved jet. Nevertheless, the mass of such jets is cleanly
measured~\cite{Aad:2015rpa, Khachatryan:2014gha} and they are
key tools for such studies. In Figure~\ref{fig:jetmW} the jet mass for
a sample of lepton plus MET plus jets with top pair enhancements
illustrates the cleanliness of the merged hadronic $W$ boson decays.

For the gauge bosons, the studies of photons are already listed
above. For $W$ bosons, the leptonic decays are studied using the
lepton (electron or muon) plus MET signature. Hadronic decays are
captured as a mass peak in the resolved dijet case at low transverse
momentum and in the boosted monojet case at high transverse
momentum. For $Z$ bosons dilepton pairs are used, both electrons and
muons~\cite{Aad:2011dm, Chatrchyan:2014mua}; $\tau$ leptons are not
included with one exception detailed in Section~\ref{sec:ZZ}. In
addition, the larger branching fraction neutrino pair decay mode is tagged using a MET
signature. Hadronic $Z$ boson decays are not fully resolved in the
(di)jet mass from $W$ boson decays.

The results described in this review combine the boson signatures at 7
and 8~TeV center of mass energy in a variety of final states detailed
in Sections~\ref{diboson} to~\ref{VBS}. Limits on anomalies in the
gauge couplings appear in Sections~\ref{aTGC} and~\ref{aQGC}, derived
by exploring the high-mass spectrum of the (multi-)bosons themselves
or by use of the transverse momentum of one of the bosons or one of
the boson decay products depending on the specific analysis.
Background processes are evaluated by using Monte Carlo models, by
extrapolating from background dominated control regions or by
data-driven methods, depending on the importance of the background
source and reliability of the available MC modeling.

\section{Diboson Production}
\label{diboson}
\subsection{$\gamma\gamma$ Production}
\label{sec:aa}
Measurement of diphoton production represents a stringent test of
higher-order perturbative QCD corrections, since beyond the LO
quark-antiquark annihilation the quark-gluon channel contributes at
NLO and the gluon-gluon channel box diagram at NNLO. This process is
also sensitive to soft fragmentation contributions where photons arise
from the fragmentation of colored partons. With the discovery of a
Higgs boson~\cite{Aad:2012tfa, Chatrchyan:2012xdj} a resonant
production mode has become available to which diphoton production
constitutes an irreducible background that needs to be
well-characterized for detailed Higgs boson studies as well as for
searches for new resonances.

Both ATLAS~\cite{Aad:2011mh, Aad:2012tba} and
CMS~\cite{Chatrchyan:2011qt, Chatrchyan:2014fsa} studied diphoton
production at 7~TeV in data samples with integrated luminosities of up
to 5~\ifb. The measured total cross sections are most
compatible with the theoretical predictions at NNLO, and partial
N$^3$LO results including the NLO corrections to the gluon-gluon
channel box diagram lead to a further 7\% increase of the total cross-section 
prediction~\cite{Campbell:2016yrh}.

Both experiments provide in addition differential cross-section
measurements as a function of, for example, the invariant mass,
transverse momentum and azimuthal separation of the diphoton
system. As illustrated in Figure~\ref{fig:aadiffxsecs}, these
measurements show better agreement with NNLO predictions
compared to NLO ones, albeit the fixed order NNLO calculation fails to
describe data in regions where fragmentation contributions (not
included in the calculation) are relevant, such as low mass or
intermediate transverse momentum of the diphoton system. Mass scales
slightly below 1~TeV are probed already with these 7~TeV data sets.

\begin{figure}[htbp]
  \begin{tabular}{c}
\includegraphics[width=0.39\textwidth]{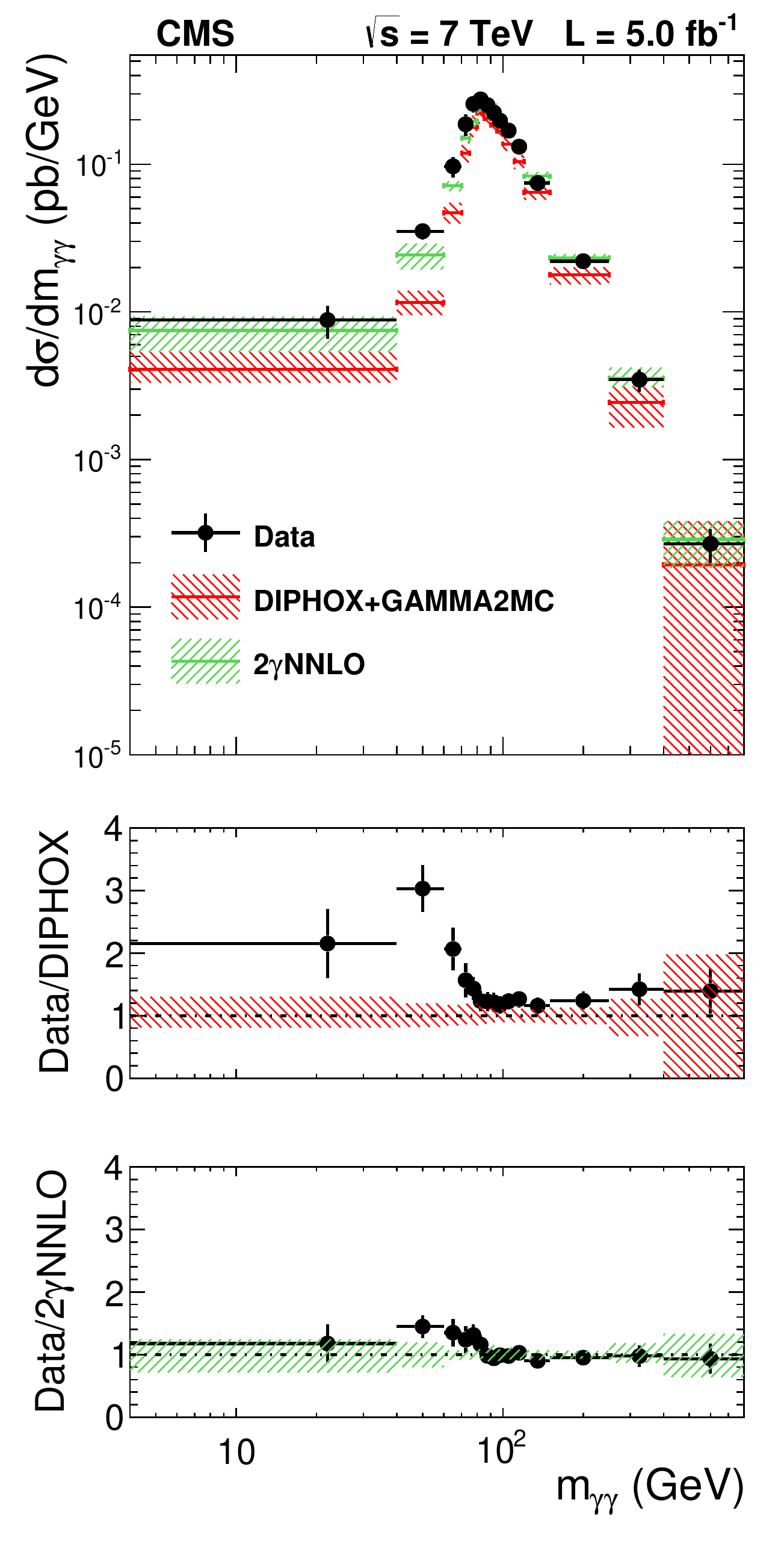}\\
    (a)\\
\includegraphics[width=0.4\textwidth]{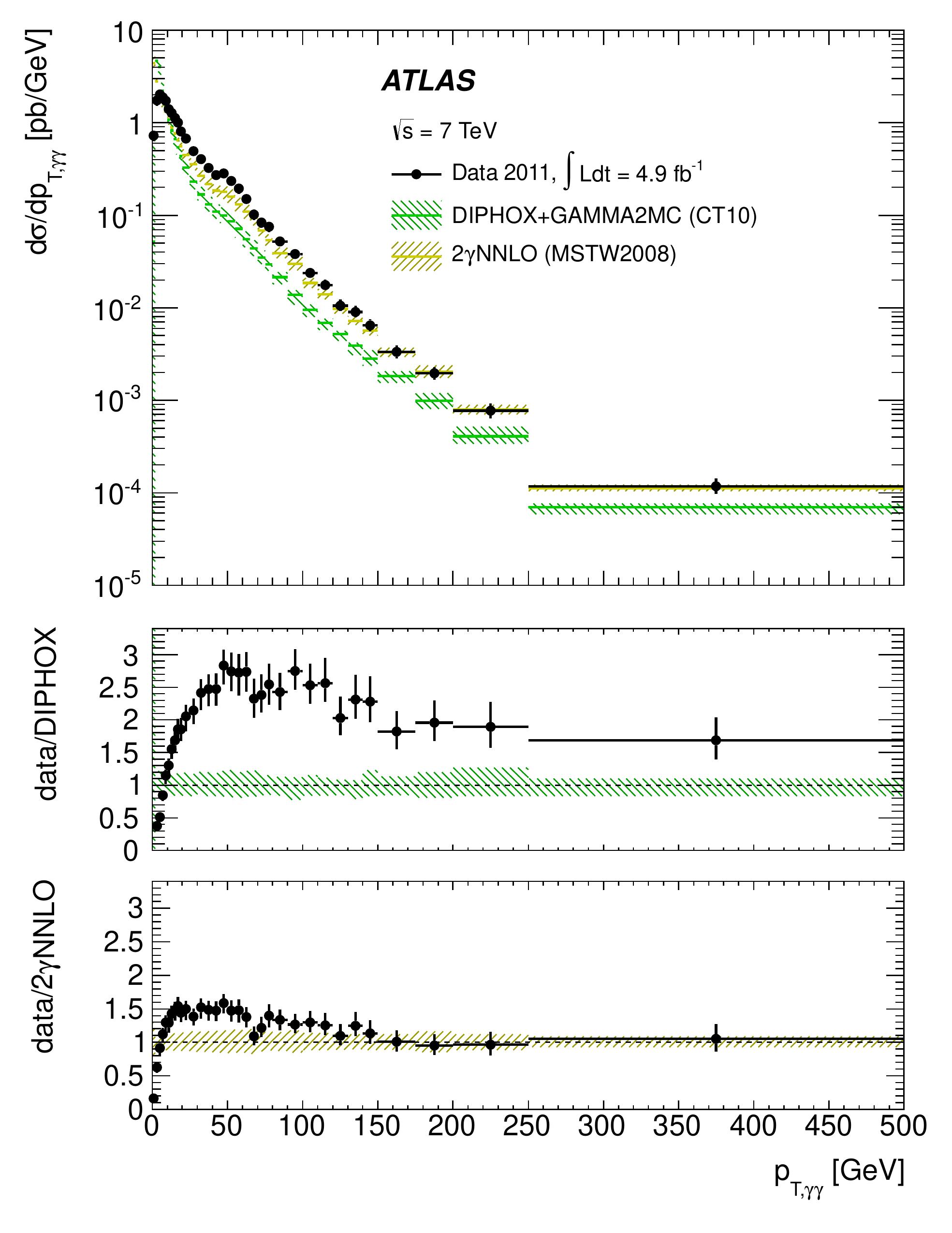}\\
    (b)\\
  \end{tabular}
  \caption{Comparison of $\gamma\gamma$ differential cross-section
    measurements as a function of (a) the invariant
    mass~\cite{Chatrchyan:2014fsa} and (b) transverse
    momentum~\cite{Aad:2012tba} of the diphoton system with NLO and
    NNLO predictions.}
  \label{fig:aadiffxsecs}
\end{figure}

\subsection{$W\gamma$ Production}
Studies of the $W\gamma$ final state have been published by
ATLAS~\cite{Aad:2011tc, Aad:2012mr, Aad:2013izg} and
CMS~\cite{Chatrchyan:2011rr, Chatrchyan:2013fya} at 7~TeV
using data samples with integrated luminosities of up to 5~\ifb, where the
$W$ boson was observed in the leptonic final state with the charged
lepton being either an electron or a muon and the photon was required
to be isolated. Both experiments provide inclusive diboson cross
sections, and ATLAS additionally provides exclusive cross sections
where central jet activity has been vetoed. As illustrated in
Figure~\ref{fig:Wgammaxsec}, CMS finds the cross section to be
compatible with the \mcfm prediction at NLO in QCD as a function of
the photon \et out to 100~GeV, while ATLAS measures inclusive cross
sections higher than the NLO prediction in the inclusive process for
high-\et photons.  NNLO corrections are found to increase the NLO
prediction by $\approx 20\%$, hence improving the agreement with the
measured cross sections~\cite{Grazzini:2015nwa}.
\begin{figure}[htbp]
  \begin{tabular}{c}
\includegraphics[width=0.33\textwidth]{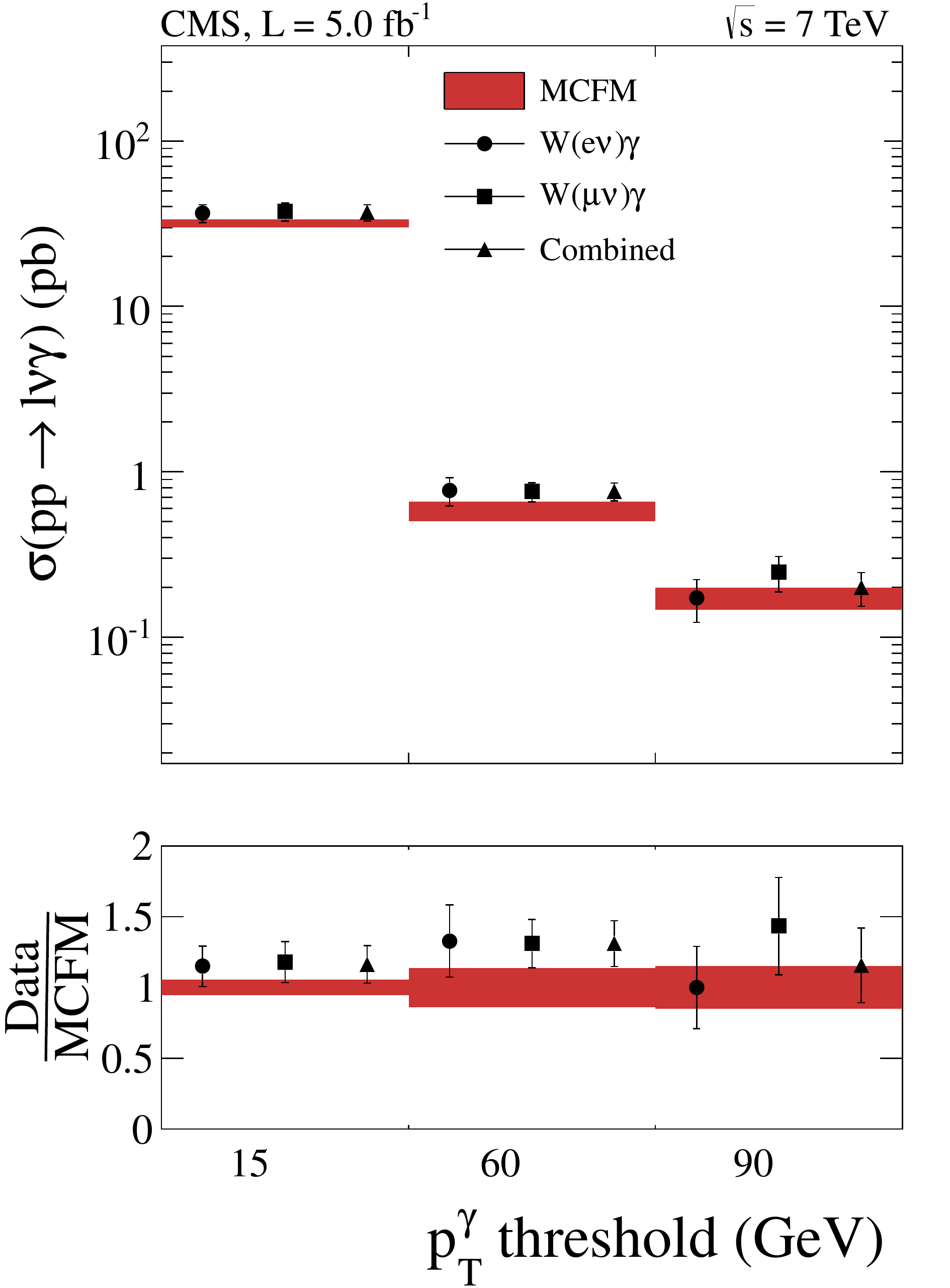}\\
    (a)\\
\includegraphics[width=0.43\textwidth]{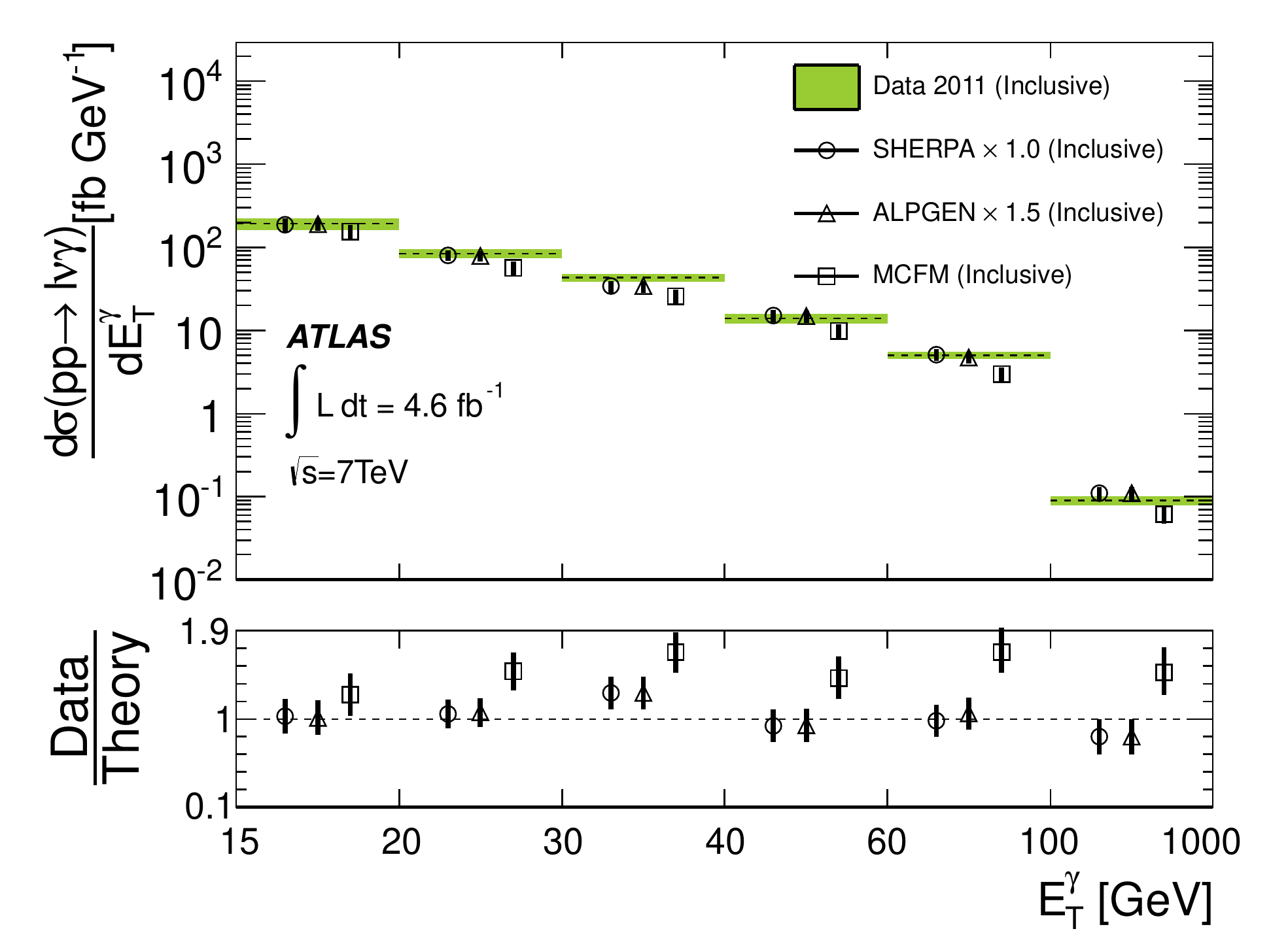}\\
    (b)\\
\end{tabular}
\caption{7 TeV $W\gamma$ inclusive cross section as a function of
  photon \et: (a) comparison of the CMS measurements with \mcfm
  predictions~\cite{Chatrchyan:2013fya} and (b) comparison of the
  ATLAS measurements with \mcfm, \sherpa and \alpgen~\cite{Mangano:2002ea} predictions,
  where the latter two have been scaled to match the total number of
  observed events in data~\cite{Aad:2013izg}.
  Note that \mcfm gives an NLO prediction, which is known
  to increase by $\approx 20\%$ when taking NNLO corrections into account.}
\label{fig:Wgammaxsec}
\end{figure}

The SM TGC of $WW\gamma$ contributes to $W\gamma$ production.  Limits
on the aTGCs $\Delta\kappa_\gamma$
and $\lambda_\gamma$ are set by comparing their effect on the photon
\et spectrum with the observed spectrum as shown in
Figure~\ref{fig:Wgammapt}.
\begin{figure}[htbp]
  \begin{tabular}{c}
\includegraphics[width=0.39\textwidth]{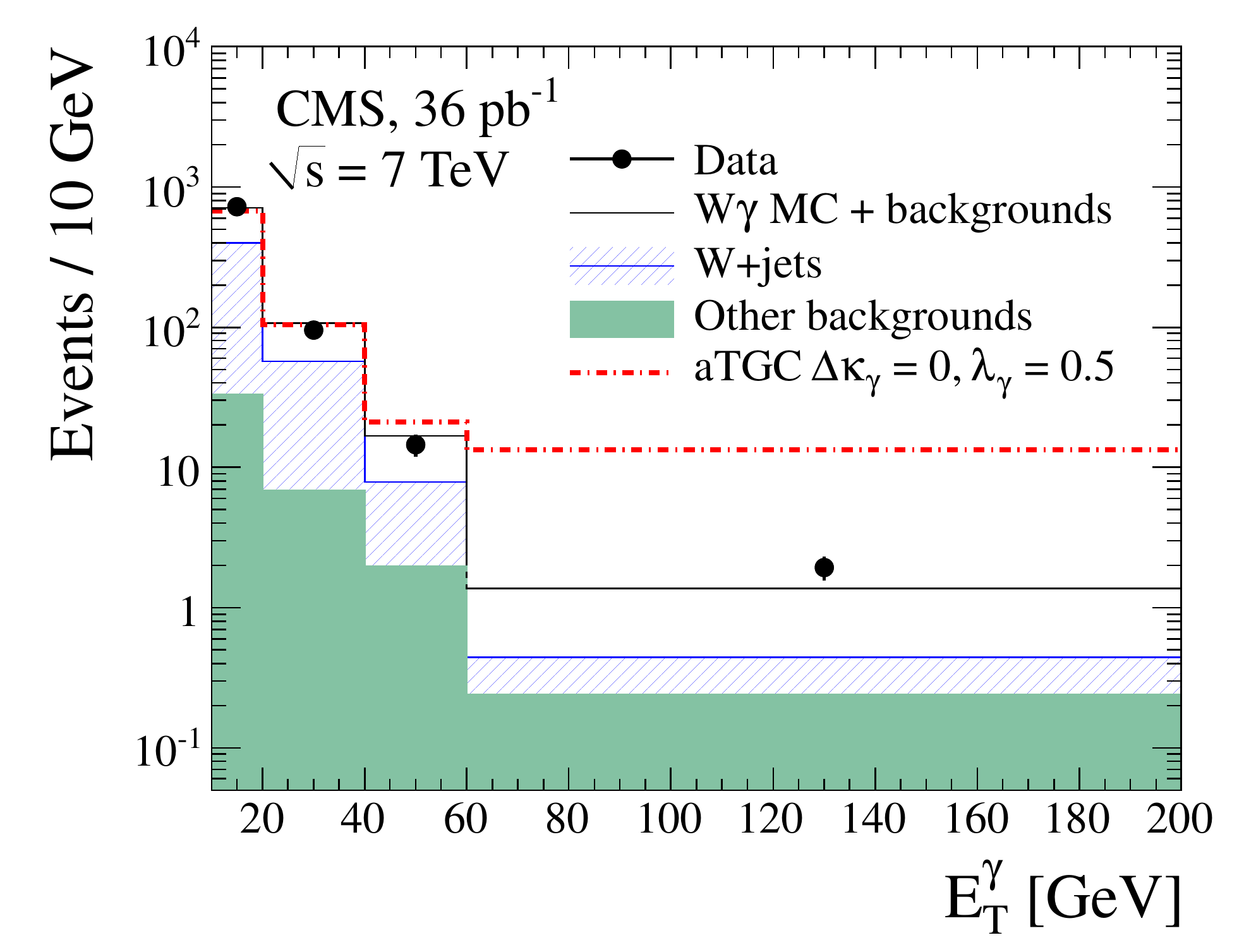}\\
    (a)\\
\includegraphics[width=0.43\textwidth]{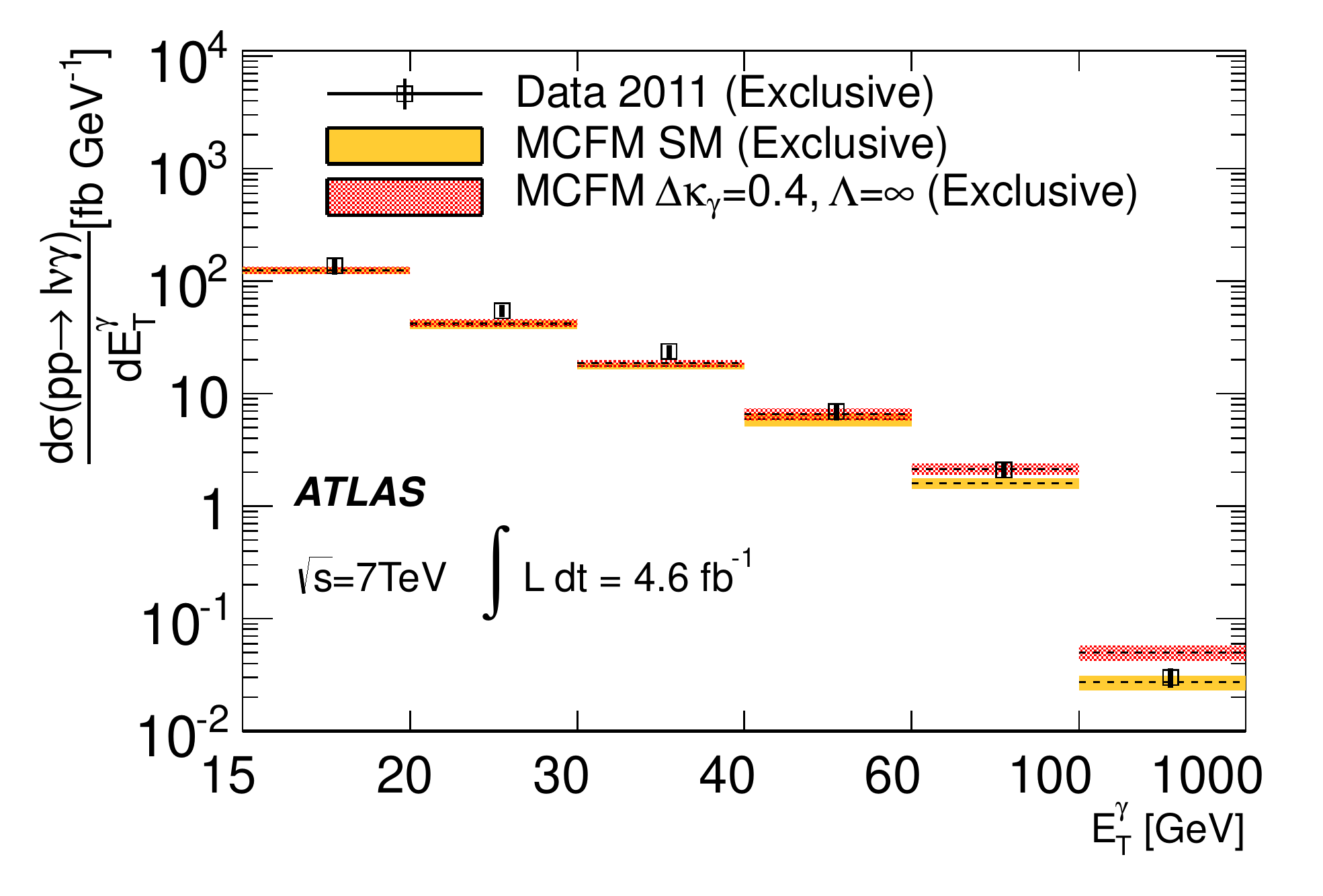}\\
    (b)\\
\end{tabular}
\caption{Effect of aTGCs in the $W\gamma$ final state at 7~TeV: (a)
  photon \et spectrum measured by CMS with an inclusive
  selection~\cite{Chatrchyan:2011rr} and (b) $W\gamma$ cross section
  measured by ATLAS as a function of photon \et with an exclusive
  selection vetoing central jets~\cite{Aad:2013izg}.}
\label{fig:Wgammapt}
\end{figure}
ATLAS uses exclusive events (vetoing central jets) to set limits on anomalous couplings in
order to increase the expected sensitivity in high \et photon events,
which otherwise also tend to exhibit more jet activity in the SM.
For CMS, no constraints are placed on additional objects in the event due to 
issues of possible systematic bias in Monte Carlo modeling of those additional 
objects.

\subsection{$Z\gamma$ Production}

The production of $Z\gamma$ pairs in final states with an oppositely
charged electron or muon pair and an isolated photon has been studied by
ATLAS~\cite{Aad:2011tc, Aad:2012mr, Aad:2013izg} and
CMS~\cite{Chatrchyan:2011rr, Chatrchyan:2013fya} at 7~TeV using data
samples with integrated luminosities of up to 5~\ifb.  Both
experiments provide inclusive diboson cross sections, and ATLAS
additionally provides exclusive cross sections where central jet
activity has been vetoed. As illustrated in
Figure~\ref{fig:Zgammaxsec7TeV}, both ATLAS and CMS find the cross
section to be compatible with the \mcfm prediction at NLO in QCD as a
function of the photon \et. NNLO corrections are found to be much
smaller compared to $W\gamma$ and increase the NLO prediction by
$\approx 8\%$~\cite{Grazzini:2015nwa}.
\begin{figure}[htbp]
  \begin{tabular}{c}
\includegraphics[width=0.33\textwidth]{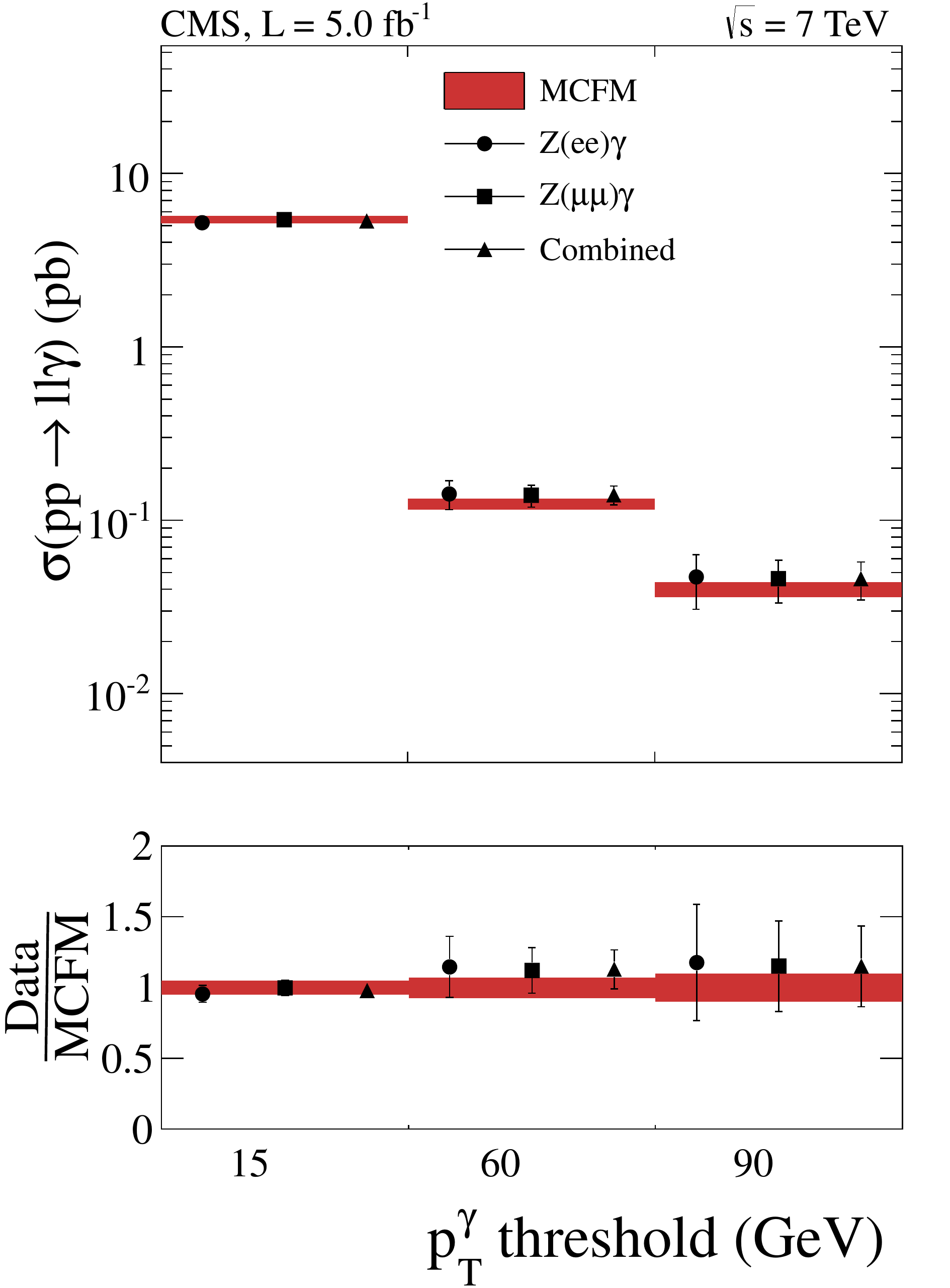}\\
    (a)\\
\includegraphics[width=0.43\textwidth]{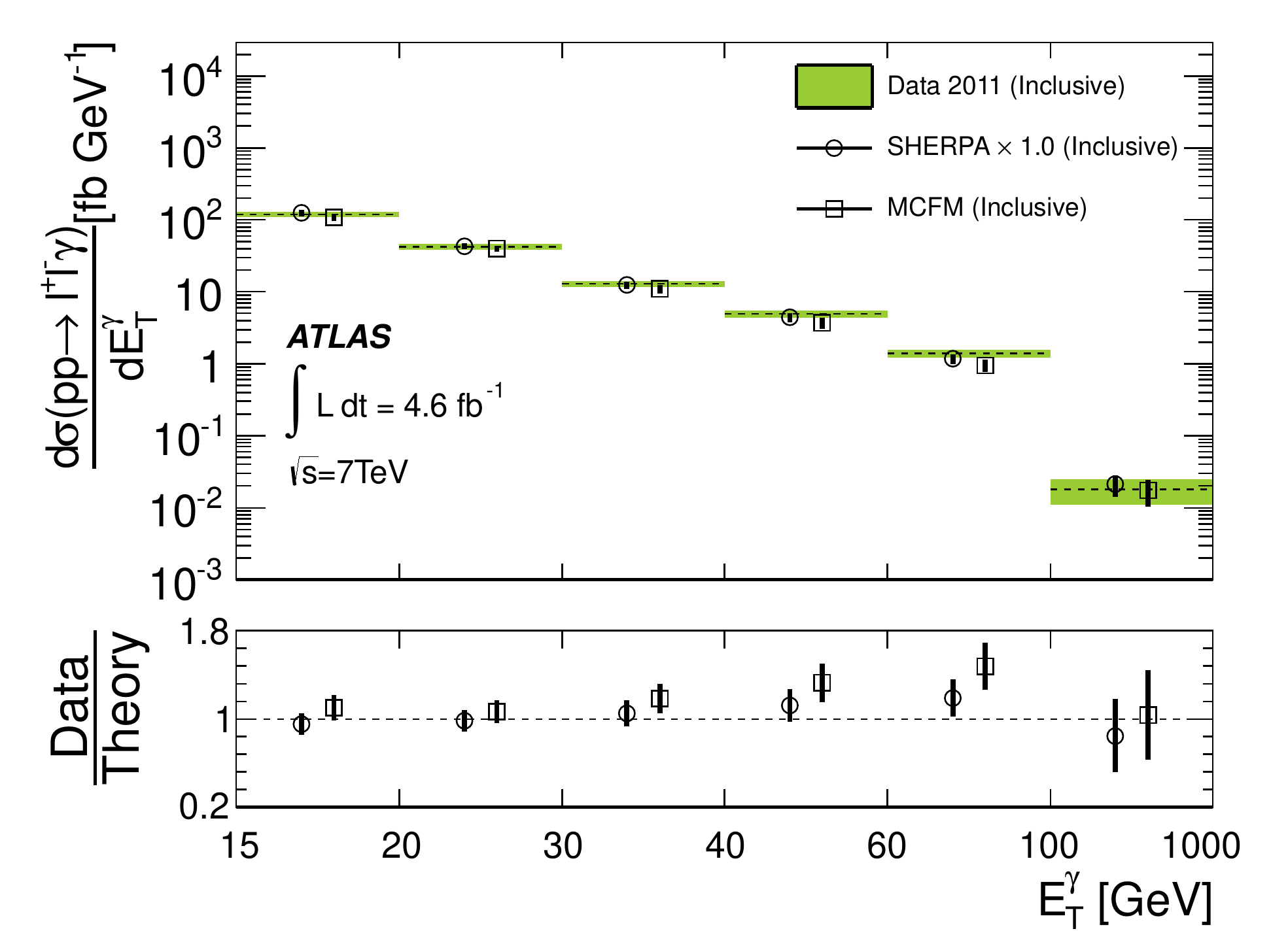}\\
    (b)\\
\end{tabular}
\caption{7 TeV $Z\gamma$ inclusive cross section as a function of
  photon \et: (a) comparison of the CMS measurements with \mcfm
  predictions~\cite{Chatrchyan:2013fya} and (b) comparison of the
  ATLAS measurements with \mcfm and \sherpa predictions,
  where the latter has been scaled to match the total number of
  observed events in data~\cite{Aad:2013izg}.
  Note that \mcfm gives an NLO prediction, which is known
  to increase by $\approx 8\%$ when taking NNLO corrections into account.}
\label{fig:Zgammaxsec7TeV}
\end{figure}
The same final state was studied by ATLAS~\cite{Aad:2016sau}
and CMS~\cite{Khachatryan:2015kea} in 8~TeV data samples with
integrated luminosities of up to 20~\ifb. Both inclusive and exclusive
production cross sections are extracted and found to be in agreement
with \mcfm and NNLO predictions. Figure~\ref{fig:Zgammaxsec8TeV} shows
the inclusive differential cross-section measurements as a function of
photon \et from both experiments.
\begin{figure}[htbp]
  \begin{tabular}{c}
\includegraphics[width=0.43\textwidth]{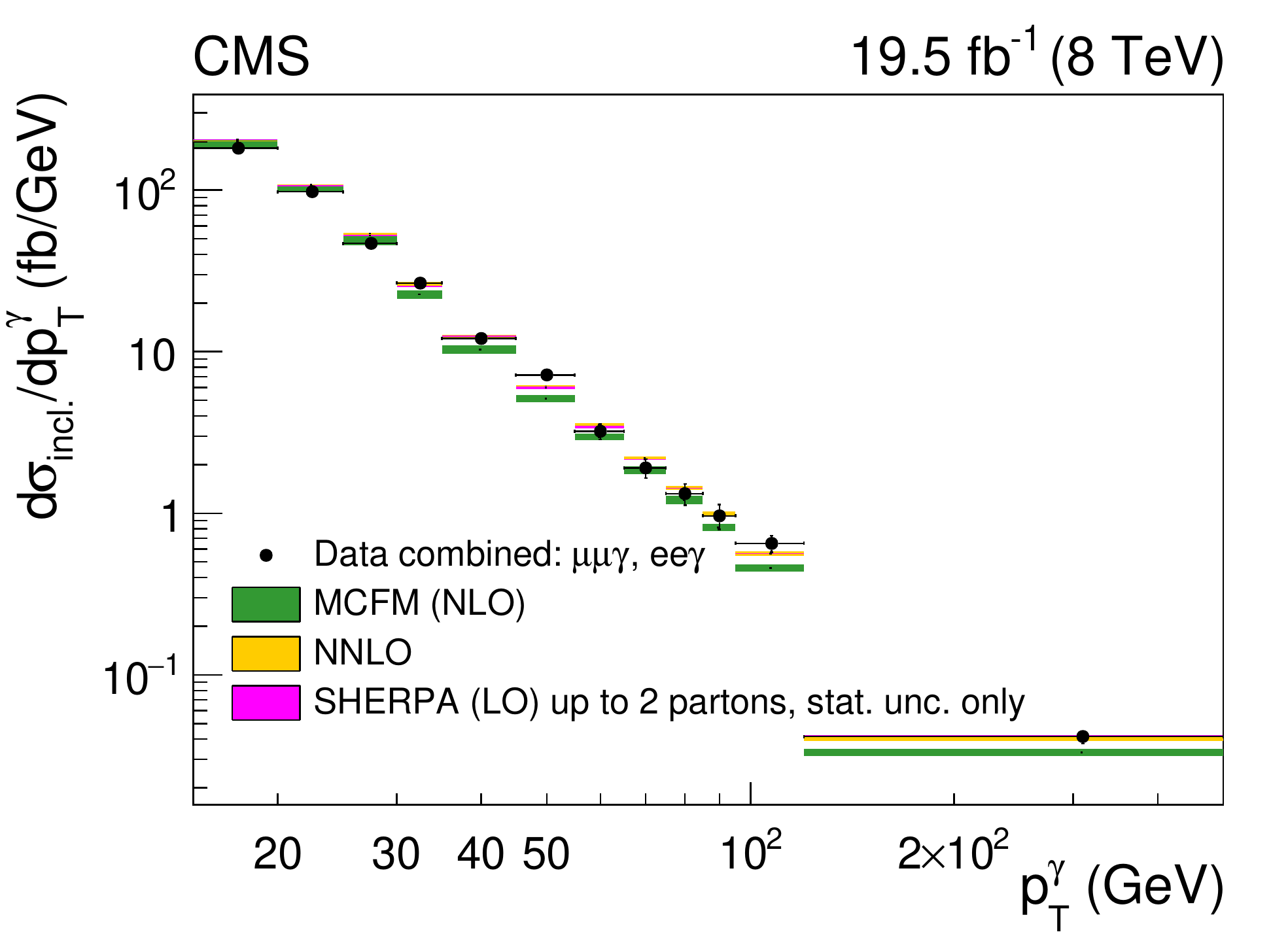}\\
(a)\\
\includegraphics[width=0.43\textwidth]{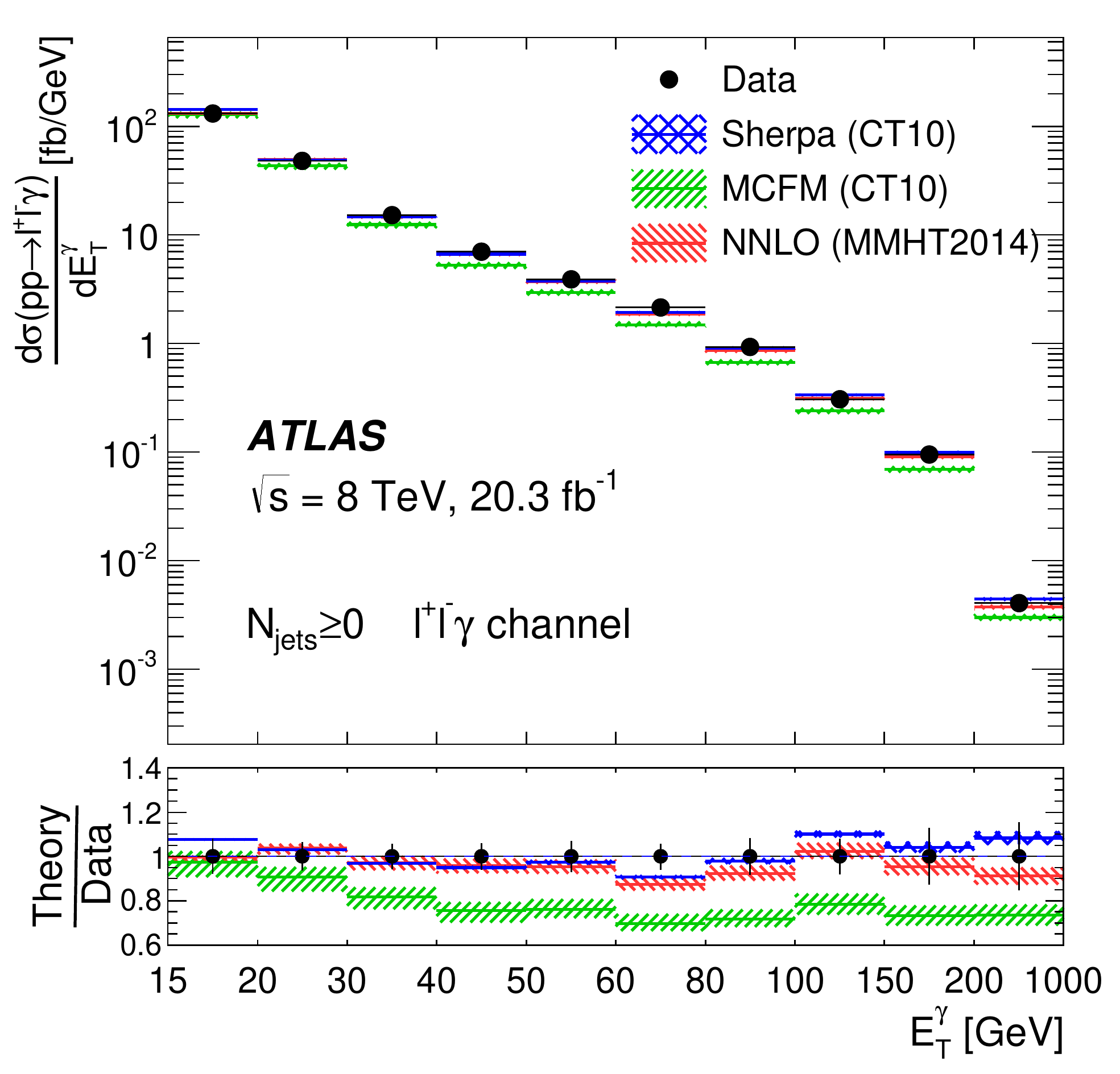}\\
(b)\\
\end{tabular}
\caption{8 TeV $Z\gamma$ inclusive cross section as a function of
  photon \et in comparison with \mcfm, \sherpa, and NNLO predictions:
  (a) CMS~\cite{Khachatryan:2015kea}, where \sherpa is normalized to
  the NNLO cross section, and (b) ATLAS~\cite{Aad:2016sau}.}
\label{fig:Zgammaxsec8TeV}
\end{figure}

SM $Z\gamma$ production arises from photons radiated from
initial state quarks or radiative $Z$ boson decays to charged leptons
as well as fragmentation of final state quarks and gluons into photons.
$Z\gamma Z$ and
$Z\gamma\gamma$ anomalous triple gauge couplings $h_3^V,h_4^V$ ($V =
Z, \gamma$) are constrained by comparing their effect on the photon
\et spectrum with the observed spectrum. The sensitivity to these
aTGCs can be significantly enhanced by studying the $Z\to\nu\bar{\nu}$
decay mode due to the 6 times larger branching fraction compared to the
charged lepton decay modes and the increased detector acceptance. Both
ATLAS~\cite{Aad:2013izg, Aad:2016sau} and
CMS~\cite{Chatrchyan:2013nda, Khachatryan:2016yro} have studied the resulting
final state of large missing transverse energy and an energetic isolated
photon in the 7 and 8~TeV data sets and observe
production rates in agreement with theoretical predictions. The photon
\et spectra extend to about 1~TeV and
are utilized to constrain aTGC contributions as
illustrated in Figure~\ref{fig:Zvvgamma8TeVaTGC}, which also serves to
set the scale for the sensitivity of the data to non-SM couplings.
\begin{figure}[htbp]
  \begin{tabular}{c}
\includegraphics[width=0.43\textwidth]{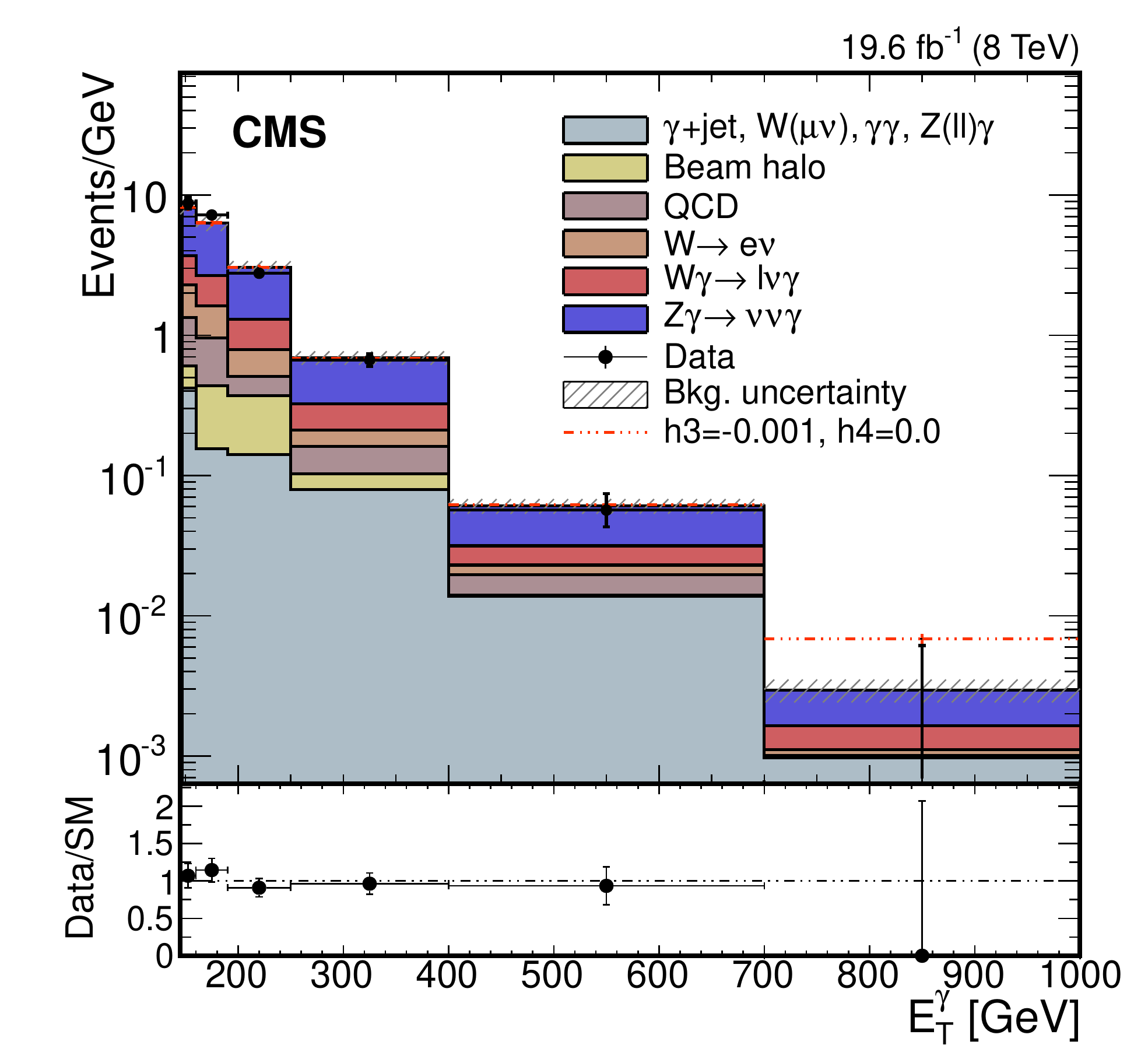}\\
    (a)\\
\includegraphics[width=0.43\textwidth]{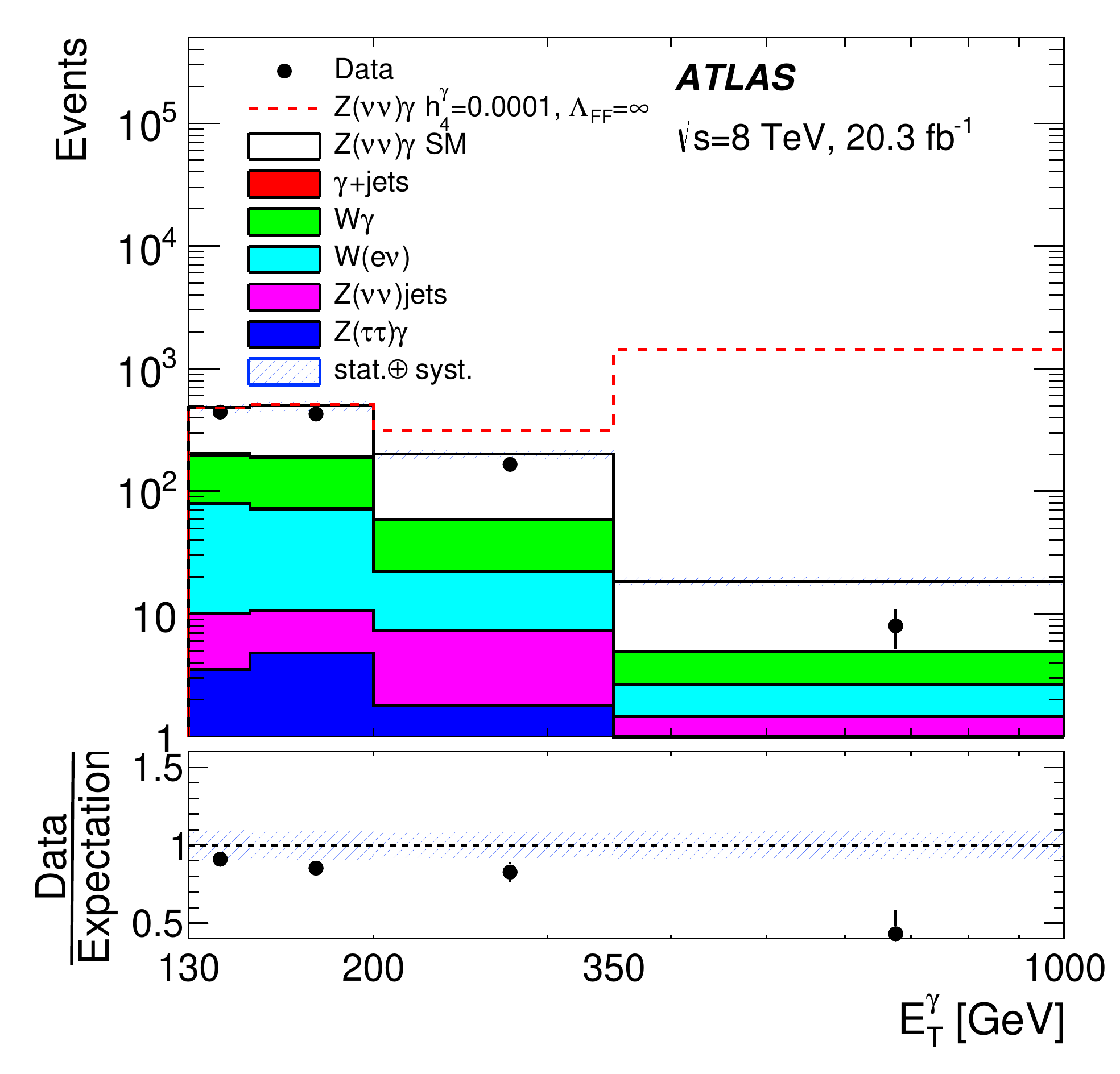}\\
    (b)\\
  \end{tabular}
  \caption{Photon \et spectrum at 8~TeV for the
    $Z\gamma\to\nu\bar{\nu}\gamma$ final state and effect of a representative aTGC on the
    spectrum for (a) CMS~\cite{Khachatryan:2016yro} and (b) ATLAS~\cite{Aad:2016sau} with an exclusive selection vetoing central jets.}
  \label{fig:Zvvgamma8TeVaTGC}
\end{figure}
Again, ATLAS uses exclusive events to set limits on anomalous couplings in
order to increase the expected sensitivity in high-\et photon events,
which otherwise also tend to exhibit more jet activity in the SM.

\subsection{$W^+W^-$ Production}
For the case of $W^+W^-$ production, two decay modes have been
studied. In the leptonic mode, both $W$ bosons decay into a charged
lepton and a neutrino (MET). In the semi-leptonic case, one $W$ boson
decays leptonically while the other decays hadronically. The leptonic
mode has less background but the branching fraction of the $W$ pair is
about 6 times smaller than in the semi-leptonic case when
considering the decay modes involving electrons and muons.
In addition, the $WW$ pair mass in a
semi-leptonic decay can be fully reconstructed up to a quadratic
ambiguity, so that the energy at the TGC vertex is directly measurable
in contrast to the leptonic decay case. However, in the semileptonic
decay mode hadronic $W$ boson decays cannot be fully distinguished from
hadronic $Z$ boson decays due to limited dijet mass resolution. The
semileptonic $WW$ decay is hence studied together with the
semileptonic $WZ$ decay in Section~\ref{sec:WVprod}. Both the
$WW\gamma$ and the $WWZ$ SM TGC contribute to $WW$ production in
distinction to $W\gamma$ production. Deviations from the SM TGC are
labeled by parameters $\lambda_V$, $\Delta\kappa_V$ ($V = Z, \gamma$)
following the nomenclature already introduced for $W\gamma$
production, and $\Delta g_1^Z$.

The production of $WW$ pairs in the fully leptonic decay mode with an
oppositely charged lepton (electron or muon) pair and missing
transverse energy in the final state has been studied by
ATLAS~\cite{Aad:2011kk, Aad:2012oea, ATLAS:2012mec} and
CMS~\cite{Chatrchyan:2011tz, Chatrchyan:2013yaa} at 7~TeV using data
samples with integrated luminosities of up to
5~\ifb. Figure~\ref{fig:WWxsec7TeV} shows the spectra of the highest
\pt lepton of the final state pair as observed by ATLAS and CMS. Also
shown are the modifications to the spectrum caused by aTGCs for which
no evidence was found.
\begin{figure}[htbp]
  \begin{tabular}{c}
\includegraphics[width=0.40\textwidth]{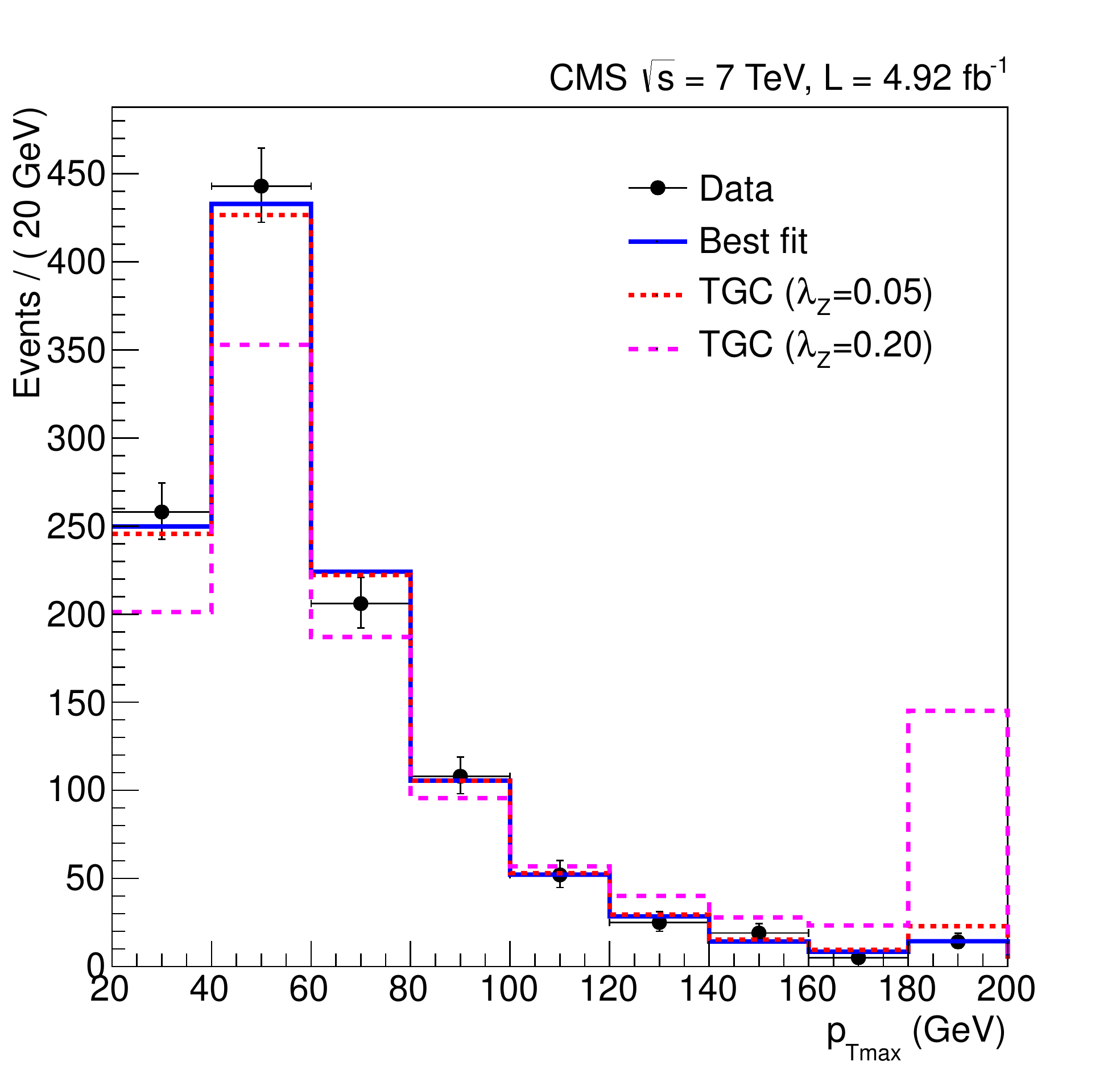}\\
(a)\\
\includegraphics[width=0.43\textwidth]{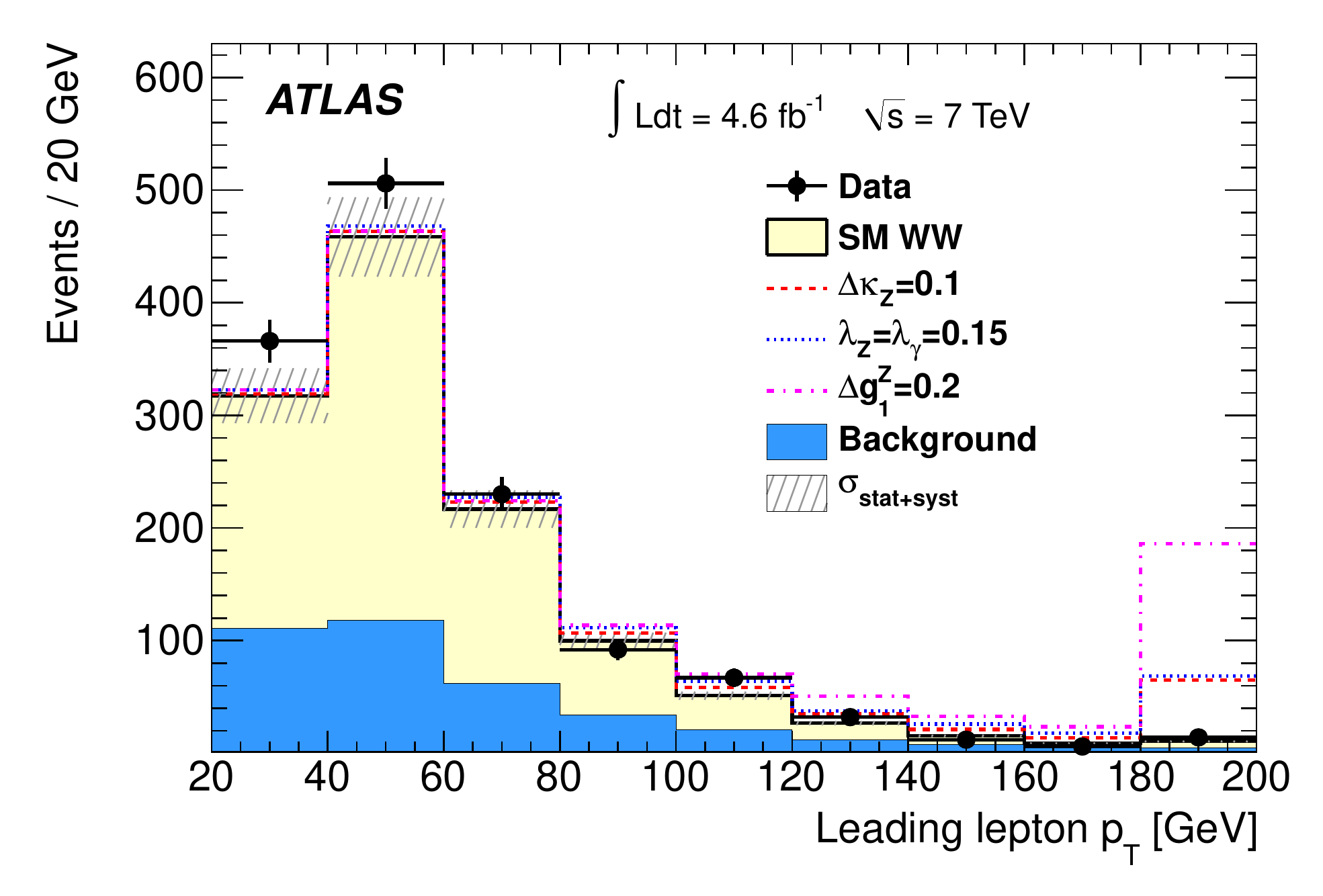}\\
(b)\\
\end{tabular}
\caption{Leading lepton \pt spectra in 7~TeV fully leptonic $WW$
  candidate events and the impact of different anomalous TGC predictions
  for (a) CMS~\cite{Chatrchyan:2013yaa} and (b)
  ATLAS~\cite{ATLAS:2012mec}. The last bin includes the overflow.}
\label{fig:WWxsec7TeV}
\end{figure}
Both experiments do not include resonant production via the Higgs
boson in their signal model and observe $WW$ production cross sections
larger than (then state-of-the-art) NLO predictions, consistent with
the significant cross-section enhancements predicted by NNLO
calculations~\cite{Gehrmann:2014fva}. Additional measurements such as
the ratio of the inclusive $WW$ cross section to the $Z$ boson cross
section~\cite{Chatrchyan:2013yaa} and normalized fiducial cross
section as function of the leading lepton \pt~\cite{ATLAS:2012mec} are
provided as well and are found to be in agreement with theory predictions.

$WW$ production in the fully leptonic decay mode has been studied by
ATLAS~\cite{Aad:2016wpd} and CMS~\cite{Chatrchyan:2013oev,
  Khachatryan:2015sga} as well in 8~TeV data samples with integrated
luminosities of up to 20~\ifb. While ATLAS includes Higgs-mediated
$WW$ production as signal, CMS subtracts the small corresponding expected
contribution. The measured fiducial and total production cross
sections are found to be consistent with NNLO predictions~\cite{Grazzini:2016ctr}, and
(normalized) differential cross sections are measured as a function of
kinematic event variables. CMS includes a measurement of the total 
$WW$ production cross section in events with exactly one jet, while 
ATLAS vetoes events with reconstructed jets.
No evidence for anomalous $WW\gamma$ and
$WWZ$ TGCs is observed and hence limits on the corresponding
parameters are set. An alternative EFT formulation of aTGC with
dimension-6 operators is introduced~\cite{Degrande:2012wf} with
corresponding coefficients $c_W$, $c_{WWW}$, and $c_B$ that can be
mapped to the LEP formulation which allows comparisons with earlier
data. Figure~\ref{fig:WWmll8TeV} shows the dilepton mass spectrum as
measured by CMS~\cite{Khachatryan:2015sga} together with the distorted
spectral shape that would result from aTGC contributions. 
\begin{figure}
  \centering
\includegraphics[width=0.40\textwidth]{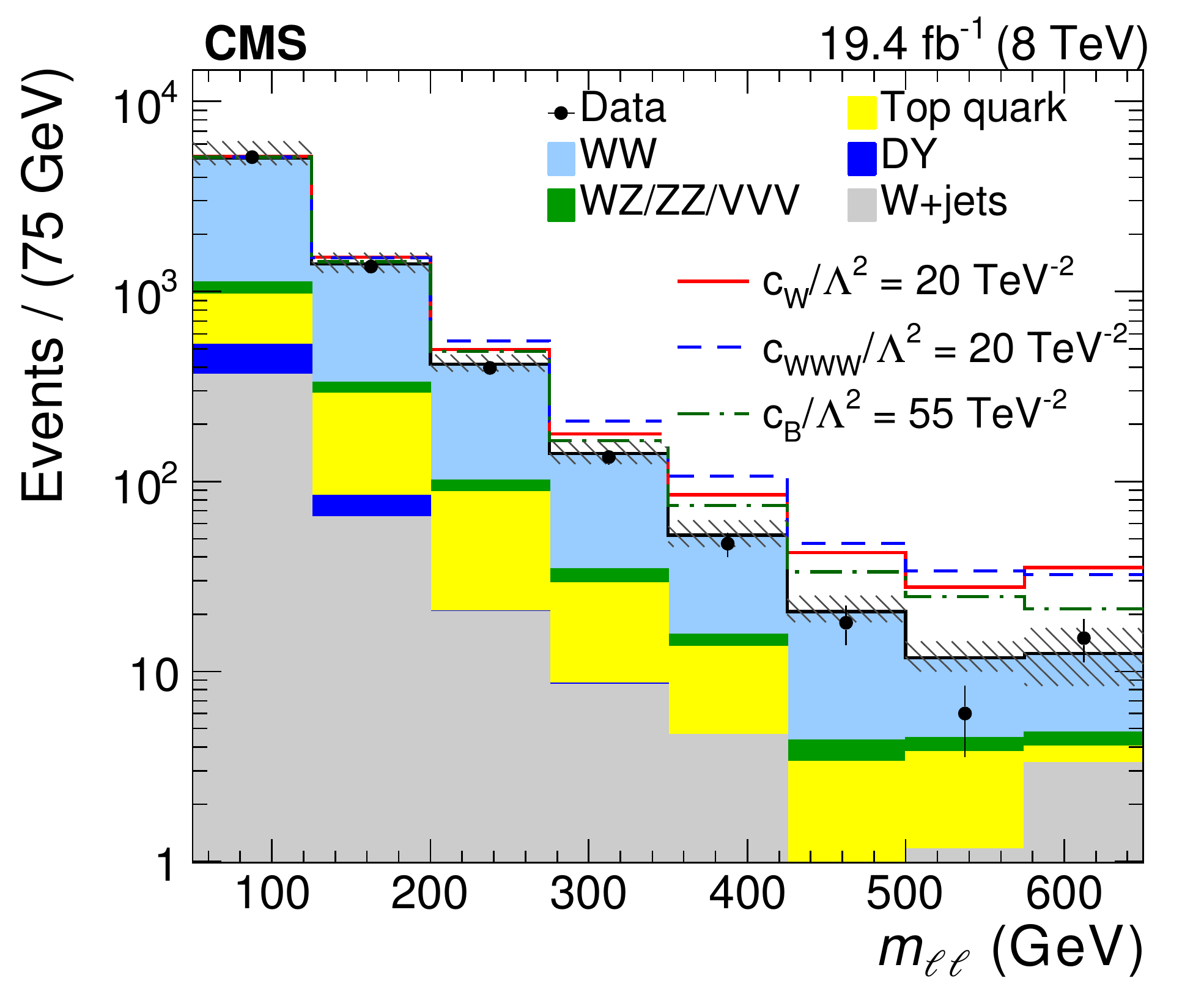}
  \caption{Dilepton mass spectrum in 8~TeV fully leptonic $WW$
    candidate events and impact of different anomalous TGC
    predictions~\cite{Khachatryan:2015sga}. The last bin includes the
    overflow.}
  \label{fig:WWmll8TeV}
\end{figure} 
Figure~\ref{fig:WWxsecsummary} gives an overview of the total $WW$
production cross sections measured at hadron colliders at different
center of mass energies in comparison with the expectations of theory.
\begin{figure}
  \centering
\includegraphics[width=0.4\textwidth]{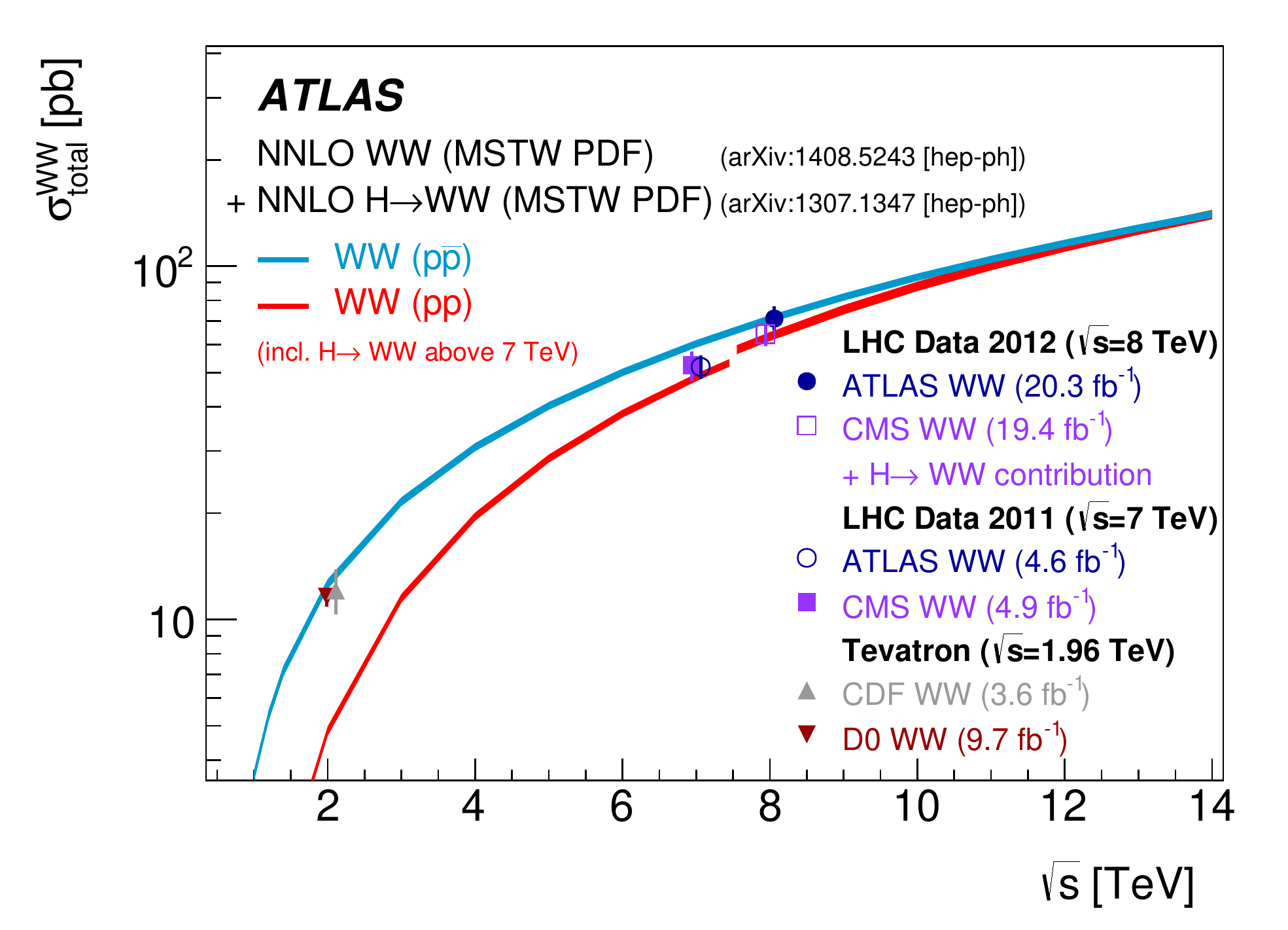}
  \caption{Comparison of measured total $WW$ production cross sections
    at hadron colliders and NNLO theory predictions as a function of
    $\sqrt{s}$~\cite{Aad:2016wpd}.}
  \label{fig:WWxsecsummary}
\end{figure}

ATLAS has studied $WWj$ production in the $e\mu$, MET, and exactly one
jet final state~\cite{Aaboud:2016mrt} in the full 8~TeV data set, where
the largest background from top quark production is suppressed with a
$b$-jet veto. Both $WW+1$ jet and $WW+\leq 1$ jets (the latter in
combination with the 0-jet analysis~\cite{Aad:2016wpd}) fiducial cross
sections are provided and in good agreement with state-of-the-art
theoretical predictions. Extrapolating the $WW+\leq 1$ jets fiducial
measurement to the total cross section, better agreement with the
theoretical prediction is observed than in the 0-jet analysis, and the
overall uncertainty improves by 12\%. The ratio of $WW+1$ jet to
$WW+0$ jets fiducial cross sections is found to be consistent with
theoretical predictions.

\subsection{$W^\pm V$ Production}
\label{sec:WVprod}

Semileptonic $WV$ decays ($V = W,Z$) with one charged lepton (electron
or muon), missing transverse energy and exactly two jets in the final
state have been studied by ATLAS~\cite{Aad:2014mda} and
CMS~\cite{Chatrchyan:2012bd} at 7~TeV using data samples with
integrated luminosities of up to 5~\ifb.  The background (dominated by
$W+$jets production) is much more important in this case compared to
the leptonic decay modes and care is needed to accurately assess the level of
background. The measured sums of the inclusive $WW$ and $WZ$
cross sections are found to be in good agreement with the NLO SM
prediction.  Both experiments constrain anomalous $WWZ$
and $WW\gamma$ couplings utilizing the \pt distribution of the
hadronically decaying $V$ in a narrow mass window 75~GeV $< m_{jj}<
95$~GeV that improves the signal-to-background ratio and 
enhances the expected contribution of $WW$ over $WZ$.
Figure~\ref{fig:WV7TeV} shows the observed dijet-\pt spectra measured by both
experiments in the muon channel together with the potential impact of
aTGCs.
\begin{figure}[htbp]
  \begin{tabular}{c}
\includegraphics[width=0.40\textwidth]{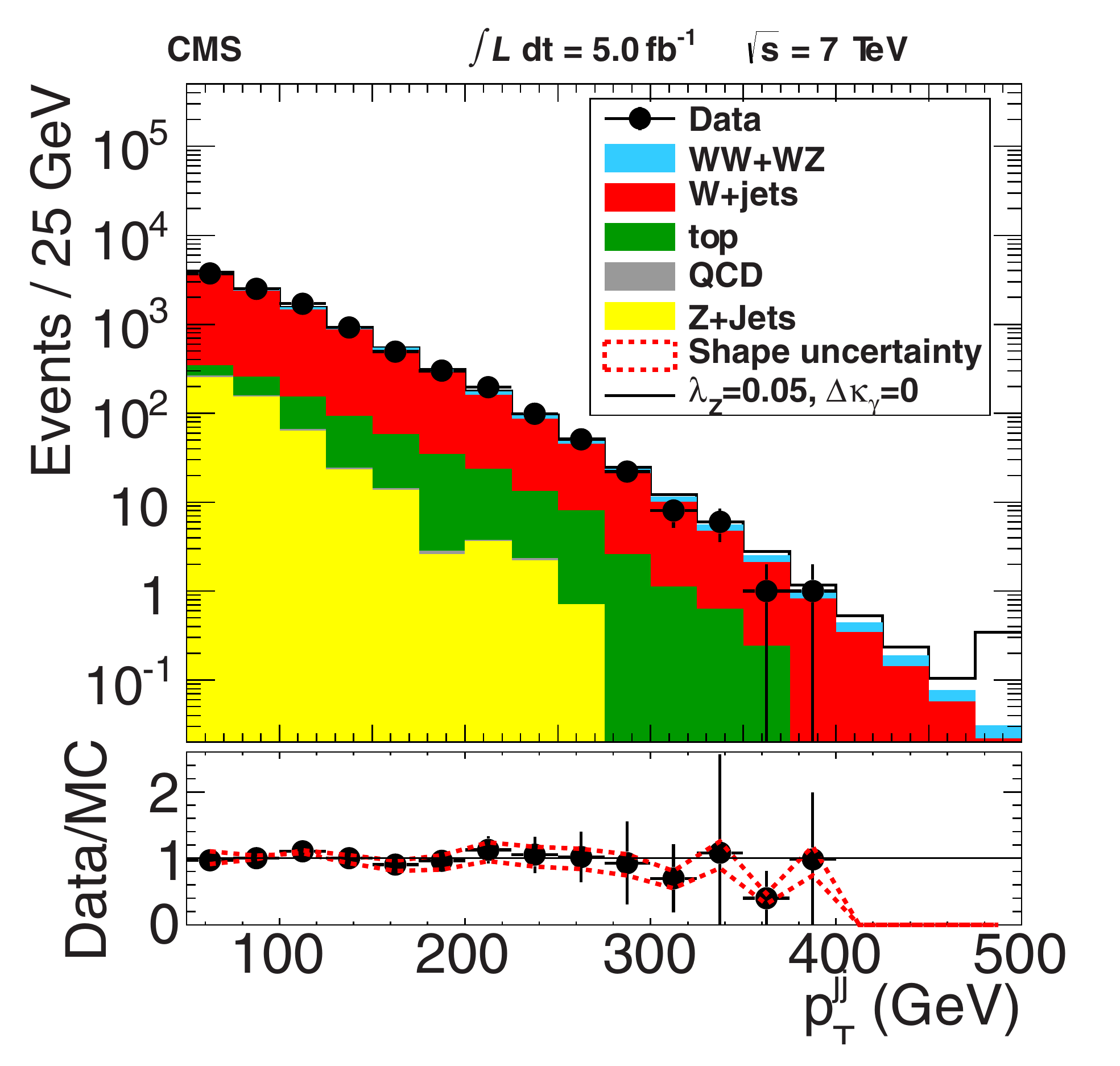}\\
(a)\\
\includegraphics[width=0.40\textwidth]{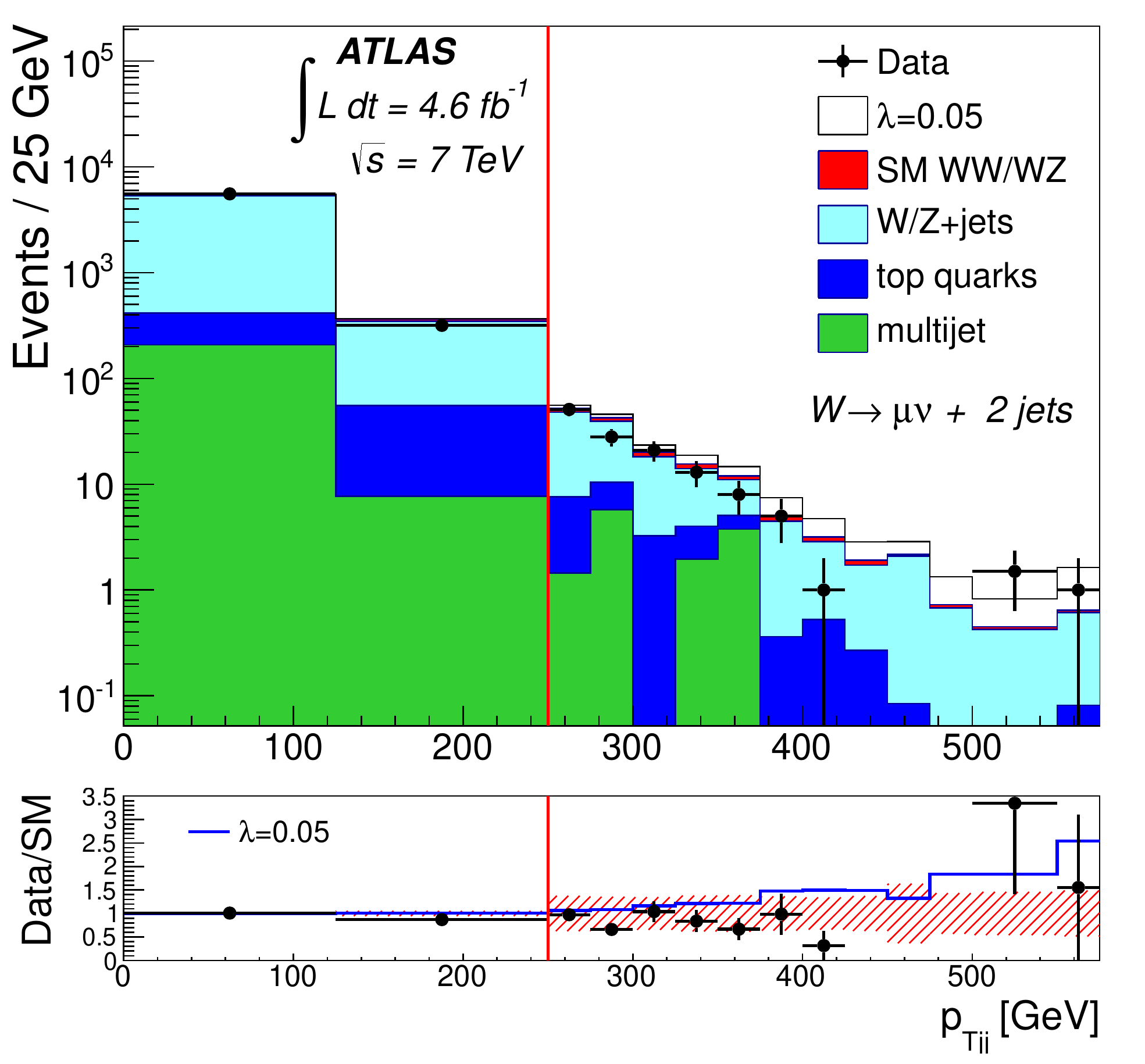}\\
(b)\\
\end{tabular}
\caption{Dijet-\pt spectra in the muon channel of 7~TeV semileptonic
  $WV$ candidate events and the impact of different anomalous TGC
  predictions for (a) CMS~\cite{Chatrchyan:2012bd} and (b)
  ATLAS~\cite{Aad:2014mda}. The last bin includes the overflow.}
\label{fig:WV7TeV}
\end{figure}

\subsection{$ZV$ Production}
CMS has studied semileptonic $ZV$ decays ($V = W,Z$), where the $Z$
boson decays into a pair of $b$-tagged jets in 18.9~\ifb $pp$ data at
8~TeV~\cite{Chatrchyan:2014aqa}. The second $V$ boson is detected
through leptonic final states giving rise to MET (mainly due to
$Z\to\nu\bar{\nu}$), one charged lepton (electron or muon) and MET
($W\to\ell\nu$), or a same-flavor, oppositely-charged lepton pair
(electrons or muons, $Z\to\ell\ell$). A significant $ZV\to b\bar{b}V$
signal is observed, and the simultaneously measured $WZ$ and $ZZ$
cross sections are found to be in agreement with their NLO
predictions, as illustrated in Figure~\ref{fig:ZV8TeV}. The fiducial
cross sections for high-\pt($V$) events are as well found to be in good
agreement with NLO theory predictions and hence give no indication
of anomalous TGC contributions.
\begin{figure}[htbp]
  \centering
\includegraphics[width=0.33\textwidth]{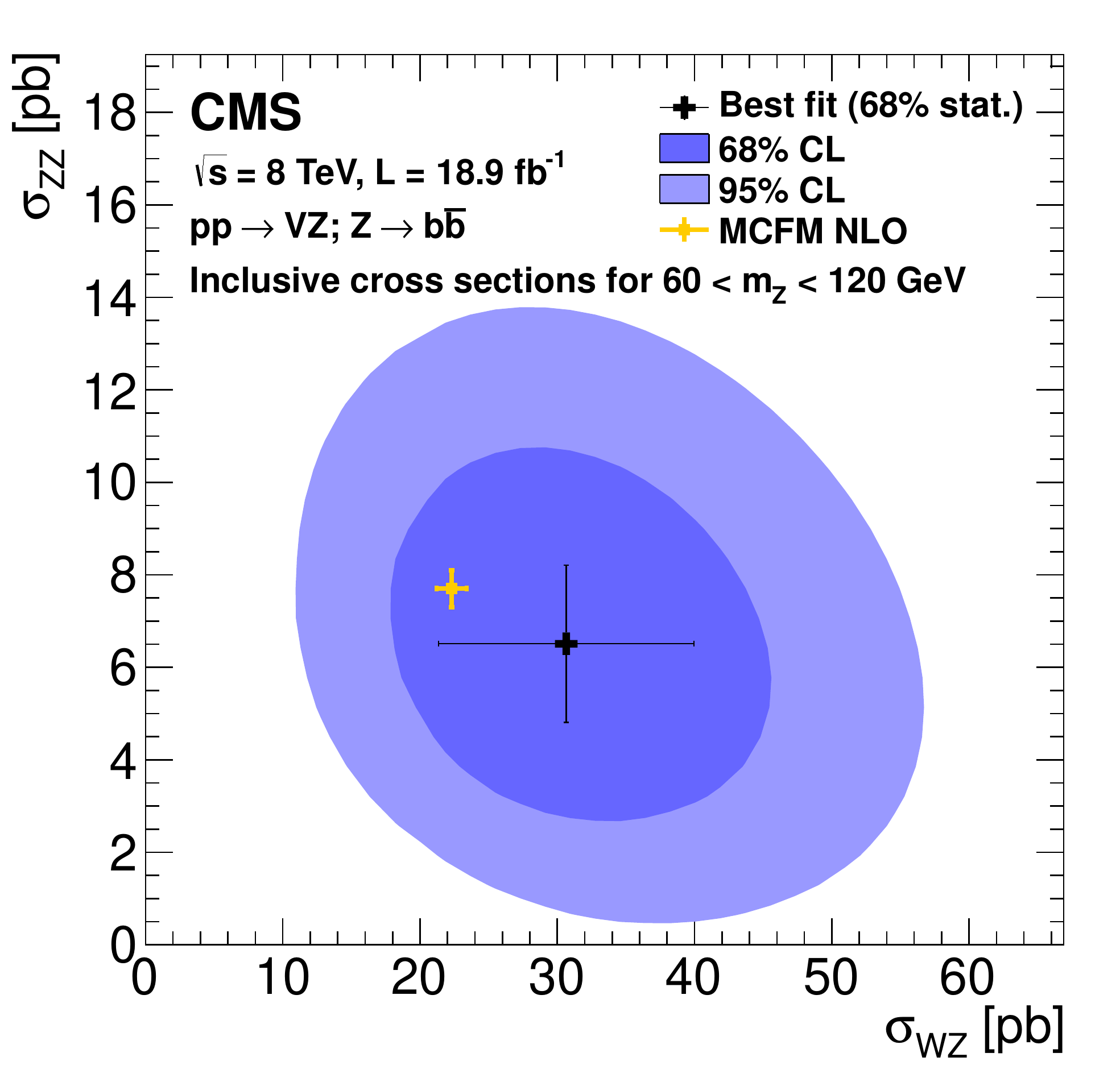}
  \caption{68\% and 95\% C.L. cross-section contours for $WZ$ and $ZZ$
    production as observed in $ZV\to b\bar{b} V$
    events~\cite{Chatrchyan:2014aqa} in comparison with NLO theory
    predictions.}
  \label{fig:ZV8TeV}
\end{figure}

\subsection{$W^\pm Z$ Production}
The production of $W^\pm Z$ boson pairs in the three lepton plus MET
final state where the $Z$ boson decays into an electron or muon pair
while the $W$ boson decays leptonically has been studied by
ATLAS~\cite{Aad:2016ett, Aad:2012twa, Aad:2011cx} and
CMS~\cite{Khachatryan:2016poo} at both 7 and 8~TeV using data
samples with integrated luminosities of up to 5 and 20~\ifb,
respectively.

The selected data sets are quite cleanly dominated by the signal
process. The measured $WZ$ cross sections are found to be consistent
with NLO SM predictions, and differential cross sections for a variety
of kinematic variables such as the transverse momentum of the $Z$ and
$W$ boson~\cite{Aad:2016ett} or leading jet \pt and jet
multiplicity~\cite{Khachatryan:2016poo} are provided.  The cross-section 
ratios of inclusive $W^+ Z$ and $W^- Z$ production are
measured as well by ATLAS and found to be in agreement with NLO theory
predictions. A first calculation of the SM cross section at
NNLO~\cite{Grazzini:2016swo} that became available only after the
ATLAS analyses were published significantly improves the agreement
between prediction and measurements as illustrated in
Figure~\ref{fig:WZxsec}.
\begin{figure}[htbp]
  \begin{tabular}{c}
\includegraphics[width=0.42\textwidth]{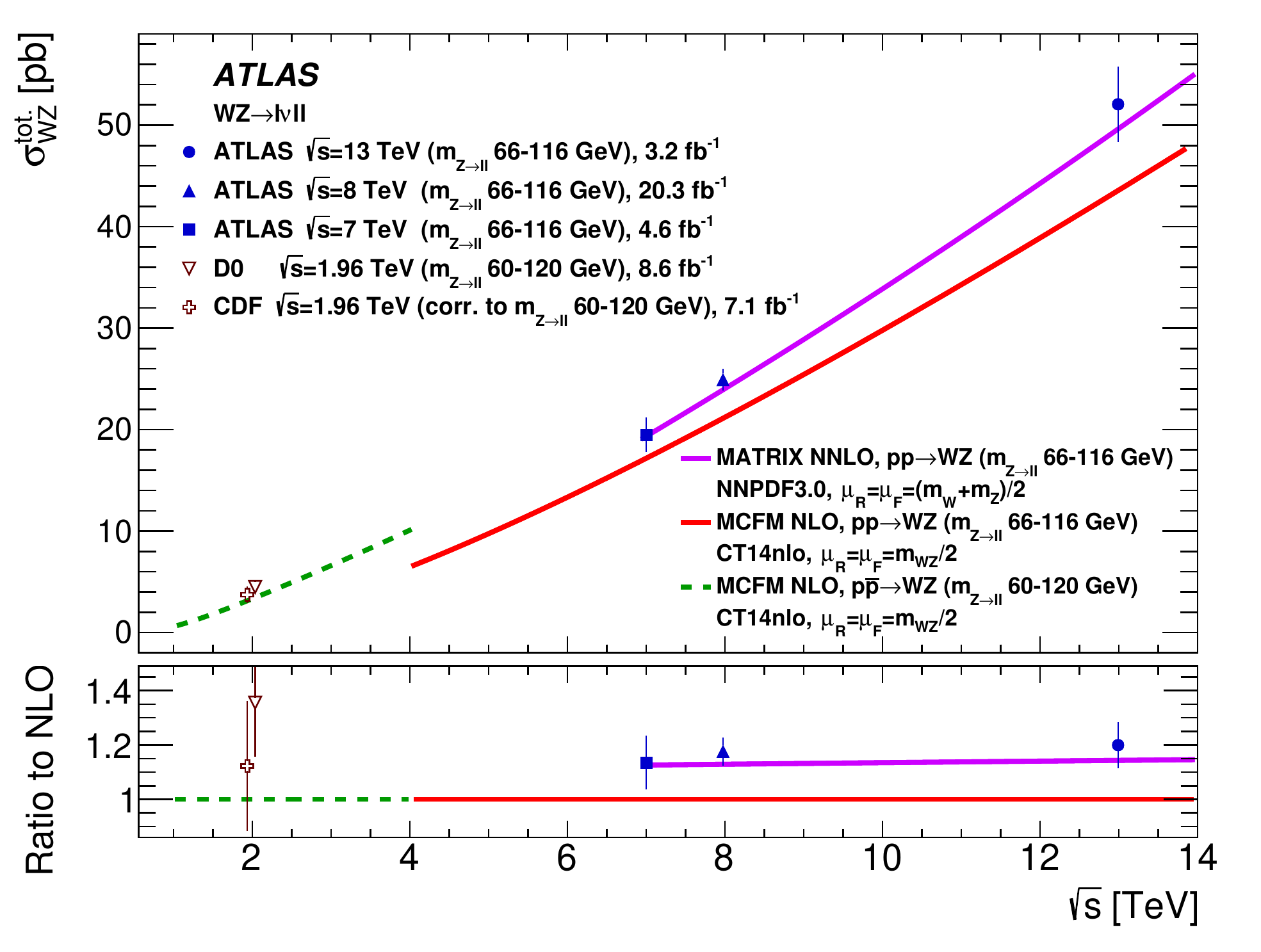}\\
    (a)\\
\includegraphics[width=0.4\textwidth]{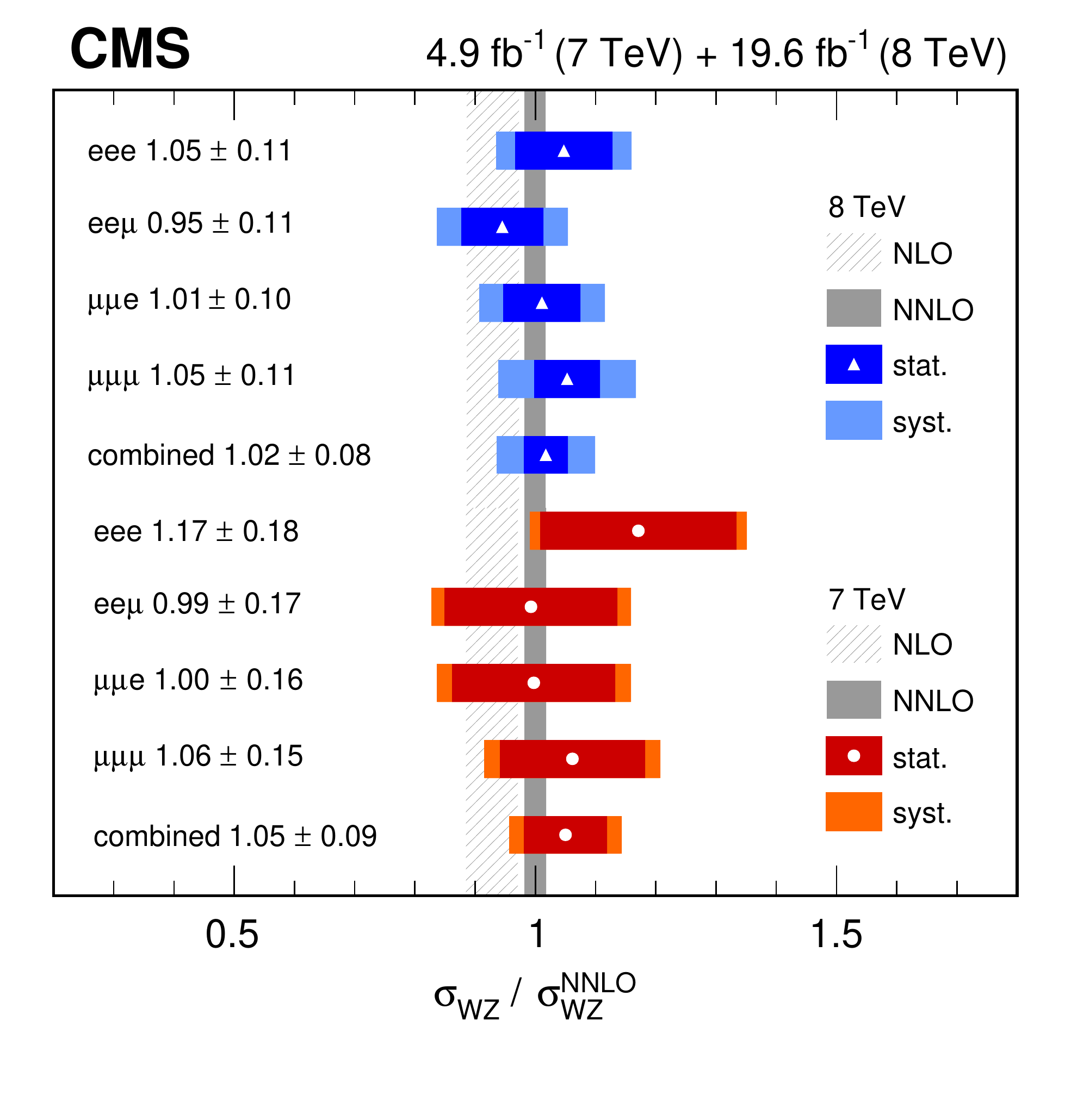}\\
    (b)\\
  \end{tabular}
  \caption{Comparison of $W^\pm Z$ production cross-section
    measurements with NLO and NNLO predictions (a) at various center
    of mass energies~\cite{Aaboud:2016yus} and (b) in the individual
    and combined channels at 7 and
    8~TeV~\cite{Khachatryan:2016poo}.}
  \label{fig:WZxsec}
\end{figure}

$WZ$ production includes only the TGC of $WWZ$ as opposed to $WW$
production which has both $WWZ$ and $WW\gamma$ SM vertices. The
variables chosen to search for aTGC are the \pt of the $Z$ boson and
the transverse mass of the $W^\pm Z$ system, shown in
Figure~\ref{fig:WZ}. As the observed spectra agree with the SM
prediction, stringent limits on aTGC contributions are derived.
\begin{figure}[htbp]
  \begin{tabular}{c}
\includegraphics[width=0.4\textwidth]{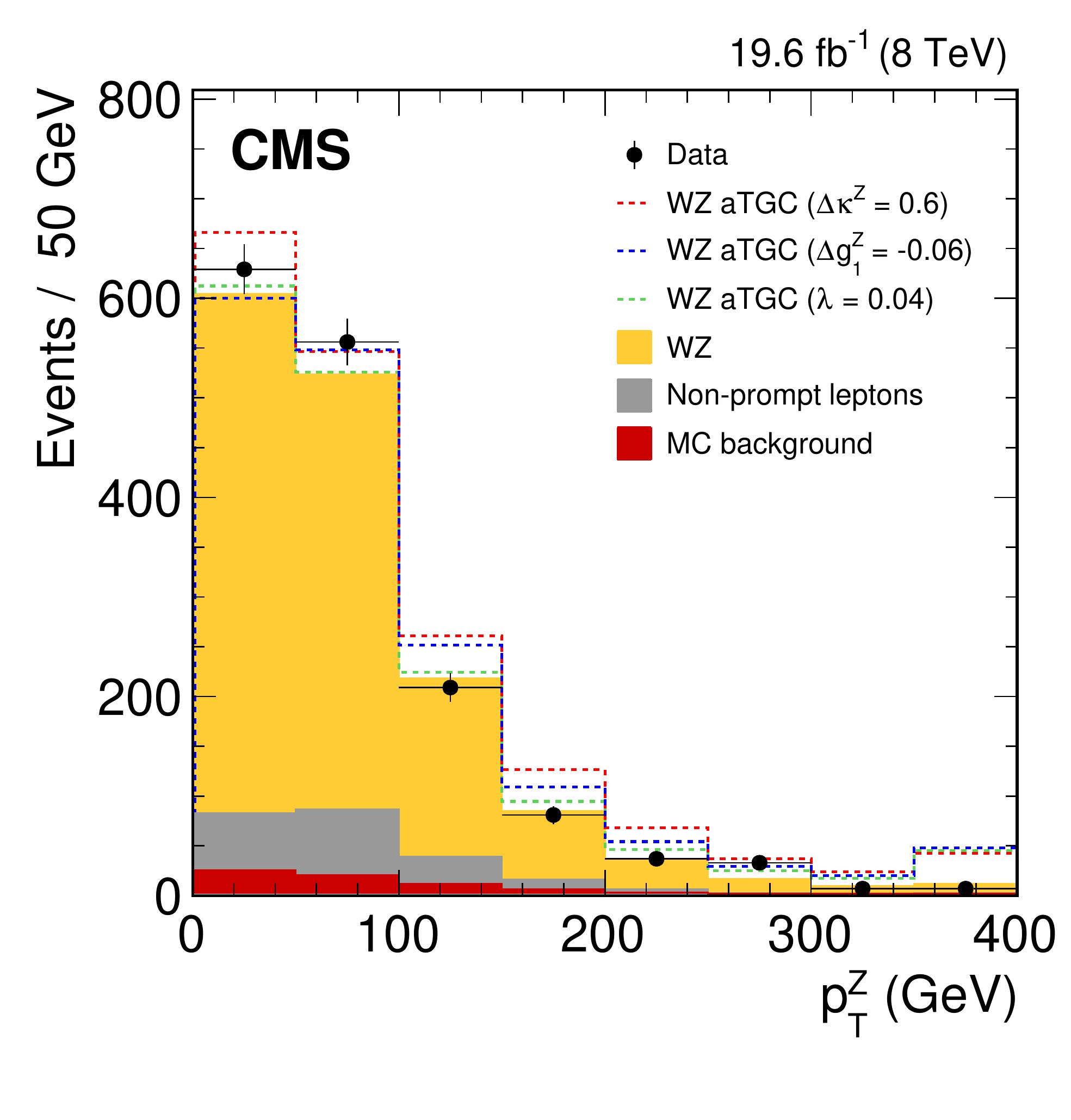}\\
    (a)\\
\includegraphics[width=0.4\textwidth]{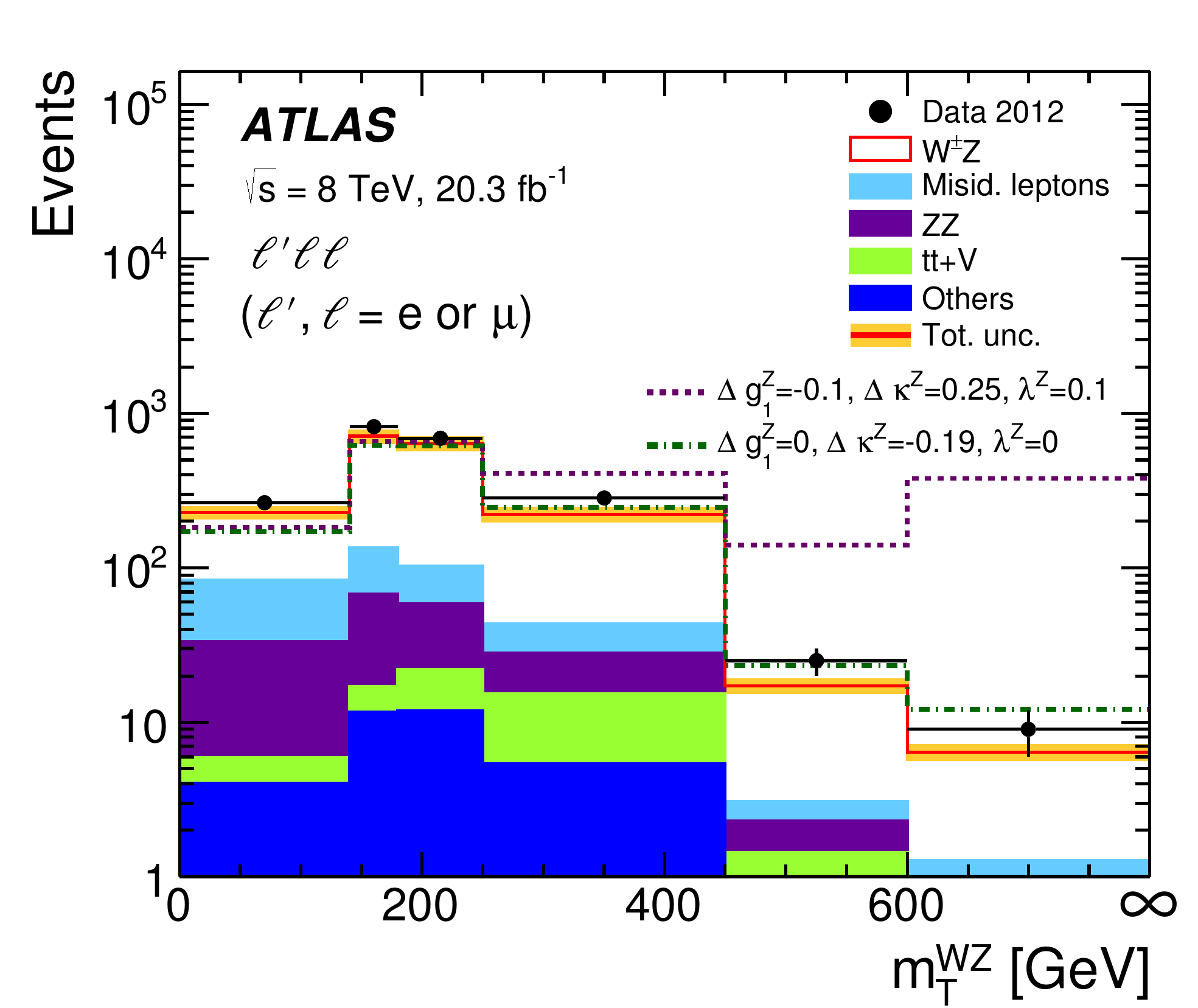}\\
    (b)\\
  \end{tabular}
  \caption{Observed (a) \pt distribution of the $Z$
    boson~\cite{Khachatryan:2016poo} and (b) transverse mass spectrum
    of the $W^\pm Z$ system~\cite{Aad:2016ett} in $W^\pm Z$ candidate
    events at 8~TeV and the impact of different anomalous TGC
    predictions.}
  \label{fig:WZ}
\end{figure}

\subsection{$ZZ$ Production}
\label{sec:ZZ}
Pairs of $Z$ bosons cannot be created at a single vertex in the SM because there
is no SM TGC available; only $WWZ$ and $WW\gamma$ exist in the SM. The
$HZZ$ vertex is here not considered to be a TGC vertex. Anomalous
$ZZ\gamma$ and $ZZZ$ couplings can be added with an effective
Lagrangian approach and parametrized using two CP-violating ($f^V_4$ )
and two CP-conserving ($f^V_5$ ) parameters ($V = \gamma,Z$) in direct
analogy to the $Z\gamma$ case, where there is also no SM TGC.

The production of $ZZ$ boson pairs has been studied in two decay
modes. In the ``$4\ell$'' mode, both $Z$ bosons decay into
same-flavor, oppositely-charged lepton pairs, resulting in a very
low-background, kinematically fully reconstructable final state that
however suffers from low statistics due to the branching
fractions involved. In the ``$2\ell 2\nu$'' mode, one $Z$ boson decays into a
same-flavor, oppositely-charged lepton pair, while the other one
decays to neutrinos, giving rise to large missing transverse energy in
the final state. While this decay mode suffers from larger background
contributions and is not kinematically fully reconstructable, it
benefits from better signal statistics due to the increased branching fraction and
detector acceptance.

Both $ZZ$ decay modes have been studied by ATLAS
($4\ell$~\cite{Aaboud:2016urj, Aad:2012awa,
  Aad:2011xj}; $2\ell 2\nu$~\cite{Aaboud:2016urj, Aad:2012awa}) and CMS
($4\ell$~\cite{CMS:2014xja, Chatrchyan:2012sga};
$2\ell 2\nu$~\cite{Khachatryan:2015pba}) at both 7 and 8~TeV
using data samples with integrated luminosities of up to 5 and
20~\ifb, respectively. CMS includes the decay of one $Z$ boson into
$\tau$ leptons in the $4\ell$ decay mode. The measured $ZZ$ cross
sections are found to be consistent with NLO SM predictions, as
illustrated in Figures~\ref{fig:ZZ4l7TeV}--\ref{fig:ZZ2l2v8TeV}.
NNLO corrections~\cite{Grazzini:2015hta} increase the expected
fiducial cross sections by about 15\% with respect to NLO predictions.
\begin{figure}[htbp]
  \begin{tabular}{c}
\includegraphics[width=0.40\textwidth]{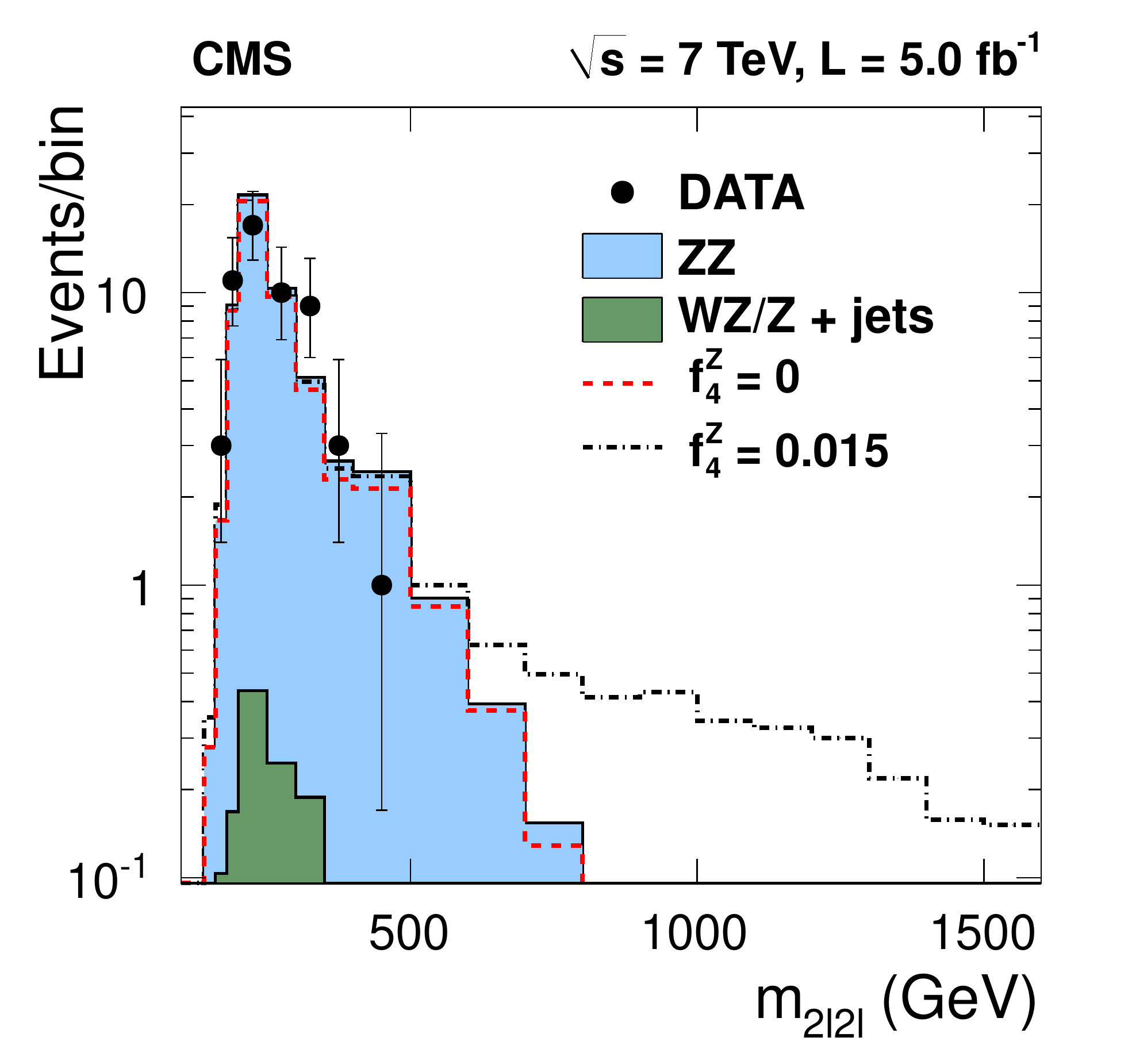}\\
    (a)\\
\includegraphics[width=0.40\textwidth]{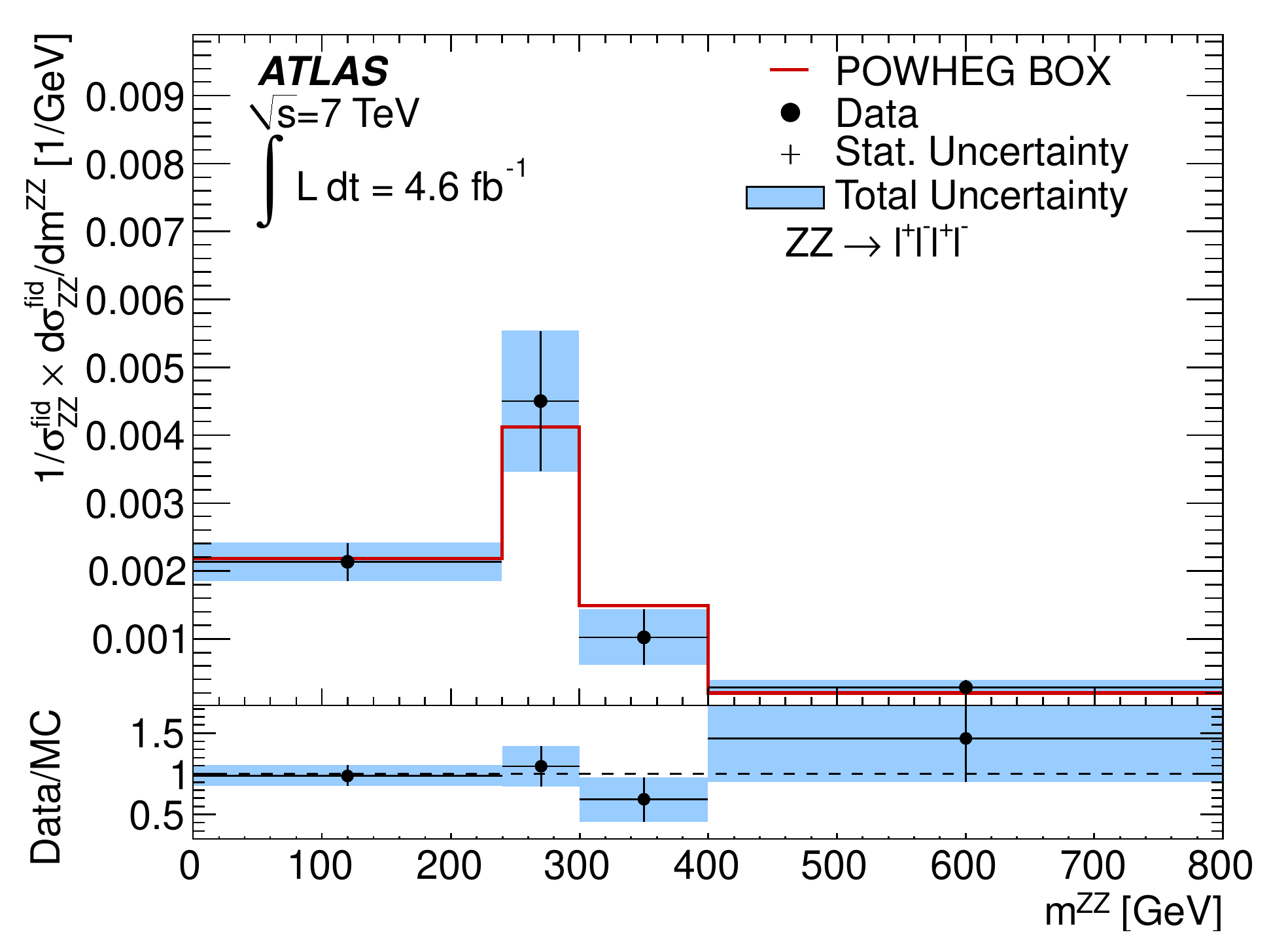}\\
    (b)\\
  \end{tabular}
  \caption{(a) $ZZ$ mass spectrum in the $4\ell$ decay channel
    ($\ell = e, \mu$) and the impact of anomalous
    TGCs~\cite{Chatrchyan:2012sga}. (b) Unfolded $ZZ$ fiducial cross
    sections in the $4\ell$ decay channel in bins of $ZZ$
    mass~\cite{Aad:2012awa}.}
  \label{fig:ZZ4l7TeV}
\end{figure}
\begin{figure}[htbp]
  \begin{tabular}{c}
\includegraphics[width=0.40\textwidth]{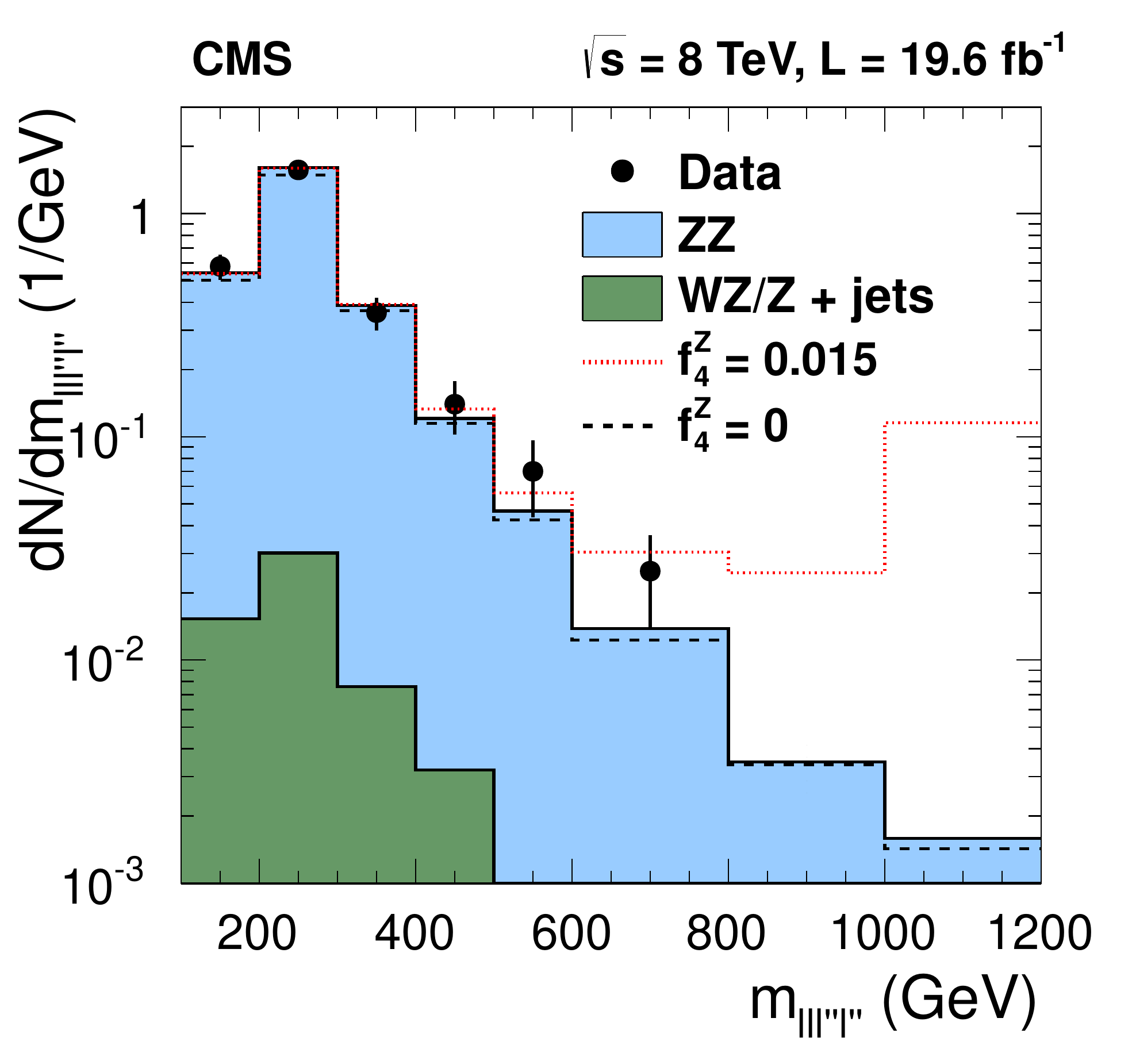}\\
    (a)\\
\includegraphics[width=0.40\textwidth]{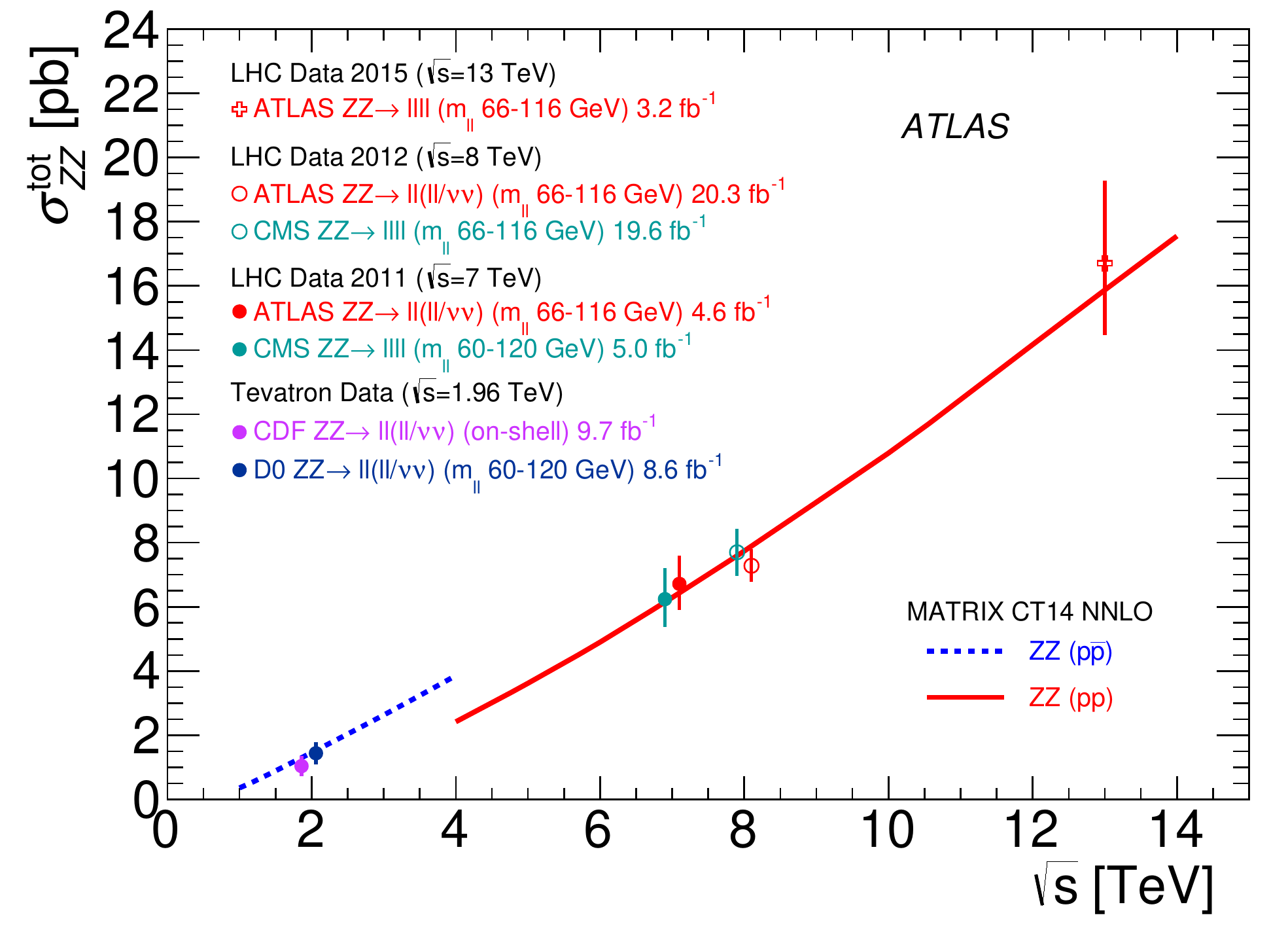}\\
    (b)\\
  \end{tabular}
  \caption{(a) $ZZ$ mass spectrum in the $4\ell$ decay channel
    ($\ell = e, \mu$) and the impact of anomalous
    TGCs~\cite{CMS:2014xja}. (b) Comparison of $ZZ$ production cross
    sections measured at hadron colliders and NNLO predictions as a
    function of center of mass energy~\cite{Aaboud:2016urj}.}
  \label{fig:ZZ4l8TeV}
\end{figure}
\begin{figure}[htbp]
  \begin{tabular}{c}
\includegraphics[width=0.40\textwidth]{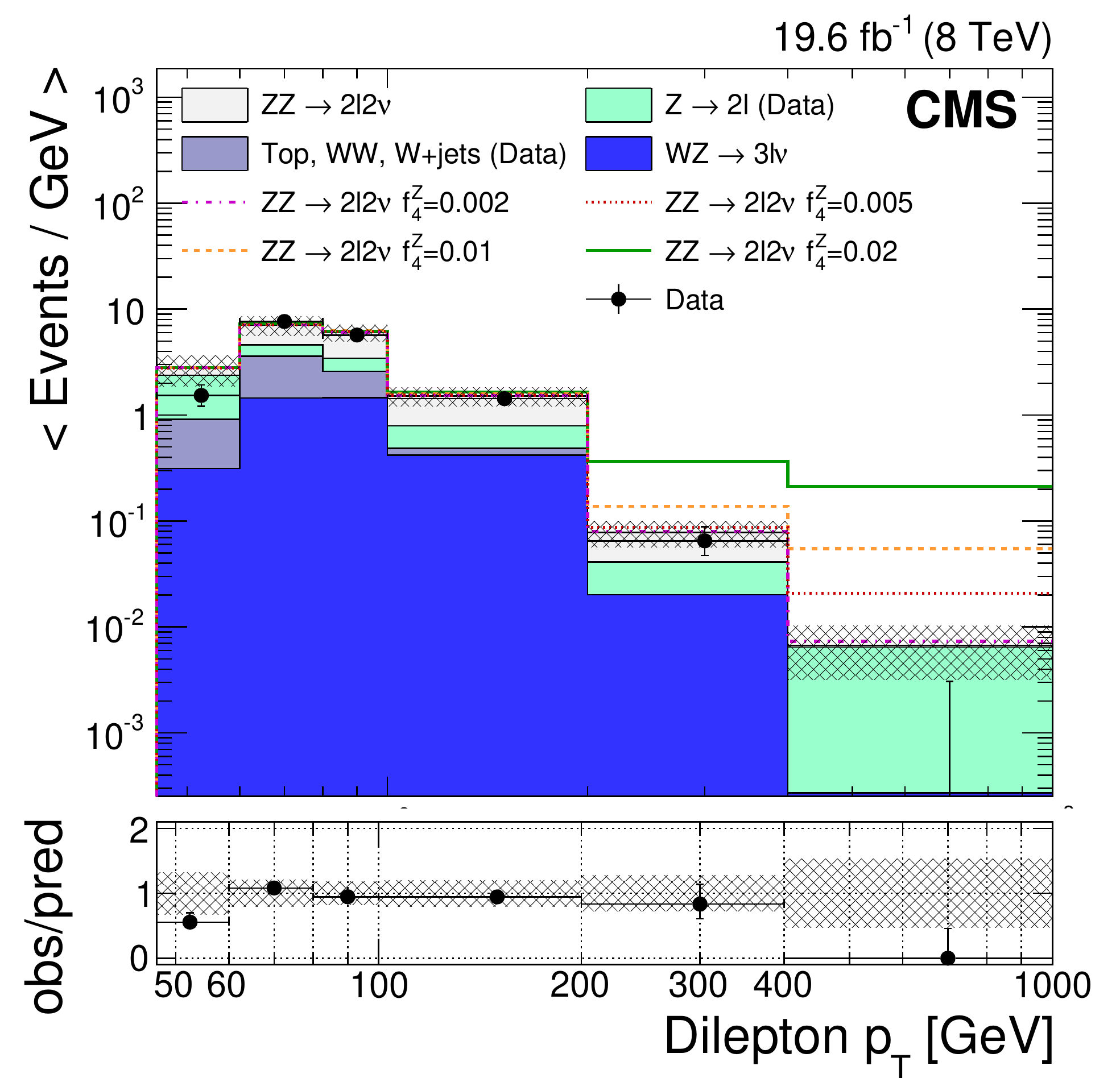}\\
    (a)\\
\includegraphics[width=0.40\textwidth]{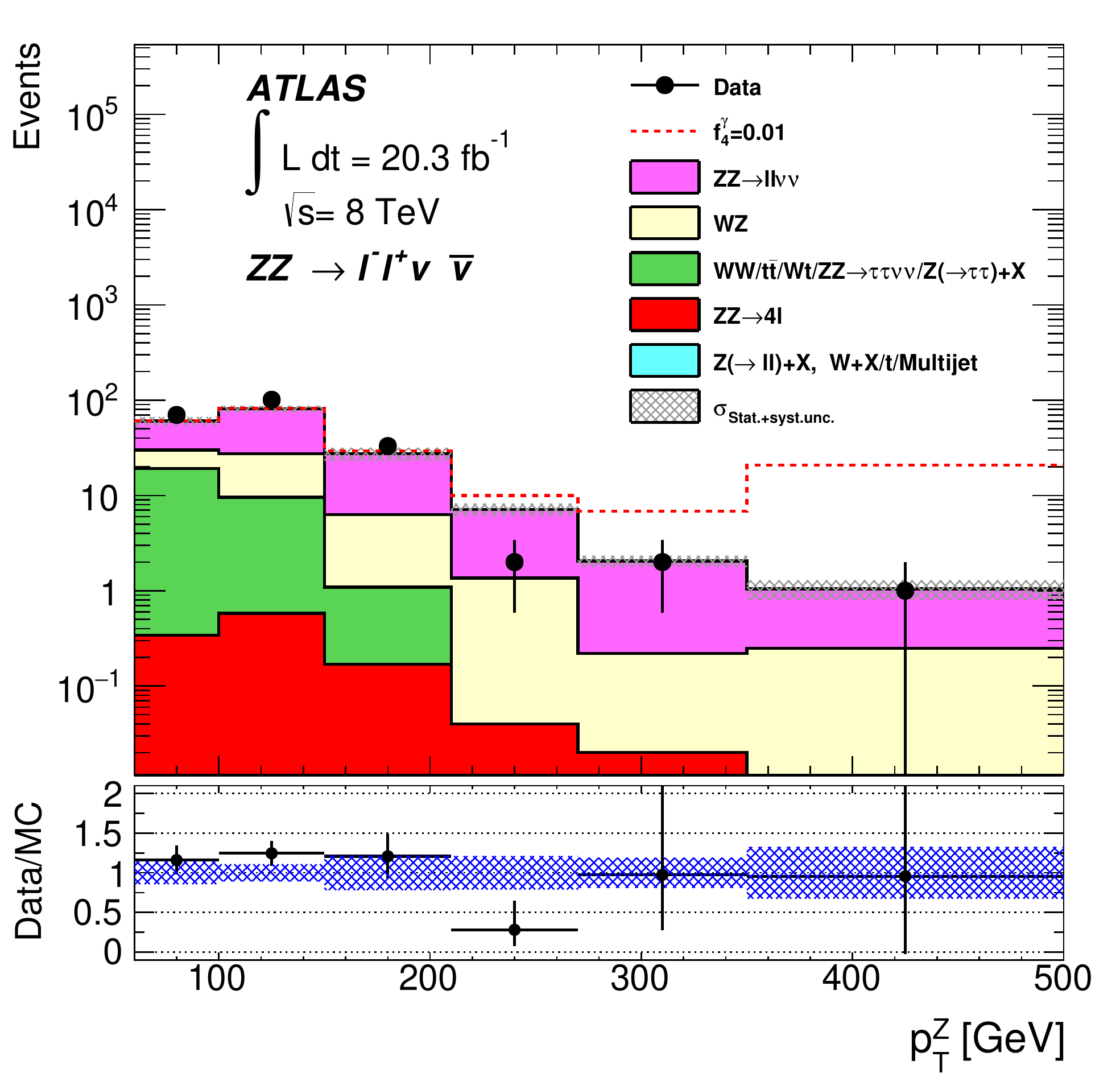}\\
    (b)\\
  \end{tabular}
  \caption{Dilepton ($Z$) \pt distributions in $ZZ$ candidate events
    at 8~TeV in the $2\ell 2\nu$ decay channel ($\ell = e, \mu$) and
    the impact of anomalous TGCs: (a)~\cite{Khachatryan:2015pba} and
    (b)~\cite{Aaboud:2016urj}.}
  \label{fig:ZZ2l2v8TeV}
\end{figure}

Figures~\ref{fig:ZZ4l7TeV} and~\ref{fig:ZZ4l8TeV} show that in the
$4\ell$ final state masses of the $ZZ$ pair up to about 0.5~TeV at
7~TeV and 0.8~TeV at 8~TeV are explored in a situation where the $ZZ$
signal dominates. The dilepton, or $Z$, \pt in the $2\ell 2\nu$ final
state at 8~TeV extends out to about 0.5~TeV as presented in
Figure~\ref{fig:ZZ2l2v8TeV}; however, here the $ZZ$ signal has large
backgrounds compared to the $4\ell$ final state.

Limits on aTGCs arise when the spectra shown are confronted with
models having deviations from the SM. As is customary, 95\% C.L. 
limits are derived for aTGCs as limits either in one
dimension or in two dimensions allowing two couplings to vary freely
from their SM values as will be shown later.

\section{Triboson Production}
\label{triboson}

The inclusive production of three gauge bosons has a much lower cross
section compared to that for the production of two gauge bosons.
Large aQGC are searched for in an EFT formulation with dimension-6
or -8 operators. The lowest dimension operators that {\em only} introduce aQGC are of dimension
eight.

\subsection{$W\gamma\gamma$ Production}

The largest inclusive triple gauge boson cross section is that for
$W\gamma\gamma$ production. The best signal-to-background ratio is
achieved when studying the leptonic $W$ boson decay modes into a
charged lepton ($e$ or $\mu$) and a neutrino (MET), leading to a final
state with one isolated lepton, MET, and two isolated photons.

ATLAS~\cite{Aad:2015uqa} has studied this final state in an 8~TeV data
sample with an integrated luminosity of 20~\ifb and observes first
evidence for the $W\gamma\gamma$ process at the level of $>3\sigma$,
with the production rate in agreement with theoretical NLO
predictions. ATLAS additionally provides exclusive cross sections
where additional jet activity has been vetoed.

\begin{figure}[htbp]
  \begin{tabular}{c}
\includegraphics[width=0.40\textwidth]{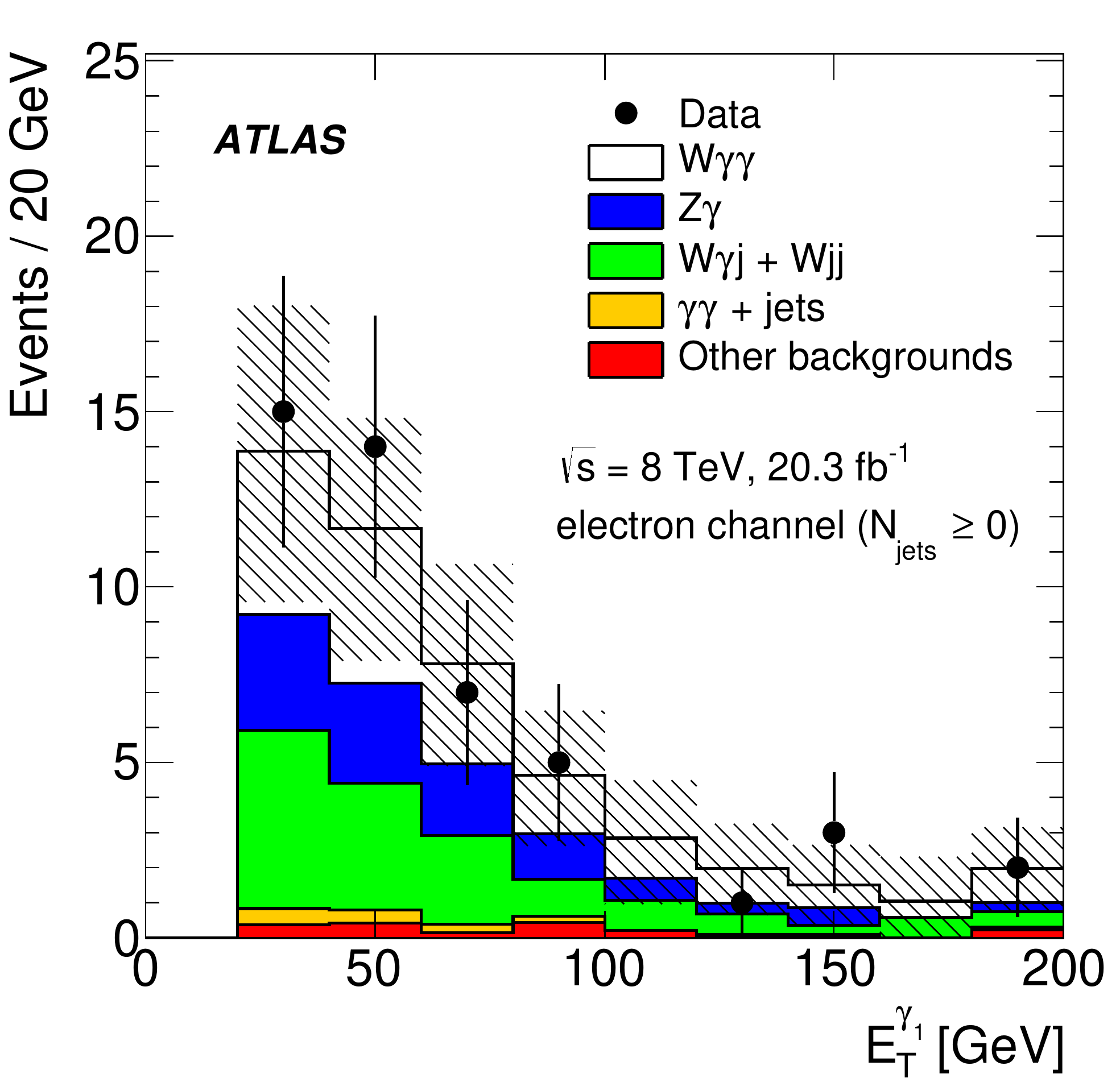}\\
    (a)\\
\includegraphics[width=0.40\textwidth]{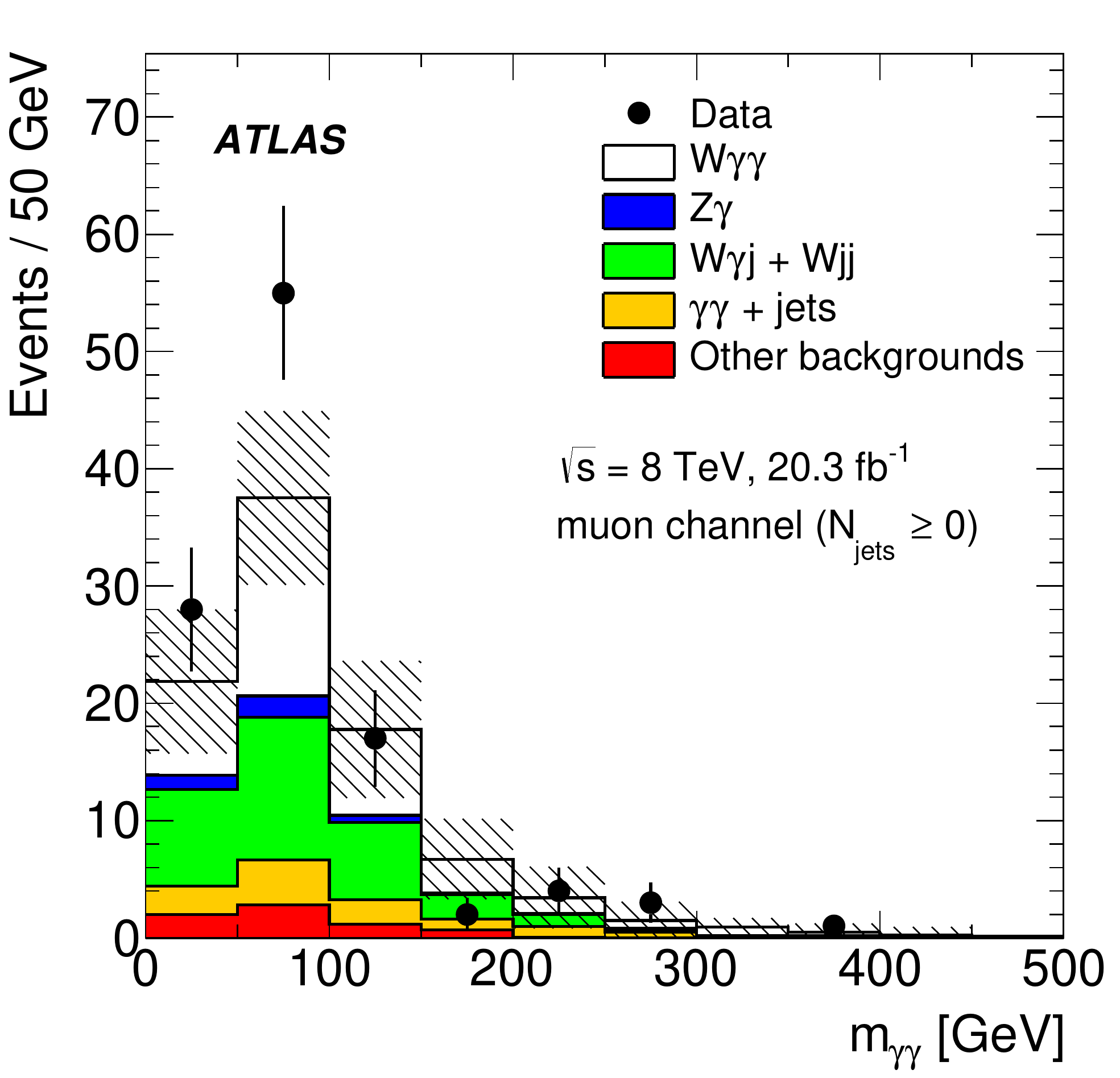}\\
    (b)\\
  \end{tabular}
  \caption{$W\gamma\gamma$ candidate events at 8~TeV in the
    $\ell\nu\gamma\gamma$ final state~\cite{Aad:2015uqa}: (a) \et
    spectrum of the leading photon in the electron channel and (b)
    diphoton invariant mass distribution in the muon channel.}
  \label{fig:Wgg8TeV}
\end{figure}

Figure~\ref{fig:Wgg8TeV} shows leading photon \et and diphoton
invariant mass distributions in $W\gamma\gamma$ candidate events which
extend out to about 0.2 and 0.4~TeV, respectively. There is no
evidence for a large non-SM contribution to the production process.
Limits on anomalous $WW\gamma\gamma$ couplings are placed using the
tail of the diphoton invariant mass distribution and vetoing
additional jet activity to constrain dimension-8 operators with
couplings $f_{T0}$, $f_{M2}$, and $f_{M3}$.

\subsection{$Z\gamma\gamma$ Production}
SM $Z\gamma\gamma$ triboson production arises from $Z$ boson
production with photons radiated off from initial state quarks or
radiative $Z$ boson decays to charged leptons as well as fragmentation
of final state quarks and gluons into photons and cannot occur in a
single vertex due to the lack of neutral $ZZ\gamma\gamma$ and
$Z\gamma\gamma\gamma$ QGCs in the SM. Such anomalous QGCs can be
introduced with EFT dimension-8 operators with couplings $f_{T0}$,
$f_{T5}$, $f_{T9}$, $f_{M2}$, and $f_{M3}$.

The production of $Z\gamma\gamma$ tribosons has been studied in two
decay modes, each of which requires two isolated photons in the final
state. In the ``$2\ell$'' mode, the $Z$ boson decays into a
same-flavor, oppositely-charged lepton (electron or muon) pair,
resulting in a low-background, kinematically fully reconstructable
final state. In the ``$2\nu$'' mode, the $Z$ boson decays into
neutrinos, giving rise to large missing transverse energy in the final
state. While this decay mode suffers from larger background
contributions and is not kinematically fully reconstructable, it
benefits from an increased branching fraction and detector acceptance 
in order to constrain anomalous QGCs.

ATLAS~\cite{Aad:2016sau} has studied the $2\ell$ and $2\nu$ decay
modes in an 8~TeV data sample with an integrated luminosity of 20~\ifb
and provides the first cross-section measurement for $Z\gamma\gamma$
production with $>5\sigma$ significance. The observed production rate
is found to be consistent with theoretical NLO predictions. ATLAS also
provides exclusive cross sections where additional jet activity was
vetoed.

\begin{figure}[htbp]
  \begin{tabular}{c}
\includegraphics[width=0.40\textwidth]{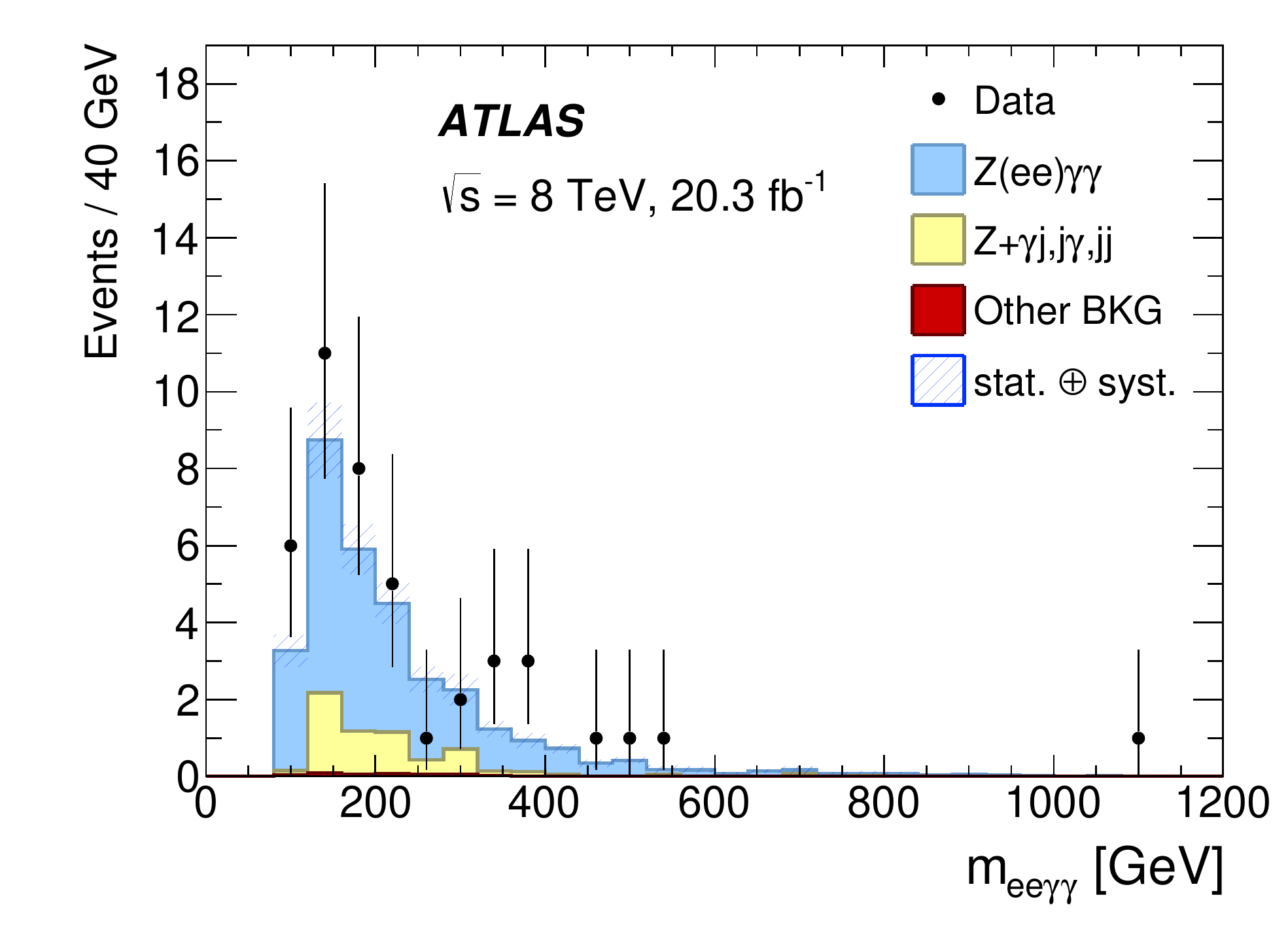}\\
    (a)\\
\includegraphics[width=0.40\textwidth]{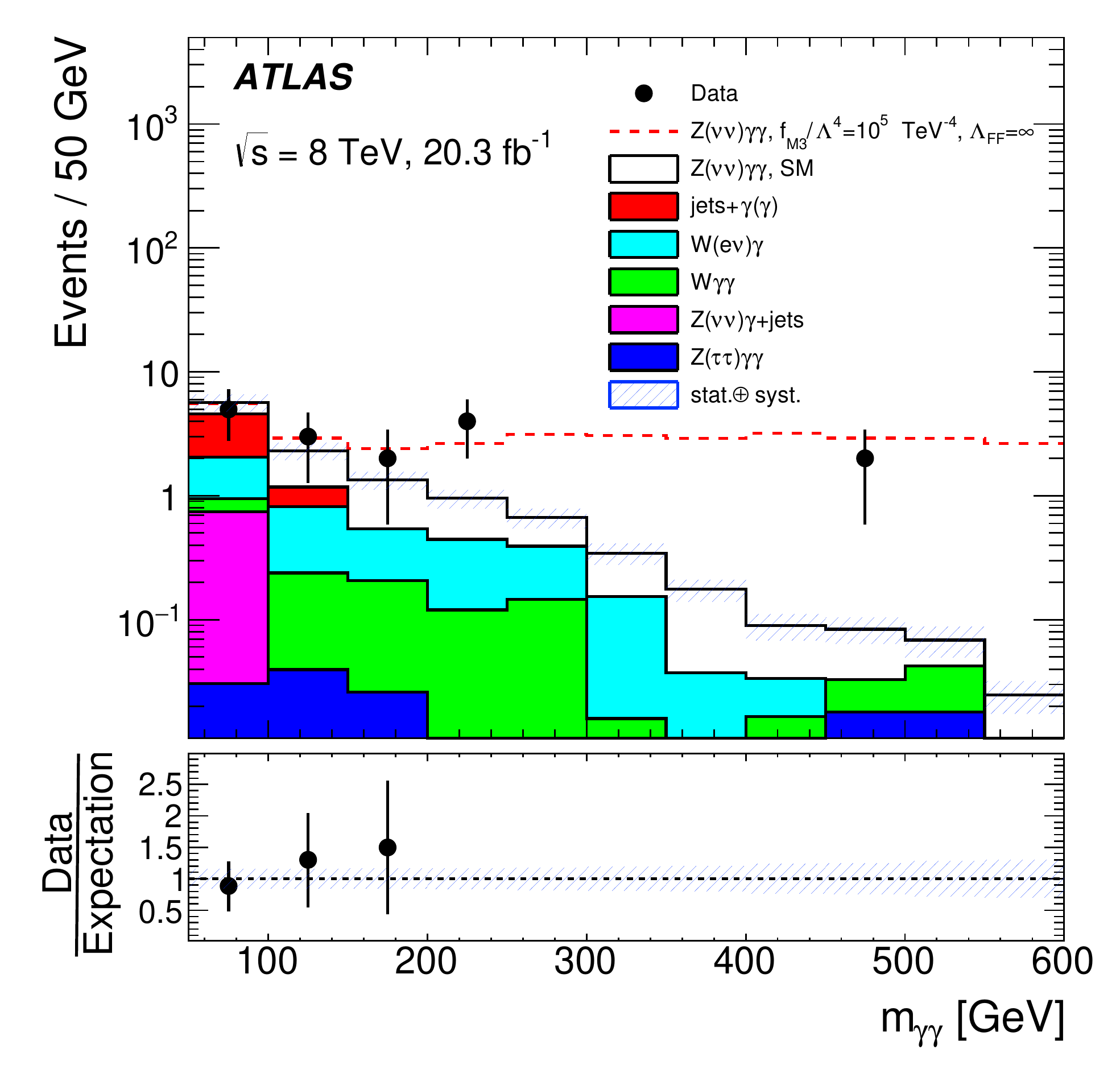}\\
    (b)\\
  \end{tabular}
  \caption{$Z\gamma\gamma$ candidate events at 8~TeV~\cite{Aad:2016sau}. 
    (a) Spectrum of the four-body invariant mass $m_{ee \gamma\gamma}$ 
    in the electron channel of the $\ell\ell\gamma\gamma$ final state.
    (b) Diphoton invariant mass distribution in the exclusive 
    $\nu\nu\gamma\gamma$ final state and potential impact of aQGCs.}
  \label{fig:Zgg8TeV}
\end{figure}

Figure~\ref{fig:Zgg8TeV} shows the four-body $ee \gamma\gamma$ and
diphoton invariant mass distributions in $Z\gamma\gamma$ candidate
events which extend out to about 1.1 and 0.5~TeV,
respectively. With no evidence found for a large non-SM contribution
to the production process, ATLAS places limits on anomalous QGCs using
exclusive fiducial cross sections with high diphoton invariant mass
requirements in the $2\ell$ and $2\nu$ decay modes.

\subsection{$WV\gamma$ Production}

Semileptonic $WV\gamma$ decays ($V = W,Z$) with one charged lepton
(electron or muon), missing transverse energy, at least two jets and
an energetic photon in the final state represent an extension of the
study of $WV$ production described in Section~\ref{sec:WVprod}. While
the large hadronic branching fraction of the $W$ or $Z$ boson makes
this triboson production mode more accessible, $W$ and $Z$ bosons
cannot be fully distinguished since the dijet mass resolution is
comparable to their mass difference. However, the $WW\gamma$ mode
dominates because the $WZ\gamma$ cross section is smaller and the
dijet mass resolution provides some discrimination. The
expected SM QGC contributions to $WV\gamma$ production are $WWZ\gamma$
and $WW\gamma\gamma$.

The production of $WV\gamma$ has been searched for by
CMS~\cite{Chatrchyan:2014bza} at 8~TeV using a data sample with
integrated luminosity of 19~\ifb. The $W\gamma$
plus jet background dominates the signal. An upper limit on $WV\gamma$
production is placed based on the observed data yields corresponding
to about $3.4$ times the SM NLO QCD theoretical
expectation. Nevertheless useful limits can be placed on large
contributions of aQGCs using
the photon \pt spectrum as shown in
Figure~\ref{fig:WVgamma8TeV}.
\begin{figure}[htbp]
  \centering
\includegraphics[width=0.4\textwidth]{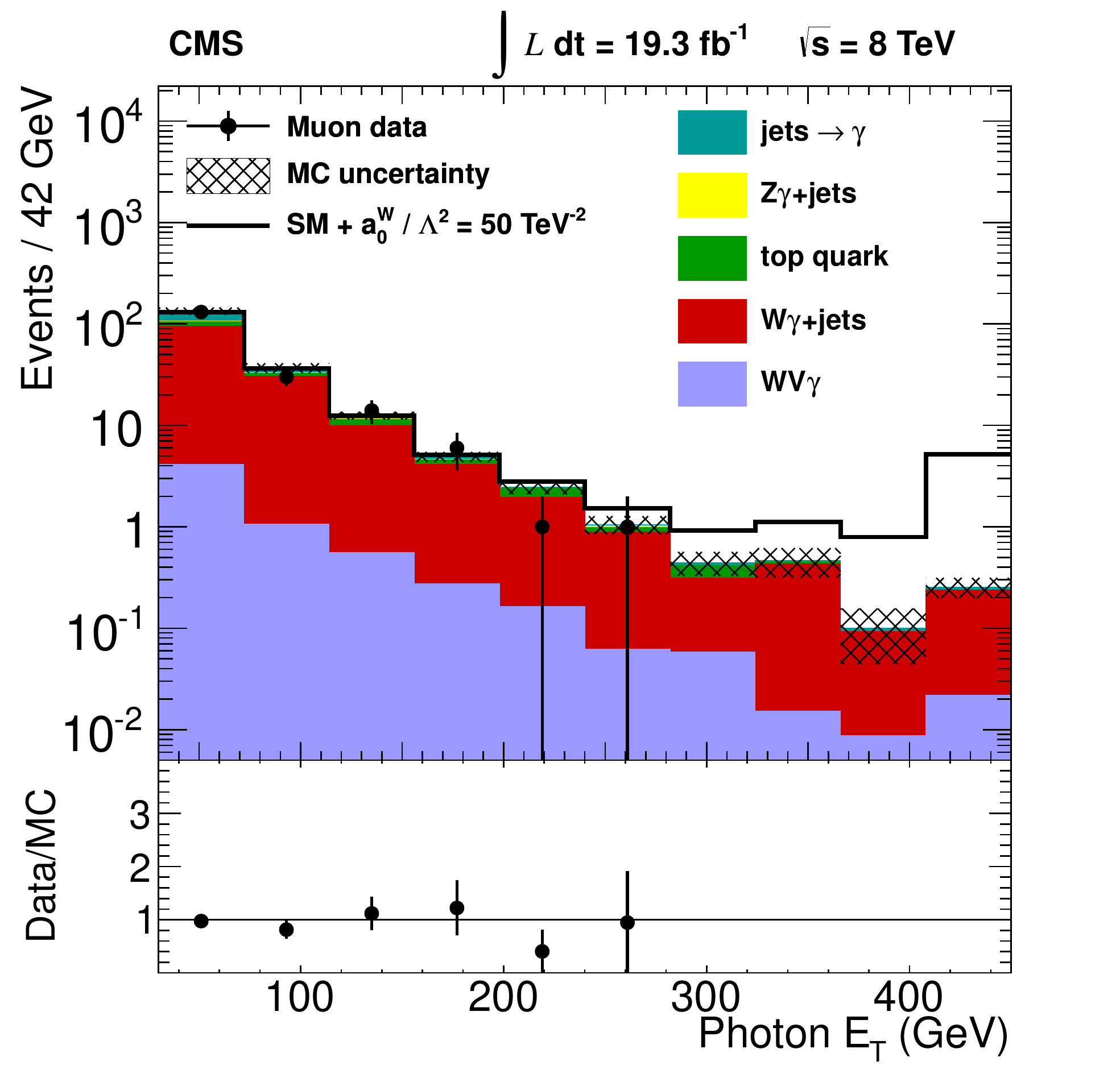}
  \caption{Photon \pt spectrum in $WV\gamma$ candidate events in the
    $\ell\nu jj\gamma$ final state at 8~TeV and potential impact of
    aQGCs~\cite{Chatrchyan:2014bza}.}
  \label{fig:WVgamma8TeV}
\end{figure}
Constraints are provided on the dimension-8 operator with coupling
$f_{T0}$ and alternatively on the dimension-6 operators with couplings $a_0^W$,
$a_C^W$ for $WW\gamma\gamma$ and $\kappa_0^W$, $\kappa_C^W$ for
$WWZ\gamma$ vertices, respectively.

\subsection{$W^\pm W^\pm W^\mp$ Production}

The production of $W^\pm W^\pm W^\mp$ constitutes the largest
inclusive triple gauge boson cross section with three massive bosons
and includes contributions from TGCs, Higgs production, and the SM
$WWWW$ QGC. The possible decay modes include the very clean fully
leptonic final state $\ell^\pm\nu\ell^\pm\nu\ell^\mp\nu$ exhibiting
three charged leptons ($e$ or $\mu$) and MET as well as a semileptonic
final state $\ell^\pm\nu\ell^\pm\nu jj$ with two leptons of the same
sign ($e$ or $\mu$), MET and two jets that -- while suffering from
larger background contributions -- benefits from a larger branching
fraction.

ATLAS has studied both of these signatures at 8~TeV using a data
sample with an integrated luminosity of 20~\ifb~\cite{Aaboud:2016ftt}.
To optimize signal sensitivity, the selection criteria are adjusted
according to the number of Same Flavor Opposite Sign (SFOS) lepton
pairs present in the leptonic final state and according to the same-sign 
lepton flavor combination in the semileptonic final state. The
latter is a ``spin-off'' from the $W^\pm W^\pm jj$ analysis described
in Section~\ref{sec:ssWWprod}, where the dijet invariant mass and
rapidity separation cuts have been modified to select $W$ boson decays
instead.

The data are described well by the signal and background model for
both final states as illustrated in Figure~\ref{fig:WWW8TeV} and the
combined signal significance is $\approx 1\sigma$. Given the current
statistical limitation to establish the signal cross section, upper
limits on $W^\pm W^\pm W^\mp$ production are placed based on the
observed data yields in good agreement with predictions from theory.

\begin{figure}[htbp]
  \begin{tabular}{c}
\includegraphics[width=0.40\textwidth]{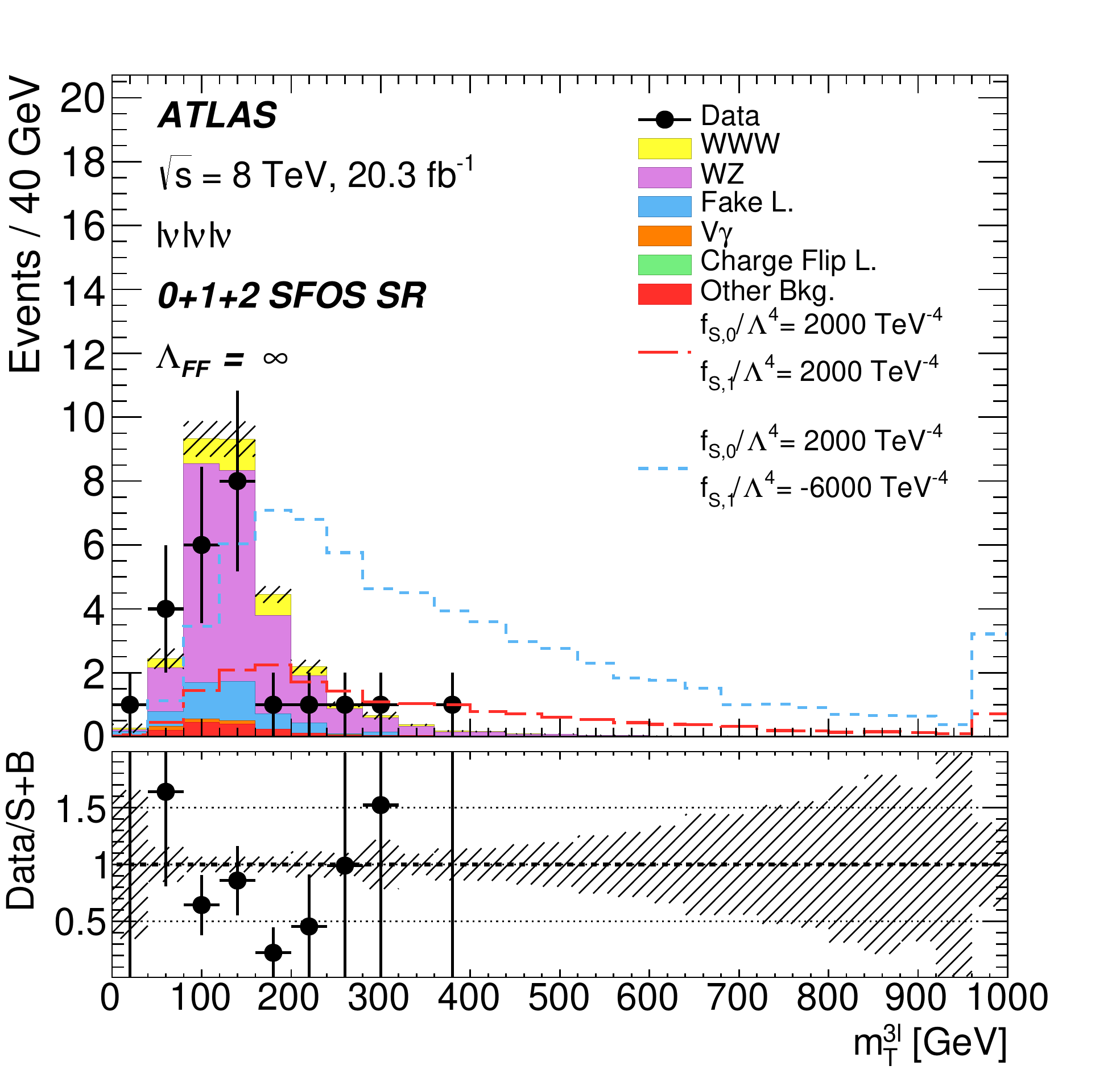}\\
    (a)\\
\includegraphics[width=0.40\textwidth]{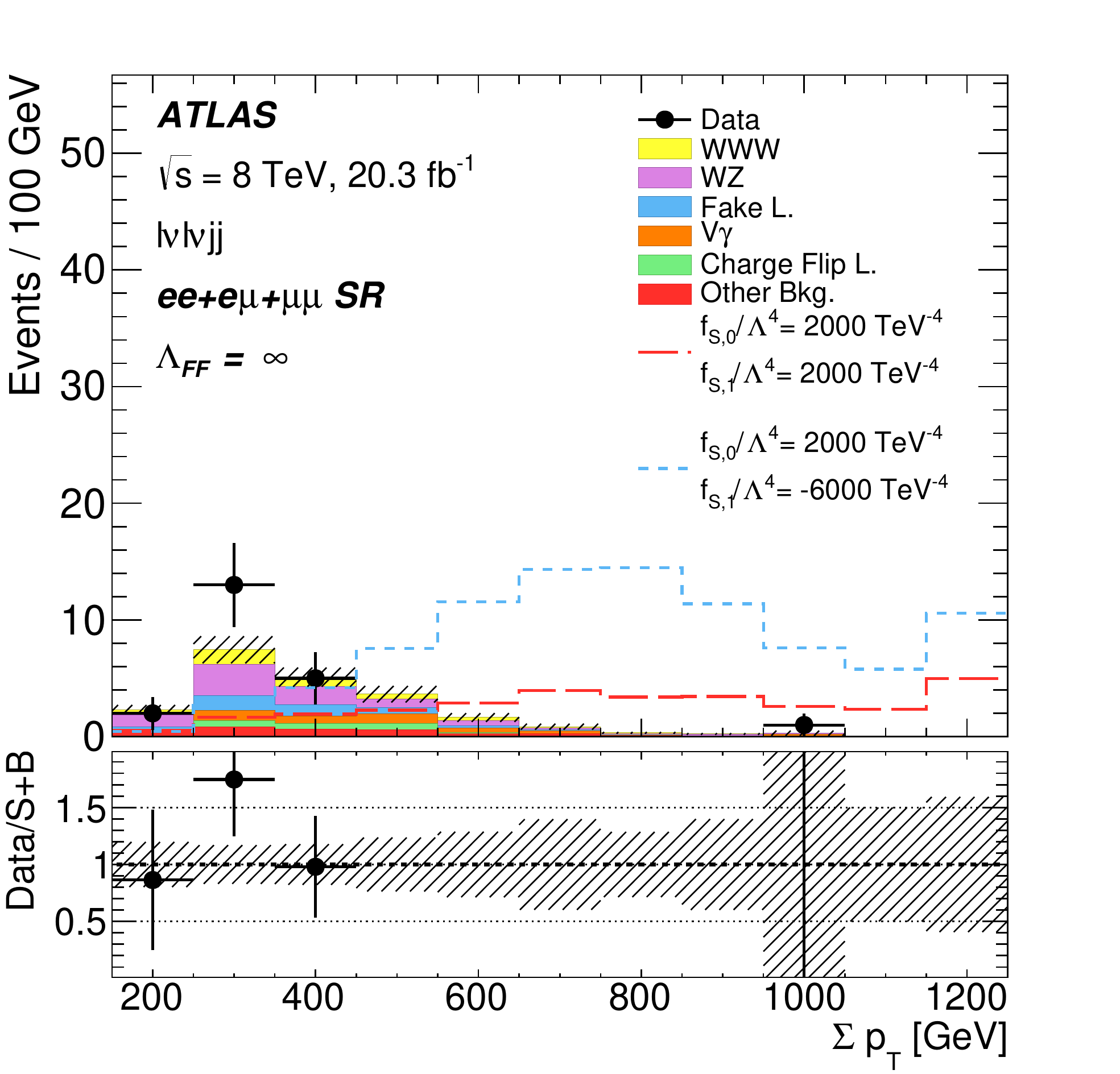}\\
    (b)\\
  \end{tabular}
  \caption{$W^\pm W^\pm W^\mp$ candidate events at
    8~TeV~\cite{Aaboud:2016ftt}: (a) spectrum of the trilepton
    transverse mass in the $\ell^\pm\nu\ell^\pm\nu\ell^\mp\nu$ final
    state and (b) sum of scalar \pt for all selected objects (leptons,
    jets, MET) distribution in the $\ell^\pm\nu\ell^\pm\nu jj$ final
    state. The potential impact of aQGCs is shown as well.}
  \label{fig:WWW8TeV}
\end{figure}

Possible aQGC contributions are constrained using the spectrum of the
trilepton transverse mass in the $\ell^\pm\nu\ell^\pm\nu\ell^\mp\nu$
final state and the sum of scalar \pt for all selected objects
(leptons, jets, MET) in the $\ell^\pm\nu\ell^\pm\nu jj$ final state,
where data extend to 1~TeV. %
Dimension-8 operators with couplings $f_{S0,1}$ are probed.

\section{Vector Boson Fusion}
\label{VBF}

Vector Boson Fusion ($VV\to V$) is an exclusive process wherein a
constituent of each proton emits a boson which then both fuse together
to form a single boson. The proton emission leads to remnant jets near
to the initial beam directions. That topology is exploited in
attempting to isolate the specific process. The emitted virtual vector
boson can be a photon, $W$ boson or $Z$ boson.

Typically the rapidity difference of the forward/backward ``tag''
jets is large as is the dijet mass. These facts are used to enhance
the VBF process. Nevertheless, the final states are also available to
other processes whose amplitudes interfere with the VBF
process, making a completely clean separation impossible, even at a
conceptual level.

The study of VBF events also constrains aTGC contributions
in a way complementary to diboson production, since in the VBF process
the two bosons radiated by the protons exhibit space-like
four-momentum transfer and not time-like four-momentum as is the case
in diboson production~\cite{Baur:1993fv}.
The sensitivity of such limits can be competitive with that from
diboson production~\cite{Eboli:2004gc}.

\subsection{$Wjj$ Production}

The largest cross-section VBF process studied at the LHC is the
production of a $W$ boson in association with two jets. The leptonic
decay of the $W$ boson is used in the examination of the lepton
($e,\mu$) plus MET plus two jet final state.

CMS has studied this signature at 8~TeV using a data sample with
integrated luminosity of 19~\ifb~\cite{Khachatryan:2016qkk}.
\begin{figure}[htbp]
  \begin{tabular}{c}
\includegraphics[width=0.40\textwidth]{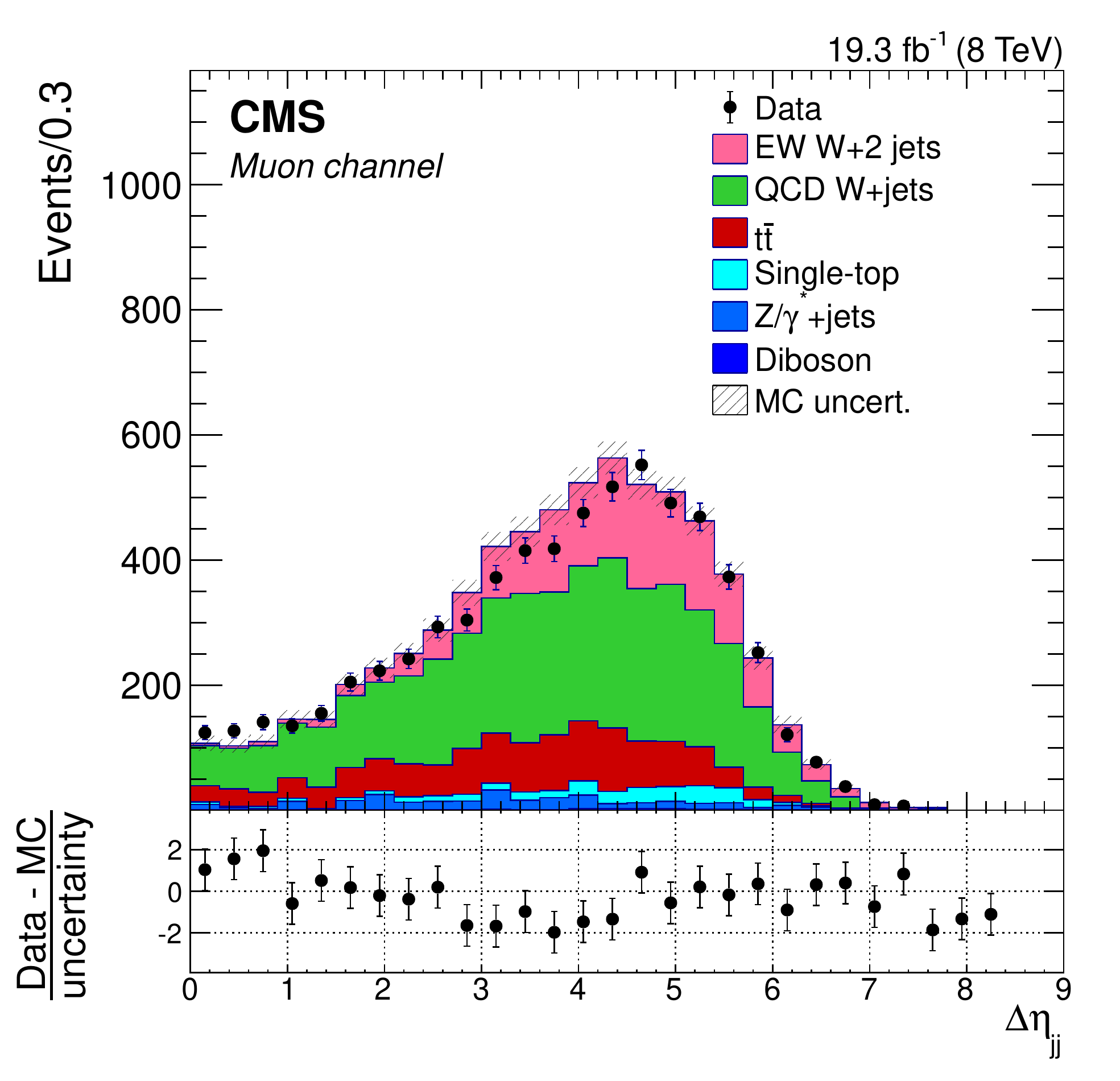}\\
    (a)\\
\includegraphics[width=0.40\textwidth]{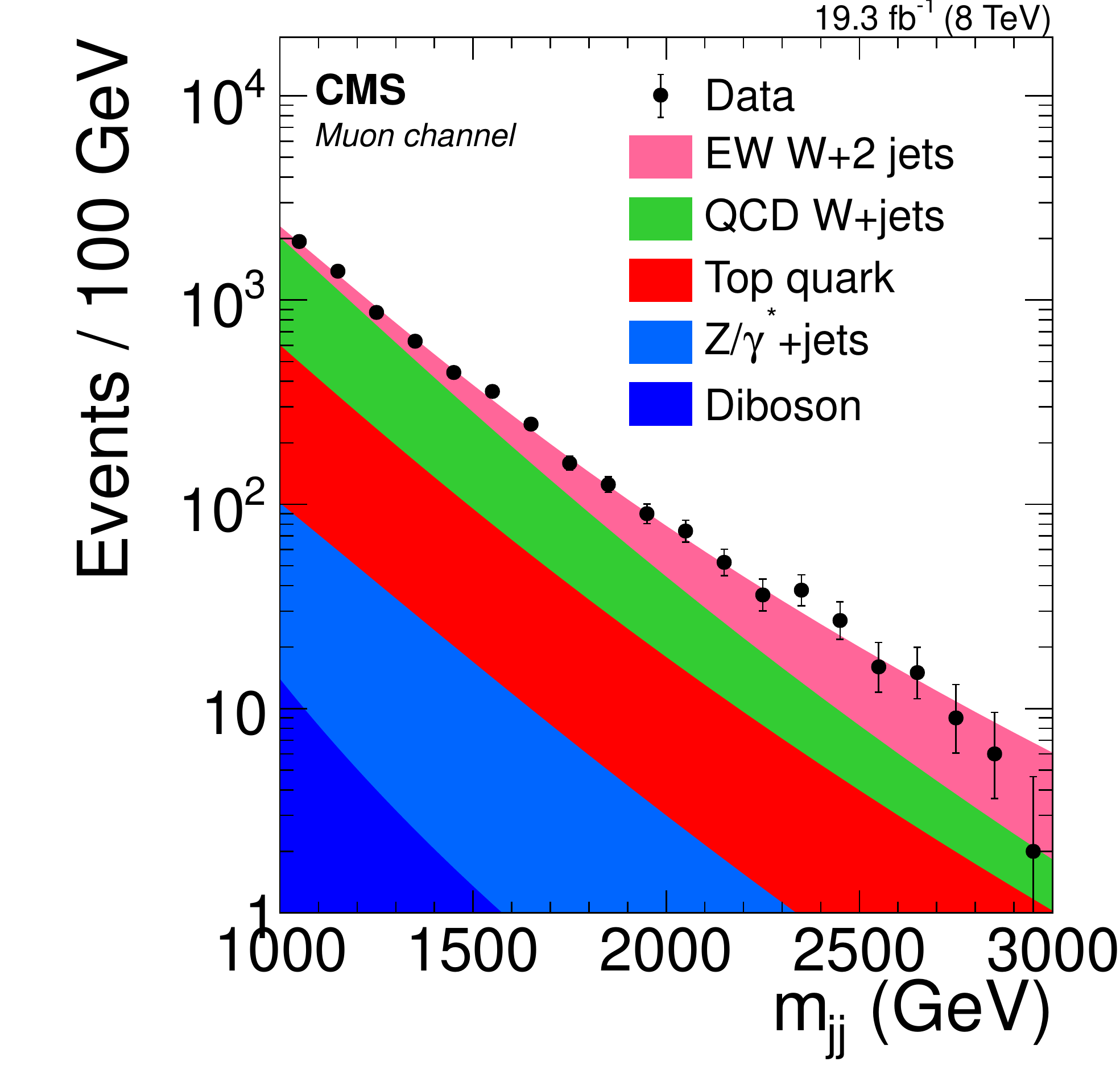}\\
    (b)\\
  \end{tabular}
  \caption{VBF-$W$ candidate events in the $\ell\nu jj$ final state at
    8~TeV~\cite{Khachatryan:2016qkk}. (a) Pseudorapidity difference of the tag
    jets. (b) Dijet mass spectrum of the tag jets.}
  \label{fig:VBFW8TeV}
\end{figure}
As seen in Figure~\ref{fig:VBFW8TeV} (a), the EW processes can be
large in carefully selected regions of phase space. Normalizing the
dominant background arising from $W$ boson plus jets production via
the strong interaction with a Boosted Decision Tree (BDT) technique, the
dijet mass tail above 1~TeV is examined as shown in
Figure~\ref{fig:VBFW8TeV} (b).
At large masses, greater than about 2~TeV, the EW processes dominate
the data sample. The largest background, QCD $W$ plus jets production,
falls with mass more rapidly than the EW signal. The fiducial
electroweak production cross section of a $W$ boson in association
with two jets is extracted and found to be consistent with the SM
prediction.

The SM $WW\gamma$ and $WWZ$ TGCs contribute to this process, but the aTGC 
limits are presently not competitive with the limits coming from inclusive $VV$ production. The VBF-$W$ production study 
shows that the EW process is well modeled and can be
enhanced in selected regions of phase space.

\subsection{$Zjj$ Production}

Electroweak production of a $Z$ boson in association with two jets
includes VBF $Z$ boson production via the $WWZ$ TGC and has been
studied in the final state with a same-flavor, oppositely-charged
lepton pair (electrons or muons) and two jets.

CMS has performed measurements~\cite{Khachatryan:2014dea, Chatrchyan:2013jya}
at both 7 and 8~TeV using data samples with integrated
luminosities of 5 and 20~\ifb, respectively. In the 8~TeV
analysis a BDT technique is used. As seen in
Figure~\ref{fig:VBFZ8TeVCMS} (a), a BDT variable selection can be used
to choose a region of phase space dominated by the EW process. 
The two major processes at high BDT values are EW and Drell-Yan
(DY). Since the fit is normalized, the two processes are
anti-correlated, as shown in Figure~\ref{fig:VBFZ8TeVCMS} (b). 
The magnitude of the EW cross section is found to be in agreement with
theoretical NLO QCD predictions.
\begin{figure}[htbp]
  \begin{tabular}{c}
\includegraphics[width=0.40\textwidth]{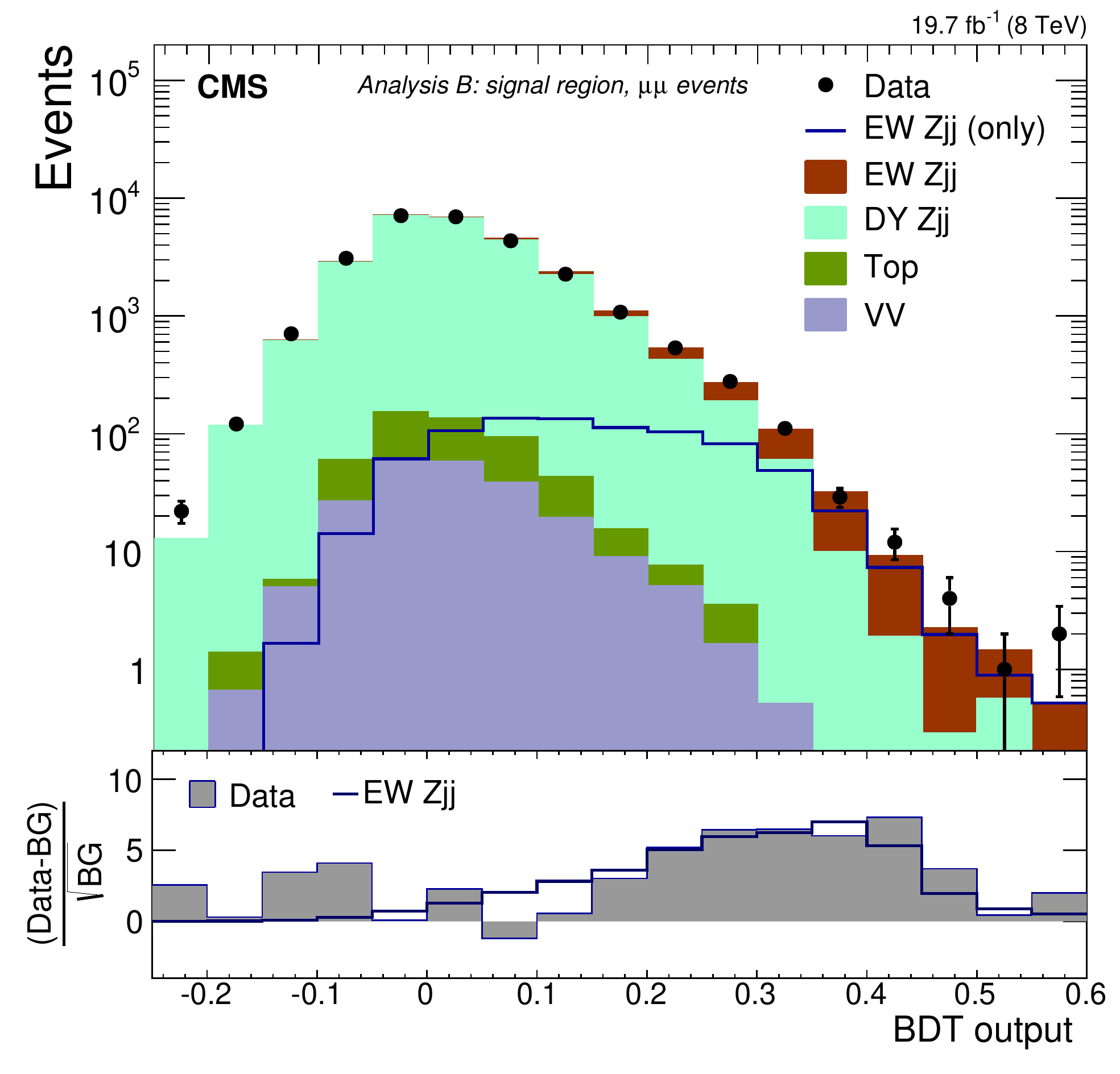}\\
    (a)\\
\includegraphics[width=0.40\textwidth]{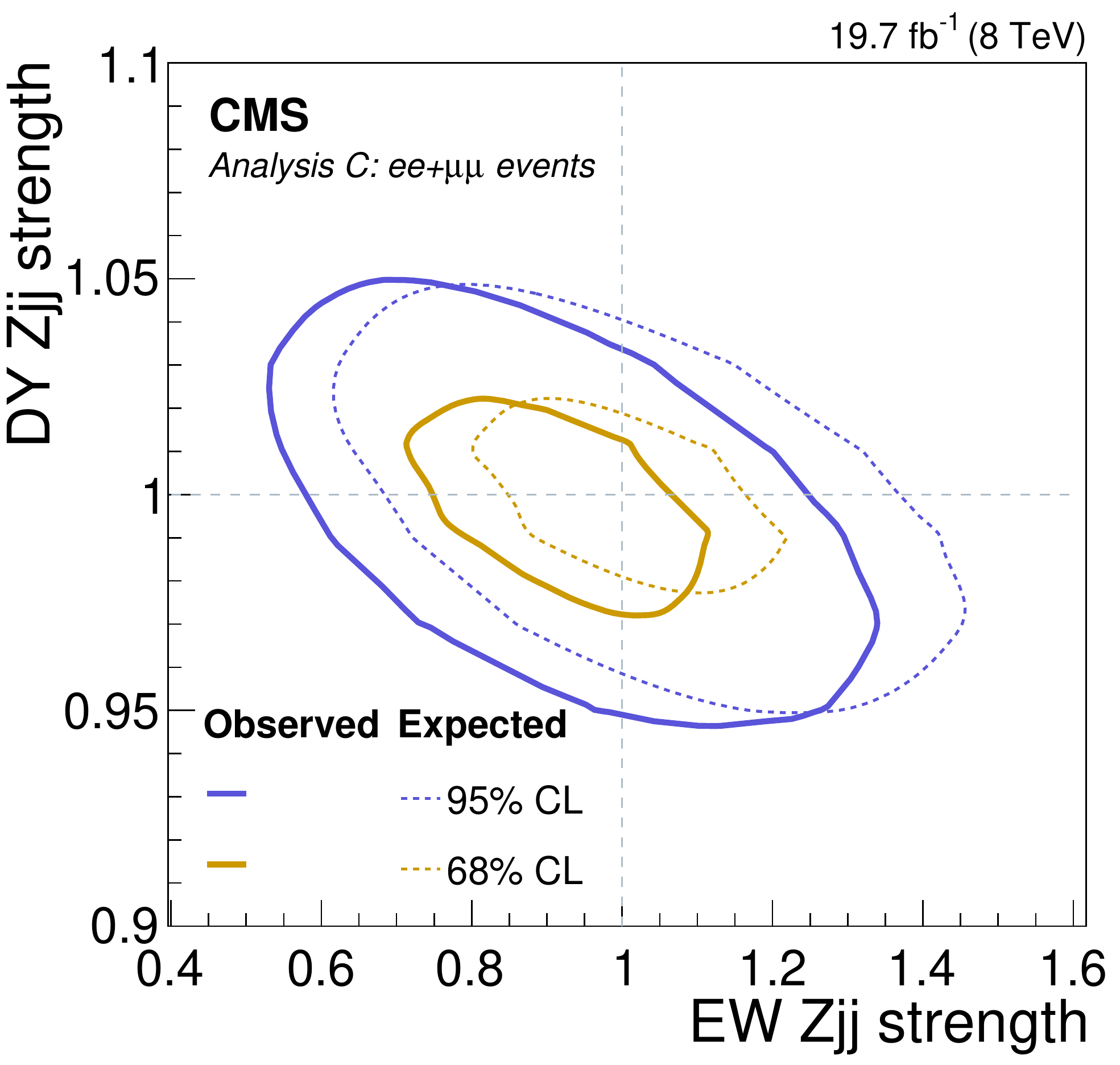}\\
    (b)\\
  \end{tabular}
  \caption{(a) BDT output and Monte Carlo expectations for VBF-$Z$
    candidate events in the $\ell\ell jj$ final state at
    8~TeV~\cite{Khachatryan:2014dea}. (b) Expected and observed 68\%
    and 95\% C.L. signal strength contours for EW and DY production of
    $\ell\ell jj$ at 8~TeV~\cite{Khachatryan:2014dea}.}
  \label{fig:VBFZ8TeVCMS}
\end{figure}

ATLAS has studied the $\ell\ell jj$ final state in 20~\ifb of 8~TeV
data~\cite{Aad:2014dta} and uses a fit of the dijet invariant mass
distribution with electroweak signal and QCD background templates to
extract the electroweak production cross section in a fiducial region
that enhances the signal contribution. The extracted signal is
established with more than $5\sigma$ significance and the production
rate is found to be in agreement with NLO SM predictions. In addition,
cross sections and differential distributions are measured in five
fiducial regions with different sensitivity to EW $Zjj$ production,
and limits on $WWZ$ aTGCs $\lambda_z$ and $\Delta g_1^Z$ are placed
based on the observed event yields in the tail of the dijet invariant
mass distribution, shown in Figure~\ref{fig:VBFZ8TeVATLAS}.
\begin{figure}[htbp]
  \begin{tabular}{c}
\includegraphics[width=0.40\textwidth]{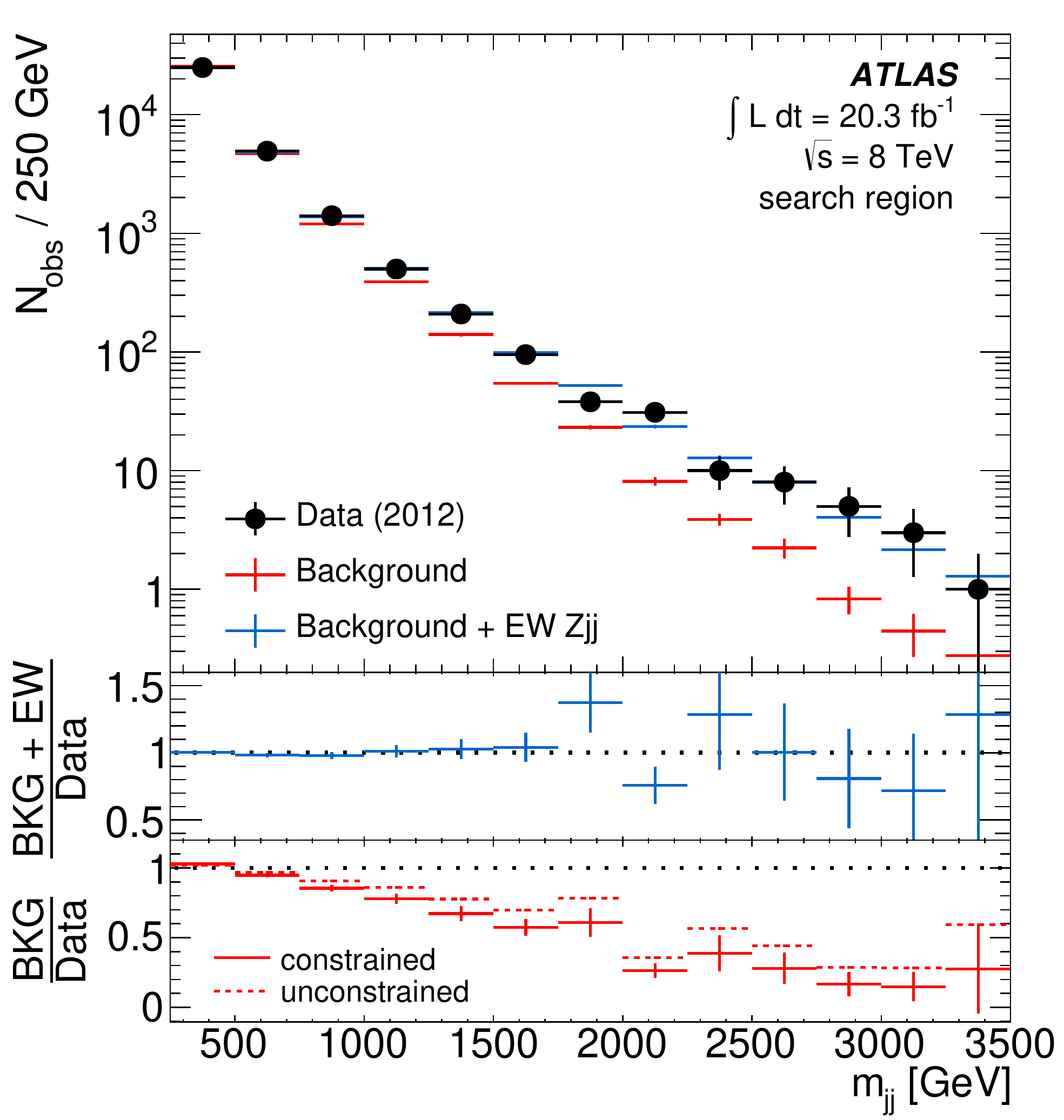}\\
    (a)\\
\includegraphics[width=0.40\textwidth]{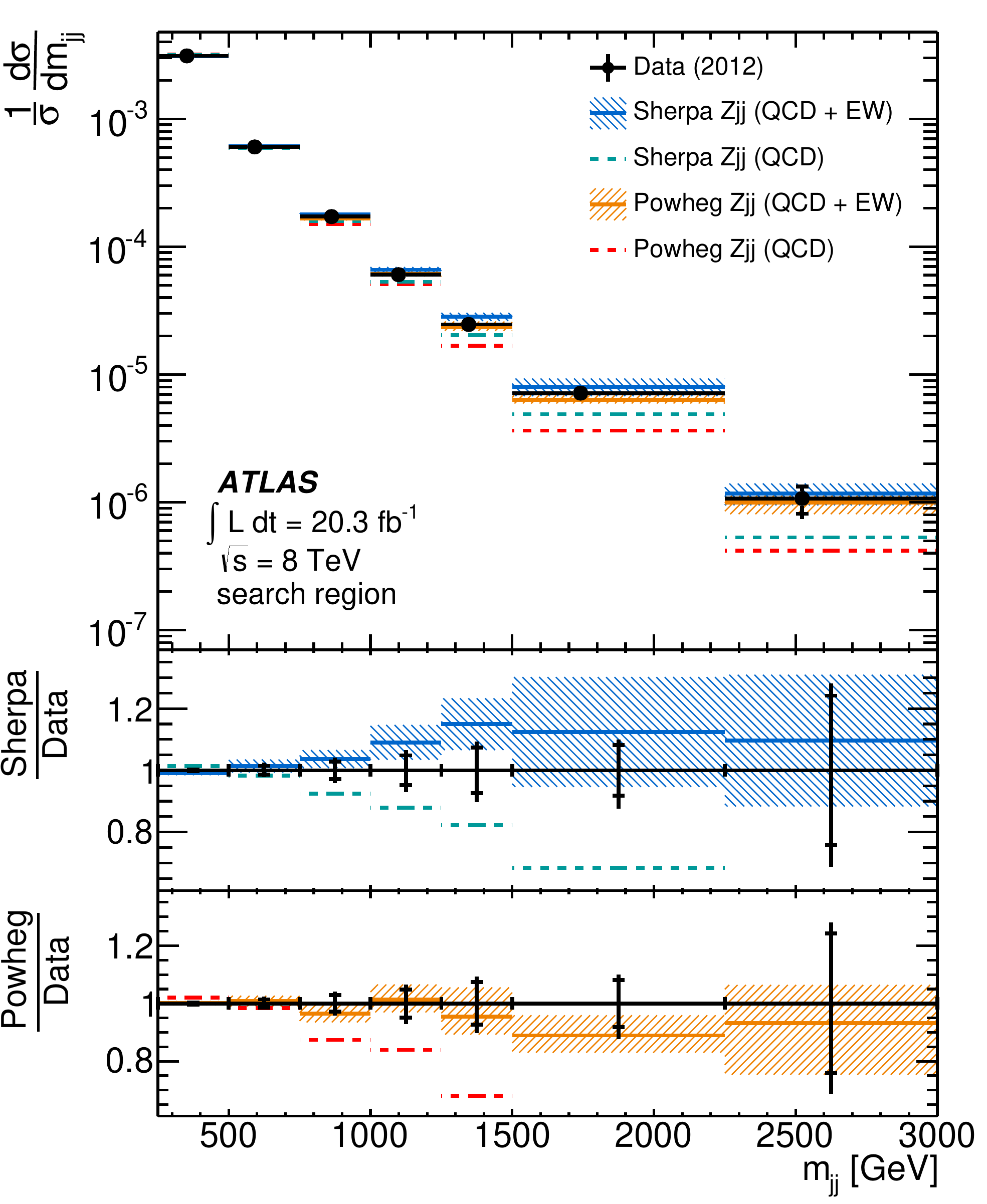}\\
    (b)\\
  \end{tabular}
  \caption{(a) Dijet invariant mass distribution of VBF-$Z$ candidate
    events in the $\ell\ell jj$ final state at
    8~TeV~\cite{Aad:2014dta}.  (b) Unfolded normalized differential
    $Z jj$ production cross section as a function of dijet invariant
    mass~\cite{Aad:2014dta}.}
  \label{fig:VBFZ8TeVATLAS}
\end{figure}

\section{Vector Boson Scattering}
\label{VBS}
Vector Boson Scattering ($VV\to VV$) is an exclusive process
wherein a constituent of each proton emits a boson which then interact
with each other causing the emission of two new bosons. As in the case
of VBF, the proton emission leads to remnant forward/backward or
tag jets near to the initial beam directions with large rapidity
difference and dijet mass. The resulting $VVjj$ final state
($V=\gamma,~W^\pm,~Z$) has contributions from both electroweak and QCD mediated
processes. The latter can be suppressed by requiring the
stated scattering topology. The electroweak processes include quartic
boson self-interactions, whose amplitudes interfere with those of
the other contributing diagrams, making a completely clean separation
impossible.

One main argument for expecting new particles and/or interactions at
the TeV scale is the linear divergence of the scattering amplitude for
longitudinally polarized weak bosons as the center of mass energy
squared increases~\cite{Lee:1977yc}, which leads to the violation of
unitarity at about 1~TeV. In the framework of the SM, this divergence
is canceled through diagrams involving the exchange of a Higgs
boson. Even if the recently discovered boson turns out to be the Higgs
boson, its role in VBS still needs to be experimentally established to
confirm the SM nature of EWSB. A wealth of models with dynamical EWSB
in lieu of or in addition to the Higgs mechanism exists, making the
measurement of VBS both a fundamental test of the SM and a window to
new physics.

\subsection{$W^\pm \gamma jj$ Production}

The largest cross-section VBS process studied at the LHC is the
production of a $W\gamma$ boson pair in association with two jets,
which includes SM QGC contributions from the $WW\gamma\gamma$ and
$WWZ\gamma$ vertices. Purely longitudinal scattering effects cannot be
studied in this channel due to the presence of the photon. 

CMS has performed a search for electroweak $W^\pm \gamma jj$
production using leptonic $W$ boson decays in final states with one
charged lepton (electron or muon), missing transverse energy, two jets
well-separated in rapidity and an energetic photon in 8~TeV data with
an integrated luminosity of 20~\ifb~\cite{Khachatryan:2016vif}. After
preliminary selections the dijet mass of the tag jets is shown in
Figure~\ref{fig:VBSWgamma}~(a). At masses greater than about 1~TeV the
electroweak signal process dominates.  An upper limit on electroweak
$W^\pm \gamma jj$ production is placed based on the observed data
yields in the tails of the dijet mass distribution, corresponding to
about $4.3$ times the SM NLO QCD theoretical expectation. The combined
electroweak and strong $W^\pm \gamma jj$ production is measured in
good agreement with theoretical expectations.

The search for aQGCs uses the shape of the \pt spectrum of the $W$
boson in events with a tightened selection, including the requirement
of a very energetic photon, as shown in Figure~\ref{fig:VBSWgamma}~(b).
The \pt values extend to about 0.25~TeV, and constrain dimension-8
operators with couplings $f_{M0..7}$ and $f_{T0..2,5..7}$. 
The notation for the subscripts indicates which operators are
considered, where dots indicate contiguous indices.
\begin{figure}[htbp]
  \begin{tabular}{c}
\includegraphics[width=0.40\textwidth]{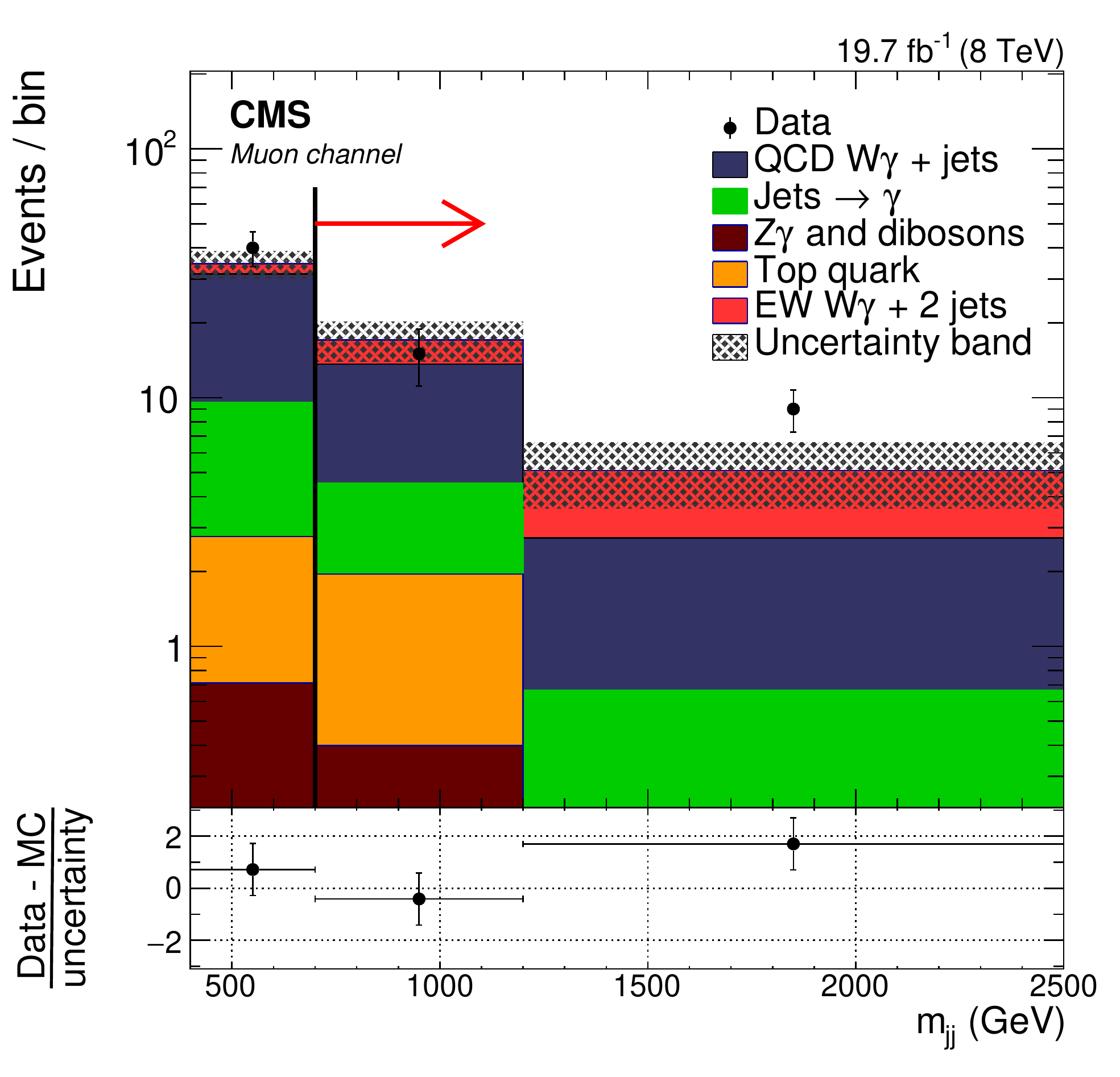}\\
    (a)\\
\includegraphics[width=0.40\textwidth]{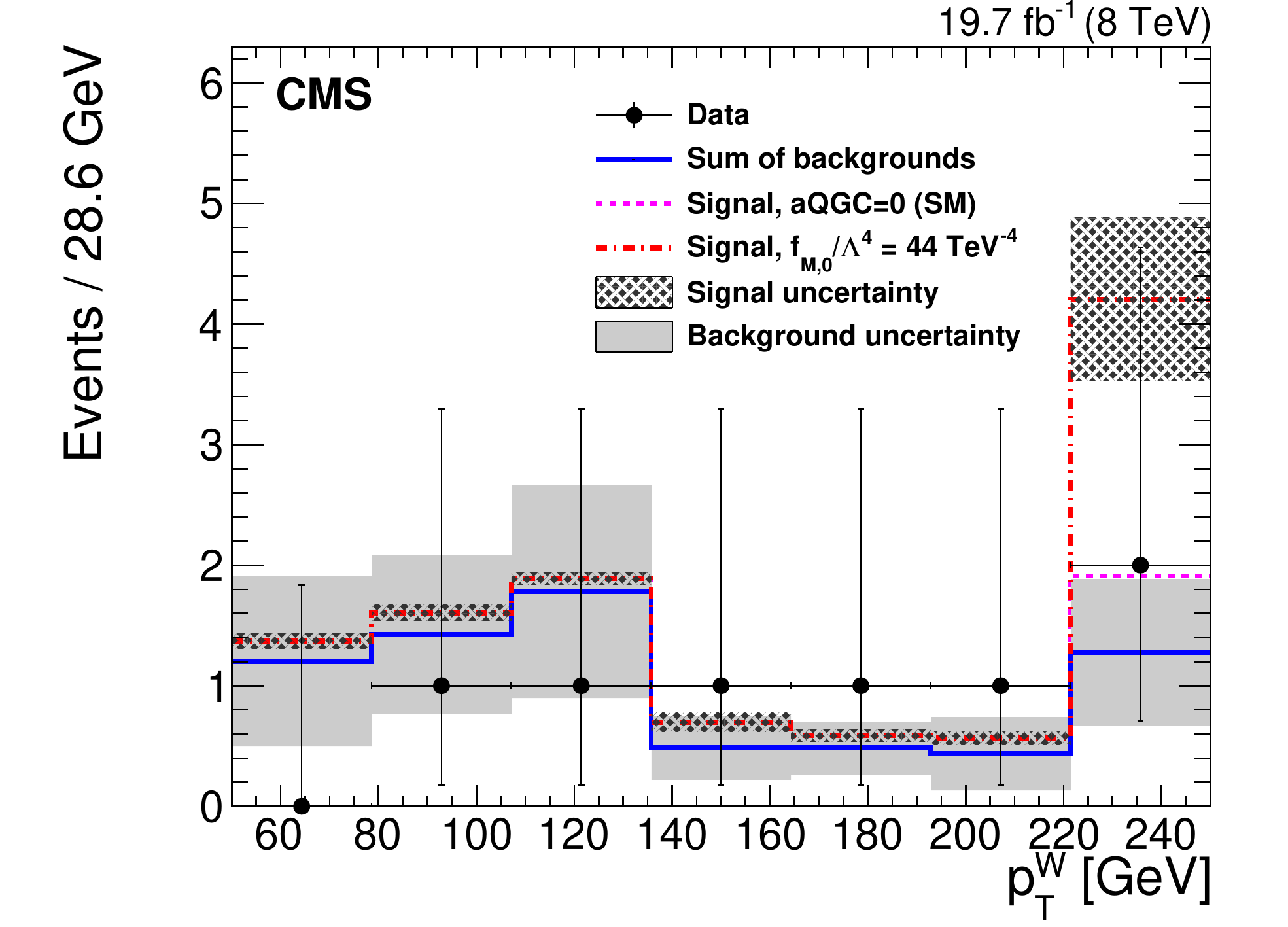}\\
    (b)\\
  \end{tabular}
  \caption{(a) Dijet mass distribution of the tag jets in the
    $\mu\nu\gamma jj$ final state at 8~TeV~\cite{Khachatryan:2016vif}. (b) \pt
    spectrum of the $W$ boson in VBS candidate events in the
    $\ell\nu\gamma jj$ final state at 8~TeV. The effect of a
    representative aQGC on the spectrum is also
    shown~\cite{Khachatryan:2016vif}.}
  \label{fig:VBSWgamma}
\end{figure}

\subsection{$W^\pm V jj$ Production}
\label{sec:WVjjprod}
The study of the semileptonic $WV$ ($V = W,Z$) VBS process benefits
from the large hadronic branching fraction of the $W$ or $Z$ boson
compared to the leptonic decays and the ability to fully reconstruct
the $WW$ contribution up to a quadratic ambiguity, resulting in
improved sensitivity to anomalous event kinematics. Searching for
anomalous quartic couplings in the high-mass tail of the $WV$ spectrum
is facilitated by the continually improving substructure
techniques to analyze boosted monojets arising from the hadronically
decaying $V$ boson (see Figure~\ref{fig:jetmW}). The $W^\pm V jj$ semileptonic final state includes
contributions from the $W^\pm W^\mp jj$, $W^\pm W^\pm jj$, and
$W^\pm Z jj$ VBS processes.

Building on the semileptonic $WV$ decay signature described in
Section~\ref{sec:WVprod} with one charged lepton (electron or muon),
missing transverse energy, and exactly two jets in the final state, the
corresponding VBS processes can be studied by requiring in addition
the presence of a tagging jet pair with large invariant mass.

ATLAS~\cite{Aaboud:2016uuk} has performed a first search for anomalous
couplings in $W^\pm V jj$ semileptonic VBS candidate events at 8~TeV
using a data sample with an integrated luminosity of 20~\ifb. While
the extraction of the SM signal cross section is not yet possible due
to large background contributions from $W+$jets and $t\bar{t}$
production, the analysis is optimized for aQGC sensitivity in a phase
space where the SM contributions are sufficiently suppressed. The
hadronic weak boson decay is reconstructed either via two jets in a
``resolved'' event category (which is split by lepton charge) or via a
large monojet in a ``merged'' event category.

No excess is observed in the data, and the transverse mass
distribution of the $WV$ system in the two resolved and one merged
event categories is used to constrain dimension-8 operators with
couplings $\alpha_4$ and $\alpha_5$. Two of the observed distributions are
shown in Figure~\ref{fig:VBSWVjj} with the data extending to about
0.9~TeV in transverse mass. The obtained limits are more stringent
than those obtained in the separate analyses of $W^\pm W^\mp jj$ and
$W^\pm Z jj$ leptonic final states described in the following two
sections. Given the largest sensitivity to aQGCs in the tail of the
transverse mass distribution, the merged event category presently improves the
expected sensitivity by about 40\% compared to the resolved categories
alone.

\begin{figure}[htbp]
  \begin{tabular}{c}
\includegraphics[width=0.40\textwidth]{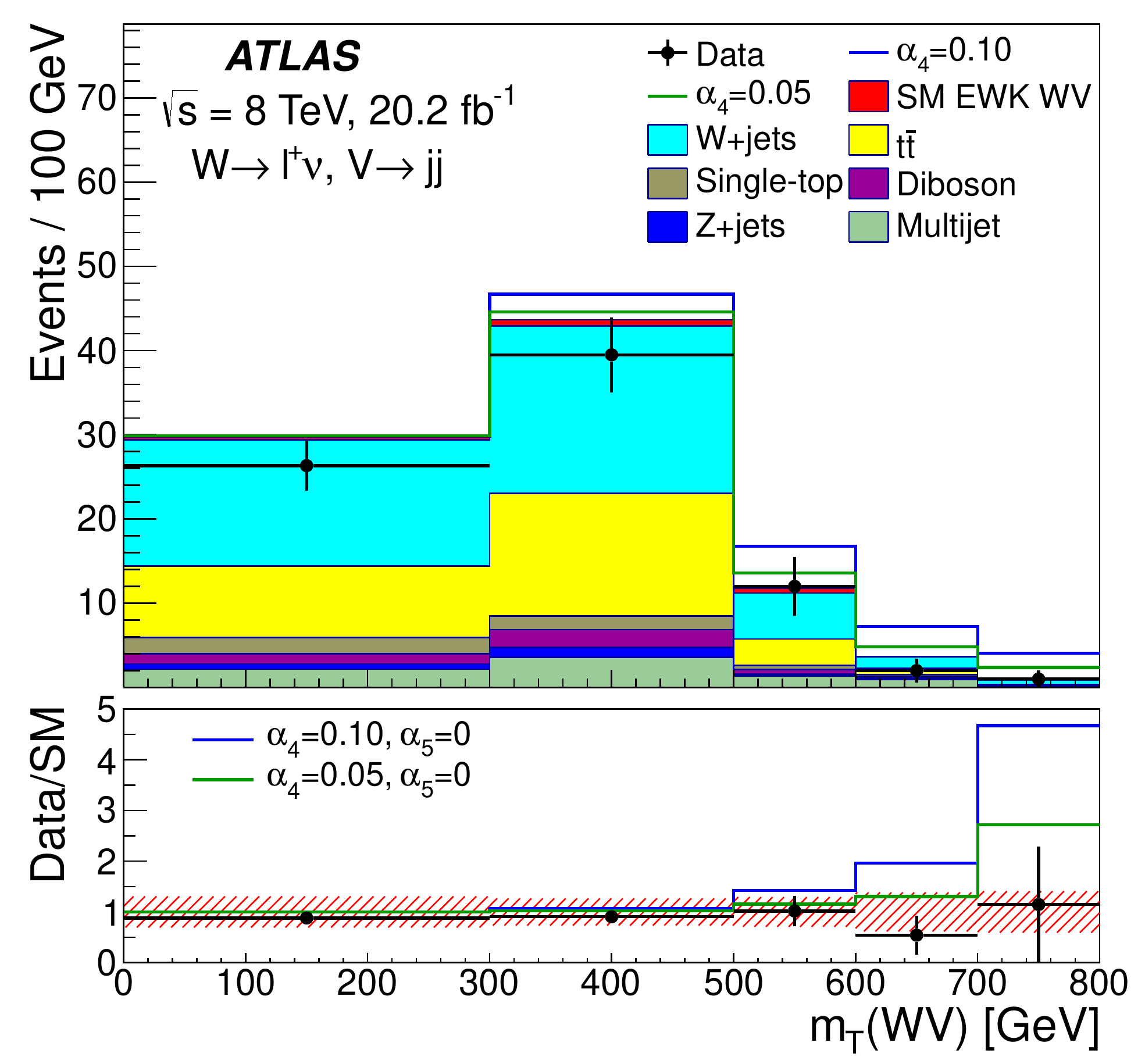}\\
    (a)\\
\includegraphics[width=0.40\textwidth]{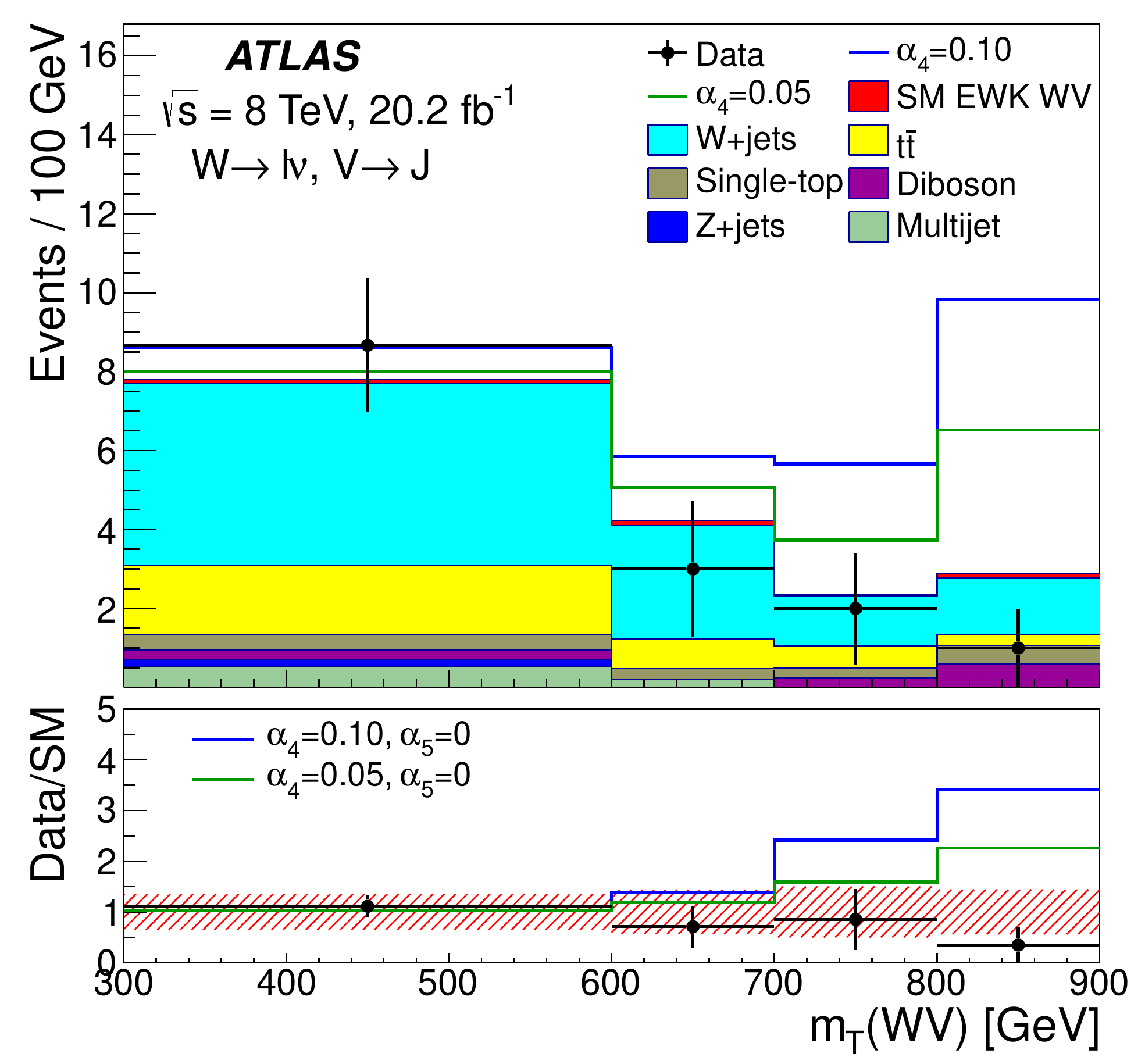}\\
    (b)\\
  \end{tabular}
  \caption{$W^\pm V jj$ candidate event transverse diboson mass
    distributions in the $\ell^\pm\nu~(jj/J)~jj$ final state at
    8~TeV~\cite{Aaboud:2016uuk}: (a) resolved ($V\to jj$) category for
    positively charged leptons and (b) merged ($V\to J$) category. The
    potential impact of aQGCs is shown as well.}
  \label{fig:VBSWVjj}
\end{figure}

\subsection{$W^\pm W^\pm jj$ Production}
\label{sec:ssWWprod}
The production of same-sign $W$ boson pairs in association with two
jets includes the SM QGC contribution from the $WWWW$ vertex and is
particularly valuable for the study of VBS processes with massive bosons since
the strong production mode does not dominate over the electroweak mode
of interest as is the case for the other $VVjj$ ($V=W,Z$) processes.
The best signal-to-background ratio is achieved when studying the
leptonic $W$ boson decays, giving rise to final states with two
leptons of the same sign ($e$ or $\mu$), MET, and two jets.

Both ATLAS~\cite{Aad:2014zda} and CMS~\cite{Khachatryan:2014sta} have
studied this final state in 8~TeV data samples with integrated
luminosities of up to 20~\ifb, requiring the two leading (tag)
jets to exhibit a large dijet invariant mass and to be well-separated
in rapidity to enhance the VBS contribution (see
Figures~\ref{fig:VBSssWWCMS}~(a) and \ref{fig:VBSssWWATLAS}). ATLAS
and CMS find evidence for electroweak $W^\pm W^\pm jj$ production
with 3.6 and 2.0~$\sigma$ significance, respectively, compatible with
SM NLO expectations.

\begin{figure}[htbp]
  \begin{tabular}{c}
\includegraphics[width=0.40\textwidth]{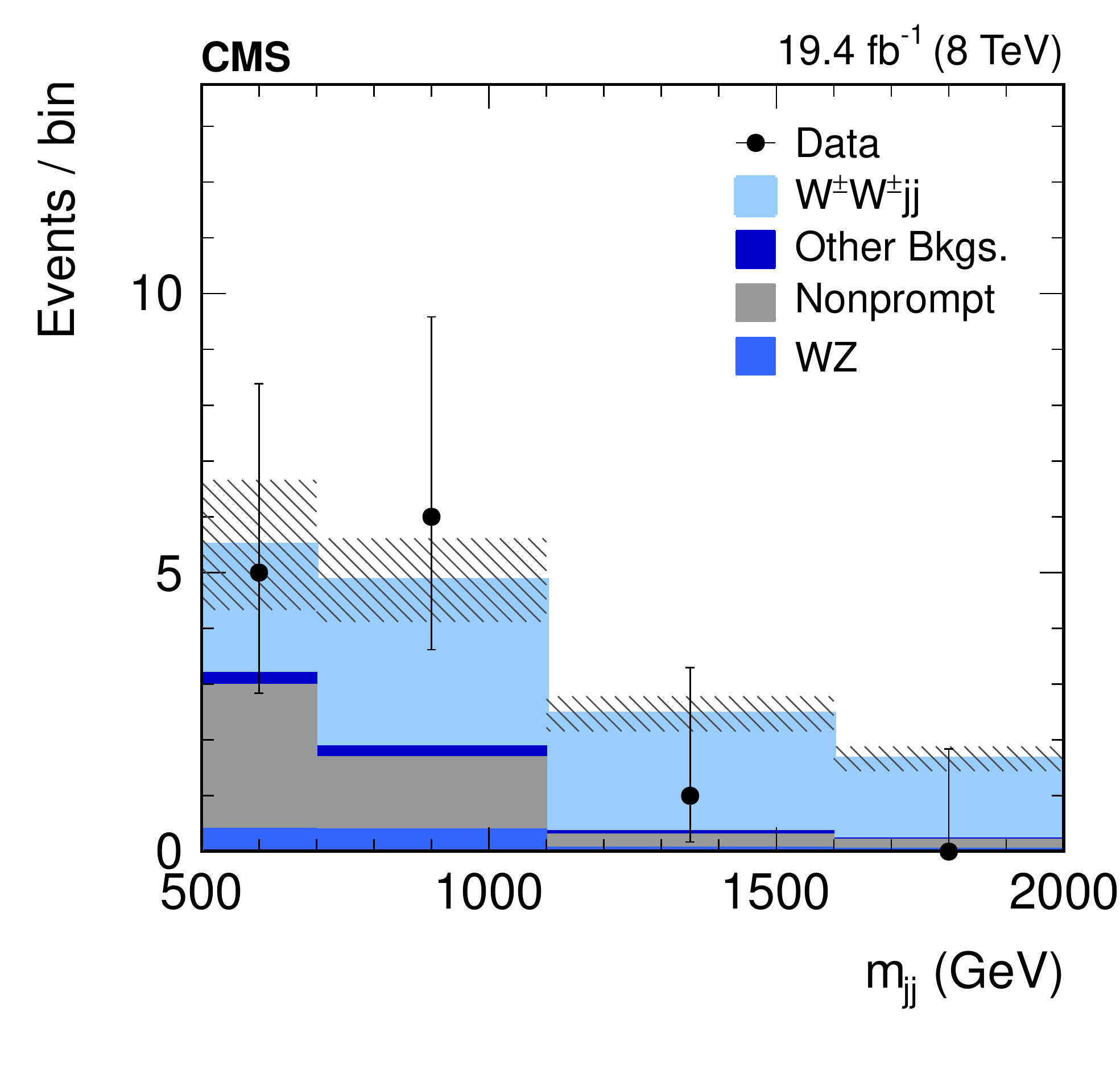}\\
    (a)\\
\includegraphics[width=0.40\textwidth]{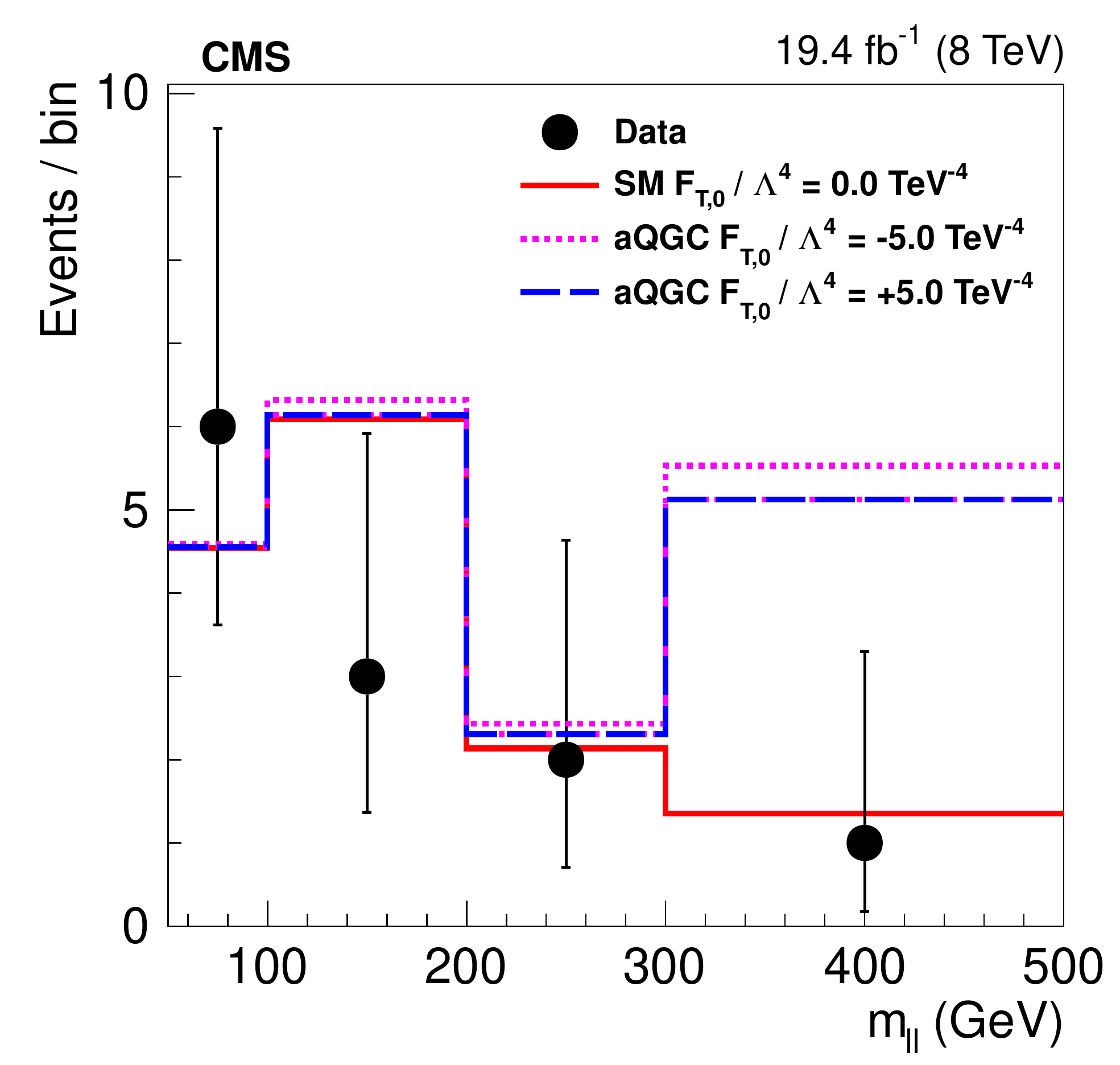}\\
    (b)\\
  \end{tabular}
  \caption{VBS-$W^\pm W^\pm$ candidate events in the
    $\ell^\pm\nu\ell^\pm\nu jj$ final state at
    8~TeV~\cite{Khachatryan:2014sta}: (a) dijet mass of the tag
    jets and (b) dilepton mass distribution, where the effect of a
    representative aQGC on the spectrum is also shown.}
  \label{fig:VBSssWWCMS}
\end{figure}
\begin{figure}[htbp]
  \begin{tabular}{c}
\includegraphics[width=0.40\textwidth]{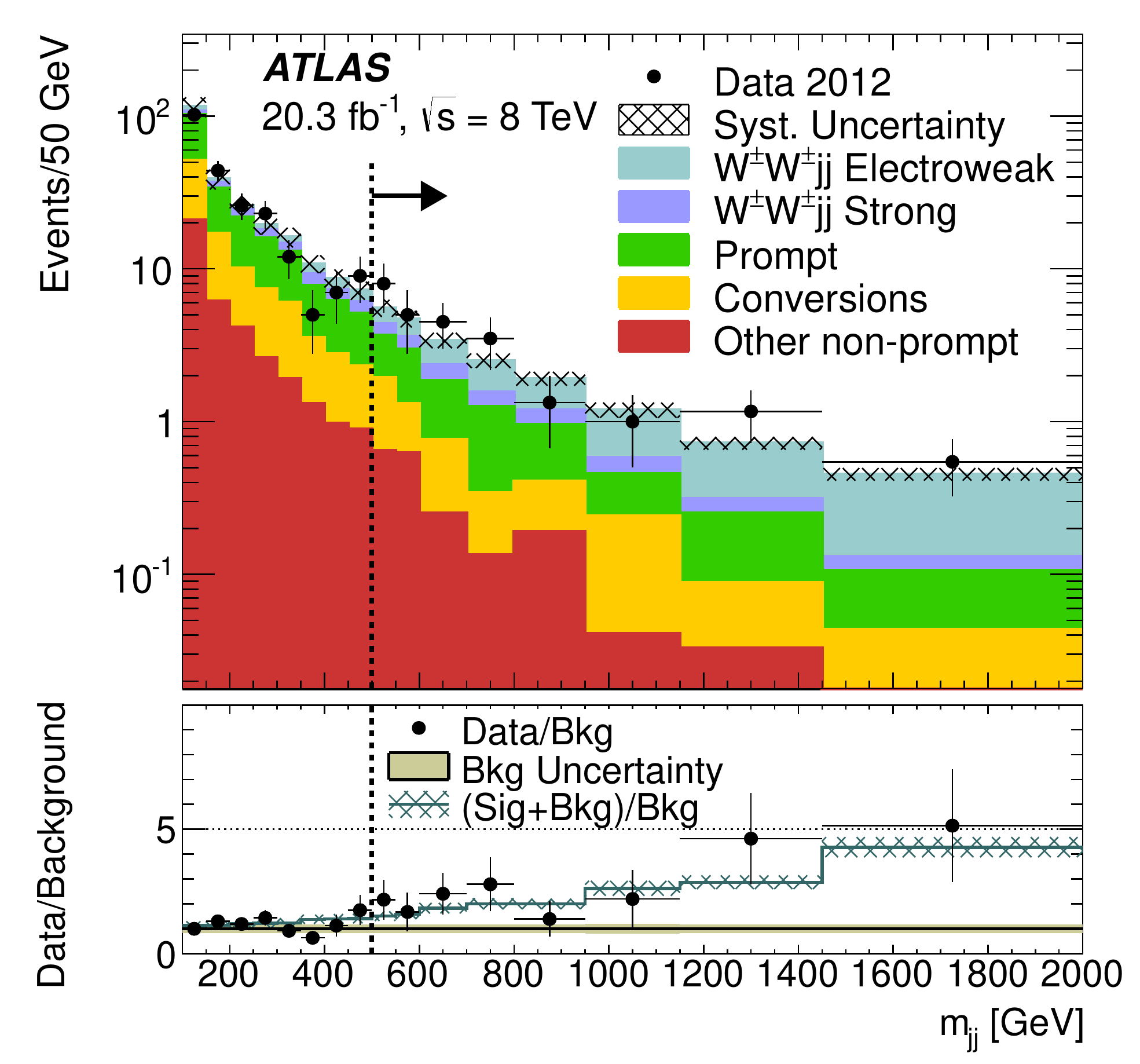}\\
    (a)\\
\includegraphics[width=0.40\textwidth]{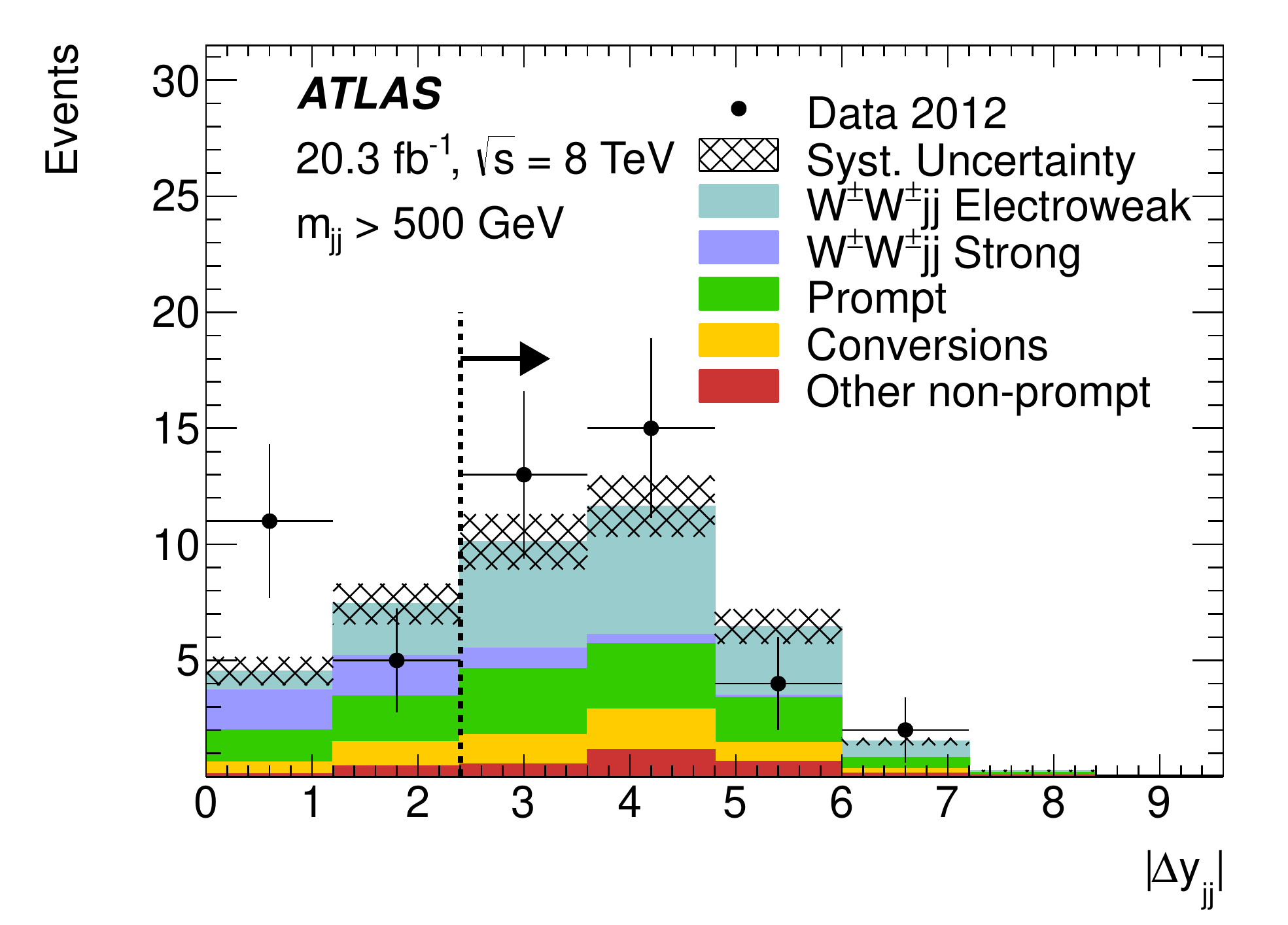}\\
    (b)\\
  \end{tabular}
  \caption{VBS-$W^\pm W^\pm$ candidate events in the
    $\ell^\pm\nu\ell^\pm\nu jj$ final state at
    8~TeV~\cite{Aad:2014zda}: (a) dijet mass and (b) rapidity
    separation of the tag jets. The applied selections are
    indicated by dotted lines.}
  \label{fig:VBSssWWATLAS}
\end{figure}

To constrain possible aQGC contributions, the measured cross section
in the VBS fiducial region (ATLAS) or the dilepton mass shape is used
(CMS, see Figure~\ref{fig:VBSssWWCMS}~(b)), where the data extend to
about 0.5~TeV in dilepton mass.  Dimension-8 operators with
couplings $\alpha_4$ and $\alpha_5$ or alternatively $f_{S0,1}$, $f_{M0,1,6,7}$
and $f_{T0..2}$ are probed.

ATLAS~\cite{Aaboud:2016ffv} has published in addition a detailed
writeup of a re-analysis of the same data set, where the sensitivity
to anomalous couplings was optimized through an additional cut on the
estimated transverse mass of the $WW$ system. As a result, the
expected $\alpha_4$ and $\alpha_5$ sensitivity is improved by 35\% with
respect to the previous analysis~\cite{Aad:2014zda}. Upper limits on
the cross section in the resulting fiducial volume are provided as
well.
\subsection{$W^\pm Z jj$ Production}
The production of $W^\pm Z$ boson pairs with two jets includes the SM
QGC contribution from the $WZWZ$ and $W\gamma WZ$ vertices. The best signal-to-background
ratio is achieved when studying the leptonic boson decay modes
involving electrons and muons, resulting in a final state with three
charged leptons, MET, and at least two jets.

ATLAS has performed a first measurement in this final state at 8~TeV
using a data sample with an integrated luminosity of
20~\ifb~\cite{Aad:2016ett}. Requiring a large invariant mass of the
two leading tag jets, 95\% C.L. limits on electroweak $W^\pm Z jj$
production are placed about a factor of 4.8 higher than the SM cross-section 
expectation at NLO in QCD in the fiducial volume under study,
consistent with the expected sensitivity.

Additional selection criteria are applied to the data in order to
optimize the expected sensitivity for aQGCs: Both a large difference
in azimuthal angle between reconstructed $W$ and $Z$ boson directions
and a large scalar sum of the \pt of the three charged leptons are
required, with the distributions prior to the cuts shown in
Figure~\ref{fig:VBSWZATLAS}. The resulting measured fiducial cross
section is used to constrain dimension-8 operators with couplings
$\alpha_4$ and $\alpha_5$ or alternatively $f_{S0,1}$.
\begin{figure}[htbp]
  \begin{tabular}{c}
\includegraphics[width=0.40\textwidth]{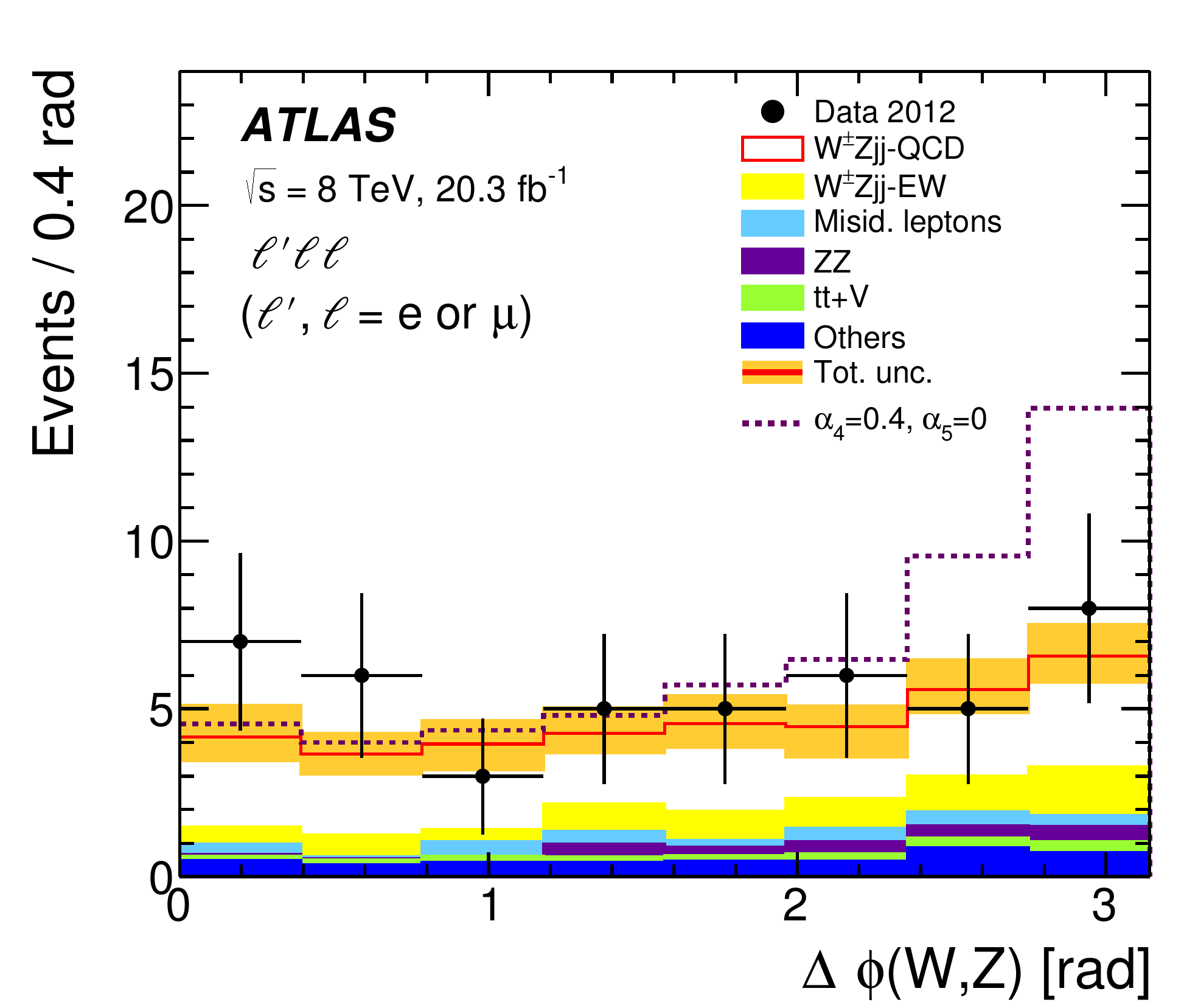}\\
    (a)\\
\includegraphics[width=0.40\textwidth]{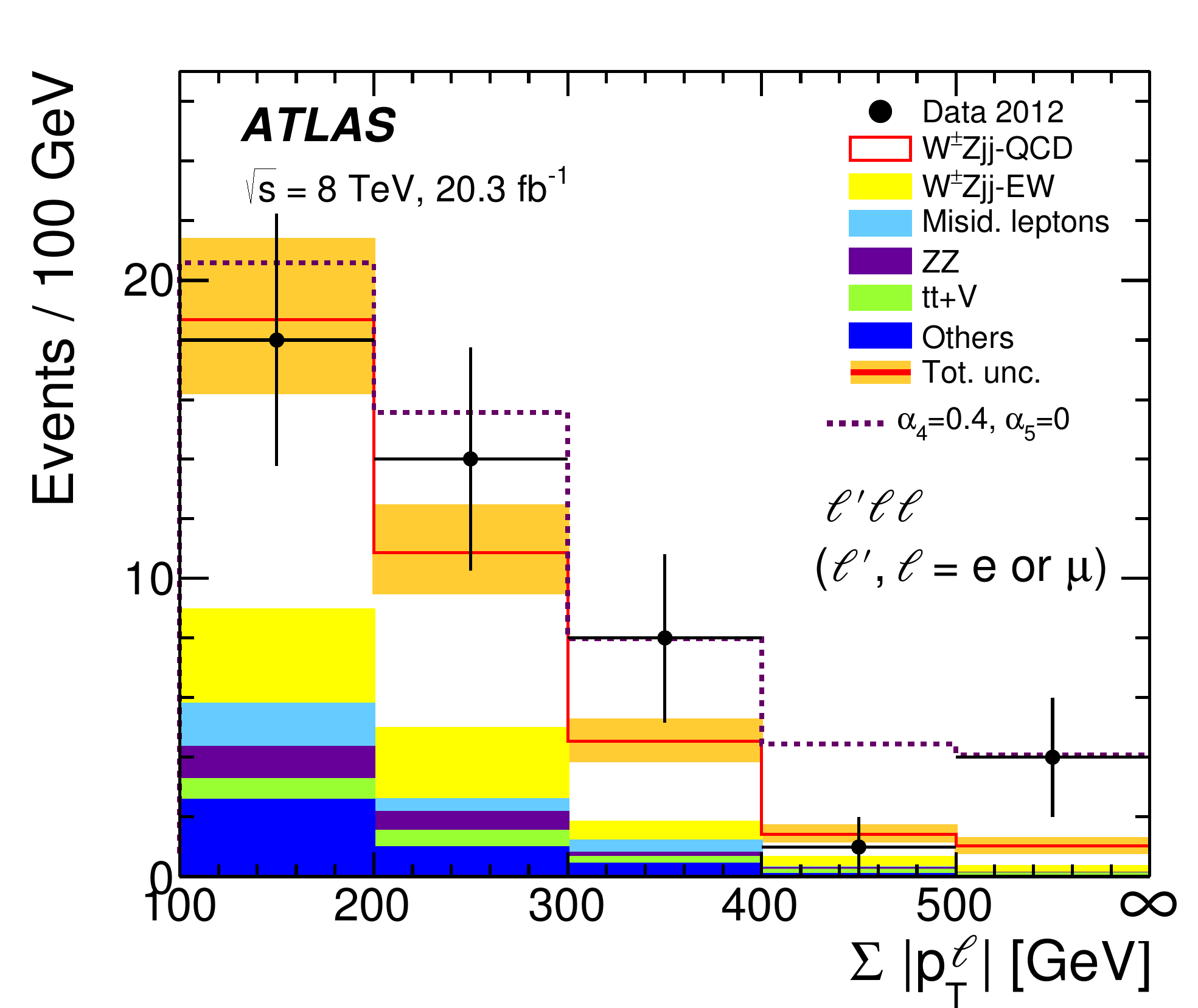}\\
    (b)\\
  \end{tabular}
  \caption{Electroweak $W^\pm Z jj$ candidate event distributions in
    the $\ell^\pm\nu\ell^\pm\ell^\mp jj$ final state at
    8~TeV~\cite{Aad:2016ett}: (a) difference in the azimuthal angle
    between reconstructed $W$ and $Z$ boson directions and (b) scalar
    sum of the \pt of the three charged leptons. The potential impact
    of aQGCs is shown as well.}
  \label{fig:VBSWZATLAS}
\end{figure}

\subsection{exclusive $WW$ Production}

Exclusive production of a $W$ boson pair, $pp\to W^+W^- pp$, proceeds
via the emission of photons from the beam protons, which then interact
to yield the $W$ boson pair: $\gamma\gamma\to W^+W^-$. In the elastic
case, both protons remain intact after the interaction, while in the
case of single (double) dissociation one (both) of the protons
dissociates. In either case, the proton (remnants) closely follows the
original beam direction and hence escape detection, leaving only the
$W$ boson decay products in the detector without the 
additional activity present in inclusive processes. The production of
$W^+W^-$ from photon scattering gives access to the SM QGC from the
$WW\gamma\gamma$ vertex. The
$WW\gamma\gamma$ coupling is the sole SM QGC contribution to the
process since no tag jets indicating beam breakup are allowed
which suppresses the $WWWW$, $WWZZ$, and $WWZ\gamma$ processes.

The best signal-to-background ratio is achieved when studying
different-flavor leptonic $W$ boson decays, giving rise to a final
state with one electron and one muon of opposite charge and MET.
ATLAS has performed a measurement~\cite{Aaboud:2016dkv} based on the
full 20~\ifb of its 8~TeV data set, while CMS utilizes both 7 and
8~TeV data samples with integrated luminosities of 5 and 20~\ifb,
respectively~\cite{Chatrchyan:2013akv,Khachatryan:2016mud}. Exclusive
events are selected by requiring no additional charged particles be
present at the $e\mu$ vertex and a large \pt of the $e\mu$ pair.

ATLAS measures the exclusive $W^+W^-$ cross section in good agreement
with SM expectation with a significance of 3.0~$\sigma$.
Figure~\ref{fig:VBSaaWWATLAS}~(a) shows the distribution of the
difference in azimuthal angle between electron and muon, clearly
indicating the need for the signal contribution to describe the
observed data. First upper limits on exclusive Higgs boson production
in the $H\to WW$ decay mode are provided as well, based on a
separately optimized selection.
\begin{figure}[htbp]
  \begin{tabular}{c}
\includegraphics[width=0.42\textwidth]{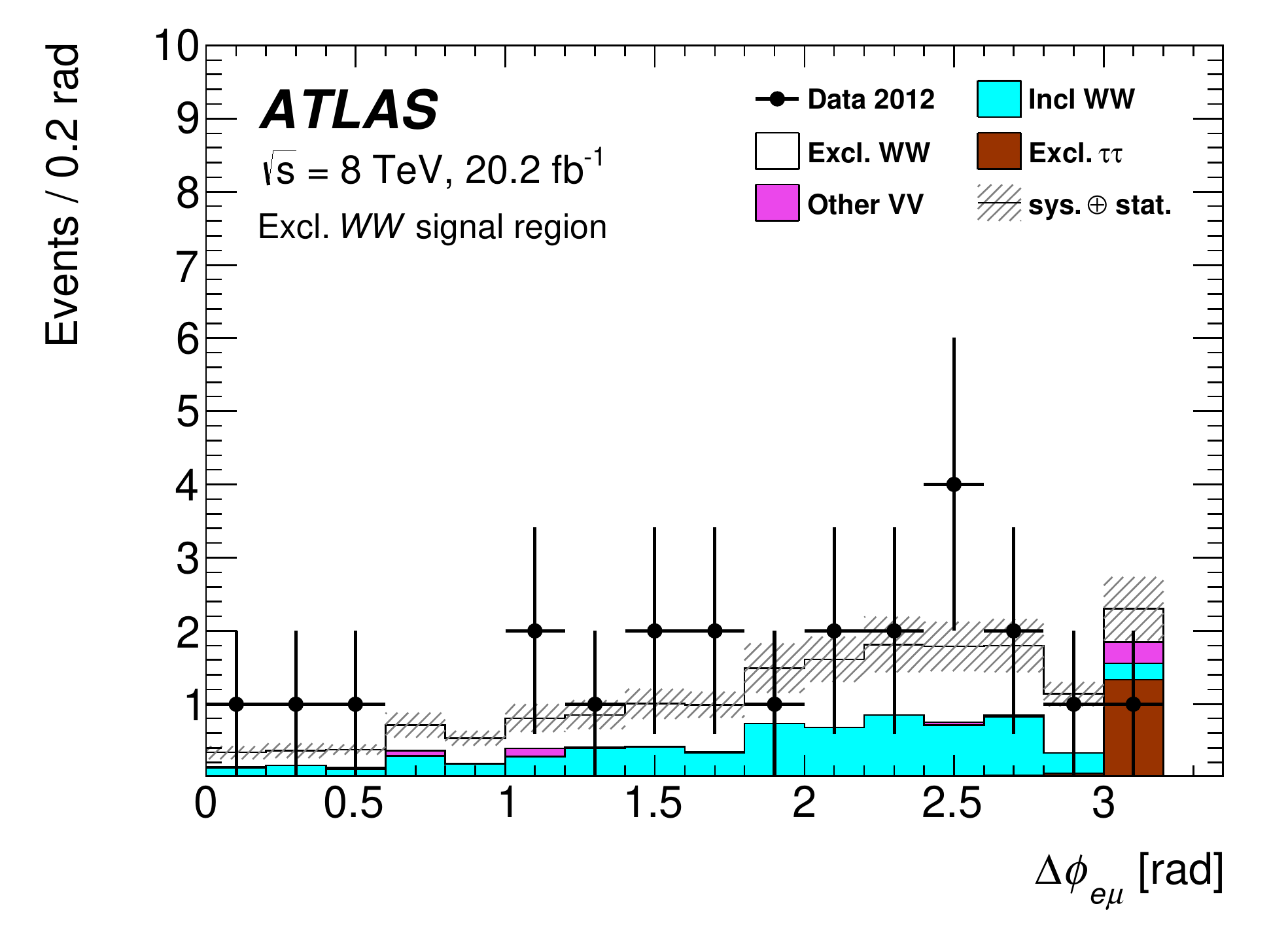}\\
    (a)\\
\includegraphics[width=0.42\textwidth]{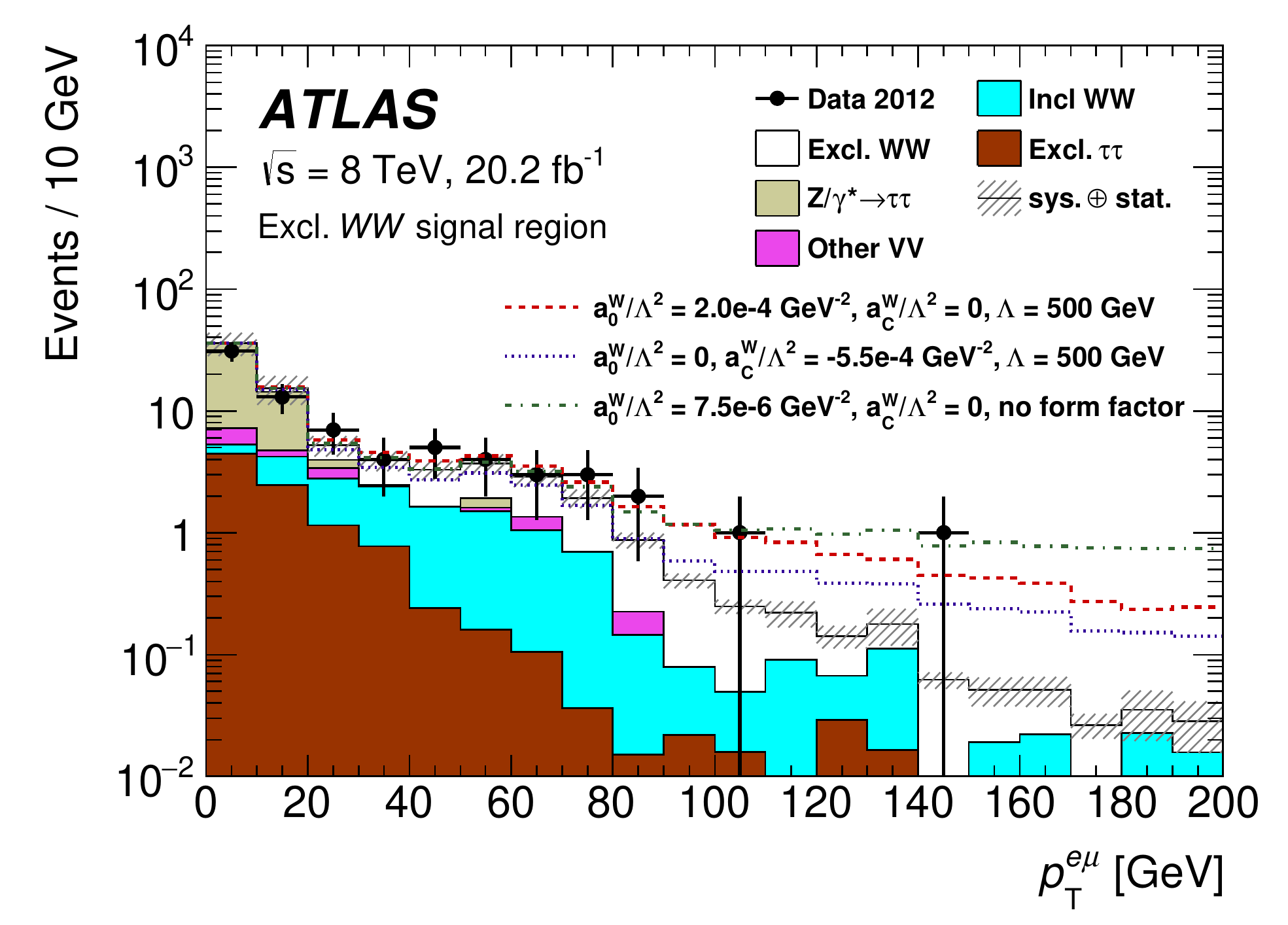}\\
    (b)\\
  \end{tabular}
  \caption{VBS-$\gamma\gamma\to W^+W^-$ candidate events in the
    $e^\pm\nu\mu^\mp\nu$ final state in 8~TeV
    data~\cite{Aaboud:2016dkv}: (a) difference in the azimuthal angle
    between the electron and muon and (b) dilepton \pt distribution before
    applying the 30~GeV selection cut. The potential impact of aQGCs
    is shown as well.}
  \label{fig:VBSaaWWATLAS}
\end{figure}

CMS places an upper limit on $\gamma\gamma\to W^+W^-$ production in
the 7~TeV analysis, corresponding to about $2.6$ times the SM
theoretical expectation at 95\% C.L., while at 8~TeV first evidence for
the signal is observed with a significance of 3.2~$\sigma$. Combining
the 7 and 8~TeV data, the signal significance increases to
3.4~$\sigma$. The $e\mu$ acoplanarity is shown in
Figure~\ref{fig:VBSaaWWCMS}~(a) in the 8~TeV data, indicating
consistent yields with respect to signal and background expectations
and a dominant VBS contribution.
\begin{figure}[htbp]
  \begin{tabular}{c}
\includegraphics[width=0.40\textwidth]{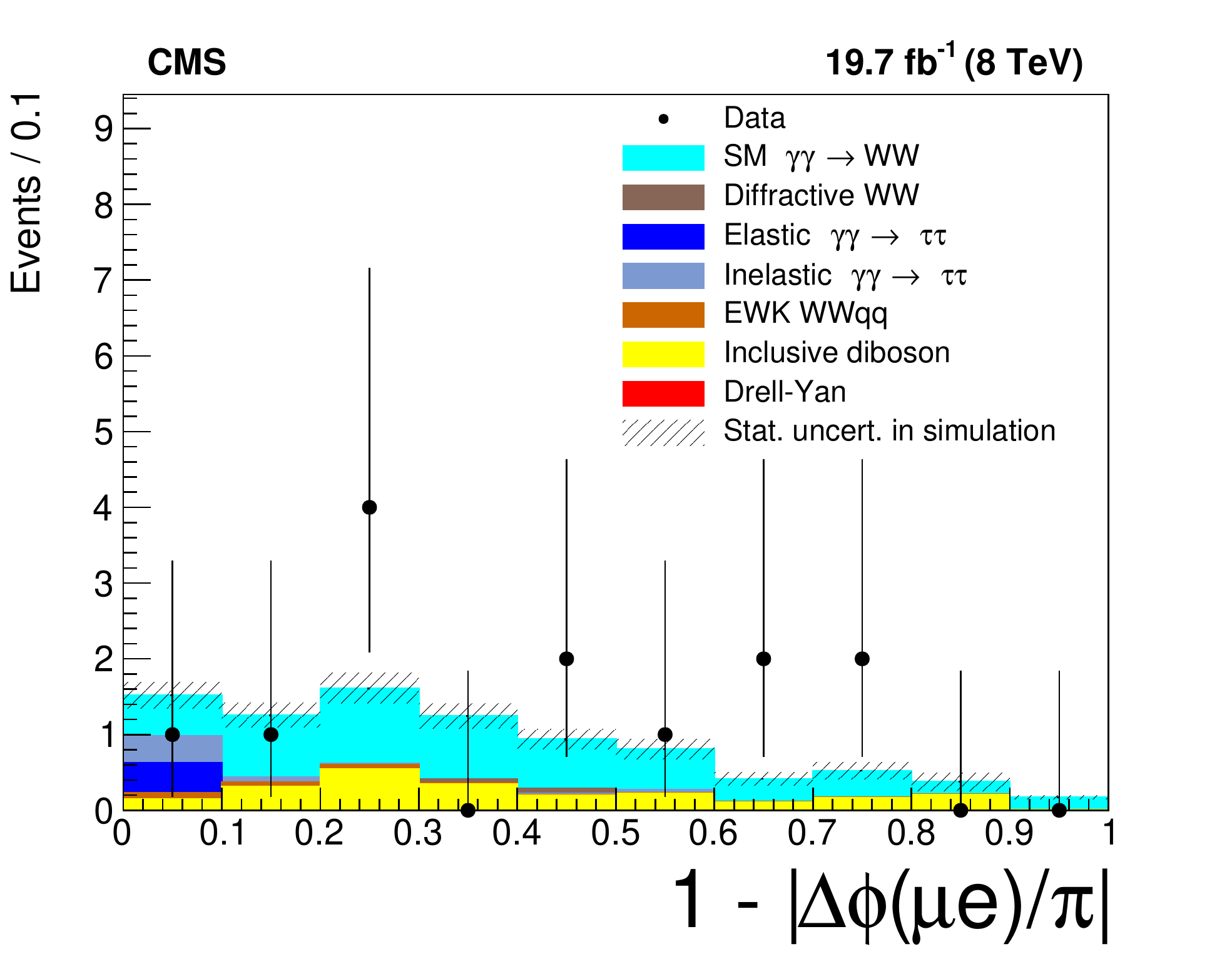}\\
    (a)\\
\includegraphics[width=0.40\textwidth]{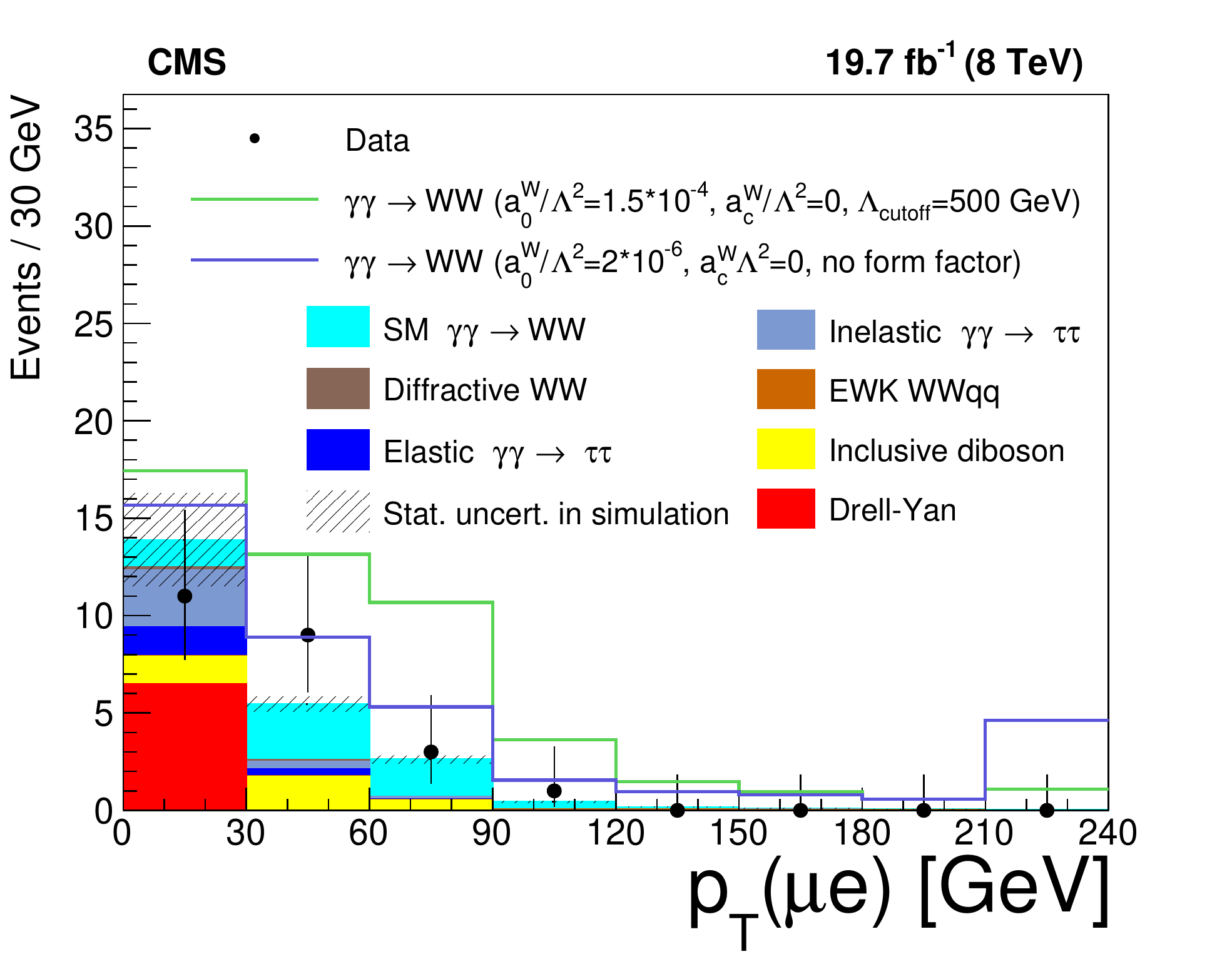}\\
    (b)\\
  \end{tabular}
  \caption{VBS-$\gamma\gamma\to W^+W^-$ candidate events in the
    $e^\pm\nu\mu^\mp\nu$ final state in 8~TeV
    data~\cite{Khachatryan:2016mud}: (a) acoplanarity of the $e\mu$
    system and (b) dilepton \pt distribution before applying the 30~GeV
    selection cut. The potential impact of aQGCs is shown as well.}
  \label{fig:VBSaaWWCMS}
\end{figure}

Both ATLAS and CMS use the shape of the dilepton \pt distribution, shown in
Figures~\ref{fig:VBSaaWWATLAS}~(b) and~\ref{fig:VBSaaWWCMS}~(b) for 
the 8~TeV data set, to
limit aQGC dimension-6 operators with couplings $a_0^W$ and $a_C^W$.
Corresponding transformed limits on dimension-8
operators with couplings $f_{M0..3}$ are provided as well. 

\section{Constraints on anomalous Triple Gauge Couplings}
\label{aTGC}

The exploration of high-$\hat{s}$ diboson and VBF events leads to
limits on possible triple gauge couplings which are differing from or
not present in the SM: anomalous triple gauge couplings. Limits
on aTGCs have been presented by experiments at LEP, the Tevatron, and
the LHC. aTGC limits arise when specific spectra of final state
particles are compared to the expectations of the SM with additional
aTGC terms in the Lagrangian. The specific spectra used at the LHC
were shown in Sections~\ref{diboson} and \ref{VBF} for aTGCs and
Sections~\ref{triboson} and \ref{VBS} for aQGCs. Note that
higher-order (NNLO QCD and NLO EW) corrections will significantly
impact the SM expectation in the tails of the utilized distributions,
and the incorporation of such corrections depends on the timing of the
corrections becoming available versus when the analysis was carried
out. The various limits made
with different diboson and VBF final states are collected here and
compared.  Typically, one-dimensional (1D) limits are quoted where
only one operator is allowed to be non-zero at a time. In a few cases
two operators are allowed to float simultaneously, a procedure which
illustrates the correlations between the effects of the operators.

In the SM there are $WW\gamma$ and $WWZ$ TGCs. They are studied in
$WW$, $WZ$, $W\gamma$, VBF-$W$, and VBF-$Z$ final states. Beyond the SM
there are $ZZZ$, $ZZ\gamma$, $Z\gamma Z$, $Z\gamma\gamma$, and
$\gamma\gamma\gamma$ couplings, where limits on the first four are
placed by exploring $ZZ$ and $Z\gamma$ final states.
In the future, with higher luminosity data taking, the $\gamma\gamma$
final state (Section~\ref{sec:aa}) can be used to explore the non-SM
$\gamma\gamma\gamma$ aTGC.

\subsection{$WW\gamma$ and $WWZ$ limits}
\label{sec:WWVaTGCs}
$WWZ$ and $WW\gamma$ limits
can be formulated with aTGC as was done in other prior experiments at
LEP (ALEPH, DELPHI, L3, and OPAL) and the Tevatron (CDF and D0). The
five independent C- and P-conserving aTGC parameters that remain after
imposing electromagnetic gauge invariance,
$\Delta g_1^Z (\equiv g_1^Z - 1)$,
$\Delta \kappa_Z (\equiv \kappa_Z - 1)$,
$\Delta \kappa_\gamma (\equiv \kappa_\gamma-1)$, $\lambda_Z$, and
$\lambda_\gamma$ are all zero in the SM, and limits on all these
parameters have been provided independently by recent LHC
publications. In order to be able to compare limits from the LHC, Tevatron,
and LEP on equal footing, results from the ``LEP
scenario''~\cite{Altarelli:1996gh, Gounaris:1996rz} are used, which
are available from all experiments. Motivated by SU(2)$\times$U(1)
symmetry, the LEP scenario assumes
$\Delta \kappa_\gamma = (\Delta g_1^Z -
\Delta\kappa_Z)/\tan^2\theta_W$, and $\lambda_\gamma = \lambda_Z$,
thereby reducing the number of independent parameters to three.

\begin{figure*}[htbp]
  \centering
\includegraphics[width=\textwidth]{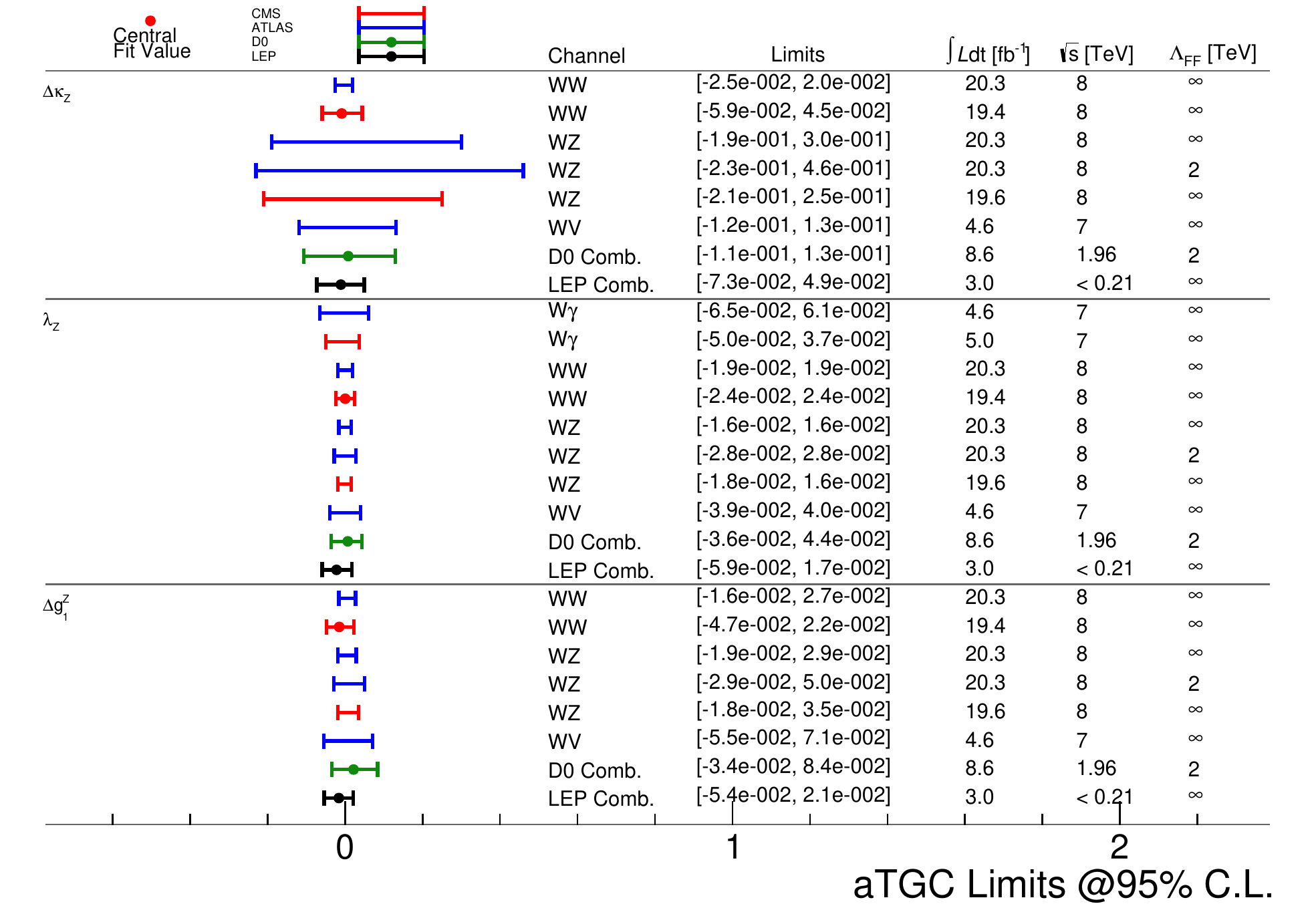}
  \caption{Comparison of the most competitive aTGC limits in the LEP
    scenario for the LHC analyses presented in this review as well as
    the combination of limits by the D0~\cite{Abazov:2012ze} and
    LEP~\cite{Schael:2013ita} experiments.}
  \label{fig:aTGCsummary}
\end{figure*}

Figure~\ref{fig:aTGCsummary} shows a comparison of the most competitive
limits derived in the LEP scenario by experiments at the LHC, Tevatron
and LEP. The impact of imposing unitarity constraints on the anomalous
couplings via a dipole form factor with a suppression scale $\Lambda_{\rm FF}$ that dampens the cross-section
increase at high $\hat{s}$ for any anomalous coupling
$\alpha$ with value $\alpha_0$ at low energies,
$\alpha(\hat{s}) = \alpha_0/(1+\hat{s}/\Lambda_{\rm FF}^2)^2$, is shown as
well. The LHC limits using $WW$ and $WZ$ final states for constraining
$WW\gamma$ and $WWZ$ couplings are already more stringent than the
combined D0 or LEP limits. Presently, the higher energy and higher
statistics data at 8~TeV give the strongest LHC limits. Increased
luminosity and center of mass energy in Run~II and beyond will further
reduce the LHC limits.

\begin{figure}[htbp]
  \begin{tabular}{c}
\includegraphics[width=0.40\textwidth]{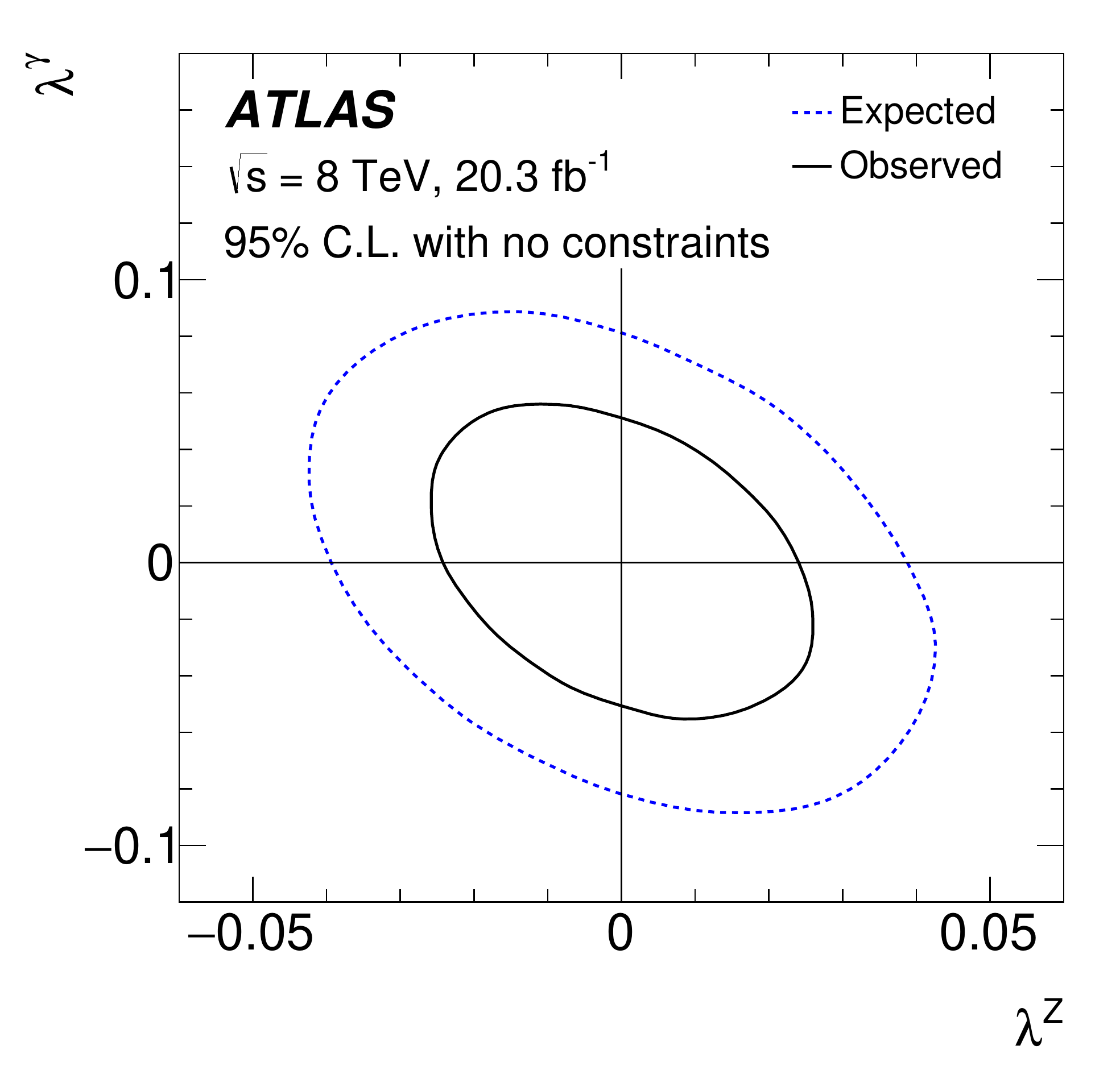}\\
    (a)\\
\includegraphics[width=0.40\textwidth]{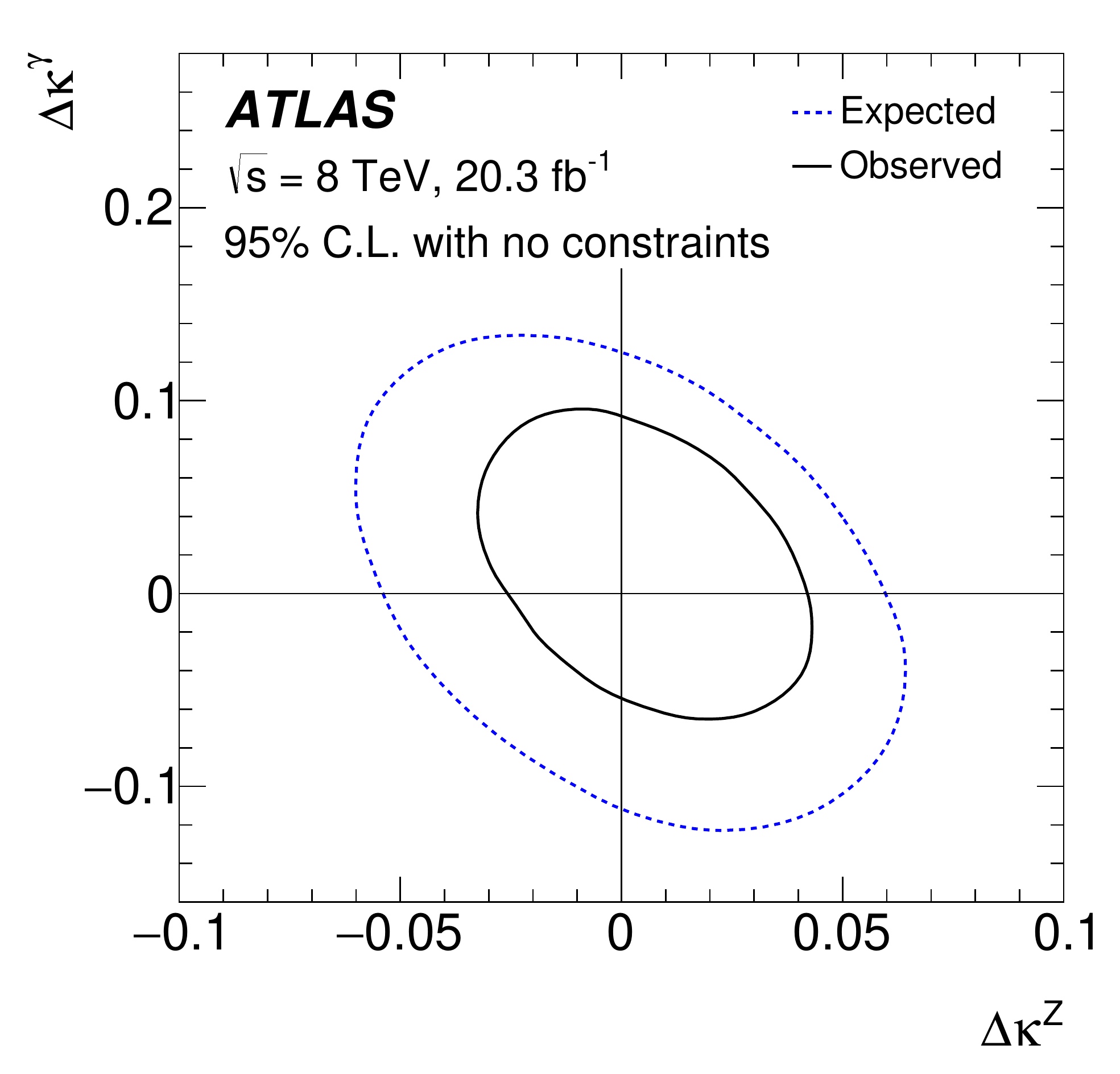}\\
    (b)\\
  \end{tabular}
  \caption{Expected and observed 95\% C.L. contours for aTGC limits
    derived from 8~TeV fully leptonic $WW$ candidate
    events~\cite{Aad:2016wpd}, illustrating the anti-correlations
    between (a) $\lambda^V$ and (b) $\Delta \kappa^V$ parameters when
    no constraints between aTGCs are assumed. The aTGCs not shown are
    set to zero.}
  \label{fig:aTGC2dnoconst}
\end{figure}

The two-dimensional limits shown in Figure~\ref{fig:aTGC2dnoconst}
illustrate the anti-correlation of the $\Delta \kappa^V$ and
$\lambda^V$ parameters when no constraints are assumed on the five
aTGC parameters. Typically, only the 1D limits are shown since the
correlations are usually small.

More recently, the EFT formulation of possible aTGC in terms of
dimension-6 operators for triple boson couplings has come into use.
A marked difference with respect to the anomalous Lagrangian vertex couplings is that
the EFT-based anomalous couplings are not valid to
arbitrary energy scales, but instead are valid only below the scale
$\Lambda$ where new physics sets in. Using the same assumptions as in
the LEP scenario and applying no unitarization, the aTGC parameters
can be directly translated into EFT coefficients $c_W$, $c_{WWW}$, and
$c_B$~\cite{Degrande:2012wf}.  

Since these dimension-6 operators are not expected to lead to
unitarity violation in diboson production at the LHC center of mass
energies~\cite{Degrande:2012wf}, the same must hold true for their
aTGC counterparts. The reason that, for example, the ATLAS $WW$
analysis~\cite{Aad:2016wpd} nevertheless gives aTGC unitarization
bounds is that the used unitarity considerations
in~\cite{Aihara:1995iq} are valid for {\em arbitrary center of mass
  energies}.

Figure~\ref{fig:TGCEFTsummary} shows a comparison of the best aTGC
limits, arising from $WW$ and $WZ$ analyses in leptonic final states
by ATLAS and CMS using the full 8~TeV data sets and converted to the
EFT formalism.
\begin{figure*}[htbp]
  \centering
\includegraphics[width=\textwidth]{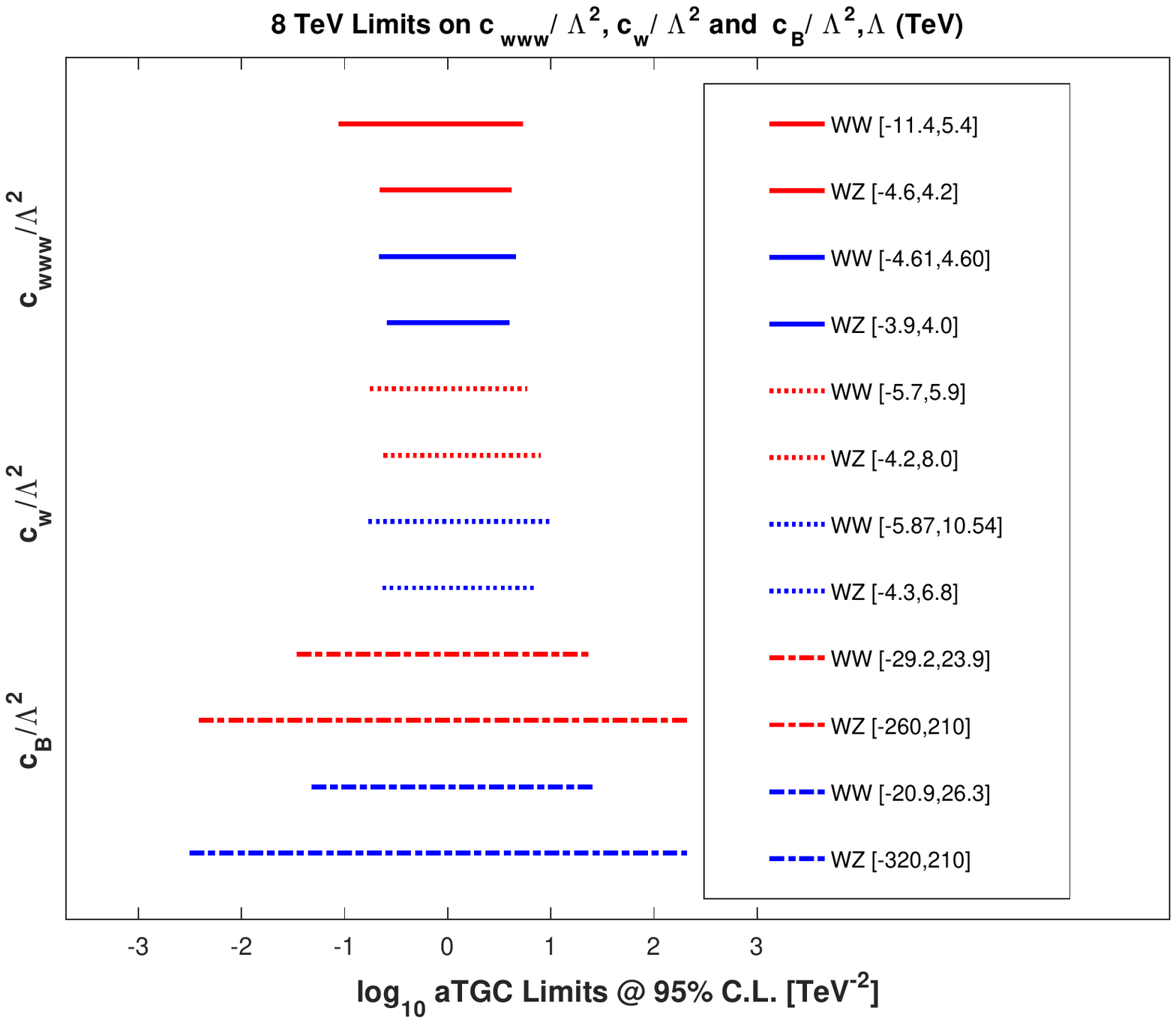}
  \caption{Comparison of the most competitive aTGC EFT limits based on
    the 8~TeV $WW$ and $WZ$ analyses in leptonic final states by ATLAS
    (blue) and CMS (red), using the full available data sets.}
  \label{fig:TGCEFTsummary}
\end{figure*}
Figure~\ref{fig:TGCEFTcorr} illustrates the weak correlations between
these EFT parameters.
\begin{figure}[htbp]
  \centering
\includegraphics[width=0.4\textwidth]{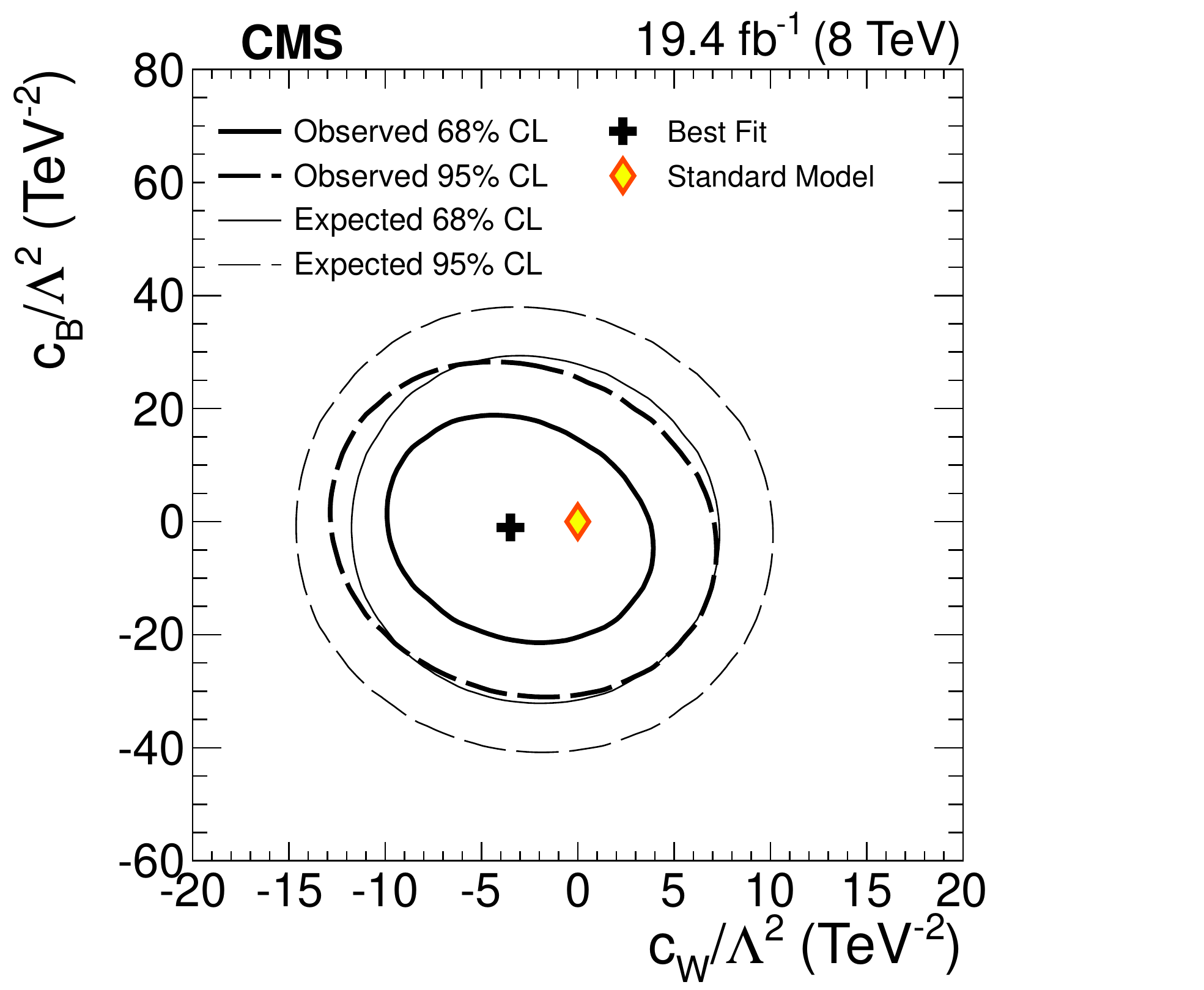}
  \caption{Expected and observed 68\% and 95\% C.L. contours for aTGC
    limits in the EFT formulation derived from 8~TeV fully leptonic
    $WW$ candidate events~\cite{Khachatryan:2015sga} illustrating the
    weak correlations between these EFT parameters, with $c_{WWW}$ set
    to zero.}
  \label{fig:TGCEFTcorr}
\end{figure}

\subsection{$Z\gamma\gamma$ and $Z\gamma Z$ limits}
Limits on the $Z\gamma\gamma$ and $Z\gamma Z$ couplings are usually
given using the CP-conserving parameters $h_3^V$ and $h_4^V$ since
there is no interference with the CP-violating couplings associated
with the $h_1^V$ and $h_2^V$ parameters and the corresponding cross
sections and sensitivities are very similar~\cite{Baur:1992cd}.
Figure~\ref{fig:nTGCAsummary} shows a comparison of the most
competitive limits, set by ATLAS and CMS. The combined limits by
LEP~\cite{Schael:2013ita} as well as the best Tevatron limits by
CDF~\cite{Aaltonen:2011zc} on $h_3^V$ and $h_4^V$ are not competitive
with those achieved at the LHC.  The impact of imposing
unitarity constraints on the anomalous couplings via a form factor with a suppression scale $\Lambda_{\rm FF}$, 
$\alpha(\hat{s}) = \alpha_0/(1+\hat{s}/\Lambda_{\rm FF}^2)^n$, is
shown as well. The form factor exponent $n$ is equal to the index $i$
of the parameter $h_i^V$ under study~\cite{Baur:1992cd}, in contrast with 
the dipole form factor ($n = 2$)  assumed in Section~\ref{sec:WWVaTGCs}. 
This illustrates the model dependence inherent in the form factor approach. 
In general, the parameter $\Lambda_{\rm FF}$ is chosen differently for different
processes and the choice of the exponent $n$ can also vary in the absence of a
definitive prediction.
\begin{figure*}[htbp]
  \centering
\includegraphics[width=\textwidth]{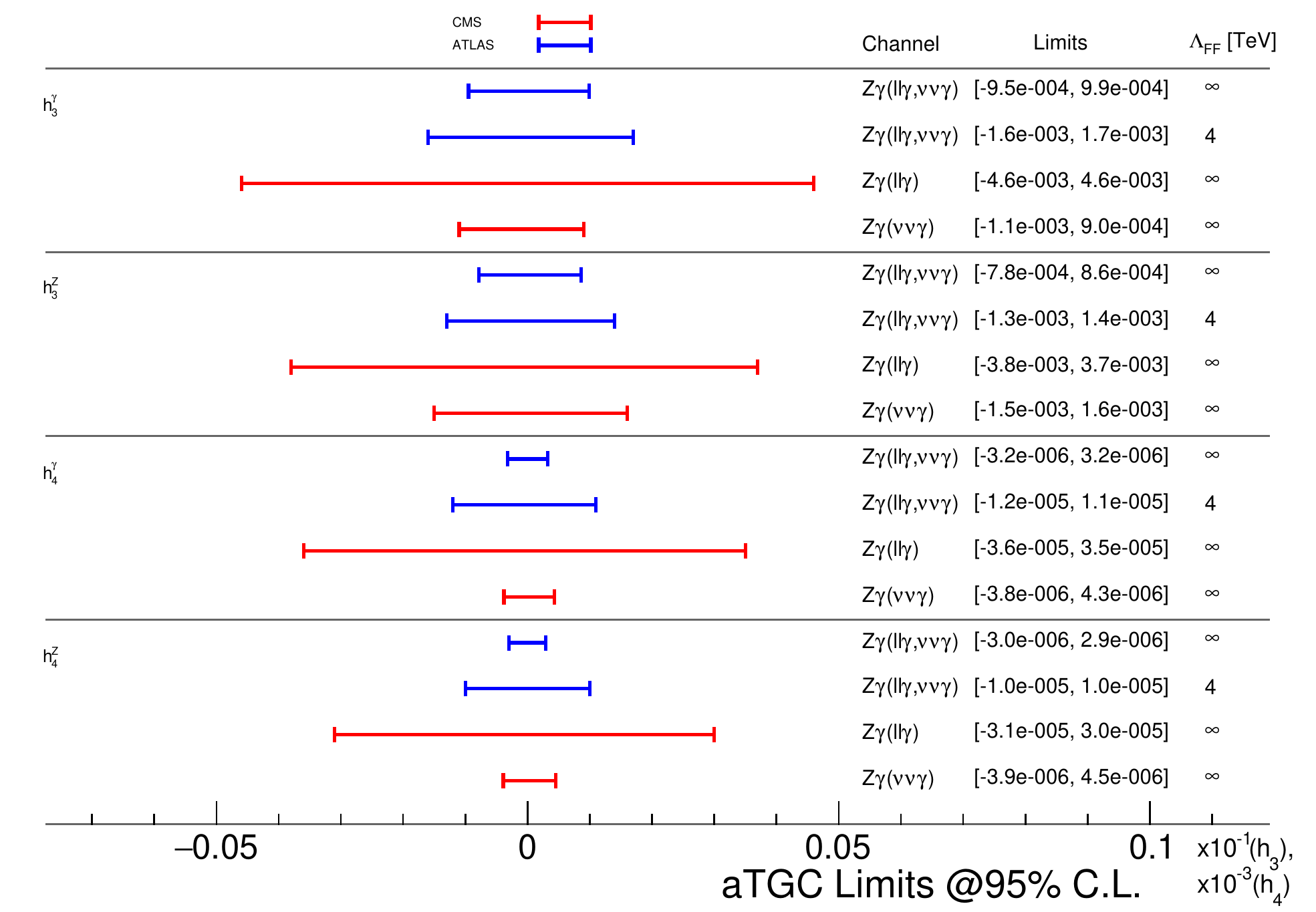}
  \caption{Comparison of the most competitive $Z\gamma\gamma$ and
    $Z\gamma Z$ limits, set by the LHC analyses presented in this review.
  All limits are based on the full 8~TeV, $\approx 20$~\ifb data sets.}
  \label{fig:nTGCAsummary}
\end{figure*}

\subsection{$ZZ\gamma$ and $ZZZ$ limits}

Turning to $ZZ$ final states, the limits on anomalous triple gauge
couplings are expressed in terms of two CP-violating ($f_4^V$) and two
CP-conserving ($f_5^V$) parameters, all of which are zero in the SM.
The limits on $f_i^V$ are negatively correlated for a given $i$ as
illustrated in Figure~\ref{fig:nTGCZ2d} which is based on 7~TeV $ZZ$
candidate events in the $4\ell$ decay mode.
\begin{figure}[htbp]
\includegraphics[width=0.40\textwidth]{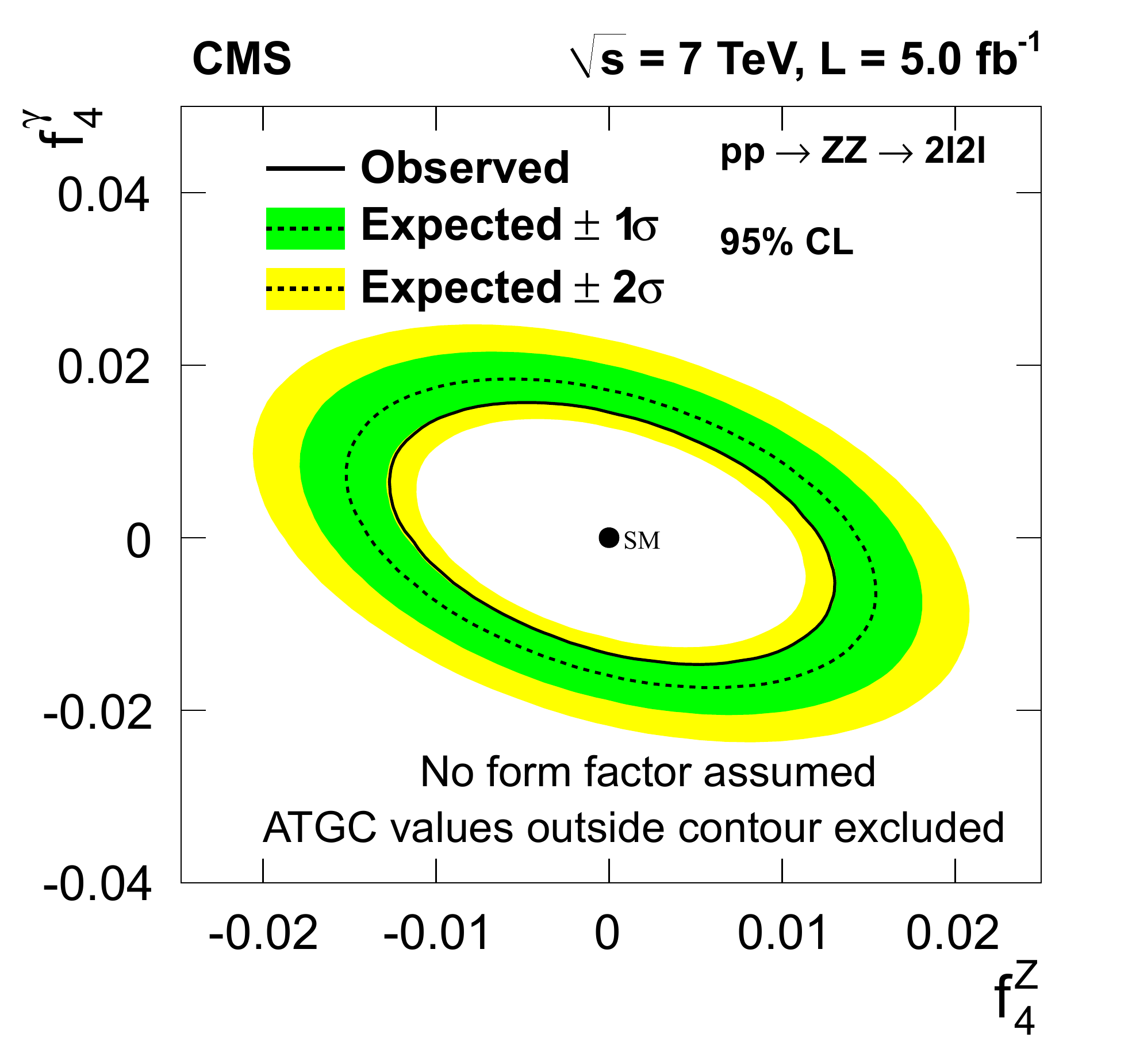}
  \caption{Expected and observed 95\% C.L. contours for aTGC limits
    derived from 7~TeV $ZZ$ candidate events in the $4\ell$ decay
    mode~\cite{Chatrchyan:2012sga} for the $f_4^\gamma$ versus $f_4^Z$
    parameters. aTGCs not shown are set to zero.}
  \label{fig:nTGCZ2d}
\end{figure}

One-dimensional limits for the $f_i^V$ parameters derived from $ZZ$
final states are shown in Figure~\ref{fig:nTGCZsummary}. The
$2\ell 2\nu$ decay mode gives the most stringent limits due to increased
branching fraction and detector acceptance. The 8~TeV data give
significantly stronger limits on the $f_i^V$ parameters, due to
larger statistics and an extended reach in $Z$ boson transverse
momentum. The combined limits by LEP~\cite{Schael:2013ita} as well as
the best Tevatron limits by D0~\cite{Abazov:2007ad} are not
competitive with those achieved at the LHC. The impact of imposing
unitarity constraints on the anomalous couplings via a form factor
$\alpha(\hat{s}) = \alpha_0/(1+\hat{s}/\Lambda_{\rm FF}^2)^3$ is
shown as well. In this specific case the exponent $n = 3$ is chosen.
Studying the sensitivity at 8~TeV, ATLAS~\cite{Aaboud:2016urj} finds
that a unitarization with a dipole form factor is no longer needed as
the aTGC limits more and more approach the SM expectation~\cite{Gounaris:2000tb}.

ATLAS and CMS have also performed a first combination of aTGC limits
based on their 7~TeV $ZZ$ analyses~\cite{ATLAS:2016hao}. With a
negligible impact due to systematic uncertainties, the combination
improves the aTGC sensitivity by about 20~\% compared to the
sensitivity of each experiment. While the resulting limits are not
competitive with the 8~TeV results presented above, this is an
important first step toward future combined LHC limits on anomalous
couplings.

\begin{figure*}[htbp]
  \centering
\includegraphics[width=\textwidth]{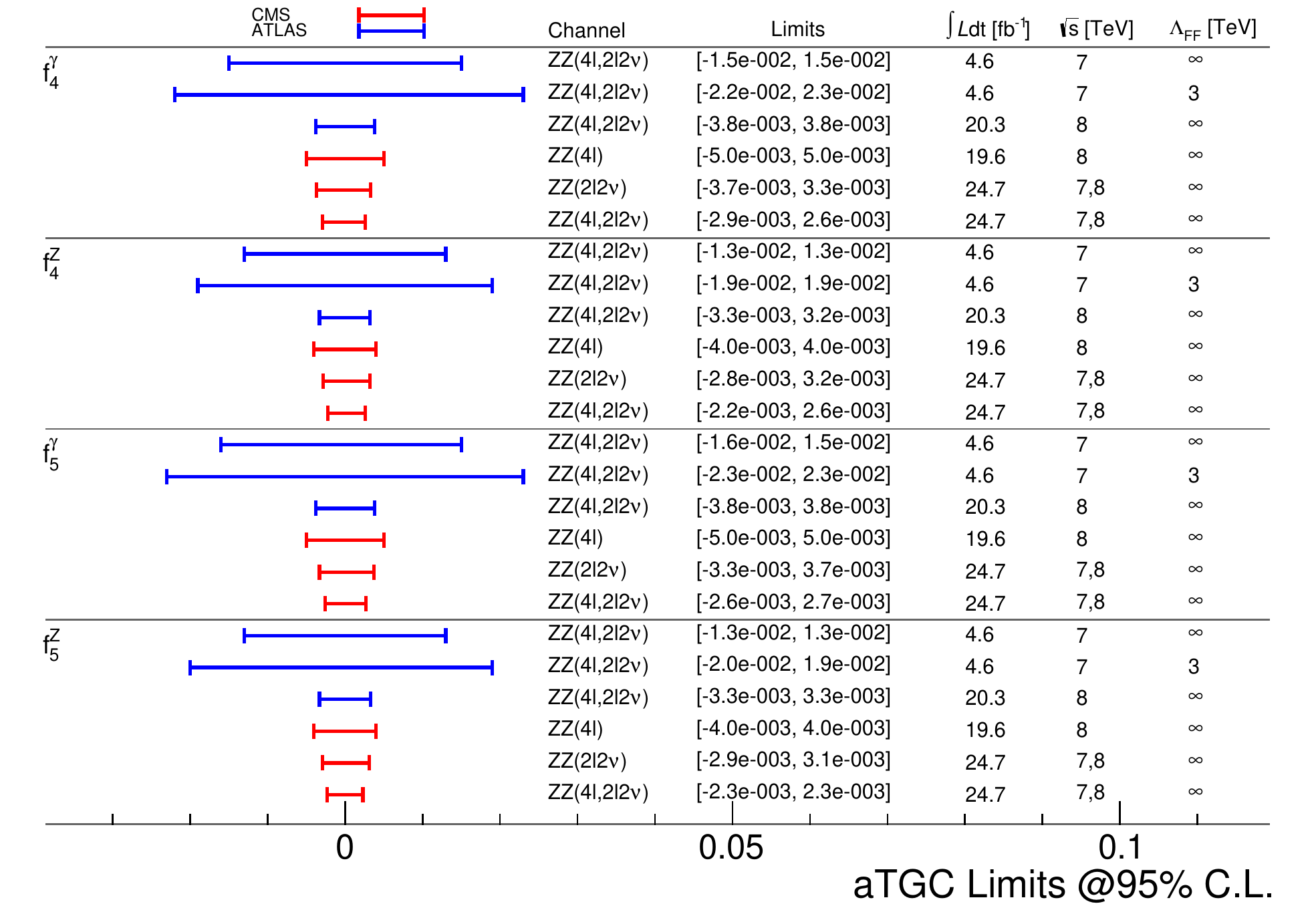}
  \caption{Comparison of the most competitive $ZZ\gamma$ and
    $ZZZ$ limits, set by the LHC analyses presented in this review.}
  \label{fig:nTGCZsummary}
\end{figure*}

\section{Constraints on anomalous Quartic Gauge Couplings}
\label{aQGC}

In the SM there are $WWWW$, $WWZZ$, $WWZ\gamma$, and $WW\gamma\gamma$
couplings. Beyond the SM there are possible $ZZZZ$, $ZZZ\gamma$,
$ZZ\gamma\gamma$, $Z\gamma\gamma\gamma$, and $\gamma\gamma\gamma\gamma$
couplings as listed in Table~\ref{tab:dim8EFT}. In Run~I ATLAS and CMS have only begun to investigate a few
of these possible couplings, with much more data planned in Run~II and
beyond.

The aQGC limits follow from the examination of the production of inclusive triple gauge
bosons, VBS dibosons, and exclusive dibosons. The limits are
generally taken to be limits on the coefficients of dimension-8
operators~\cite{Eboli:2006wa} $f$, although with
assumptions~\cite{Chatrchyan:2014bza} some of these are related to an
equivalent set $a$ of dimension-6 operators~\cite{Belanger:1992qi,
  Eboli:1993wg, Stirling:1999ek},
commonly used in Tevatron and LEP analyses. Table~\ref{tab:dim8EFT}
lists the 18 different dimension-8 operators and which quartic
vertex they affect. Note that these operators do not include TGCs.
\begin{table*}[]
  \centering
  \begin{tabular}{|c|c|c|c|c|c|c|c|c|c|} \hline
    & $WWWW$            & $WWZZ$            & $WW\gamma Z$ & $WW\gamma\gamma$ & $ZZZZ$            & $ZZZ\gamma$ & $ZZ\gamma\gamma$ & $Z\gamma\gamma\gamma$ & $\gamma\gamma\gamma\gamma$ \\ \hline
    ${\cal O}_{S,0}$, ${\cal O}_{S,1}$                                                                         & \checkmark & \checkmark & ~                       & ~                                    & \checkmark & ~                      & ~                                    & ~                                                  & ~                                                                \\ \hline
    ${\cal O}_{M,0}$, ${\cal O}_{M,1}$,${\cal O}_{M,6}$ ,${\cal O}_{M,7}$      & \checkmark & \checkmark & \checkmark     & \checkmark                  & \checkmark & \checkmark    & \checkmark                  & ~                                                  & ~                                                                \\ \hline
    ${\cal O}_{M,2}$ ,${\cal O}_{M,3}$,     ${\cal O}_{M,4}$ ,${\cal O}_{M,5}$ & ~                   & \checkmark & \checkmark     & \checkmark                  & \checkmark & \checkmark    & \checkmark                  & ~                                                  & ~                                                                \\ \hline
    ${\cal O}_{T,0}$ ,${\cal O}_{T,1}$ ,${\cal O}_{T,2}$                                       & \checkmark & \checkmark & \checkmark     & \checkmark                  & \checkmark & \checkmark    & \checkmark                  & \checkmark                                & \checkmark                                              \\ \hline
    ${\cal O}_{T,5}$ ,${\cal O}_{T,6}$ ,${\cal O}_{T,7}$                                       & ~                   & \checkmark & \checkmark     & \checkmark                  & \checkmark & \checkmark    & \checkmark                  & \checkmark                                & \checkmark                                              \\ \hline
    ${\cal O}_{T,8}$ ,${\cal O}_{T,9}$                                                                         & ~                   & ~                   & ~                       & ~                                    & \checkmark & \checkmark    & \checkmark                  & \checkmark                                & \checkmark                                              \\ \hline
  \end{tabular}
  \caption{Dimension-8 operators and the quartic vertices they affect~\cite{Degrande:2013rea}. The first four columns show the only QGC vertices which exist in the SM.}
  \label{tab:dim8EFT}
\end{table*}

The LEP L3 and OPAL experiments have set their best aQGC limits by
combining $W^+W^-\gamma$,
$\nu\bar{\nu}\gamma\gamma$~\cite{Achard:2001eg} and $W^+W^-\gamma$,
$\nu\bar{\nu}\gamma\gamma$ and
$q\bar{q}\gamma\gamma$~\cite{Abbiendi:2004bf} analyses, respectively.
These limits are surpassed by Tevatron's D0 experiment using the
exclusive VBS process, $\gamma\gamma\to WW$~\cite{Abazov:2013opa}.
Since the early LHC results are already considerably more restrictive
than the LEP and Tevatron limits, they are not shown in the following
comparisons.

The $W\gamma\gamma$ data are affected only by the SM $WW\gamma\gamma$
coupling, while $WV\gamma$ and VBS $W\gamma jj$ data have contributions owing to
$WW\gamma\gamma$ and $WWZ\gamma$ couplings. The VBS $WZ jj$ data are
affected by the SM $WWZZ$ and $WWZ\gamma$ couplings. The $WWW$ and
same-sign $WW$ VBS data select only the SM $WWWW$ coupling while the
exclusive $\gamma\gamma\to WW$ data select only the SM
$WW\gamma\gamma$ coupling. Finally, the VBS $WV jj$ data are affected
by all SM quartic couplings.

The one-dimensional limits on the EFT coefficients $f_{T,i}$ for
dimension-8 operators containing just the field strength tensors
are shown in Figure~\ref{fig:aQGCfTsummary}.
The VBS diboson channels yield similar limits, which are better than the
triple boson production limits.

The ATLAS $W\gamma\gamma$ and $Z\gamma\gamma$ results are derived with
\vbfnlo MC samples, which use a different convention for the
dimension-8 operators than the corresponding CMS results derived
with \mg MC samples. To be able to compare the results 
with CMS, the ATLAS results were converted using the redefinition
of operator coefficients outlined in~\cite{Degrande:2013rea}.

The impact of imposing unitarity constraints on the anomalous
couplings via a dipole form factor with a suppression scale $\Lambda_{\rm FF}$ is shown as
well. Note that the impact of unitarization is much
larger than in the case of the aTGCs.  Limits
without unitarization hence clearly probe a regime where unitarity is
violated at the scales probed and are more a benchmark than physically
meaningful.

The analogous plot of limits for the $f_{M,i}$ coefficients for
``mixed'' dimension-8 operators containing covariant derivatives
and the field strength tensors are shown in
Figure~\ref{fig:aQGCfMsummary}.
\begin{figure*}[htbp]
  \centering
\includegraphics[width=0.82\textwidth]{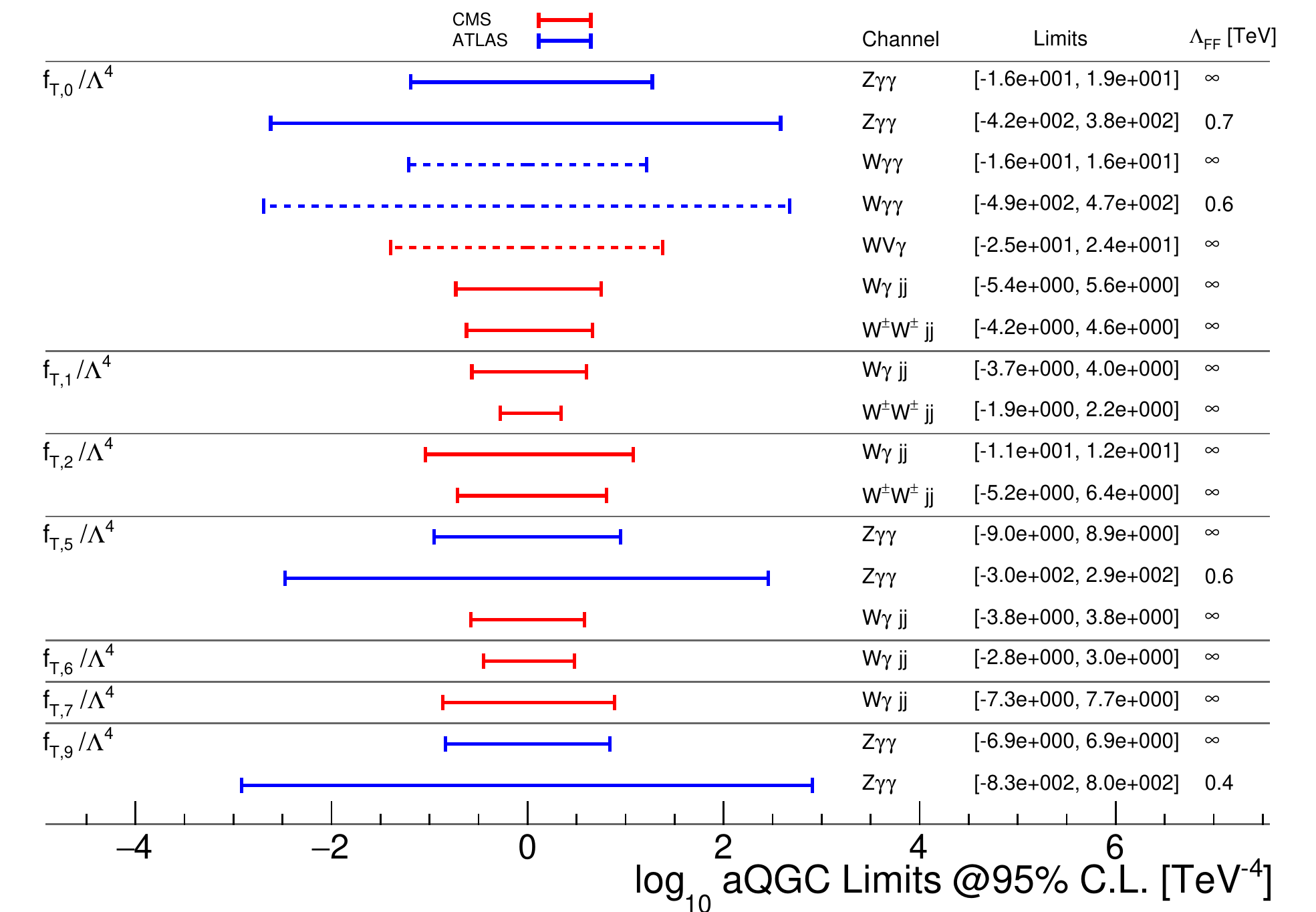}
  \caption{Comparison of the most competitive limits involving
    $f_{T,i}$ coefficients, set by the LHC analyses presented in this
    review. All limits are based on the full 8~TeV, $\approx 20$~\ifb
    data sets.}
  \label{fig:aQGCfTsummary}
\end{figure*}
\begin{figure*}[htbp]
  \centering
\includegraphics[width=0.82\textwidth]{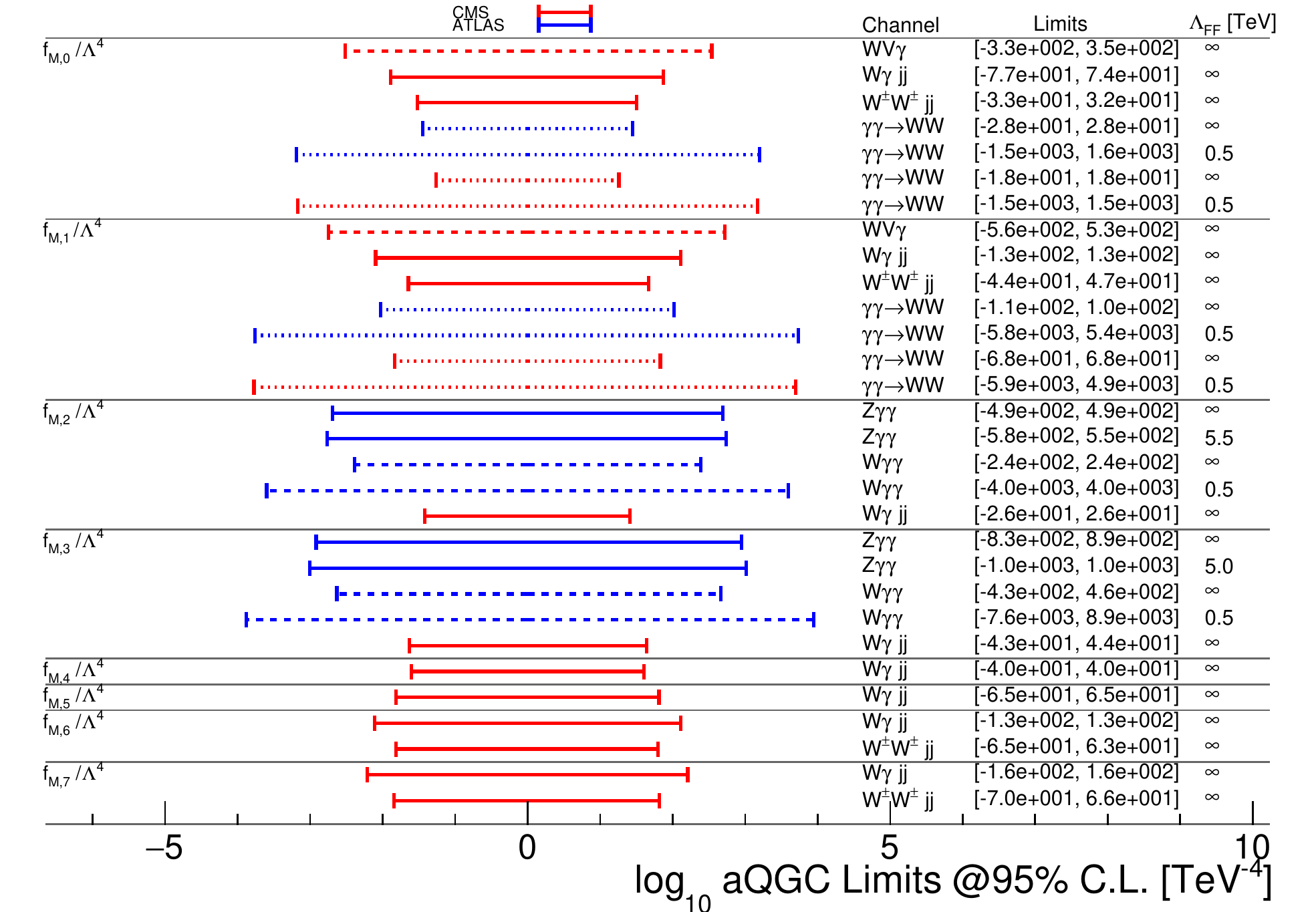}
  \caption{Comparison of the most competitive limits involving
    $f_{M,i}$ coefficients, set by the LHC analyses presented in this
    review. All limits are based on the full 8~TeV, $\approx 20$~\ifb
    data sets, except for the $\gamma\gamma\to WW$ analysis by CMS,
    using in addition the full $\approx 5$~\ifb of the 7~TeV
    data set.}
  \label{fig:aQGCfMsummary}
\end{figure*}
Again, the VBS diboson channels are all comparable and yield the
tightest limits although generally the same-sign $WW$ limits are the
most stringent. Where the exclusive process $\gamma\gamma\to WW$ is
used to set a limit it is the most stringent, because the signal is so
clean that it dominates the final selected data. The sensitivity of
ATLAS and CMS to anomalous couplings is generally very similar. 
Limits on $f_{M,2}$ and
$f_{M,3}$ were not included in the summary when they are trivially related to
$f_{M,0}$ and $f_{M,1}$ by a factor of 2 under the assumption of a
vanishing anomalous $WWZ\gamma$ coupling~\cite{Khachatryan:2016mud}.
The ATLAS $W\gamma\gamma$ and $Z\gamma\gamma$ results are again
converted to the convention employed by CMS, using the relations given
in~\cite{Degrande:2013rea}. The $WV\gamma$ and $\gamma\gamma\to WW$
results are based on the dimension-6 operators with coefficients
$a_{0,C}^W$ which are then converted to dimension-8 operators with
coefficients $f_{M,0}$ and $f_{M,1}$. The conversion conventions
employed by ATLAS~\cite{Degrande:2013rea} and
CMS~\cite{Belanger:1999aw} differ because CMS implemented their
own Lagrangians in \mg for $WV\gamma$ and $\gamma\gamma\to WW$.
To enable comparisons, the results from these two analyses
and the ATLAS $\gamma\gamma\to WW$ analysis are derived from their 
$a_{0,C}^W$ results using the conversion in~\cite{Degrande:2013rea}
to give results following the standard \mg convention.

The impact of imposing unitarity constraints on the anomalous
couplings via a dipole form factor with a suppression scale $\Lambda_{\rm FF}$  is shown as
well. Again, unitarization can change some of the limits by orders of
magnitude for these dimension-8 operators, indicating that such
limits without unitarization are driven by unphysical parameter
regions where unitarity is violated.

Limits on the $f_{S,i}$ coefficients whose operators affect the
scattering of longitudinal vector bosons have been placed by the CMS
same-sign $WW$ and ATLAS $WWW$ analyses and are summarized in
Figure~\ref{fig:aQGCfSsummary}, illustrating the impact of applying unitarization with
a form factor exponent $n=1$.

The $W^\pm V jj$, $W^\pm W^\pm jj$ and $W^\pm Z jj$ results by ATLAS
are not included in the above summary plots since they set limits on
the parameters $\alpha_{4,5}$ in an electroweak chiral Lagrangian
model~\cite{Appelquist:1980vg, Longhitano:1980iz, Longhitano:1980tm,
  Appelquist:1993ka} with K-matrix
unitarization~\cite{Alboteanu:2008my, Kilian:2014zja} applied. While vertex-dependent
conversions to $f_{S,0}$ and $f_{S,1}$ exist~\cite{Degrande:2013rea},
for the $W^\pm V jj$ analysis multiple vertices
contribute. Consequently, the resulting limits are summarized
separately in Figure~\ref{fig:a4a5summary}.
\begin{figure}[htbp]
  \centering
\includegraphics[width=0.4\textwidth]{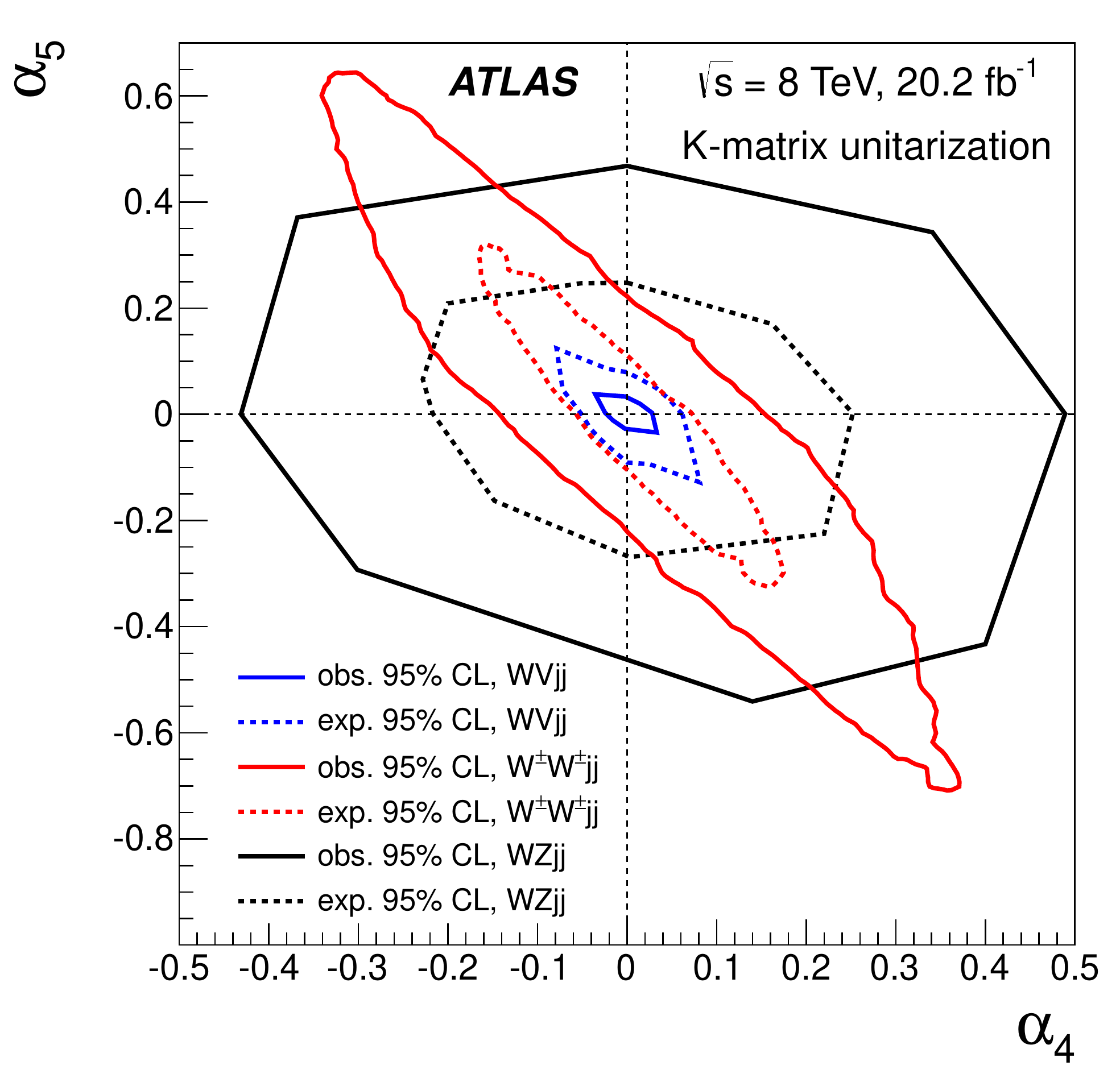}
  \caption{Expected and observed 95\% C.L. contours for aQGC limits in
    the $\alpha_{4}$ and $\alpha_{5}$ plane derived from 8~TeV
    $W^\pm V jj$, $W^\pm W^\pm jj$, and $W^\pm Z jj$ results by
    ATLAS~\cite{Aaboud:2016uuk}.}
  \label{fig:a4a5summary}
\end{figure}

The introduction of an additional dimension-8 operator
${\cal O}_{S,2}$~\cite{Eboli:2016kko} enables the vertex-independent
conversion to the parameters $\alpha_{4,5}$ when considering quartic
gauge-boson vertices only, through the study of a linear combination
of ${\cal O}_{S,2}$ and ${\cal O}_{S,0}$~\cite{Rauch:2016pai}. This would
be highly desirable for future studies, and note that
these resulting conversions are also applicable {\em after} K-matrix
unitarization~\cite{Sekulla:2016yku}.

In order to be able to compare the analyses on equal footing, we use
the conversion for the $WWWW$ vertex for the CMS same-sign $WW$ and
ATLAS $WWW$ analyses, omitting the results unitarized with a
form-factor as the applicability of the conversion is then
questionable. The resulting comparison is given in
Figure~\ref{fig:aQGCa4a5summary}, with the relative sensitivities to
be taken with a grain of salt when no unitarization is applied.
\begin{figure*}[htbp]
  \centering
\includegraphics[width=0.7\textwidth]{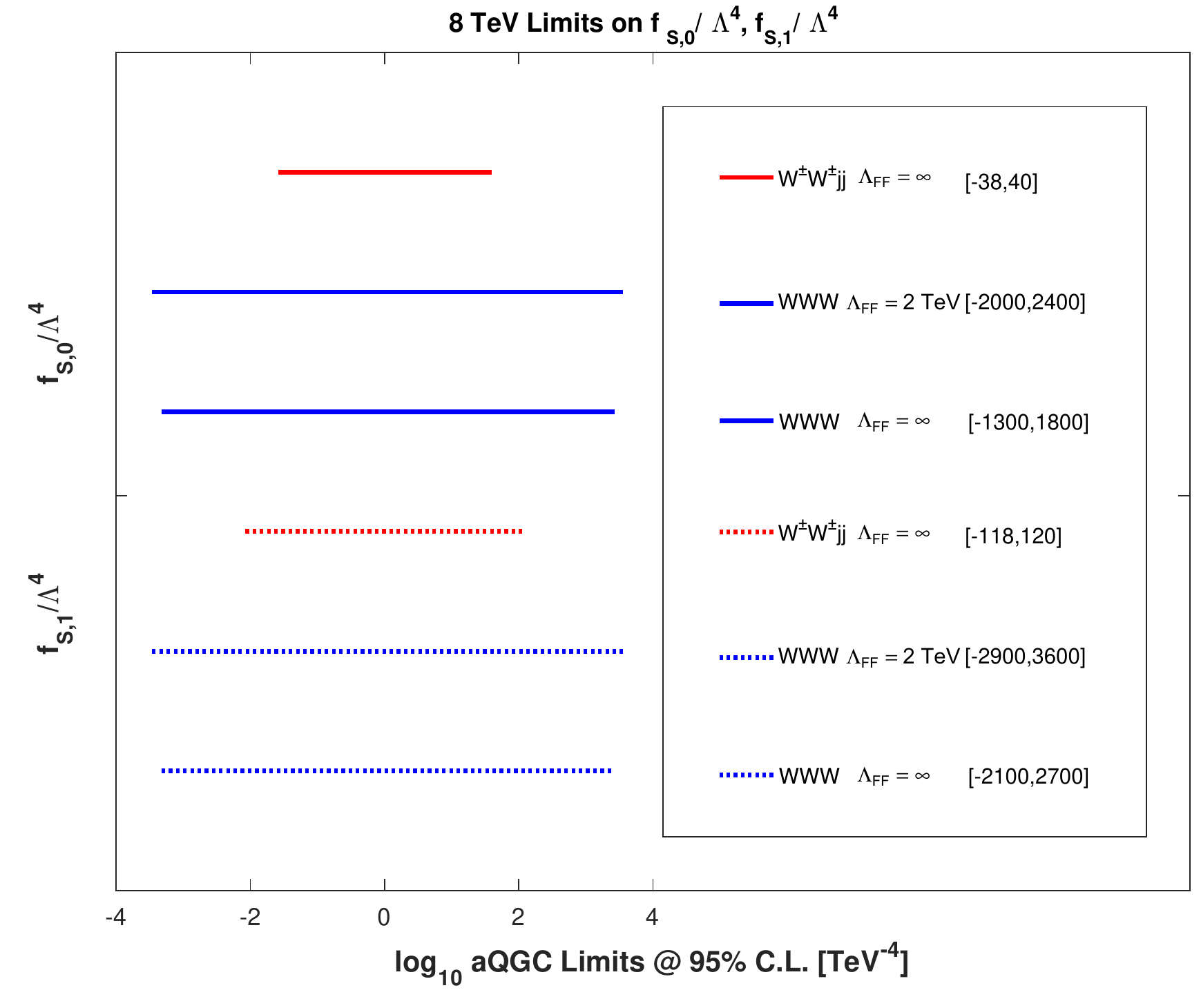}
  \caption{Comparison of the available limits involving $f_{S,i}$
    coefficients, set by the LHC analyses presented in this
    review. All limits are based on the full 8~TeV, $\approx 20$~\ifb
    data sets.}
  \label{fig:aQGCfSsummary}
\end{figure*}
\begin{figure*}[htbp]
  \centering
\includegraphics[width=0.8\textwidth]{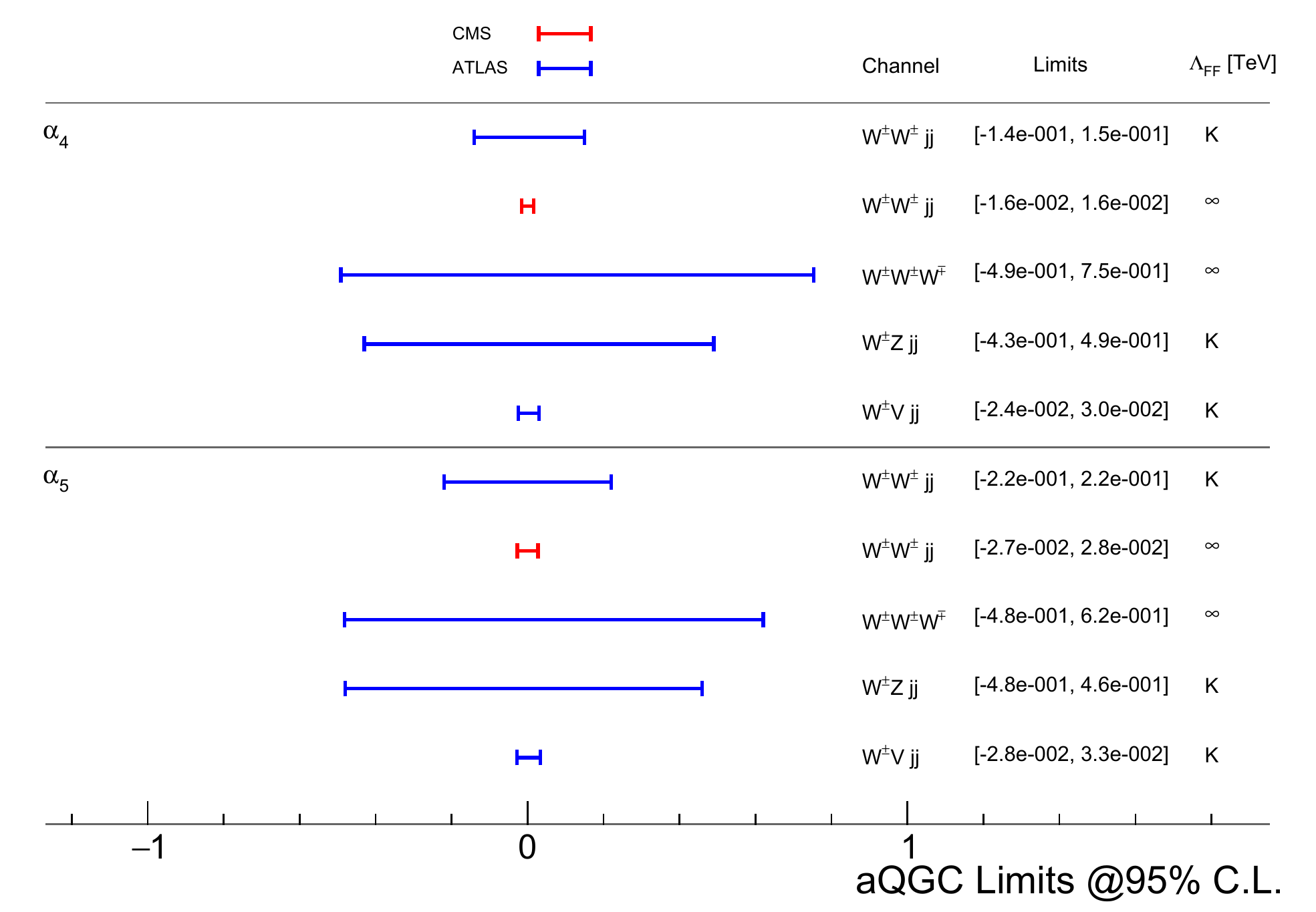}
  \caption{Comparison of the available limits involving $\alpha_{4,5}$
    coefficients, set by the LHC analyses presented in this
    review. All limits are based on the full 8~TeV, $\approx 20$~\ifb
    data sets. A ``K'' in the $\Lambda_{\rm FF}$ column indicates that
    K-matrix unitarization was applied.}
  \label{fig:aQGCa4a5summary}
\end{figure*}

Note that the $W^\pm V jj$ analysis is the first
semi-leptonic VBS analysis and exhibits a high sensitivity to
anomalous couplings due to the larger branching fraction as well as
probing the three distinct processes $W^\pm W^\mp jj$, $W^\pm W^\pm jj$, and $W^\pm Z jj$. 
It yields the most stringent unitarized limits thus far of
$-0.024 < \alpha_4 < 0.030$ and $-0.028 < \alpha_5 < 0.033$ at 95\%
C.L., corresponding to a new physics scale above $\approx 1.4$
TeV when assuming $\alpha_i = v^2/\Lambda^2$~\cite{Reuter:2013gla},
where $v$ is the Higgs vacuum expectation value ($v \approx 246$~GeV).

\section{Sensitivity prospects at the HL-LHC}
\label{HL-LHC}
Both the updated European Strategy for Particle Physics~\cite{ESC13}
and the Particle Physics Project Prioritization Panel (P5)
report~\cite{P5:2014pwa} outlining a ten-year strategic plan for HEP
in the U.S. emphasize the use of the full LHC potential through a
high-luminosity upgrade (HL-LHC) as top priority for the field. An
important ingredient for the physics case is the detailed studies of
multi-boson interactions which enable the test of the EWSB mechanism
 as well as the search for extensions beyond the SM.

The Physics Briefing Book~\cite{EuropeanStrategy:2013fia} for the
European Strategy for Particle Physics points out that studies of
longitudinal VBS to explore the EWSB mechanism in detail will not be
possible without the HL-LHC data set. As an example for the enhanced
sensitivity achievable with the HL-LHC to unveil new phenomena, a
study by ATLAS~\cite{ATL-PHYS-PUB-2012-005} using simplified detector
performance parametrizations is shown, where a new physics VBS $ZZ$
resonance could be discovered only using the HL-LHC data set. 

For the Snowmass community study preceding the P5 formation, both
ATLAS~\cite{ATLAS:2013hta} and CMS~\cite{CMS:2013xfa} provided
contributions which outline the physics program as well as sensitivity
improvements at the HL-LHC. With the enhanced data set, diboson
differential cross-section measurements in the high-$\hat{s}$ tails of
distributions will be possible as well as detailed studies of VBF, VBS,
and triboson production, from establishing the signals to measuring
differential cross sections with high precision. The discovery reach
for new higher-dimensional operators studied in $W^\pm W^\pm$, $WZ$, and
$ZZ$ VBS processes and $Z\gamma\gamma$ is at least doubled in the
HL-LHC data set~\cite{ATL-PHYS-PUB-2013-006}. The proceedings of the
Snowmass community study~\cite{snowmassbook} also quantify the
dimension-8 operator sensitivity increase due to the HL-LHC to be a
factor of 2 to 3 over the LHC, based on independent
studies~\cite{Baak:2013fwa, Degrande:2013yda} of $W^\pm W^\pm$, $WZ$, and
$ZZ$ VBS processes and $WWW$ and $Z\gamma\gamma$ triboson production.

Both ATLAS and CMS are preparing major detector upgrades for the HL-LHC
and have used VBS interactions as a benchmark for the anticipated
performance~\cite{CERN-LHCC-2015-010, CERN-LHCC-2015-020}.  Extended
tracking systems will enable improved lepton identification also in
the forward detector regions as well as crucial suppression of pileup
jet contributions to the tagging jet signature.

Measuring the polarization fractions in VBS is a crucial experimental
test of the predicted unitarization of the longitudinal VBS cross
section by the SM Higgs boson. CMS~\cite{CERN-LHCC-2015-010} has
evaluated the expected sensitivity to measure the longitudinal
fraction in $W^\pm W^\pm$ scattering using a two-dimensional template
fit of the $\Delta\Phi$ between the two tagging jets and the \pt of
the leading lepton and expect a significance of $\approx 2.4~\sigma$ in
the HL-LHC data set as illustrated in
Figure~\ref{fig:ssWWHLLHC}~(a). Combining this with a corresponding
$WZ$ VBS analysis, the longitudinal VBS significance increases to
$2.75~\sigma$. Note that a significant increase in
sensitivity should still be possible using a deep machine learning
technique instead of a ``simple'' variable fit, as demonstrated for
the $W^\pm W^\pm$ VBS case in~\cite{Searcy:2015apa}.
\begin{figure}[htbp]
  \begin{tabular}{c}
\includegraphics[width=0.40\textwidth]{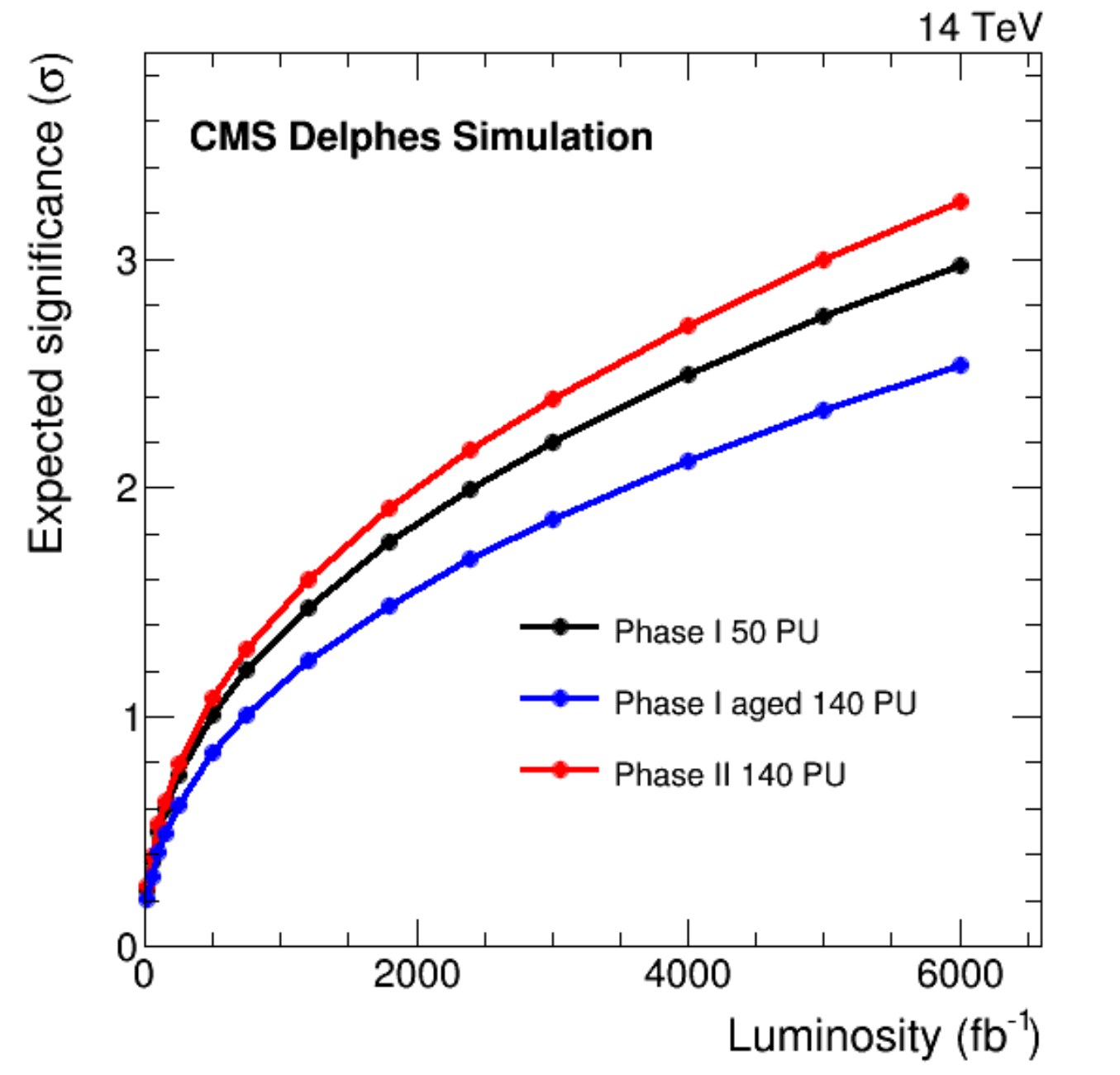}\\
    (a)\\
\includegraphics[width=0.40\textwidth]{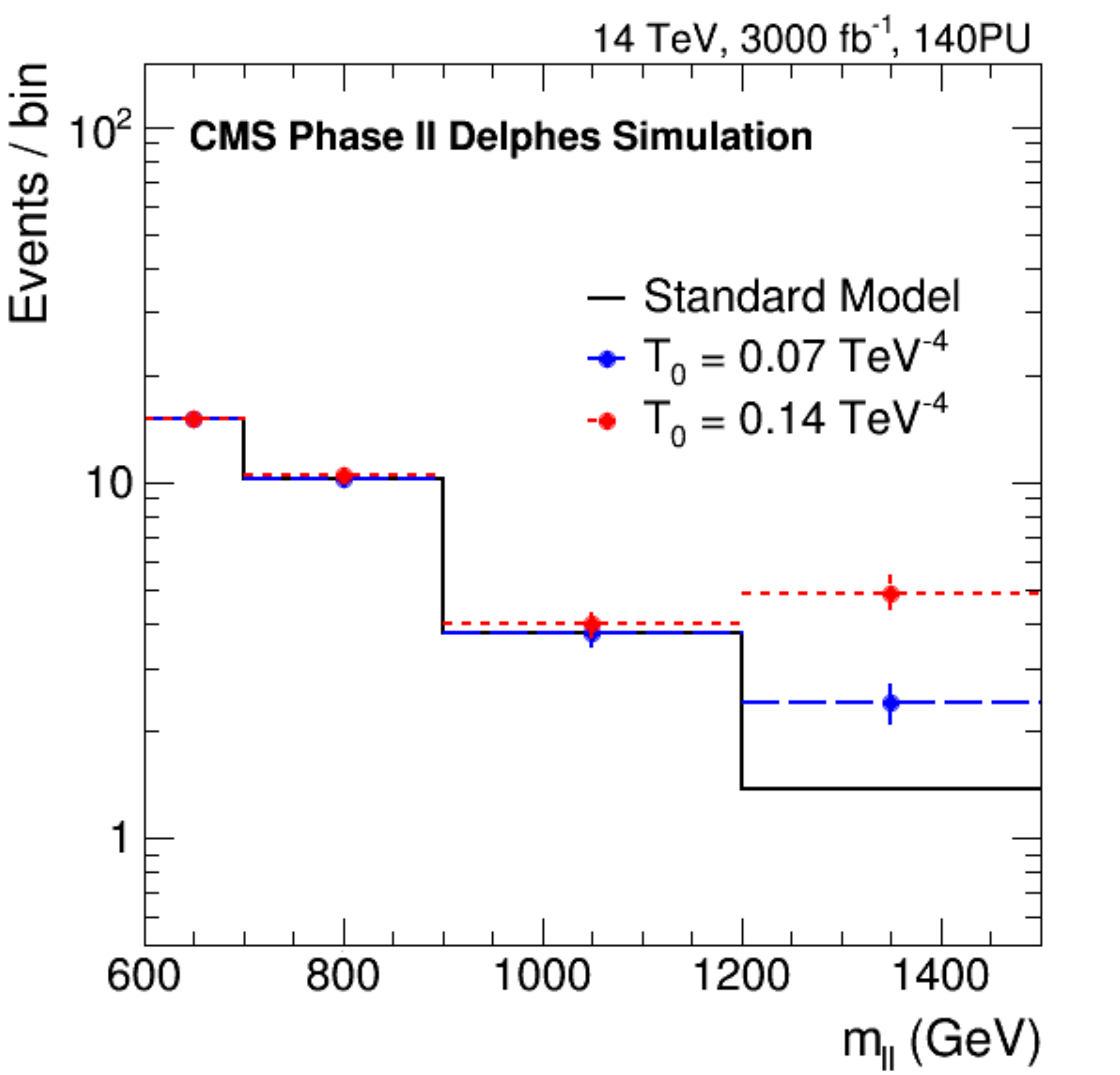}\\
    (b)\\
  \end{tabular}
  \caption{Expected CMS performance for $W^\pm W^\pm$ VBS measurements
    at the HL-LHC~\cite{CERN-LHCC-2015-010}: (a) significance of
    measuring the longitudinal $W^\pm W^\pm$ VBS cross section as a
    function of integrated luminosity and (b) dilepton mass
    distribution, where the effect of aQGCs on the spectrum is also
    shown.}
  \label{fig:ssWWHLLHC}
\end{figure}

The HL-LHC data set will also greatly enhance the sensitivity to
anomalous couplings. Figure~\ref{fig:ssWWHLLHC}~(b) shows the expected
dilepton mass distribution in $W^\pm W^\pm$ VBS candidate events with
3000~\ifb, where the data extend to about 1.5~TeV. This is a factor of
3 higher than accessible in Run~I and will improve the 
aQGC limits by about a factor of 50 for the aQGC example shown.

The search for new physics contributions in multi-boson interactions
either indirectly through anomalous couplings or directly through
resonance searches will greatly benefit from the increased
exploitation of final states with hadronically decaying $W/Z$ bosons.
To probe the high-mass tail of the $VV$ spectrum, this means the
identification of merged dijets into boosted monojets (the
hadronically decaying $V$ boson) will be of crucial
importance~\cite{Aad:2015ipg, Khachatryan:2014gha}, in particular,
at the HL-LHC (see Figure~\ref{fig:jetmW}).

With the advent of the first analyses in the Higgs sector employing the
EFT approach~\cite{Aad:2015tna} rather than the $\kappa$
framework~\cite{Heinemeyer:2013tqa} which allows the modification of
Higgs couplings without affecting its kinematics, combined constraints
from the Higgs and multi-boson analyses will be possible. Such analyses
will properly reflect the interconnectedness of multi-boson and Higgs
interactions in EFTs and yield improved constraints, as 
demonstrated by external global fits~\cite{Corbett:2013pja,
  Pomarol:2013zra, Ellis:2014jta, Falkowski:2014tna, Butter:2016cvz}.

To benefit from the tremendous progress in the theoretical predictions
it is important to perform measurements of carefully chosen
observables that can be studied with high precision and exhibit small
theoretical uncertainties. Ratios of diboson production rates have
been proposed~\cite{Frye:2015rba} that could enable precision tests of
the theoretical predictions, potentially to the level of being
sensitive to electroweak corrections.

\section{Conclusions}
\label{conclusions}

The LHC has enabled studies of multi-boson interactions at an
unprecedented level. Previously unobserved SM processes including
vector boson fusion, triboson production, and vector boson scattering
were established or at least observed with first evidence. In
particular, processes involving quartic gauge boson couplings were
probed for the first time, allowing the test of uncharted territory in
the SM. The SM signal is modeled in most analyses with MC generators
implementing NLO QCD calculations. Higher-order corrections at NNLO
QCD and NLO EW generally tend to improve the agreement with the data.
Such corrections are sizable (and of {\em opposite sign}), in particular, in 
the high-energy tails of distributions and need to be incorporated where
available to set more accurate anomalous coupling limits.

The data taken at the LHC through 2012 yielded many limits on
aTGCs and aQGCs which have confirmed the SM gauge couplings at the
level of accuracy which was accessible given the integrated luminosity
and the LHC energy. The limits for aTGCs arise largely from inclusive
diboson production properties at high diboson mass. The aQGC limits
are determined from the inclusive production of three bosons and from
the exclusive VBS production of boson pairs, also at high tri-boson and diboson
masses. These limits are presently the most stringent, exceeding those
found at LEP and the Tevatron both in the numerical limits themselves
and in the breadth of the processes explored in the search for
anomalous behavior.

Limits on aQGCs prove to be very sensitive to the application of a
unitarization procedure due to the higher dimensionality (eight) of
the operators involved, indicating that limits without unitarization
are driven by unphysical parameter regions where unitarity is
violated. On the other hand, introducing any unitarization turns the
EFT ansatz into a specific model, defeating the original purpose of
model independence.  Limits on aTGCs in an EFT framework can also suffer from inconsistencies if a generic power counting implies that the scale of new physics is parametrically below the mass scale probed by the LHC.  Both the aQGC and aTGC EFT issues stem from using the EFT in a region that is not self-consistent.  Eventually when the LHC has obtained sufficient precision, and if no deviation from the SM is found, EFTs will be able to be generically used to set precision bounds.  However, as emphasized in Section~\ref{s.theory} this is not the case as of yet. 
One way to avoid these issues is to provide upper
limits on fiducial cross sections as an alternative to EFT
interpretations or unfolded differential cross-section distributions
in sensitive variables that can be confronted with any new physics
model of interest.

The LHC experiments have already gone well beyond previous limits on
triple gauge couplings and have advanced into limits on quartic gauge
couplings by using the new energy frontier opened up by the LHC.
The LHC has since run at enhanced energy going from 8~TeV in 2012 to 13~TeV
starting in 2015. The luminosity has also risen substantially. Thus,
the limits given in this review will be improved upon in the near
future. In the more distant future the high-luminosity LHC will yet
again substantially improve on the aTGC and aQGC limits. These
data will serve to explore the mutual couplings of the gauge bosons
and ascertain if the SM is the correct description of those
non-Abelian couplings.

\section{Acknowledgments}
We thank CERN and the LHC Accelerator team for making this fantastic
machine a reality and for operating it so well. 
In addition, we thank the ATLAS and CMS Collaborations in general, and, in particular, the electroweak and standard model
working group members, past and present, for their help and
cooperation in advising and encouraging us and for providing
experimental details of the results quoted in this review. We also
gratefully acknowledge the tremendous effort and progress on the
theoretical side, enabling the predictions and MC tools utilized in
the measurements presented in this review and those forthcoming.

Finally we thank Matt Herndon and Harikrishnan Ramani for useful comments on the draft. 
The work of M.-A.P. was supported by the DOE Contract No. DE-SC0012704. The work of P.M. was supported in part by the NSF CAREER Awards No. NSF-PHY-1056833 and No. NSF-PHY-1620628.
\clearpage
\bibliography{VV}

\end{document}